\documentclass[preprint,showpacs,preprintnumbers,amsmath,amssymb,rmp]{revtex4}
\usepackage{graphicx}
\usepackage{dcolumn}
\usepackage{bm}
\usepackage{epsfig}
\newcommand{\geqnew}{\stackrel{>}{\!\ _{\sim}}}
\newcommand{\leqnew}{\stackrel{<}{\!\ _{\sim}}}
\def\single_space{\baselineskip 12pt plus 1pt minus 1pt}
\begin{document}
\preprint{To appear in   {\bf Physics Reports}}

\title{Quantum wave packet revivals}

\author{R. W. Robinett} \email{rick@phys.psu.edu}
\affiliation{
Department of Physics\\
The Pennsylvania State University\\
University Park, PA 16802 USA \\
}

\date{\today}

\begin{abstract}

The numerical prediction, theoretical analysis, and experimental verification 
of the phenomenon of wave packet revivals in quantum systems has flourished 
over the last decade and a half. 
Quantum revivals are characterized by initially localized quantum states 
which have a short-term, quasi-classical time evolution, which then can
spread significantly over several orbits, only to reform later in the form 
of a quantum revival in which the spreading reverses itself, the wave packet 
relocalizes, and the semi-classical periodicity is once again evident.
Relocalization of the initial wave packet into a number of smaller copies 
of the initial packet (`minipackets' or `clones') is also possible,
giving rise to fractional revivals.  
Systems exhibiting such behavior are a fundamental realization of 
time-dependent interference phenomena for bound states with quantized
energies in quantum mechanics 
and are therefore of wide interest in the physics and chemistry communities.

We review the theoretical machinery of quantum wave packet construction
leading to the existence of revivals and fractional revivals, in systems with
one (or more) quantum number(s), as well as discussing how information on
the classical period and revival time is encoded in the energy eigenvalue
spectrum. We discuss a number of one-dimensional model systems
which exhibit revival behavior, including the infinite well, the 
quantum bouncer, and others, as well as several two-dimensional integrable 
quantum billiard systems. Finally, we briefly review the experimental 
evidence for wave packet revivals in atomic, molecular, and other systems,
and related revival phenomena in condensed matter and optical systems.

\end{abstract}


\maketitle

\tableofcontents

\newpage


\single_space

\section{Introduction}
\label{sec:intro}

The study of localized, time-dependent solutions to bound state problems 
in quantum mechanics has been of interest to both researchers and students 
of the subject alike since the earliest days of the development of the 
field. Schr\"{o}dinger \cite{schrodinger} and others \cite{kennard} - 
\cite{debroglie} discussed the connections between the quantum and classical 
descriptions of nature by exhibiting explicit wave packet solutions to many 
familiar problems, including the cases of the free-particle, 
uniform acceleration (constant electric or gravitational field), 
harmonic oscillator (forerunner of coherent and squeezed states), 
and uniform magnetic field. Many such examples then appeared in 
early textbooks \cite{kemble} - \cite{rojansky}  only a decade later, 
discussing both wave packet spreading and periodic time-dependence
in a way which  is easily accessible to students even today.

Despite Schr\"{o}dinger's hope \cite{schrodinger} that 
``...{\it wave groups can be constructed
which move round highly quantised Kepler ellipses and are the representations
by wave mechanics of the hydrogen electron}...'' (without spreading, as with 
the constant width harmonic oscillator packet he derived), 
early investigators soon found \cite{darwin}, \cite{lorentz} that such 
dispersion was a natural feature of wave packets for the Coulomb problem. 

Attempts at constructing localized semi-classical solutions 
(of the coherent-state type) for the 
Coulomb problem, following up on Schr\"odinger's suggestions, 
continued with theoretical results \cite{lowell_brown} - \cite{ghosh}
appearing in the literature much later. 
It was, however, the development of modern experimental techniques,
involving the laser-induced excitation of atomic Rydberg wave packets,
including the use of the `pump-probe' \cite{generation}
or  `phase-modulation' \cite{ramsey_fringes} techniques 
to produce,  and then monitor the subsequent time-development of,
such states which led to widespread interest in the physics of
wave packets. (For reviews of the subject, 
see \cite{rydberg_review} - \cite{new_fielding}.)
Updated proposals for the production of such states in the context of
Rydberg atoms, where one could study the connections between localized 
quantum mechanical solutions and semi-classical notions 
of particle trajectories,  led first to the creation of such spatially
localized states \cite{yeazell_big_one}, to experiments which 
observed their return to near the atomic core \cite{noordam_return}, 
and then the observation of the classical Kepler periodicity \cite{wolde}, 
\cite{yeazell_classical} of Rydberg wave packets, over only a few cycles 
in the early experiments.

However, this interest also led to the prediction of 
qualitatively new features in the long-term time development 
in such bound state systems, such as quantum wave 
packet revivals. Parker and Stroud \cite{parker} were the first to find
evidence for this behavior in numerical studies of Rydberg atoms, while 
Yeazell and Stroud \cite{yeazell_2}, \cite{yeazell_fractional} 
and others \cite{meacher}, \cite{wals} soon confirmed their predictions 
experimentally.

The phenomenon of wave packet revivals, which has now been observed in 
many experimental situations, arises when a well-localized wave packet is 
produced and initially exhibits a short-term time evolution with almost 
classical periodicity ($T_{cl}$) and then spreads significantly after a 
number of orbits, entering a so-called collapsed phase where the probability 
is spread (not uniformly) around the classical trajectory.
On a much longer time scale after the initial excitation, however, 
called the revival time (with $T_{rev} >> T_{cl}$),
the packet relocalizes, in the form of a quantum revival, 
in which the spreading reverses itself and the classical
periodicity is once again apparent. Even more interestingly, many 
experiments have since observed additional temporal structures,  
with smaller  periodicities (fractions of $T_{cl}$), 
found at times equal to rational fractions
of the revival time ($pT_{rev}/q$). These have been elegantly interpreted 
\cite{perelman_1}  as the
temporary formation of a number of `mini-packets' or `clones' of the
original packet, often with $1/q$ of the total probability, exhibiting
local periodicities $T_{cl}/q$, and have come to be known as fractional 
revivals. 
Observations of fractional revivals have been made  in a number of
atomic \cite{yeazell_fractional} - \cite{wals} and molecular 
\cite{vrakking} systems. 

A simple picture \cite{yeazell_american}
of the time-dependence of the quantum state leading to these behaviors, 
modeling the individual energy eigenstates and their 
exponential ($\exp(-iE_nt/\hbar))$ time-dependent factors 
as an ensemble of
runners or race-cars on a circular track, is often cited. The quantum 
mechanical spreading arises from the differences in speed, while the
classical periodicity of the system is observable over a number of
revolutions (or laps.)
For longer times, however, the runners/race-cars spread out and  no
correlations (or clumpings) are obvious, while after the fastest 
participants have lapped their slower competitors (once or many times), 
obvious patterns can return, including smaller `packs' of racers, clumped 
together, which model fractional revivals. 

A different metaphor involves the (deterministic) shuffling of an initially
highly ordered deck of playing cards. One shuffling method involves splitting
the deck into two equal halves, and then alternately placing the bottom card 
from each half into a pile, reforming and reordering  the deck. After only 
a few such shuffles, the original order is seemingly completely lost and the 
cards appear to have randomized. After only a few more turns, however, clear 
patterns of ordered subsets of suits and ranks appear, increasingly so until 
after only eight such shuffles the deck has returned to its original highly 
ordered state. Only the simplest of mathematical concepts is required to 
describes these analogs of fractional and full revivals.

The {\it Dynamics of wave packets of highly-excited states of atoms
and molecules}, including a discussion of wave packet revivals and 
fractional revivals, and descriptions of the experimental observations of 
these phenomena, were  discussed in the excellent 1991 review 
by Averbukh and Perelman \cite{averbukh_review}, while a nicely accessible 
general discussion by Bluhm and Kosteleck\'{y} \cite{bluhm_ajp} has 
also appeared. There have been developments in the field since then, and
many of the basic quantum mechanical concepts behind revival behavior 
have also begun to appear in the pedagogical literature, so it seems 
appropriate to provide a review of some of the fundamental ideas behind the 
short- and long-term behavior of quantum wave packets, describing both the 
classical periodicity and revival behavior of wave packets in many model 
systems, and their experimental realizations. 

The theoretical machinery required to understand many aspects of 
revival behavior is sufficiently accessible, and potentially
of enough general interest, that our review of the subject will contain
many tutorial aspects, such as
\newcounter{temp}
\begin{list}
{(\roman{temp})}{\usecounter{temp}}
\item the use of familiar model systems (such as the infinite well and others)
as illustrative of the fundamental concepts,
\item an emphasis on examples in which exact revival behavior is 
found (to be used as benchmarks for more realistic systems), 
\item a focus on semi-classical methods, including the WKB approximation,
which are appropriate for many wave packet systems constructed from large $n$
energy eigenstates,
\item general discussions of how information on both the classical periodicity
and quantum revival times are encoded in the energy eigenvalue spectrum,
\item and references not only to the original research literature, but to
many of the pedagogical papers appearing on the subject.
\end{list}
In these areas, and others, we hope to extend the reviews in 
Refs.~\cite{averbukh_review} and \cite{bluhm_ajp} in useful ways.

We start in Sec.~\ref{sec:general}  with a general introduction to the 
theoretical tools required for
understanding revival behavior, including the autocorrelation function.
We then turn to discussions of many model systems 
(in Sec.~\ref{sec:model_systems}) 
which illustrate various aspects of the quantum mechanical
time development of localized wave packets, including background material
on unbound systems (free-particle and uniform acceleration cases) and
for ones exhibiting only periodic behavior (the harmonic oscillator). 
The infinite well is then discussed in great detail, as are other familiar
one-dimensional problems. Two-dimensional quantum systems, especially
quantum billiard geometries,  are studied in 
Sec.~\ref{sec:2d_and_3d_systems}. We then briefly review experimental
evidence for revival behavior in Sec.~\ref{sec:experimental_results} 
in atomic (Coulomb) and molecular (vibrational, rotational) systems, 
as well as in situations where the quantum revivals are due to the 
quantized nature of the electromagnetic field in two-state atom-field systems,
or in the excitation spectrum of Bose-Einstein condensates, and finally we
discuss related revival phenomena in a variety of optical systems.

\section{General analysis}
\label{sec:general}

\subsection{Autocorrelation function}
\label{subsec:autocorrelation_function}

The study of the time-development of wave packet solutions of
the Schr\"{o}dinger equation has a long history and often makes use of 
the concept of the overlap ($\langle \psi_t|\psi_0\rangle$) 
of the time-dependent quantum state ($|\psi_t\rangle$) with the initial 
state ($|\psi_0\rangle$). For example,
early work by Mandelstam and Tamm \cite{mandelstam}
on the time-development of isolated quantum systems led to the inequality
\begin{equation}
|\langle \psi_{t}|\psi_{0}\rangle|^2
 \geq
\cos^2\left(\frac{\Delta H t}{\hbar}\right)
\quad
,
\quad
\mbox{valid for}
\qquad
0 \leq t \leq \frac{\pi \hbar}{2\Delta H}
\, , 
\label{mandelstam}
\end{equation}
where $\Delta H = \sqrt{\langle H^2\rangle - \langle H\rangle^2}$
is the uncertainty in the free-particle energy of the wave packet.
These ideas were used to study energy-time uncertainty relations,
as well as the minimum time required to reach an orthogonal quantum state.

This overlap is most often referred to as the {\it autocorrelation function} 
and can be evaluated in either position- or momentum-space to give
\begin{equation}
A(t) \equiv \langle \psi_t | \psi_0 \rangle
=  \int_{-\infty}^{+\infty} \,\psi^{*}(x,t)\, \psi(x,t)\,dx 
=  \int_{-\infty}^{+\infty} \,\phi^{*}(p,t)\, \phi(p,t)\,dp 
\label{autocorrelation_general_definition}
\end{equation}
which emphasizes that in order for $|A(t)|$ to be large, the wave function
at later times must have significant overlap with the initial state
in both $x$- and $p$-space.

For one-dimensional bound state systems, where a wave packet is expanded 
in terms of energy eigenfunctions, $u_n(x)$, with quantized energy 
eigenvalues, $E_n$, in the form
\begin{equation}
\psi(x,t) = \sum_{n=0}^{\infty} a_n u_n(x)\, e^{-iE_nt/\hbar}
\label{general_expansion_in_eigenstates}
\end{equation}
with 
\begin{equation}
a_n = \int_{-\infty}^{+\infty} [u_n(x)]^{*}\, \psi(x,0)\,dx
\end{equation}
the most useful form for the autocorrelation function
\cite{nauenberg} is
\begin{equation}
A(t) = \sum_{n=0}^{\infty} |a_n|^2 e^{+iE_nt/\hbar}
\end{equation}
and the evaluation of $A(t)$ in this form for initially 
highly localized wave packets will be one of the main topics of this review.
Besides being of obvious theoretical value in the analysis of 
time-dependent systems, more physically, the autocorrelation function
is important because it is very directly related to the observable
ionization signal \cite{generation}, \cite{yeazell_2} 
in the pump-probe type experiments where 
such behavior is studied experimentally.

Studies of the long-term time development of quantum states is also
a topic with a long history, often being discussed from a more formal
point of view. For example, Bocchieri and Loinger \cite{bocchieri}
produced statements of a `{\it quantum recurrence theorem}' 
which attempted to generalize Poincar\'{e}'s theorem
to show that systems with discrete energy eigenvalues would eventually 
return arbitrarily closely  to the initial state, 
in the sense that the norm $||\psi_t - \psi_0||$ could be made smaller
than any arbitrarily small number. (Such ideas have also been examined
\cite{baltz} from a pedagogical point of view.) 
Similar studies in the mathematical
literature often deal with analyses of {\it almost periodic functions}
\cite{almost_periodic_functions}, \cite{percival}. Such long-term
time behavior has also been compared \cite{fermi_pasta_ulam_1}
to Fermi-Pasta-Ulam recurrences \cite{fermi_pasta_ulam_2} in the
dynamical behavior of non-linearly coupled oscillators. 
Investigations of recurrence (revival-like) phenomena in the context
of non-stationary (time-dependent) Hamiltonians \cite{hogg} have resulted in 
similar theorems.

In this context, Gutschick and Nieto \cite{nieto_3} noted that 
``{\it ...if one
waits long enough any state which has significant overlap with a 
finite number of states will eventually return to ``almost''
its original shape...it amounts to having a totally dispersed wave
packet suddenly regenerate itself after very long times.}
Peres \cite{peres} also pointed out that very
long time scale recurrences of this type would occur in systems with
hydrogenic energy levels. However, both analyses focused on time
scales which are much longer than those ultimately
observed in wave packet experiments. 
For the laser-induced wave packets produced
in modern experiments, the structure of the expansion in eigenstates
in Eqn.~(\ref{general_expansion_in_eigenstates}) 
is such that a quite general systematic analysis of
the time scales involved in the problem is possible and that is the
topic of the next section.

\subsection{Time-dependence of one-dimensional localized wave packets}
\label{subsec:general_time_dependence}

In general, the time-dependence of an arbitrary time-dependent bound state
wavefunction, $\psi(x,t)$, with the expansion in eigenstates, $u_n(x)$,
of the form in Eqn.~(\ref{general_expansion_in_eigenstates}) 
can be quite complex. However, in many experimental realizations, 
a localized wave packet is excited with an energy spectrum which is
tightly spread around a large central value of the quantum number, $n_0$,
so that $n_0 >> \Delta n >> 1$. In that case, it is appropriate to expand 
the individual energy eigenvalues, $E(n) \equiv E_n$, about this value, 
giving
\begin{equation}
E(n) \approx E(n_0) + E'(n_0)(n-n_0) + \frac{E''(n_0)}{2}(n-n_0)^2
+ \frac{E'''(n_0)}{6}(n-n_0)^3 + \cdots
\end{equation}
where $E'(n_0) = (dE_n/dn)_{n=n_0}$ and so forth.
This gives the time-dependence of each individual quantum eigenstate 
through the factors
\begin{eqnarray}
e^{-iE_nt/\hbar} & = & \exp\left( -i/\hbar\left[E(n_0)t + (n-n_0) E'(n_0)t +
\frac{1}{2} (n-n_0)^2 E''(n_0) t \right. \right. \nonumber \\
& & 
\qquad \qquad \qquad 
\left. \left. 
+ \frac{1}{6}(n-n_0)^3 E'''(n_0)t  + \cdots 
\right]\right) \nonumber \\
& \equiv  & \exp \left( -i\omega_0 t - 2\pi i(n-n_0) t/T_{cl}
- 2\pi i(n-n_0)^2t/T_{rev} \right. \\
& & \qquad \qquad \qquad \left. - 2\pi i(n-n_0)^3t/T_{super} + \cdots \right)
\nonumber
\label{expansion_in_time}
\end{eqnarray}
where each term in the expansion (after the first) defines an important
characteristic time scale, via
\begin{equation}
T_{cl} = \frac{2\pi \hbar}{|E'(n_0)|}
\quad
,
\quad
T_{rev} = \frac{2\pi \hbar}{|E''(n_0)|/2}
\quad
, 
\quad
\mbox{and}
\quad
T_{super} = \frac{2\pi \hbar}{|E'''(n_0)|/6}
\, . 
\label{all_relevant_times}
\end{equation}

The first ($\omega_0 = E(n_0)/\hbar$) term is an unimportant,
$n$-independent  overall phase, common to all
terms in the expansion, and which therefore induces no interference between 
them; it is similar to the time-dependent phase for a single stationary 
state solution and has no observable effect in $|\psi(x,t)|^2$. 

The second term in the expansion is familiar from  
correspondence principle arguments \cite{correspondence}
as being associated with the classical
period of motion in the bound state, so that
\begin{equation}
T_{cl} \equiv \frac{2\pi\hbar}{|E'(n_0)|}
\, .
\label{classical_period}
\end{equation}
This connection is perhaps most easily seen using a semi-classical
argument and the WKB quantization condition, familiar variations on
classical action-angle methods \cite{action_angle_1} - \cite{action_angle_2}.

For a particle of
fixed energy $E$ in a 1D bound state potential, $V(x)$, we have
$E = m v(x)^2/2 + V(x)$ 
and the short time, $dt$, required to traverse a distance $dx$ 
can be obtained from this and integrated over the range defined by the
classical turning points; this then gives {\it half} the classical period as 
\begin{equation}
dt = \frac{dt}{v(x)}
= 
\sqrt{\frac{m}{2}} \frac{dx}{\sqrt{E-V(x)}}
\qquad
\longrightarrow
\qquad
\frac{\tau}{2} = \int_{a}^{b}\,dt =
\sqrt{\frac{m}{2}}\int_{a}^{b} \frac{dx}{\sqrt{E-V(x)}}
\, .
\label{classical_period_definition}
\end{equation}
The WKB quantization condition in this same potential 
(with the same classical turning points, $a,b$) can be written in the form 
\begin{equation}
\sqrt{2m } \int_{a}^{b} \sqrt{E_n -V(x)}\, dx
 =   (n +C_{L} + C_{R}) \pi \hbar 
\label{wkb_quantization_condition}
\end{equation}
in terms of the matching coefficients $C_L,C_R$. 
We recall that the appropriate values of these constants 
\cite{wkb_approximation} are given
by $C_{L,R} = 1/4$ at `linear' or `smooth' turning points,
where functions are matched smoothly onto Airy function solutions
(and $\psi(x)$ effectively penetrates roughly $1/8$ of a wavelength
into the classically disallowed region \cite{saxon}), 
while $C_{L,R} = 1/2$ at `infinite wall' type boundaries
where the wavefunction must vanish. 
This expression can then be differentiated implicitly
with respect to the quantum number, $n$, to obtain
\begin{equation}
\sqrt{2m} \int_{a}^{b} \frac{|dE_n/dn|\,dx}{2\sqrt{E_n-V(x)}} = 
\pi \hbar
\, . 
\end{equation}
This, in turn, can be related to the classical period in 
Eqn.~(\ref{classical_period_definition}) to give
\begin{equation}
T_{cl} \equiv \tau =  \sqrt{2m}\int_{a}^{b} \frac{1}{\sqrt{E_n - V(x)}}\,dx
= \frac{2\pi \hbar}{|dE_n/dn|}
\, . 
\label{classical_period_from_wkb}
\end{equation}
The most obvious example of such a connection is for the harmonic oscillator,
where the WKB condition gives the exact eigenvalues, $E_n = (n+1/2) \hbar
\omega$, and the classical period from Eqn.~(\ref{classical_period}) is
\begin{equation}
T_{cl}= 
\frac{2\pi \hbar}{|dE/dn|} =
\frac{2\pi \hbar}{\hbar \omega} = \frac{2\pi}{\omega}
\end{equation}
as expected. For the special case of the oscillator, all wave packets
are exactly periodic and all higher order  derivatives 
($E'', E''',...$) vanish, so no other longer time scales are present. 
Nauenberg \cite{nauenberg} has also provided an elegant argument
connecting $E'(n_0)$, the classical periodicity, and the structure of
the autocorrelation function, which we review in
Appendix~\ref{appendix:time_scales}.

For future reference, even in the
presence of higher-order time scales, we define the {\it classical}
component of the wave packet to be
\begin{equation}
\psi_{cl}(x,t) 
\equiv 
\sum_{n=0}^{\infty} a_n u_n(x) e^{-2\pi i (n-n_0)E_n't/\hbar}
\equiv 
\sum_{k} a_k u_k(x) e^{-2\pi i kt/T_{cl}}
\label{classical_wave_packet}
\end{equation}
where we define $k \equiv n-n_0$. 
This component can be 
used to describe the short-term ($t \approx T_{cl} << T_{rev}$) 
time-development and is especially helpful in discussing fractional revivals.
We note that here, and elsewhere, we will henceforth ignore the overall 
$\exp(-iE(n_0)t/\hbar)$ phase factor.

The next term in the expansion in Eqn.~(\ref{expansion_in_time})
is given by 
\begin{equation}
T_{rev} = \frac{2\pi \hbar}{|E''(n_0)|/2}
\label{quantum_revival}
\end{equation}
which will be associated with the quantum revival time scale. 
This time scale determines the relative importance of the $(n-n_0)^2$
term in the exponent for $t>0$. It is responsible for the long-term
($t >>T_{cl}$) spreading of the wave packet in the same way that the 
difference in the $p$-dependence of the $px$ and $p^2t/2m$ terms in 
$\exp(i(px-p^2t/2m)/\hbar)$
in the plane wave expansion of free particle wave packets 
gives rise to dispersion; it can be shown 
quite directly (Ref.~\cite{nauenberg}
and Appendix~\ref{appendix:time_scales}) that the spreading time is 
proportional to $T_{rev}$ for $t<< T_{rev}$. More interestingly, 
for times of the order of $T_{rev}$, 
the additional $\exp(2\pi i (n-n_0)^2t/T_{rev})$
phase terms all return to unity, giving the $t\approx 0$ time-dependence
in Eqn.~(\ref{classical_wave_packet}) and a return to 
approximate semi-classical behavior.

The anharmonic oscillator provides an example of the calculation of
the revival time of relevance to physically realizable systems such as the
vibrational motion of molecules.  For example, the addition of a cubic term, 
$\lambda \hbar \omega x^3$, to the usual oscillator Hamiltonian yields 
(in second-order perturbation theory \cite{cohen}) the energy eigenvalues 
\begin{equation}
E_n = \hbar \omega
\left[
(n+1/2)  - \frac{15 \lambda^2}{4} (n+1/2)^2 
         - \frac{7}{16} \lambda^2 \right]
= \hbar \omega \left[
\gamma + n(1-\alpha) - \alpha n^2 \right]
\end{equation}
where $\gamma \equiv 1/2-11\lambda^2/8$ and $\alpha \equiv 15\lambda^2/4$.
The effect of the anharmonicity (for either sign of $\lambda$) is to
increase the classical period ($T_{cl} = 2\pi/(1-\alpha)\omega$), 
but more importantly it also introduces  a finite revival time given by
\begin{equation}
T_{rev} = \frac{2\pi}{\omega \alpha} \approx  \frac{T_{cl}}{\alpha}
>> T_{cl}
\qquad
\mbox{for $\alpha << 1$}
\, . 
\end{equation}

As another example of the hierarchy of  relative magnitudes of $T_{rev}$ and 
$T_{cl}$ which
is possible, we consider a family of (symmetric) one-dimensional
power-law potentials \cite{sukhatme} -- \cite{robinett_power_law}
defined by
\begin{equation}
V_{(k)}(x) \equiv V_0 \left|\frac{x}{L}\right|^k
\, . 
\label{power_law_potential}
\end{equation}
This family includes both the harmonic oscillator (when $k=2$ and
$V_0/L^2 = m \omega^2/2$) and the infinite well of width $2L$ 
(when $k\rightarrow +\infty$, for any $V_0$). The same WKB approximation
considered above (see Appendix~\ref{appendix:power_law_wkb} for details) 
shows that the energy eigenvalues scale (for large $n$) as
\begin{equation}
E_n{(k)} \propto n^{2k/(k+2)}
\label{power_law_scaling}
\end{equation}
which does indeed give the correct large $n$ behavior for the oscillator 
($E_n \propto n$) and the infinite well ($E_n \propto n^2$); it even
gives the appropriate dependence on quantum number  for the Coulomb problem 
($E_n \propto n^{-2}$ for $k=-1$) so that these simple arguments are
useful for the more realistic case of Rydberg atoms. (Many similar scaling
laws can be derived for truly three-dimensional systems in different
ways,  as shown in Ref.~\cite{quigg_rosner}.)

For this general class of potentials, we find that the ratio of the
revival time to the classical period is given by
\begin{equation}
\frac{T_{rev}^{(k)}}{T_{cl}^{(k)}}
= \left(\frac{4\pi \hbar}{|E''(n_0)|}\right)
  \left(\frac{|E'(n_0)|}{2\pi \hbar}\right) 
= 2\left|\frac{k+2}{k-2}\right| n_0
\label{power_law_ratio}
\, . 
\end{equation}
The revival time diverges for the special case of the oscillator 
($k=2$), as do all of the other higher-order time scales, 
while for the general ($k\neq 2$) case one can quite generally 
 have $T_{rev} >> T_{cl}$ for $n_0 >> 1$, which is typical for the 
localized wave packets studied 
experimentally. We note that for the experimentally important case
of the Coulomb potential ($k=-1$), the hierarchy is given by 
$T_{rev}^{(Coul)} = (2n_0/3) T_{cl}^{(Coul)}$. 

We note that several important model systems, including the infinite well,
and the 2D ($E_m \propto m^2$) and 3D ($E_l \propto l(l+1)$) free rotors
\cite{bluhm_ajp} have energy spectra with no higher than quadratic 
dependence on the 
quantum number, implying that higher order derivatives vanish and longer
time scales are sent to infinity. The interaction energy of the
particles in Bose-Einstein condensates can be modeled in the form
$E_n = U_0n(n-1)/2$ \cite{bec_revivals} providing another highly realizable
physical example with a quadratic energy spectrum, and revival
behavior. 

Continuing in this fashion, we note that the next term in the expansion is
typically an even longer time scale, called the {\it superrevival} time, 
defined by 
\begin{equation}
T_{super} = \frac{2\pi \hbar}{|E'''(n_0)|/6}
\label{quantum_superrevival}
\end{equation}
with
\begin{equation}
\frac{T_{super}^{(k)}}{T_{rev}^{(k)}} = \frac{3(k+2)}{4} n_0
\end{equation}
for the power-law family of potentials; we discuss superrevivals
in more detail in Sec.~\ref{subsec:superrevivals}.

Since this type of behavior is generic to localized wave packets, we
will first exemplify the short-term semi-classical periodicity, then the
approach to a collapsed (incoherent sum) state, as well as the structure of  
quantum revivals and fractional revivals using a simple model energy
eigenvalue spectrum and standard Gaussian distribution of eigenstates.
We will assume that the expansion coefficients are given by
\begin{equation}
a_n = \frac{1}{\sqrt{\Delta n \sqrt{2\pi}}} 
\exp\left[-\frac{(n-n_0)^2}{4\Delta n^2}
\right]
\label{gaussian_components}
\end{equation}
which gives the required normalization
\begin{equation}
\sum_{n=0}^{\infty} |a_n|^2 \approx 1
\end{equation}
to exponential accuracy for $n_0 >> \Delta n >> 1$. For the examples
in the rest of this section we use $n_0 = 400$ which is typically larger
than that found in experimental realizations, but useful for making the
visualized examples we present more transparent. We often 
use $\Delta n = 6$, but also will examine the effect of varying 
$\Delta n$ as well; several notable experiments on Rydberg atoms 
\cite{yeazell_2}, \cite{wals} have made use of wave packets constructed 
from $5-10$ states, corresponding to $\Delta n \approx 2-5$ in our 
notation.

For the energy eigenvalues, we choose, for simplicity,  a generalized 
anharmonic oscillator spectrum of the form 
\begin{equation}
E_n = 2\pi(n - \alpha n^2/2 +  \beta n^3/6)
\label{anharmonic}
\end{equation}
and use two sets of parameters for comparison, namely
\begin{center}
\begin{tabular}{|c|c|c|} \hline
Case       &    A            & B \\ \hline
$\alpha$   &       $1/800$   & $1/800$  \\ \hline 
$\beta$    &        $0$      & $2\cdot10^{-6}$ \\ \hline
$T_{cl}$   &       $2$       & $1.515$ \\ \hline
$T_{rev}$  &   $1600$        &  $4444.4$ \\ \hline
$T_{rev}/T_{cl}$ &  $800$    & $2933.3$  \\ \hline
\end{tabular}
\\
\vskip 0.5cm
Table I.
\end{center}

For Case A, with only linear and quadratic terms, we expect exact
revival behavior, and this case is arranged to be even more special
by having $T_{rev}/T_{cl}$ be an integer and so can be considered as an
ideal case.  The introduction of an additional
small $n^3$ ($\beta \neq 0$) term in Case B will exemplify systems 
containing  higher order terms, 
longer time scales, and hence only approximate revival behavior.

\subsection{Classical periodicity and approach to the collapsed state}
\label{subsec:period_and_collapse}

Even in the absence of a specific physical system, one can still examine, 
in great detail, the time-dependence of a generic wave packet using the 
autocorrelation function in the form
\begin{equation}
A(t) = \sum_{n=0}^{\infty} |a_n|^2 e^{iE_n t/\hbar}
\, . 
\end{equation}
Nauenberg \cite{nauenberg}, for example, has shown how to elucidate 
the approximate
periodicity apparent in $|A(t)|$ for bound state systems 
in a quite general way, and we discuss
his approach in Appendix~\ref{appendix:time_scales}, but we first
present some numerical examples. 

Using the values for Case A in Table I, we examine the behavior of
$A(t)$ over the first $100$ classical periods in 
Fig.~\ref{fig:collapse}, corresponding in this case to $1/8$ of the
entire revival time. In that plot (upper half), the initial 
classical periodicity
is clear, as is the effect of the wave packet spreading, shown by the
decreasing value of $A(t)$ at integral multiples of the classical period
(also seen in the bottom half), as well as the increasing `width' of
the (decreasingly small) peaks in $A(t)$. In more specific model systems
where we can examine the dynamics, we can elucidate the nature of the loss
of coherence to the collapsed phase in more detail, as in 
Sec.~\ref{subsec:infinite_classical}.

For times longer than approximately $50T_{cl}$ (in this case), 
the $|A(t)|$ oscillates (more or less rapidly) about a constant 
value, indicated by the horizontal dashed line. To understand this, we 
examine the general structure of $|A(t)|^2$ and note that 
\begin{equation}
|A(t)|^2 = \left|\sum_{n}|a_n|^2 e^{iE_n t/\hbar}\right|^2
= \sum_{n} |a_n|^4 
+ 
2\sum_{n\neq m} |a_n|^2 |a_m|^2 \cos\left(\frac{(E_n-E_m)t}{\hbar}\right)
\, . 
\label{general_coherence}
\end{equation}
For time scales of the order of the classical period, the 
$\cos((E_n-E_m)t/\hbar)$ terms are still highly correlated and 
reproduce the  approximate classical periodicity. For longer times, the 
oscillatory components become increasingly out of phase and can lead to
high frequency ($f >> 2\pi/T_{cl}$) excursions around the constant value 
given by the first term in Eqn.~(\ref{general_coherence}), namely
\begin{equation}
|A_{inc}|^2 \equiv \sum_{n=0}^{\infty} |a_n|^4 
\end{equation}
which we will label the {\it incoherent} limit of $|A(t)|^2$. For the
Gaussian expansion coefficients in Eqn.~(\ref{gaussian_components}),
this sum can also be done approximately (to the same accuracy as the 
normalization) giving
\begin{equation}
|A_{inc}|^2 \equiv \sum_{n=0}^{\infty} |a_n|^4 
= \frac{1}{\Delta n 2 \sqrt{\pi}}
\approx 0.047
\qquad
\mbox{for $\Delta n = 6$}
\label{incoherent_value}
\end{equation}
which is included  in Fig.~\ref{fig:collapse} as the dotted horizontal lines
and indicated by the horizontal arrows.
The larger the number of states contained in the expansion, the smaller the
resulting incoherent value during the collapsed state; because of the
discrete nature of the bound state spectrum, the autocorrelation function
does not asymptotically approach zero as it would in the free particle
case, for example, but oscillates about  this generic plateau  value.

In the bottom half of Fig.~\ref{fig:collapse}, we also show the values of 
$A(t)$ at multiples of the classical period (squares) and at times a half 
period away  ($t = (n+1/2)T_{cl}$, stars) 
showing how the `in-phase' (`out-of-phase') components first 
shrink (grow) until they are of the same order as the incoherent value of 
$|A_{inc}|^2$, and then clearly exhibit highly correlated behavior, which 
is the first hint of fractional revivals; some evidence of this is the
appearance of highly oscillatory behavior at fractional multiples of
$T_{rev}$ (note the $1/9$ - $1/11$ labels in 
Fig.~\ref{fig:collapse}(a).)

We note, for future reference, that the form of the probability density,
$|\psi(x,t)|^2$ during the collapsed or incoherent phase of the
time-development can be similarly written in the form
\begin{equation}
|\psi(x,t)|^2 
= \sum_{n}|a_n|^2 |u_n(x)|^2
+
2\Re
\left[\sum_{n\neq m} a_n^{*} a_m u_n(x)u_m^{*}(x)e^{i(E_n-E_m)t/\hbar}\right]
\label{collapsed_wave_function}
\, . 
\end{equation}
The probability density will then oscillate  around a `static' value
determined by the incoherent sum of the probability densities for each 
eigenstate, $|u_n(x)|^2$. In the large $n$ limit which is applicable here, 
WKB methods can be used to approximate the energy eigenstates,  and one 
can write
\begin{equation}
P_{n}(x) = |u_n(x)|^2 \approx \frac{2}{\tau_n} \sqrt{\frac{m}{E_n -V(x)}}
\end{equation}
where $E_n,\tau_n$ are the quantized WKB energies and classical periods
in Eqns.~(\ref{wkb_quantization_condition}) and
 (\ref{classical_period_from_wkb}).
Similar statements can be made about the momentum-space probability 
densities.

\subsection{Revivals and fractional revivals}
\label{subsec:revivals_and_fractional}

\subsubsection{Revival time}
\label{subsubsec:revival_time}

We now turn our attention to the longer term behavior of the autocorrelation
function over an entire revival time and plot $|A(t)|^2$ over the
interval ($0,T_{rev})$ in Fig.~\ref{fig:new_three} for three cases. 
The top two plots, (a) and (b), 
correspond to the parameters of Case A in Table I, 
showing the effect of changing $\Delta n$;  for larger (smaller)
values of $\Delta n$, the spread in momentum values is larger (smaller),
so that the quantum mechanical spreading time, $t_0$, is smaller (larger) 
and the rate at which the initial classical periodicity is lost is
corresponding faster (slower). The initial size of the wave packet (and, as
we will see, the subsequent size of the `mini' wave packet components 
at fractional revivals) is smaller (larger) for bigger (smaller) values 
of $\Delta n$ and this correlation is also obvious in that finer details 
are apparent for the larger $\Delta n$ case. 

In the bottom plot in Fig.~\ref{fig:new_three}(c), we show  the results 
corresponding to case B in Table I, where one includes contributions
from terms of order $(n-n_0)^3$. The overall pattern is similar, and 
many of the same individual distinct features are seen, but without the 
obvious symmetry or near 
exact revivals as in the case of purely quadratic $n$-dependence where 
no higher order time scales are present. 
The rich structure of features at rational fractions of $T_{rev}$ which
is apparent in both cases, and which are obviously highly correlated with 
fractional values of $|A(t=0)|^2=1$, is seemingly a robust feature 
of all the examples, and we systematically explore their general structure
next.

To examine the generic behavior of the wave packet near the revival time, 
$T_{rev}$, we use only the first two derivative terms in the expansion in 
Eqn.~(\ref{expansion_in_time}) and write
\begin{equation}
\psi(x,t\approx T_{rev})
=
\sum_{k} a_k u_k(x) e^{-2\pi i kt/T_{cl}} e^{-2\pi i k^2}
\qquad
\mbox{where $k=n-n_0$}
\end{equation}
and we focus on the effect of the ${\cal O}(t/T_{rev}$) or 
$k^2$ terms on the exponential time development. The additional phases 
arising from such terms all give unity and the wave packet is said to have 
revived since
\begin{equation}
\psi(x,t\approx T_{rev}) = 
\sum_{k} a_k u_k(x) e^{-2\pi i kt/T_{cl}}
= \psi_{cl}(x,t)
\end{equation}
has returned to something like its initial form, exhibiting
the classical periodicity. In the special case when $T_{rev}/T_{cl}$
is an integer (as in Case A in Table I,
and the top two plots in Fig.~\ref{fig:new_three})
the revival occurs exactly in phase with the original time-development,
and is exact (in that $|A(t)|$ returns to exactly unity).

The robust 
prediction that the wave packet exhibits approximately the  classical
periodicity (in phase or not) near $t = T_{rev}$ is also apparent
in Fig.~\ref{fig:compare} (where we now plot $|A(t)|^2$ near several
full or fractional revival times) even for admixtures of higher order terms,
such as the Case B values in Table I where $|A(t\approx T_{rev})| <1$,
but the return to the initial periodicity is apparent.

\subsubsection{Fractional revivals}
\label{subsubsec:fractional_revivals}

We next examine $\psi(x,t)$ near half a revival time, and find that
\begin{equation}
\psi(x,t\approx T_{rev}/2)
=
\sum_{k} a_k u_k(x) e^{-2\pi i k t/T_{cl}} e^{-\pi i k^2}
\label{one_half_revival}
\end{equation}
with additional $\exp(-\pi i k^2) = \pm 1$ factors. 
We can compare this to the semi-classical time evolution by 
noting that
\begin{equation}
\psi_{cl}(x,t + T_{cl}/2)
= 
\sum_{k} a_k u_k(x) e^{-2\pi i k (t+T_{cl}/2)/T_{cl}}
=
\sum_{k} a_k u_k(x) e^{-2\pi i k t/T_{cl}}\, e^{-\pi i k}
\label{half_revival_classical}
\end{equation}
and the additional $\pm$ phase factors in Eqn.~(\ref{one_half_revival})
can be written in the same form since both the
\begin{equation}
\mbox{(Eqn.~(\ref{one_half_revival}))}
\qquad
e^{-\pi i k^2} = (-1)^{k^2}  =  (-1)^{k} = e^{-\pi i k}
\qquad
\mbox{(Eqn.~(\ref{half_revival_classical}))} 
\label{first_trick}
\end{equation} 
terms give the same result for even and odd values of $k$. We thus see that
\begin{equation}
\psi(x,t\approx T_{rev}/2) = \psi_{cl}(x,t+T_{cl}/2)
\label{one_half_revival_function}
\end{equation}
and the wavepacket also reforms near the half revival time, with the original
classical periodicity, but half a period of phase with the initial wave form
(at least for integral $T_{rev}/T_{cl}$) and we see this behavior in
Fig.~\ref{fig:compare} for both the A and B cases. The approximate revival
at $T_{rev}/2$ seen for Case B is, in fact, somewhat better than at the full
revival time $T_{rev}$ since the higher order anharmonicities have had
less time to effect the phase structure of the revivals.

We then urn our attention to the quarter-revival time where we can write
\begin{equation}
\psi(x,t\approx T_{rev}/4)
=
\sum_{k} a_k u_k(x) e^{-2\pi i k t/T_{cl}} e^{-\pi i k^2/2}
\end{equation}
and the additional phase factor beyond the classical terms is given
by
\begin{subequations}
\begin{eqnarray}
\mbox{$k=2l$ even} & \qquad & e^{-i \pi (2l)^2/2}  =  e^{-2\pi i} = 1 \\
\mbox{$k=2l+1$ odd}& \qquad &  e^{-i\pi (2l+1)^2/2} = 
e^{-2\pi i l^2}\, e^{-2\pi i l}\, e^{-i\pi/2} = -i
\end{eqnarray}
\end{subequations}
or 
\begin{equation}
      e^{-i \pi k^2/2} = \left\{ \begin{array}{ll}
               +1 & \mbox{for $k$ even} \\
               -i & \mbox{for $k$ odd}
                                \end{array}
\right.
\, .
\end{equation}
This factor can be written in the forms
\begin{equation}
 e^{-i \pi k^2/2}
= \frac{(1-i)}{2} + \frac{(1+i)}{2} (-1)^k
= 
\frac{1}{\sqrt{2}}
\left( e^{-i\pi/4} + e^{+i\pi /4} e^{-i\pi k}\right)
\label{second_trick}
\end{equation} 
and, using the result in Eqn.~(\ref{half_revival_classical}), we can
see that
\begin{eqnarray}
\psi(x,t\approx T_{rev}/4)
& =& 
\sum_{k} a_k u_k(x) e^{-2\pi i kt/T_{cl}} e^{-\pi i k^2/2} \nonumber \\
& = & 
\sum_{k} a_k u_k(x) e^{-2\pi i kt/T_{cl}} 
\left[\frac{1}{\sqrt{2}}
\left( e^{-i\pi/4} + e^{+i\pi /4} e^{-i\pi k}\right)\right] 
\label{one_quarter_revival_function} \\
& = &
\frac{1}{\sqrt{2}}
\left[
e^{-i\pi/4}\, \psi_{cl}(x,t) +
e^{+i\pi/4}\, \psi_{cl}(x,t+T_{cl}/2)
\right] \nonumber 
\, . 
\end{eqnarray}
Thus, the wave packet near $T_{rev}/4$ (and $3T_{rev}/4$ as well) consists 
of two, high-correlated copies of the original packet, in- and out-of-phase, 
each containing half of the probability. This structure is similar to
that proposed in early suggestions made for Schr\"odinger cat type states
\cite{yurke_milburn},  and experimental work on Rydberg atom
realizations of this state \cite{noel} has been discussed in just that 
context.

The signal for such behavior is the presence of  peaks in $|A(t)|^2$ with
half the magnitude, and half  the classical periodicity, as each 
`mini-packet' or `clone' or `fractional revival' state now approaches 
the same location in phase space as the initial wave form twice during 
each classical period; this behavior is indeed
seen in Fig.~\ref{fig:compare}, with the exact revival structure and 
phase relation to the initial state for Case A (with a maximum value of
$|A(t)|^2 = (1\sqrt{2})^2 = 1/2$ as shown by the horizontal dashed line), 
and in a more approximate manner as for Case B.

\subsubsection{General structure of fractional revivals}
\label{subsubsec:general_structure}

A very general pattern of well-defined revivals, characterized by 
temporally localized structures in $A(t)$,  of local periodicity $T_{cl}/q$
and with magnitude $|A(t)|^2 = 1/q$,
at various rational multiples of the revival time given
by $t=pT_{rev}/q$ (where $p,q$ are mutually prime) is obvious from 
Fig.~\ref{fig:new_three} and leads to the term {\it fractional revivals}.
Averbukh and Perelman \cite{perelman_1} 
were the first to analyze in detail the mathematical
structure of the additional phase factors arising from the 
$\exp(-2\pi ik^2 t/T_{rev})$ terms at such times to discuss
 the ``{\it Universality in the long-term evolution of quantum wave packets
beyond the correspondence principle limit}'' and we reproduce here, in part,
their elegant arguments, for completeness. (We note that it has been
pointed out \cite{schleich_book}, \cite{schleich_factorize_numbers} that 
this problem, especially the calculation of the autocorrelation function at 
fractional revival times, is similar to that of the evaluation of Gauss 
sums \cite{gauss_sum_textbook}
which has a long history in the mathematical literature 
\cite{original_gauss}.) The case of odd $q$ is most 
straightforward and we consider the more general arguments for even $q$ in 
Appendix~\ref{appendix:fractional_revivals}.

We are interested in the structure of the additional phase terms 
at times $t = pT_{rev}/q$ of the form 
\begin{equation}
e^{-2\pi i k^2 t/T_{rev}}
= 
e^{-2\pi i (pk^2/q)} 
\equiv 
e^{-2\pi i \Theta_k} 
\label{phase_factor}
\end{equation}
and especially how to write these phases with $k^2$ in the exponential
in terms of similar factors, but with linear dependence on $k$, as in
Eqns.~(\ref{first_trick}) and (\ref{second_trick}). We first note that
the phase factor in Eqn.~(\ref{phase_factor}) will be periodic in $k$, 
namely
\begin{equation}
e^{-2\pi i \Theta_k}        =  e^{-2\pi i \Theta_{k+l}}
\qquad
\quad
\mbox{or}
\quad
\qquad 
\frac{p}{q} k^2 = \Theta_k  =  \Theta_{k+l} = \frac{p}{q} (k+l)^2
\qquad 
\mbox{(mod $1$).}
\label{periodic_condition}
\end{equation}
For this to occur, one requires that
\begin{equation}
\frac{2pk+pl^2}{q} = 0
\qquad
\qquad 
\mbox{(mod $1$)}
\end{equation}
which, for odd $q$, is satisfied if $l=q$. One can then expand any
periodic function (periodic in $k$) in terms of the basis states
\begin{equation}
e^{-2\pi i sk/l}
\qquad
\mbox{with}
\qquad
s = 0,1,...,l-1
\end{equation}
and write
\begin{equation}
e^{-2\pi i \Theta_k}
=
e^{-2\pi i p k^2/q} 
= \sum_{s=0}^{l-1}\,
b_s \, e^{-2\pi i sk/l}
\label{fourier_expansion}
\end{equation}
which can be described as a generalized trigonometric identity,
finite Fourier series, or discrete Fourier transform.
 This is already a useful result as we can then write
the wavefunction near a fractional ($p/q$) revival in the form 
\begin{eqnarray}
\psi(x,t\approx pT_{rev}/q)
& = &  
\sum_{k} a_k u_k(x) e^{-2\pi i kt/T_{cl}} e^{-\pi i pk^2/q} \nonumber \\
& = & 
\sum_{s=0}^{l-1} b_s
\left[\sum_{k} a_k u_k(x) e^{-2\pi i kt/T_{cl}} e^{-2 \pi i sk/l} 
\right] \\
& = & 
\sum_{s=0}^{l-1} b_s
\, \psi_{cl}(x,t+sT_{cl}/l)
\nonumber
\, . 
\end{eqnarray}
This form implies that there will be as many as $l$ `clones' of the
original wave packet, with amplitude $b_s$ and probability $|b_s|^2$,
differing in phase from the original packet by fractions ($T_{cl}/l$)
of the classical period. 

The expansion in
Eqn.~(\ref{fourier_expansion}) can be inverted by multiplying
both sides by $\exp(2\pi i rk/l)$, summing over $l$ possible states,
and using the periodicity of the $\Theta_k$ to obtain
\begin{equation}
\sum_{k=0}^{l-1} \, e^{2\pi irk/l - 2\pi i p k^2/q}
 = 
\sum_{s=0}^{l-1}b_s
\left[
\sum_{k=0}^{l-1} e^{2\pi i k(r-s)/l}
\right]  
= 
\sum_{s=0}^{l-1}b_s\, \left[l \delta_{r,s}\right]  =  l b_r
\end{equation}
or
\begin{equation}
b_r = \frac{1}{l} \sum_{k=0}^{l-1} e^{2\pi i(rk/l - pk^2/q)}
\, . 
\label{direct_summation}
\end{equation}
Using the arbitrariness of the summation index (due to the periodicity
of the exponentials), we can formally relabel this relation using
$k \rightarrow \overline{k} - 1$ and write
\begin{eqnarray}
b_{r} & = & \frac{1}{l} \sum_{\overline{k}} e^{2\pi i(r(\overline{k}-1)/l 
- p(\overline{k}-1)^2/q}  \nonumber \\
& = & 
\left[e^{-2\pi ir/l}\, e^{-2\pi ip/q}\right]
\frac{1}{l}
\sum_{\overline{k}}
e^{2\pi i(r\overline{k}/l + 2\overline{k}p/q - p \overline{k}^2/q}
\\ \nonumber
& = & \left[e^{-2\pi ir/l}\, e^{-2\pi ip/q}\right]
\left[
\frac{1}{l}
\sum_{\overline{k}}
e^{2\pi i(\overline{k}/l(r+2pl/q) - p \overline{k}^2/q}
\right] \nonumber \\
& = & 
\left[e^{-2\pi i(r/l+p/q)}\right]\, b_{r'}
\nonumber 
\end{eqnarray}
or
\begin{equation}
b_{r'} = e^{2\pi i(r/l+p/q)} \, b_{r}
\qquad
\mbox{where}
\qquad
r' \equiv r+ \frac{2pl}{q}
\label{recursion_relation}
\end{equation}
which recursively relates the expansion coefficients. This implies that all
of the (non-zero) $|b_r|$ have the same magnitude and,
for odd values of $q$ (and hence $l$), it gives
\begin{equation}
|b_{r}|^2 = \frac{1}{q}
\qquad
\mbox{for all $r=0,..,q-1$}
\end{equation}
and each of the $l=q$ `mini-packets' contains $1/q$ of the total
probability, overlapping with the original $t=0$ packet in $A(t)$ with a 
periodicity $T_{cl}/q$.

As an example, consider the case of $T_{rev}/3$ where $p=1$ and
$q=l=3$. In that case we use Eqn.~(\ref{recursion_relation}),
once with $r=0$ and $r'=2$ to obtain
\begin{equation}
b_2 = e^{2\pi i/3}\, b_0
\end{equation}
and again with $r=2$ and $r'=4 \longleftrightarrow r'=1$ 
(recall the periodicity) to find
\begin{equation}
b_{1} \longleftrightarrow  b_{4} = b_2
\, . 
\end{equation}
The explicit value of $b_0$ is obtained by direct summation of
Eqn.~(\ref{direct_summation}) giving
\begin{equation}
b_0 = \frac{1}{3} \sum_{k=0}^{2} e^{-2\pi i k^2/3}
=
\frac{1}{3}\left[1 + e^{-2\pi i/3} + e^{-8\pi i/3}\right]
= \frac{1}{3} \left[1 + 2e^{-2\pi i/3}\right]
= -\frac{i}{\sqrt{3}}
\, . 
\end{equation}
Thus, near $t\approx T_{rev}/3$, we find that
\begin{eqnarray}
\psi(x,t\approx T_{rev}/3)
& = & 
b_0 \psi_{cl}(x,t)
+ 
b_1 \psi_{cl}(x,t+T_{cl}/3)
+ 
b_2 \psi_{cl}(x,t+2T_{cl}/3) 
\label{one_third_revival_function}
\\
& = & 
-\frac{i}{\sqrt{3}}
\left[ \psi_{cl}(x,t)
+
e^{2\pi i/3}\left\{ 
\psi_{cl}(x,t+T_{cl}/3)
+
\psi_{cl}(x,t+2T_{cl}/3)
\right\}\right]
\nonumber 
\end{eqnarray}
as first observed by Averbukh and Perelman in Ref.~\cite{perelman_1}. 
The autocorrelation function near $T_{rev}/3$ is also shown in
Fig.~\ref{fig:compare} for our two model systems, where the exact and 
approximate realizations of this behavior are apparent.

The arguments for even $q$ are similar, but differ slightly depending
on whether $q$ is a multiple of $4$,  as discussed in 
Appendix~\ref{appendix:fractional_revivals}. The result however, is quite
similar in that one finds $r=q/2$ copies of the classical wave packet,
separated by multiples of $T_{cl}/r$, each with maximum $|A(t)|^2
= 1/r$. The cases of $q=2$ and $q=4$ are considered again explicitly in 
Appendix~\ref{appendix:fractional_revivals} and are shown to reproduce
the results in Eqns.~(\ref{one_half_revival_function}) 
and (\ref{one_quarter_revival_function}) respectively.

Averbukh and Perelman used their results to correctly analyze
the numerical calculations  of Parker and Stroud \cite{parker}, 
identifying the revival at $T_{rev}/2$ and fractional revivals of 
order $1/4,1/6,1/8$ which had earlier been described only as 
``{\it ...a complex pattern of quantum beats.}'' We reproduce the results of
their analysis of the (simulated) data in Fig.~\ref{fig:perelman}. 
Fractional revival structures of order up to $1/7$ have been observed 
in other systems \cite{wals},  as shown in Fig.~\ref{fig:wals}.

The observability of higher-order revivals, which would have $r$
distinct features in $|A(t)|^2$ separated by $T_{cl}/r$ (for $r=q$
($r=q/2$) for odd (even) $q$) depends on the level of the incoherent
`background' in Eqn.~(\ref{incoherent_value}),  so that in general such
features would not be observable if
\begin{equation}
\frac{1}{q}  = |A(t\approx pT_{rev}/q)|^2 < 
|A_{inc}|^2 = \frac{1}{\Delta n 2\sqrt{\pi}} 
\qquad
\mbox{or}
\qquad
q \geqnew \Delta n 2\sqrt{\pi}
\label{resolve_features}
\end{equation}
so that larger $\Delta n$ wave packets can `resolve' higher-order fractional
revivals, as noted in Fig.~\ref{fig:new_three}.
The behavior at such times during the collapsed or incoherent phase
is shown in Fig.~\ref{fig:compare} for our two model systems 
for a multiple of $T_{rev}/37$
where the dashed horizontal line indicates the value of $1/q$, while the
$|A(t)|^2$ values are seen to oscillate instead about the incoherent value,
shown by the dotted line.

We note that a number of authors have considered generalizations of 
these ideas \cite{leichtle}, applications to specific model systems 
\cite{example_6} - \cite{example_10}, and extensions to Hamiltonians
that are time-dependent through a slow change in a parameter
\cite{polavieja}.

\subsection{Superrevivals}
\label{subsec:superrevivals}

For systems with purely quadratic energy dependence on a single quantum
number (such as the infinite well, rigid rotor, and others), there are
no independent time scales longer than the revival time, and the pattern 
of fractional and full revivals will repeat itself indefinitely with the 
$T_{rev}$ time scale. For more realistic systems, with higher order terms 
in the expansion in Eqn.~(\ref{expansion_in_time}), 
the superrevival time, $T_{super}$ becomes
important, and Bluhm and Kosteleck\'{y} \cite{bluhm_long_term},
have exhaustively analyzed
the long-term time-dependence of wave packets, with emphasis on
Rydberg atom applications. They find qualitatively new patterns
of revival behavior, with periodicities in the motion of the packet
characterized by periods which are fractions of $T_{rev}$, giving a
self-similar structure to the auto-correlation function plots for
$t > T_{rev}$ and $t< T_{rev}$. The wave packet behavior on the
$T_{super}$ time scale is similar to that of the fractional revival
structures seen on the $T_{rev}$ scale, with integral multiples of
$T_{super}/3$ appearing prominently, and smaller time scale 
periodicities of $3T_{rev}$ appearing, explicitly due to the 
presence of the third-derivative term in Eqn.~(\ref{expansion_in_time}). 
In performing their analyses, they necessarily had to generalize and 
extend the results of Averbukh and Perelman 
\cite{perelman_1}  to include the contributions of
the $e^{-2\pi ip(n-n_0)^3/q}$ phase terms. Extensions to even more 
time scales \cite{leichtle}, \cite{schmidt}, with possible applications,
have also been discussed.

\subsection{Revivals in systems with two or more quantum numbers}
\label{subsec:two_quantum_numbers}

The generalization of the discussion of classical periodicity and revivals
(fractional or otherwise) to systems with more than one quantum number 
has been presented by several groups
\cite{nauenberg}, \cite{bluhm_2d}, \cite{agarwal_2d}.

For the case of two quantum numbers, for example, one first assumes a 
time-dependent quantum state of the form
\begin{equation}
\psi(t) = \sum_{n_1,n_2} a_{(n_1,n_2)} u_{(n_1,n_2)} e^{-iE(n_1,n_2)t/\hbar}
\end{equation}
where the coordinate labels have been suppressed, and could be either 
two position-space (e.g., $(x,y)$, $(r,\theta)$) or 
two momentum-space ($(p_x,p_y)$) values. The energy eigenvalues
are not assumed to factorize, but for localized wave packets, we do assume
that they can be expanded about a central value 
$(\overline{n}_1,\overline{n}_2)$, with the resulting expression 
(to second order) given by
\begin{eqnarray}
E(n_1,n_2) & = &
E(\overline{n}_1,\overline{n}_2)
+ 
(n_1-\overline{n}_1)
\left(\frac{\partial E(n_1,n_2)}{\partial n_1}\right)_{(\overline{n}_1,\overline{n}_2)}
+ 
(n_2-\overline{n}_2)
\left(\frac{\partial E(n_1,n_2)}{\partial n_2}\right)_{(\overline{n}_1,\overline{n}_2)}
\nonumber \\ 
 &  &
+
\frac{1}{2}(n_1-\overline{n}_1)^2
\left(\frac{\partial^2 E(n_1,n_2)}{\partial n_1^2}\right)_{(\overline{n}_1,\overline{n}_2)}
+
\frac{1}{2}(n_2-\overline{n}_2)^2
\left(\frac{\partial^2 E(n_1,n_2)}{\partial n_2^2}\right)_{(\overline{n}_1,\overline{n}_2)} \\
& & + 
(n_1-\overline{n}_1)(n_2-\overline{n}_2)
\left(\frac{\partial^2 E(n_1,n_2)}{\partial n_1 \partial n_2}\right)_{(\overline{n}_1,\overline{n}_2)}
+ \cdots
\,\,\, .
\nonumber
\end{eqnarray}
In the spirit of Eqn.~(\ref{expansion_in_time}), 
we can use this expansion to define 
important time scales. Two separate classical periods are given by
\begin{equation}
T_{cl}^{(n_1)} \equiv \frac{2\pi \hbar}{|\partial E/\partial n_1|}
\qquad
\mbox{and}
\qquad
T_{cl}^{(n_2)} \equiv \frac{2\pi \hbar}{|\partial E/\partial n_2|}
\label{two_periods}
\end{equation}
where we will henceforth suppress the subscripts indicating that the partial
derivatives are evaluated at $(\overline{n}_1,\overline{n}_2)$. The
corresponding longer-term, revival times are defined by 
\begin{equation}
T_{rev}^{(n_1)} = \frac{2\pi \hbar}{(1/2)|\partial^2 E(n_1,n_2)/\partial n_1^2|}
\qquad
,
\qquad
T_{rev}^{(n_2)} = \frac{2\pi \hbar}{(1/2)|\partial^2 E(n_1,n_2)/\partial n_2^2|}
\label{two_revival_times}
\end{equation}
and the mixed term 
\begin{equation}
T_{rev}^{(n_1,n_2)} = \frac{2\pi \hbar}{|\partial^2 E(n_1,n_2)/\partial n_1 \partial n_2|}
\, . 
\label{third_revival_time}
\end{equation}

There can be recognizable periodicities in the short-term semi-classical 
time-development, with the two classical periods beating against each
other, with the most obvious case being when the two periods are
commensurate with each other, namely when
\begin{equation}
\frac{T_{cl}^{(n_1)}}
     {T_{cl}^{(n_2)}} = \frac{a}{b}
\label{two_classical_periods}
\end{equation}
where $a,b$ are relatively prime integers. (Similar results hold for
action-angle variables in classical systems 
\cite{action_angle_1}, \cite{action_angle_2}.)
A simple example of such behavior
is for a two-dimensional oscillator with differing frequencies, namely
\begin{equation}
V(x,y) = \frac{m}{2} \left( \omega_x^2 x^2 + \omega_y^2 y^2\right)
\end{equation}
with quantized energies
\begin{equation}
E(n_x,n_y) = 
(n_x + 1/2) \hbar \omega_x
+ 
(n_y + 1/2) \hbar \omega_y
\end{equation}
where the condition in Eqn.~(\ref{two_classical_periods}) reduces to
\begin{equation}
\frac{T_{cl}^{(n_1)}}{T_{cl}^{(n_2)}} = \frac{\omega_y}{\omega_x} = 
\frac{a}{b}
\end{equation}
which is the familiar condition for the existence of Lissajous figures.
Similar analyses will be shown below to lead to the appropriate conditions
on closed or periodic orbits in simple two-dimensional quantum billiard
systems such as the square (or rectangular) case in 
Sec.~(\ref{subsec:2d_infinite_well}), 
the equilateral triangle (Sec.~\ref{subsec:equilateral_triangle}), 
and circular (Sec.~\ref{subsec:circular_billiard}) 
footprints being the most familiar. Nauenberg \cite{nauenberg}
has also considered the patterns of classical periodicities in systems
with several quantum numbers. We note that this connection of the
quantized energy eigenvalue spectrum to the structure of classical
recurring orbits is quite different from that found in another semiclassical
approach, namely periodic orbit theory \cite{gutzwiller}, \cite{brack}.

The extension of these ideas to central potentials in three dimensions,
where the energy eigenstates do not depend on the azimuthal quantum number
($E(n_r,l,m) = E(n_r,l)$ only), implies that one of the 
corresponding classical periods, $T_{cl}^{(m)} \rightarrow \infty$, 
is irrelevant, related to the fact that classical orbits are planar for all 
times. In a similar context, certain special 3D central potentials depend 
only on special combinations of quantum numbers with important semi-classical
connections. For example, for the 3D isotropic harmonic oscillator, defined by
\begin{equation}
V({\bf r}) = V(r) = 
\frac{m\omega^2}{2} (x^2 + y^2 + z^2)
= \frac{m\omega^2}{2} \, r^2
\, , 
\end{equation}
the quantized energies can be written in the form
\begin{equation}
E(N) = (N+3/2)\hbar \omega
\end{equation}
where $N$ can be written in terms of Cartesian quantum numbers in the
form $N = n_x+n_y+n_z$ or in cylindrical ($n_r,m_{\phi},n_z$) or 
spherical language ($n_r,l$) language. 
The fact that there is only one effective quantum number is related
to the symmetry properties of the oscillator potential 
(and its factorizability in many coordinate systems) and implies that
all classical periods are the same (or commensurate). This fact, in turn,
is related to the result of Bertrand's theorem \cite{rosner_bertrand}
which states that the only
3D power-law potentials for which all orbits are closed are given by
$k=2$ (the oscillator) and $k=-1$ (the Coulomb potential). For the latter
problem, the quantized energies also exhibit special patterns of 
degeneracy (dependent on $S0(4)$ symmetries beyond the 
simple $SO(3)$ rotational invariance) so that the Coulomb energy spectrum
(with $Z$ the nuclear charge and $\mu$ the reduced mass of two-body
system)
\begin{equation}
E(n_r,l) = 
-\frac{(Z\alpha)^2 \mu c^2}{2 (n_r+l)^2}
= 
-\frac{(Z\alpha)^2 \mu c^2}{2 (n)^2}
= E(n)
\label{coulomb_energies}
\end{equation}
depends only the principal quantum number, $n$, and not on 
the radial and angular quantum numbers, $n_r,l$, separately. The
factorizability of this problem in parabolic coordinates
\cite{bethe_landau} leads to the same result for the quantized
energies, and is useful for problems involving the addition of an 
external electric field (the Stark effect.)

The general time-development of wave packets which depend on two
independent quantum numbers, on longer time scales,  is then 
determined by the interplay of the three revival times, and the
corresponding commensurability condition required to observe revivals
is 
\begin{equation}
T_{rev}^{(n_1)}
=
\left(\frac{c}{d}\right)T_{rev}^{(n_2)}
= 
\left(\frac{e}{f}\right)T_{rev}^{(n_1,n_2)}
\end{equation}
where $c,d$ and $e,f$ are again pairs of relatively prime integers.
Fractional revivals are also possible if hierarchies such as
\begin{equation}
T_{frac}
= 
\left(\frac{p_1}{q_1}\right) T_{rev}^{(n_1)}
=
\left(\frac{p_2}{q_2}\right) T_{rev}^{(n_2)}
=
\left(\frac{p_{12}}{q_{12}}\right) T_{rev}^{(n_1,n_2)}
\end{equation}
exist and Bluhm, Kosteleck\'{y}, and Tudose \cite{bluhm_2d} 
have examined the structure 
of such fractional revivals, extending the results of 
Ref.~\cite{perelman_1} to the case of additional quantum numbers. 
We will provide examples of
systems described by these time scales in the context of several
2D quantum billiard systems below.

\section{Model systems}
\label{sec:model_systems}

The time-dependence of localized quantum wave packets, including 
possible quantum revival behavior, has been discussed for
a large number of pedagogically familiar, and physically relevant,
one-dimensional model systems. We briefly review several such cases, 
and then focus attention on the infinite well as a benchmark case where
exact quantum revival behavior is found.

\subsection{Free particle wave packets}
\label{subsec:free_particle}

The analysis of Gaussian free particle wave packets goes back
at least to Darwin \cite{darwin} and is a staple of introductory
textbooks. While such systems do not exhibit quantum revivals (or
even classical periodicity), it is useful to briefly review
the basic formalism of Gaussian wave packet solutions to the free-particle
Schr\"{o}dinger equation in one dimension. It helps establish notation
for later use,  as well as providing an example  for comparison with more 
realistic systems, especially illustrating important aspects of wave
packet spreading in explicit form.

A general Gaussian free-particle wave packet, with arbitrary initial
values of $\langle x \rangle_0 = x_0$ and $\langle p \rangle_0 = p_0$, 
can be written in  momentum space as
\begin{equation}
\phi(p,t) = \phi_0(p)\,e^{-ip^2t/2m\hbar} 
= 
\left[
\sqrt{\frac{\alpha}{\sqrt{\pi}}}
\, e^{-\alpha^2(p-p_0)^2/2}
\, e^{-ipx_0/\hbar}
\right]
\, e^{-ip^2t/2m\hbar} \, .
\label{gaussian_free_particle_momentum}
\end{equation}
The various expectation values related to momentum are given by
\begin{equation}
\langle p \rangle_{t} = p_0
\, , 
\qquad
\quad
\langle p^2 \rangle_{t} = p_0^2 + \frac{1}{2\alpha^2}
\, ,
\qquad
\mbox{and}
\qquad
\Delta p_t = \Delta p_0 = \frac{1}{\alpha \sqrt{2}}
\end{equation}
so that the momentum-space probability density 
 does not disperse in time,
and the expectation value of kinetic energy, given by 
$\langle p^2 \rangle_t/2m$, is obviously constant,
as it should for a free particle solution.

The corresponding solution in position-space is given by
\begin{equation}
\psi(x,t)  =   \frac{1}{\sqrt{\sqrt{\pi} \alpha \hbar (1+it/t_0)}}
\,
e^{ip_0(x-x_0)/\hbar}
\, e^{-ip_0^2t/2m\hbar}
\,
e^{-(x-x_0-p_{0}t/m)^2/2(\alpha \hbar)^2(1+it/t_0)}
\label{gaussian_free_particle_position}
\end{equation}
where 
\begin{equation}
t_0 \equiv m\hbar \alpha^2
= \frac{m \hbar}{2\ (\Delta p_0)^2}
= \frac{2m (\Delta x_0)^2}{\hbar}
\label{spreading_time}
\end{equation}
is the spreading time. The corresponding 
position-space probability density is
\begin{equation}
P(x,t) = |\psi(x,t)|^2 = \frac{1}{\sqrt{\pi} \beta_t}
e^{-(x-x(t))^2/\beta_t^2}
\end{equation}
where
\begin{equation}
x(t) \equiv x_0 + p_0t/m
\qquad
\mbox{and}
\qquad
\beta_t \equiv \alpha \hbar \sqrt{1+ t^2/t_0^2}
\end{equation}
and the time-dependent expectation values of position are 
\begin{equation}
\langle x \rangle_t = x(t) = x_0 + p_0t/m 
\, ,
\qquad
\quad
\langle x^2 \rangle_t = [x(t)]^2 + \frac{\beta^2_t}{2}
\, ,
\qquad
\mbox{and}
\quad
\Delta x_t = \frac{\beta_t}{\sqrt{2}}
\, ,
\end{equation}
which exhibits the classical constant rate of change for the
expectation value (consistent with Ehrenfest's theorem) and the
standard spreading.

For comparison to later examples, we note that the autocorrelation
function for this class of solutions is easily obtained in either
$p$- or $x$-space from Eqn.~(\ref{autocorrelation_general_definition}), 
and is given by
\begin{equation}
A(t) = \frac{1}{\sqrt{1-it/2t_0}} 
\exp\left[
\frac{i\alpha^2 p_0^2t}{2t_0(1-it/2t_0)}
\right]
\label{free_particle_A}
\end{equation}
or
\begin{equation}
|A(t)|^2 = \frac{1}{\sqrt{1+(t/2t_0)^2}}
\exp\left[-2\alpha^2 p_0^2 \frac{(t/2t_0)^2}{(1+(t/2t_0)^2)}\right]
\, .
\label{free_particle_autocorrelation}
\end{equation}
As expected, the only relevant time scale is the spreading time 
(actually $2t_0$) and we note that while for times satisfying $ t << 2t_0$  
there is an increasing exponential suppression of the overlap between $\psi_t$
and $\psi_0$, the exponential factor  does `saturate' for long times, giving
\begin{equation}
|A(t>>2t_0)|^2 \longrightarrow \frac{2t_0}{t} 
\exp\left[-\frac{p_{0}^2}{\Delta p_0^2}\right]
\quad
\mbox{since}
\quad
\alpha = \frac{1}{\sqrt{2} \Delta p_0}
\, . 
\end{equation}
The asymptotic form of the exponential factor can perhaps be best
understood by noting that the `distance in position space'  between the 
initial
`peak' at $\langle x \rangle_0 = x_0$, 
and that at later times when $\langle x \rangle_t = x_0 + p_0t/m$,  grows
linearly with $t$, while for long times the position spread,
\begin{equation}
\Delta x_t = \Delta x_0 \sqrt{ 1 + (t/t_0)^2}
\quad
\longrightarrow
\quad
\Delta x_0 \frac{t}{t_0}
\, , 
\end{equation}
increases  in the same way. This leads to factors in the exponent of the
form
\begin{equation}
\frac{(x(t) - x(0))^2}{(\Delta x_t)^2} 
\longrightarrow
\frac{(p_0t/m)^2}{(\Delta x_0 (t/t_0))^2}
\approx
\left(\frac{p_0 t_0}{m \Delta x_0}\right)^2
\approx (p_0 \alpha)^2
\quad
\mbox{since}
\quad
\Delta x_0 = \frac{\alpha \hbar}{\sqrt{2}}
\, .
\end{equation}
We distinguish the exponential suppression factor in 
Eqn.~(\ref{free_particle_autocorrelation}) (which we can describe as 
`dynamical' as it depends on the initial wave packet parameter, $p_0$) 
from the more intrinsic pre-factor term (containing only the spreading time) 
which is due to the natural dispersion of the wave packet (which we can 
therefore describe as `dispersive'.)

We note here that the Gaussian packet in 
Eqns.~(\ref{gaussian_free_particle_momentum}) or 
(\ref{gaussian_free_particle_position}) clearly
satisfies the general bound on $A(t)$ in Eqn.~(\ref{mandelstam}) 
in Sec.~\ref{subsec:autocorrelation_function}, 
saturating it to order ${\cal O}(t^2)$ for short times. In this case, 
one has
\begin{equation}
\langle H \rangle  = \frac{1}{2m} \left( p_0^2 + \frac{1}{2\alpha^2}\right)
\qquad
\quad
\mbox{and}
\qquad
\quad
\langle H^2 \rangle  =  \left(\frac{1}{2m}\right)^2
\left(  p_0^4 + \frac{3p_0^2}{\alpha^2} + \frac{3}{4\alpha^4}\right) 
\end{equation}
which give
\begin{equation}
(\Delta H)^2  =  \left(\frac{1}{2m}\right)^2\frac{2}{\alpha^2}
\left( p_0^2 + \frac{1}{4\alpha^2}\right)
\, . 
\end{equation}

\subsection{Wave packets and the constant force or uniform acceleration
problem}
\label{subsec:accelerating}

The problem of a particle under the influence of a constant
force is one of the most familiar in classical mechanics, but is less 
often treated in introductory quantum mechanics texts, especially in terms of
time-dependent solutions, despite the fact that closed form solutions
have been known since the time of Kennard \cite{kennard}. 
 For that reason, we briefly review the most
straightforward momentum-space approach to this problem. This system also
provides another example of the behavior of $A(t)$, this time
in a more dynamical system. 

In this case,
where the potential is given by $V(x) = -Fx$, we can write the
time-dependent Schr\"odinger equation in momentum-space as
\begin{equation}
\frac{p^2}{2m}\phi(p,t) - 
F\cdot \left[i\hbar \frac{\partial}{\partial p}\right] \phi(p,t)
= i\hbar \frac{\partial \phi(p,t)}{\partial t}
\label{pspace}
\end{equation}
or 
\begin{equation}
i\hbar\left(F\frac{\partial \phi(p,t)}{\partial p} 
+ \frac{\partial \phi(p,t)}{\partial t}\right) = \frac{p^2}{2m}\phi(p,t)
\, . 
\label{accelerating_equation}
\end{equation}
We note that the simple combination of derivatives guarantees that a
function of the form $\Phi(p-Ft)$ will make the left-hand side vanish,  so
we assume a solution of the form $\phi(p,t) = \Phi(p-Ft)\tilde{\phi}(p)$,
with $\Phi(p)$ arbitrary and $\tilde{\phi}(p)$ to be determined.
Using this form, Eqn.~(\ref{accelerating_equation}) reduces to
\begin{equation}
\frac{\partial \tilde{\phi}(p)}{\partial p} = 
-\frac{i p^2}{2m\hbar F}\tilde{\phi}(p)
\end{equation}
with the solution
\begin{equation}
\tilde{\phi}(p) = e^{-ip^3/6mF\hbar}
\, . 
\end{equation}
We can then write the general solution as
\begin{equation}
\phi(p,t) = \Phi(p\!-\!Ft)e^{-ip^3/6mF\hbar}
\end{equation}
or, using the arbitrariness of $\Phi(p)$, as
\begin{equation}
\phi(p,t) = \phi_0(p\!-\!Ft) e^{i((p-Ft)^3-p^3)/6mF\hbar )}
\label{pacc}
\end{equation}
where now $\phi_0(p)$ is some initial momentum distribution, since
$\phi(p,0) = \phi_0(p)$. (We note that because the $p^3$ terms cancel
in the exponential, we will be able to explicitly integrate
Gaussian type initial momentum-space waveforms.) The momentum-space
probability density now clearly satisfies
\begin{equation}
|\phi(p,t)|^2 = |\phi_{0}(p-Ft)|^2
\label{accelerating_momentum_shape}
\end{equation}
which demonstrates that  the momentum distribution simply
translates uniformly in time, with no change in shape.

For any general initial $\phi_{0}(p)$ we now have the time-dependent
expectation values
\begin{subequations}
\begin{eqnarray}
\langle p \rangle_t &= & \langle p \rangle_0 + Ft 
\label{accelerating_momentum_values}\\
\langle p^2 \rangle_t &= & \langle p^2 \rangle_0 + 
2\langle p \rangle_0 Ft +(Ft)^2 \\
\Delta p_t = \sqrt{\langle p^2\rangle_t - \langle p \rangle_t^2}
& = & \sqrt{\langle p^2\rangle_0 - \langle p \rangle_0^2} =
\Delta p_0 \\ 
\langle \hat{x} \rangle_t & = &  \langle \hat{x} \rangle_0
+ \frac{\langle p\rangle_0 t}{m} + \frac{Ft^2}{2m} 
\end{eqnarray}
\end{subequations}
giving the expectation value
\begin{equation}
\langle H \rangle 
=
\left \langle \frac{p^2}{2m} + V(x) \right\rangle = 
\frac{\langle p^2 \rangle_0}{2m}
- F\langle \hat{x} \rangle_0
\end{equation}
which, in turn, also  agrees with a similar calculation of 
$\langle \hat{E} \rangle_t$
using $\hat{E} = i\hbar (\partial/\partial t)$,  all of which are consistent
with a particle undergoing uniform acceleration.

Using the standard initial Gaussian momentum-space wavefunction,
$\phi_0(p)$ in Eqn.~(\ref{gaussian_free_particle_momentum}), 
as the $\phi_0(p-Ft)$ in Eqn.~(\ref{pacc}), 
we can evaluate the position-space solution
using the Fourier transform to obtain
\begin{eqnarray}
\psi(x,t) & = & \left[e^{iFt(x_0-Ft^2/6m)/\hbar}\, e^{i(p_0+Ft)(x-x_0 - p_0t/2m)/\hbar}\right] \left(
\frac{1}{\sqrt{\sqrt{\pi}\alpha \hbar (1+it/t_0)}}
\right) \nonumber \\ 
& & \,\,\,
\times
\, e^{-(x-(x_0+p_0t/m+Ft^2/2m))^2/2(\alpha \hbar^2)^2(1+it/t_0)}
\, . 
\label{accelerating_solution}
\end{eqnarray}
The corresponding probability density is given by
\begin{equation}
P(x,t) = \frac{1}{\sqrt{\pi} \beta_t}
e^{-(x-\tilde{x}(t))^2/\beta_t^2}
\end{equation}
where 
\begin{equation}
\tilde{x}(t) \equiv x_0 + \frac{p_0t}{m} + \frac{Ft^2}{2m}
\end{equation}
and 
\begin{equation}
\qquad
\langle x \rangle_t = \tilde{x}(t)
\, ,
\qquad
\langle x^2 \rangle_t = [\tilde{x}(t)]^2 + \frac{\beta_t^2}{2}
\, ,
\quad
\mbox{and}
\qquad
\Delta x_t = \frac{\beta_t}{\sqrt{2}} 
\label{accelerating_values}
\end{equation}
so that this accelerating wave packet spreads in the same manner as the
free-particle Gaussian example.

The calculation of the autocorrelation function can once again 
be done in either $p$- or $x$-space using 
Eqn.~(\ref{autocorrelation_general_definition}) to give
\begin{equation}
A(t)  =  \frac{1}{\sqrt{1-it/2t_0}} 
\exp\left[\frac{(2ip_0^2t/m\hbar - (\alpha F t)^2 (1+(t/2t_0)^2))
}{4(1-it/2t_0)} 
\right] 
\, 
e^{-iFt(x_0-Ft^2/6m)/\hbar} 
\end{equation}
and  the same factors of $1-it/2t_0$ as in Eqn.~(\ref{free_particle_A})
are obtained; this expression also reduces to that case in the free particle
limit  when $F \rightarrow 0$,  as it must. The modulus-squared is given by 
\begin{equation}
|A(t)|^2 = \frac{1}{\sqrt{1+(t/2t_0)^2}}
\,
\exp\left[-2\alpha^2 (p_0^2 +(Ft_0)^2(1+(t/2t_0)^2)
\left(\frac{(t/2t_0)^2}{1+(t/2t_0)^2}\right) 
\right]
\label{accelerating_autocorrelation}
\end{equation}
and we note that this result can be obtained from 
Eqn.~(\ref{free_particle_autocorrelation}) by the simple substitution
\begin{equation}
p_0^2 \longrightarrow
p_0^2 +(Ft_0)^2(1+(t/2t_0)^2)
\, . 
\end{equation}
For this case of uniform acceleration, the wave packet spreading is
identical (same $\Delta x_t$) as in 
the free-particle case, which can be  understood
by noting that the distance between two classical particles starting at the 
same initial location, undergoing the same force, but with slightly different
initial velocities (or momenta, $p_0^{(A)} - p_0^{(B)} =
\Delta p_0$) would be
\begin{equation}
x_A(t) - x_B(t) = (x_0 + p_0^{(A)}t/m + Ft^2/2m) 
                - (x_0 + p_0^{(B)}t/m + Ft^2/2m) 
= \frac{\Delta p_0t}{m}
\end{equation}
which increases linearly with time, in exactly the same way
as for the  free-particle solutions (when $F=0$).
The `distance' between the peaks in $\psi_0$ and $\psi_t$, however,
eventually grows as $t^2$ so that the exponential (`dynamical') 
suppression in $A(t)$ does not saturate, while the `dispersive' 
pre-factor is exactly the same as for the free-particle case.

\subsection{Harmonic oscillator}
\label{subsec:harmonic_oscillator}

The harmonic oscillator provides the most straightforward example of
a bound state system for which the periodic motion of wave packet
solutions (especially Gaussian) is easily derivable. 
In this case,
the initial value problem is perhaps most easily solved, especially for
Gaussian wave packets, by propagator techniques \cite{saxon}.
In this approach, one writes
\begin{equation}
\psi(x,t) = \int_{-\infty}^{+\infty} \,dx'\, \psi(x',0)\, K(x,x';t,0)
\label{propagator}
\end{equation}
where the propagator can be derived in a variety of ways 
\cite{holstein}
and can be written in the form
\begin{equation}
K(x,x';t,0) = \sqrt{\frac{m \omega}{2\pi i \hbar \sin(\omega t)}}
\exp
\left[
\frac{im \omega}{2\hbar \sin(\omega t)} ((x^2 + (x')^2)\cos(\omega t)
- 2x x')
\right]
\, .
\end{equation}

Using the initial position-space wave function
\begin{equation}
\psi(x,0) = \frac{1}{\sqrt{\beta \sqrt{\pi}}}
\, e^{ip_0x/\hbar}
\,
e^{-(x-x_0)^2/2\beta^2}
\, ,
\label{just_initial_gaussian}
\end{equation}
where $\beta \equiv \alpha \hbar$, one can evaluate the time-dependent
wave function in closed form as 
\begin{equation}
\psi(x,t) = \frac{1}{\sqrt{L(t)\sqrt{\pi}}}
\exp\left[
\frac{S(x,t)}{2\beta L(t)}
\right]
\label{general_sho_solution}
\end{equation}
where
\begin{equation}
L(t) \equiv \beta \cos(\omega t)
+ \frac{i \hbar}{m\omega \beta} \sin(\omega t)
\end{equation}
and
\begin{eqnarray}
S(x,t) & \equiv &
-x_0^2\cos(\omega t)
+2xx_0
- x^2\left[\cos(\omega t) 
+\frac{im\omega \beta^2 \sin(\omega t)}{\hbar}\right] \\
& &
\,\,\,\,\,\,\,\,\,\,\,\,\,
- \frac{ 2x_0p_0 \sin(\omega t)}{m \omega}
+ \frac{2i\beta^2 p_0x}{\hbar}
- \frac{i\beta^2 p_0^2 \sin(\omega t)}{m\omega \hbar} 
\nonumber 
\, .
\end{eqnarray}
The corresponding position-space probability density can be written as 
\begin{equation}
|\psi(x,t)|^2
= \frac{1}{\sqrt{\pi} |L(t)|}
\exp\left[
-\frac{(x-x_0\cos(\omega t) - p_0 \sin(\omega t)/m\omega)^2}{|L(t)|^2}
\right]
\label{sho_general_case}
\end{equation}
with
\begin{equation}
\langle x \rangle_t = x_0 \cos(\omega t) + \frac{p_0 \sin(\omega t)}{m\omega}
\end{equation}
and
\begin{equation}
\Delta x_t = \frac{|L(t)|}{\sqrt{2}}
= \frac{1}{\sqrt{2}} \sqrt{ \beta^2 \cos^2(\omega t) + (\hbar/m\omega \beta)^2
\sin^2(\omega t)}
\,.
\end{equation}
Thus, the expectation value evolves in accordance with 
classical expectations \cite{styer_2}, while the width in position-space
oscillates  (from wide to narrow, or 
vice versa.) The momentum-space variables behave in a similarly 
correlated manner with
\begin{eqnarray}
\langle p \rangle_t & = &
 - m\omega x_0 \sin(\omega t) + p_0 \cos(\omega t) \\ 
\Delta p_t & = & \frac{1}{\sqrt{2}} \sqrt{ (\hbar/\beta)^2 \cos^2 (\omega t)
+ (m\omega \beta)^2 \sin^2(\omega t)}
\, . 
\end{eqnarray}

For the special case of the `minimum uncertainty' wave packet,
where 
\begin{equation}
\beta^2 = \frac{\hbar}{m\omega} \equiv \beta_0^2
\, , 
\label{minimum_uncertainty}
\end{equation}
the width of the packet is fixed as
\begin{equation}
\Delta x_t = \Delta x_0 = \frac{\beta_0}{\sqrt{2}}
\end{equation}
which is the same as the ground state oscillator energy eigenvalue state,
but simply oscillates at the classical frequency, which is similar
to the famous example first cited by Schr\"{o}dinger \cite{schrodinger}
or to later examples of coherent states \cite{coherent_states}.

The total energy of the general solution can be written in the form
\begin{equation}
\langle \hat{E} \rangle = \left(\frac{p_0^2}{2m} + \frac{1}{2}
m \omega^2 x_0^2\right) +
\frac{\hbar \omega}{4} \left(\frac{\beta_0^2}{\beta^2} + 
\frac{\beta^2}{\beta_0^2}\right)
\, . 
\label{sho_total_energy}
\end{equation}
Since this is also a bound state problem, one can expand the solution
in Eqn.~(\ref{general_sho_solution}) in eigenstates as
\begin{equation}
\psi(x,t) = \sum_{n=0}^{\infty} a_n \, u_n(x)\, e^{-iE_nt/\hbar}
\qquad
\mbox{where $E_n =  (n+1/2)\hbar \omega$}
\, .
\label{new_sho_expansion}
\end{equation}
We can then note that
\begin{equation}
\langle \hat{E} \rangle = \langle (n+1/2) \rangle \hbar\omega
= 
\left(\langle n \rangle + 1/2\right)\hbar \omega 
\end{equation}
which for $x_0,p_0=0$ and $\beta = \beta_0$ gives
$\langle n \rangle = 0$ as expected, since in this limit 
we recover the ground state energy. The spread in $n$ values is useful 
in discussions of wave
packet behavior, and we note that
\begin{equation}
(\Delta E)^2
= \langle \hat{E}^2 \rangle
- \langle \hat{E} \rangle^2 
= \langle (n+1/2)^2 \rangle - \langle (n+1/2)\rangle^2
=
 (\Delta n)^2 (\hbar \omega)^2
\end{equation}
and we find that
\begin{equation}
(\Delta n)^2
= \frac{1}{2}\left[\frac{p_0^2}{(m\omega \beta)^2}
+ x_0^2 \left(\frac{m\omega \beta}{\hbar}\right)^2\right]
+ \frac{1}{8} \left( \frac{\beta_0^4}{\beta^4}
+ \frac{\beta^4}{\beta_0^4} -2\right)
\label{sho_deltan}
\end{equation}
which will be useful. In the $x_0,p_0=0$ and $\beta = \beta_0$ limit,
we find $\Delta n =0$,  as expected for an eigenstate.

All wave packet solutions (Gaussian or not) of the harmonic oscillator
can be shown 
(using, for example \cite{saxon}, the expansion 
in Eqn.~(\ref{new_sho_expansion})), to satisfy 
\begin{equation}
\psi(x,t+mT_{cl}) = (-1)^{m}\, \psi(x,t)
\label{sho_standard}
\end{equation}
with a similar result for $\phi(p,t)$ as well. This implies that
the autocorrelation function will be periodic as well, with
\begin{equation}
A(t+kT_{cl}) = (-1)^{k} A(t)
\, . 
\end{equation}

The evaluation of $|A(t)|^2$ for general values of $\beta$, $x_0$, and
$p_0$ is straightforward enough, but the resulting expressions 
are somewhat cumbersome, so we will focus on several special cases
as illustrative.

\vskip 0.7cm
\noindent
{\bf Case I}: Minimum uncertainty wave packets, $\beta = \beta_0$.

In this case the evaluation of $A(t)$ gives
\begin{equation}
A(t) = \sqrt{\cos(\omega t) + i \sin(\omega t)}
\exp\left[
- \left(\frac{x_0^2}{2\beta_0^2} + \frac{p_0^2}{2m\omega \hbar}\right)
[(1-\cos(\omega t)) - i\sin(\omega t)]
\right]
\label{sho_autocorrelation}
\end{equation}
where great simplifications have been made by noting that
\begin{equation}
\frac{1}{\cos(\omega t) -i \sin(\omega t)}
= \cos(\omega t) + i \sin(\omega t)
\, . 
\end{equation}
(We note that a very similar expression arises in analyses of the
macroscopic wavefunction for Bose-Einstein condensates \cite{bec_first}
and the collapse and revival of the matter wave field has been observed
experimentally \cite{bec_revivals}.)
Once again, the two important parameters appear together in quadrature,
as in the uniform acceleration case, and we have
\begin{equation}
|A(t)|^2
= \exp
\left[
-\left(
\frac{x_0^2}{\beta_0^2} + \frac{p_0^2}{m\omega \hbar}\right)
\left\{1-\cos(\omega t)\right\}\right]
\label{sho_case}
\end{equation}
which clearly exhibits the expected periodicity. All of the suppression
can be attributed to the `dynamical' factors (those in the exponential,
containing $x_0$ and $p_0$) as there is no `dispersive' pre-factor
component for this constant width packet.

\vskip 0.7cm

\noindent
{\bf Case II}: Arbitrary $\beta$, but $x_0, p_0 = 0$. For this case, the
wave packet does not oscillate, but only `pulsates' \cite{sho_mirror}, 
and the time-dependent wave function simplifies to 
\begin{equation}
\psi(x,t) = 
\frac{1}
{\sqrt{(\beta \cos(\omega t) +(i\hbar/m \omega \beta)\sin(\omega t))}}
\, 
\exp
\left[\frac{-(x^2 (\cos(\omega t) +(im\omega \beta^2/\hbar)\sin(\omega t))}
{2\beta^2 (\cos(\omega t) +(i\hbar/m \omega \beta)\sin(\omega t))}
\right] 
\, . 
\end{equation}
It is convenient to define the parameters
\begin{equation}
r \equiv \frac{\hbar}{m \omega \beta^2} = \frac{ \beta_0^2}{\beta^2}
\qquad
\mbox{so that}
\qquad
\frac{1}{r} = \frac{\beta^2}{\beta_0^2}
\end{equation}
in terms of which the resulting autocorrelation function in this case
has the very simple form
\begin{equation}
A(t) = \sqrt{\frac{2}{2\cos(\omega t) -i(r+1/r)\sin(\omega t)}}
\label{special_case}
\end{equation}
or
\begin{equation}
|A(t)|^2 = \frac{1}{\sqrt{\cos^2(\omega t) + (r+1/r)^2 \sin^2(\omega t)/4}}
\end{equation}
all of which can be attributed to a `dispersive' (but in this case
periodic) pre-factor.

We first note that in this case 
$A(t)$ is invariant under the transformation
$ r \rightarrow 1/r$, in other words, the time-dependence is the same
for both initially wide ($\beta > \beta_0$) or narrow ($\beta < \beta_0$)
packets. 
Plots of $|A(t)|^2$ over one classical period are shown in 
Fig.~\ref{fig:pulsate_auto}, where
it is clear that the larger the deviation from the `minimum uncertainty'
wavepacket, the faster the wavepacket `pulsates' away from its initial
shape. It is also noteworthy that in this case $|A(T_{cl}/2)|=1$ so that
the wave packet returns to its initial form (up to a constant complex
phase) {\it twice} each classical period. This can be understood from 
the expansion
of this wave form in terms of energy eigenstates. In this case,
where the parameters $x_0,p_0$ both vanish, one is expanding an even-parity
function. In the eigenstate expansion,
\begin{equation}
\psi(x,t) = \sum_{n=0}^{+\infty} a_n u_n(x) e^{-iE_nt/\hbar}
= e^{-i\omega t/2} \sum_{n=0}^{+\infty} a_n u_n(x) e^{-in\omega t}
\, , 
\label{sho_expansion}
\end{equation}
the $u_n(x)$ have parity $(-1)^n$, so that for an even-parity state,
only the even ($a_{2n}$) terms are nonvanishing and 
the $n$-dependent exponential factors in 
Eqn.~(\ref{sho_expansion}) therefore oscillate twice as rapidly
as in the general case. 

Finally, for the very special case  where $\beta = \beta_0$ ($r=1$) as well, 
we recover the ground state energy eigenstate of the oscillator, with its 
trivial stationary-state time-dependence
($\psi_0(x,t) = u_0(x)\exp(-iE_0t/\hbar)$) and 
Eqn.~(\ref{special_case}) indeed reduces to
\begin{equation}
A(t) \stackrel{r\rightarrow 1}{\longrightarrow}
\sqrt{\frac{2}{2\cos(\omega t) - 2i\sin(\omega t)}}
= \sqrt{e^{i\omega t}} = e^{+i\omega t/2}
\end{equation}
as expected. 

Because of the specially symmetric nature of the potential, besides
the relation in Eqn.~(\ref{sho_standard}), we also have 
\begin{equation}
\psi(-x,t+T_{cl}/2) = (-i) \, \psi(x,t)
\qquad
\mbox{and}
\qquad
\phi(-p,t+T_{cl}/2) = (-i) \, \phi(p,t)
\label{sho_anticorrelation}
\end{equation}
so that half a period later, the wave-packet is also reproduced, but at
the opposite `corner' of phase space, namely with $x \leftrightarrow
-x$ and $p \leftrightarrow -p$; note that two applications of
Eqn.~(\ref{sho_anticorrelation}) reproduce Eqn.~(\ref{sho_standard}).
One can also show these connections using propagator techniques, 
provided one properly identifies the complex pre-factors,
as described in detail in Ref.~\cite{saxon}.

This type of `mirror' behavior can be diagnosed using a 
variation on the standard autocorrelation function, namely 
\begin{equation}
\tilde{A}(t) 
\equiv
\int_{-\infty}^{+\infty}\, \psi^{*}(-x,t) \, \psi(x,0)\,dx
=
\int_{-\infty}^{+\infty}\, \phi^{*}(-p,t) \, \phi(p,0)\,dp
\label{anticorrelation_function_definition}
\end{equation}
which measures the overlap of the initial state with the 
`across-phase-space' version of itself at later times. 
Given the simple connections in
Eqn.~(\ref{sho_anticorrelation}), we can immediately write, for 
Case I considered above, 
\begin{equation}
|\tilde{A}(t)|^2 
= \exp
\left[
-\left(
\frac{x_0^2}{\beta_0^2} + \frac{p_0^2}{m\omega \hbar}\right)
\left\{1+\cos(\omega t)\right\}\right]
\end{equation}
which is exponentially suppressed at integral multiples of $T_{cl}$,
but unity at $t = (2k+1)T_{cl}/2$. This type of {\it anti-correlation}
function finds use in the study of wave packet revivals where quantum
wave packets may reform near $t = T_{rev}/2$, as 
in Eqn.~(\ref{one_half_revival_function}), 
but out of phase with the original packet.

Finally, we note that these results can be extended to the case of the 
`inverted' oscillator (corresponding to a particle in unstable
equilibrium, i.e., at a local maximum, instead of a local minimum 
of the potential) defined by 
\begin{equation}
\tilde{V}(x) \equiv  - \frac{1}{2} m\tilde{\omega}^2 x^2
\end{equation}
One can make the substitutions
\begin{equation}
\omega^2 \rightarrow -\tilde{\omega}^2
\, ,
\qquad
\omega \rightarrow i \tilde{\omega}
\, ,
\qquad
\sin(\omega t) \rightarrow i \sinh(\tilde{\omega} t)
\, ,
\qquad
\mbox{and}
\qquad
\cos(\omega t) \rightarrow \cosh(\tilde{\omega} t)
\label{runaway_identifications}
\end{equation}
and we briefly discuss this in Appendix~\ref{appendix:inverted_oscillator}.

\subsection{The infinite well}

\subsubsection{General comments}
\label{subsec:infinite_general}

The one-dimensional infinite potential well is the most frequently
encountered example of a 1D bound state system, finding
its way into every introductory quantum mechanics text. Given its
important pedagogical role, and familiar solutions,
it is not surprising that many aspects of wave packet propagation 
in general, and quantum revivals in particular, in this system
have been studied \cite{bluhm_ajp}, \cite{segre} - \cite{force_paper}.
(Interestingly, the time development of quantum states in
the infinite well was used as a debating point by Einstein
and Born \cite{born}, \cite{born_2} in more general discussions about the 
nature of quantum mechanics.)

In this review, we will
examine both the short-term quasi-classical time-evolution of Gaussian
wave packets, the spreading to a collapsed state, as well as the 
structure of the revivals, mirror or 'anti`-revivals, and fractional
revivals using both the standard auto-correlation function, $A(t)$ and a 
related `anti-correlated' $\overline{A}(t)$. We also focus on visualizing 
the same phenomena through the time-development of expectation values (and
uncertainties), in both position- and momentum-space, as well as 
emphasizing the visualization of such effects in a variety of ways, 
including the Wigner quasi-probability (phase-space) distribution and
the use of what have been dubbed {\it quantum carpets}. It is worth 
noting that bound-state wave packets in this system were examined in detail
by Segre and Sullivan \cite{segre} in 1976 where the existence of exact 
wave packets revivals was  discussed very explicitly, 
with the authors also noting in a footnote that
``{\it We suspect that this almost periodic behavior of packet width is
a general property of bound-state packets.}''

We begin by defining the familiar problem of a particle of mass $m$ 
confined by the potential
\begin{equation}
      V(x) = \left\{ \begin{array}{ll}
               0 & \mbox{for $0<x<L$} \\
               \infty & \mbox{otherwise}
                                \end{array}
\right.
\label{infinite_well_potential}
\end{equation}
which has energy eigenvalues and stationary state solutions given by
\begin{equation}
E_n = \frac{p_n^2}{2m} = \frac{\hbar ^2 \pi^2 n^2}{2mL^2}
= n^2 E_0
\qquad
\mbox{where}
\qquad
p_n \equiv \frac{n \hbar \pi}{L}
\qquad
\mbox{for $n=1,2,...$}
\end{equation}
and
\begin{equation}
u_n(x) = \sqrt{\frac{2}{L}} \sin\left(\frac{n\pi x}{L}\right)
\, .
\label{1d_eigenfunctions}
\end{equation}
We note that the energy eigenfunctions have a generalized parity
property (about the midpoint of the well), namely 
\begin{equation}
u_{n}(L-x) = 
\sqrt{\frac{2}{L}} \sin\left(\frac{n\pi (L-x)}{L}\right)
= 
-\sqrt{\frac{2}{L}} \cos(n\pi) \sin\left(\frac{n\pi x}{L}\right)
=
(-1)^{n+1} \, u_{n}(x)
\end{equation}
so that for $n$ odd (even), the eigenfunctions are even (odd) about
the center ($L/2$) of the well. 
This connection suggests that we generalize the notion of the
$\tilde{A}(t)$ `anti-correlation' function in 
Eqn.~(\ref{anticorrelation_function_definition}) 
to the geometry of the standard infinite well by defining
\begin{equation}
\overline{A}(t) \equiv
\int_{0}^{L} \psi^*(L-x,t)\, \psi(x,0)\,dx
\, . 
\label{other_autocorrelation_function}
\end{equation}

The classical period for this system,
obtained from Eqn.~(\ref{classical_period}),  is
\begin{equation}
T_{cl} = \frac{2\pi\hbar}{|E_{n}'|}
= \frac{2mL^2}{\hbar \pi n}
= \frac{2L}{[(\hbar \pi n/L)/m]}
= \frac{2L}{v_n}
\qquad
\quad
\mbox{where}
\quad
\qquad
v_n \equiv \frac{p_n}{m}
\label{infinite_well_period}
\end{equation}
and $v_n$ is the analog of the classical speed, giving a result
in agreement with classical expectations. The revival time is then given by
\begin{equation}
T_{rev} \equiv \frac{2\pi \hbar}{|E_{n}''|/2}
= \frac{2\pi \hbar}{E_0}
= \frac{4mL^2}{\hbar \pi}
= (2n) T_{cl}
\label{infinite_well_revival}
\end{equation}
which is also consistent with Eqn.~(\ref{power_law_ratio}) 
for general 1D power law potentials. 
Thus, for $n>>1$, one can easily arrange to have $T_{rev} >> T_{cl}$.
The superrevival and higher order terms in the expansion in 
Eqn.~(\ref{expansion_in_time})
all vanish, making this an ideal system to study exact revival behavior.

The quantum revivals in this case are in a sense exact, since we have
\begin{equation}
\psi(x,t+T_{rev}) =
\sum_{n=1}^{\infty} a_n u_n(x) e^{-iE_n t/\hbar}\, e^{-i E_n T_{rev}/\hbar}
= \sum_{n=1}^{\infty} a_n u_n(x) e^{-iE_n t/\hbar}\, e^{-i 2\pi n^2}
= \psi(x,t)
\end{equation}
for all times. 
The wave packet will also, however, reform itself at $t=T_{rev}/2$ 
(a so-called mirror revival or 'anti-revival') at a (possibly) different 
location in the well. To see this we note that
\begin{eqnarray}
\psi(L-x,t+T_{rev}/2) 
& = &
\sum_{n=1}^{\infty} 
a_{n} u_{n}(L-x) e^{-iE_n t/\hbar} e^{-iE_n T_{rev}/2\hbar}
\nonumber \\
& = & 
\sum_{n=1}^{\infty} a_n u_{n}(x) e^{-iE_n t/\hbar}\left[ (-1)^{n+1} e^{-in^2\pi}\right]
\\
& = &
-\psi(x,t)
\nonumber 
\end{eqnarray}
which implies that
\begin{equation}
|\psi(x,t+T_{rev}/2)|^2 = |\psi(L-x,t)|^2
\label{infinite_mirror_position}
\end{equation}
so that at half a revival time later, any initial wave packet will
reform itself (same shape, width, etc.), but at a location mirrored
\cite{1d_fractional}
about the center of the well.
 The use of the anti-correlation
function, $\overline{A}(t)$, defined in 
Eqn.~(\ref{other_autocorrelation_function}), will then be useful
not only for documenting the short-term, semi-classical periodicity
of such packets, but especially in establishing this additional type 
of revival structure. This behavior is a special case of the more general
$T_{rev}/2$ revival discussed in Eqn.~(\ref{one_half_revival_function}).

The behavior of the wave packet in momentum space for such mirror packets
is also easily derived using the Fourier transform connection, in this
case
\begin{equation}
\phi(p,t) = \frac{1}{\sqrt{2\pi \hbar}}
\int_{0}^{L} \psi(x,t) \, e^{ipx/\hbar}\, dx
\, .
\end{equation}
We simply note that
\begin{eqnarray}
\phi(p,t+T_{rev}/2) & = & \frac{1}{\sqrt{2\pi \hbar}} 
\int_{0}^{L}\, \psi(x,t+T_{rev}/2) \, e^{ipx/\hbar} \, dx  \nonumber  \\
& = &  - \frac{1}{\sqrt{2\pi \hbar}}
\int_{0}^{L}\, \psi(L-x,t) \, e^{ipx/\hbar} \, dx \nonumber \\
& = &  -e^{ipL/\hbar} \left[\frac{1}{\sqrt{2\pi \hbar}}
\int_{0}^{L} \, \psi(y,t) \, e^{-ipy/\hbar} \, dy\right] \\
& = &  -e^{ipL/\hbar} \phi(-p,t) \nonumber 
\end{eqnarray}
so that 
\begin{equation}
|\phi(p,t+T_{rev}/2)|^2 = |\phi(-p,t)|^2
\label{infinite_mirror_momentum}
\end{equation}
and half a revival time later the initial momentum profile is 
also reproduced,
except flipped in sign ($p \rightarrow -p$), so that the particle 
is also moving in the 'other direction'. 

It has been noted several times that the dynamical time dependence
of wave packets in this system can be described by the free-space
evolution of an infinite sequence of appropriately displaced
initial wave functions \cite{1d_fractional}, \cite{particle_german},
which can be derived using a `method of images' technique
\cite{kleber}.

\subsubsection{Gaussian wave packets}
\label{subsec:infinite_gaussian}

For definiteness, one typically considers Gaussian-like wave
packets, corresponding approximately to initial momentum- or
position-space wave functions as in the $t=0$ limits of
Eqns.~(\ref{gaussian_free_particle_momentum}) and
 (\ref{gaussian_free_particle_position}). 
The expansion coefficients corresponding to such initial states
can be well-approximated in an analytic form if we assume that the initial
position-space wavefunction,
\begin{equation}
\psi_{G}(x,0)
= 
\frac{1}{\sqrt{b\sqrt{\pi}}}
\,
e^{-(x-x_0)^2/2b^2}
\,
e^{ip_0(x-x_0)/\hbar}
\end{equation}
(where $b \equiv  \alpha \hbar$)
is sufficiently contained within the well so that we make an exponentially 
small error by neglecting any overlap with the region outside the walls,
and may thus also ignore any problems associated with possible
discontinuities at the wall. In practice, this condition
only requires the wave packet to be a few times
$\Delta x_0 = b/\sqrt{2}$ away from an infinite wall boundary.

With these assumptions, we can then extend the integration region from the 
finite $(0,L)$ interval to the entire 1D space, giving the (exponentially
good) approximation for the expansion coefficients
\cite{blueprint} 
\begin{eqnarray}
a_{n} & \approx & \int_{-\infty}^{+\infty} 
\left[u_{n}(x)\right]\, \left[\psi_{G}(x,0)\right]\,dx 
\nonumber \\
& = & 
\left(\frac{1}{2i}\right)
\sqrt{\frac{4b\pi}{L\sqrt{\pi}}}
[
e^{in \pi x_0/L}  e^{-b^2(p_0 + n\pi \hbar/L)^2/2\hbar^2}
-
e^{-in \pi x_0/L} e^{-b^2(p_0 - n\pi \hbar/L)^2/2 \hbar^2}],
\label{approximate_expansion}
\end{eqnarray}
because we can write 
\begin{equation}
\sin\left(\frac{n \pi x}{L}\right) = \frac{1}{2i}
(e^{in \pi x/L} - e^{-i n \pi x/L}),
\end{equation}
and we can perform Gaussian integrals such as
\begin{equation}
\int_{-\infty}^{+\infty} e^{-ax^2 -bx}\,dx 
= \sqrt{\frac{\pi}{a}} e^{b^2/4a}
\end{equation}
in closed form. This explicit expression goes back to at least
Born \cite{born} and in our context is very useful because it can speed
up numerical calculation involving the expansion coefficients, 
such as the evaluation of $A(t)$. It also accurately encodes 
the sometimes delicate interplay between the oscillatory pieces of 
the Gaussian ($e^{-ip_0 x/\hbar}$)
and the bound state ($e^{\pm i n \pi x/a}$) wavefunctions, which can
be difficult to reproduce in a purely numerical evaluation, and it
does so in a way that is valid for arbitrarily large values of
$p_0$, where the integrand would be highly oscillatory, as well as 
being valid even for small values of $p_0$ (or $n_0$.)
Finally, Eqn.~(\ref{approximate_expansion}) 
nicely illustrates how $\psi_{G}(x,0)$ 
and the $u_n(x)$ must not only have an appropriate overlap in position
space, but also must have an appropriate phase relationship
between their oscillatory terms. This phase connection leads 
to the $\exp(-b^2(p_0 \pm n \pi \hbar /a)^2/2 \hbar^2)$ terms, 
which can be understood from a complementary overlap in momentum space.

With this approximation ($x_0, L-x_0 >> b$), the normalization
and energy expectation value conditions
\begin{subequations}
\begin{eqnarray}
\sum_{n=1}^{\infty} |a_n|^2 &=& 1 \\
\sum_{n=1}^{\infty} |a_n|^2\,E_n&=&
\langle \hat{E} \rangle = 
\frac{1}{2m}\Bigl( p_0^2 + \frac{\hbar^2}{2b^2}\Bigr)
\end{eqnarray}
\end{subequations}
can be satisfied to arbitrary accuracy. For the illustrative numerical
results presented here, we use the nominal values
\begin{equation}
L=1, 
\qquad
\hbar=1,
\qquad
\mbox{and}
\qquad
2m =1
\label{physical_parameters}
\end{equation}
and we will often denote $p_0 = n_0 \pi \hbar /L$ to define the
central value of $n_0$ used in the eigenstate expansion, which in the
large $n$ limit will also give
\begin{equation}
\frac{1}{\alpha \sqrt{2}} = 
\Delta p_0 = \frac{\Delta n \pi \hbar}{L}
\qquad
\quad
\mbox{so that}
\qquad
\quad
\Delta n = \frac{L}{(\alpha \hbar) \pi \sqrt{2}}
= \frac{L}{\Delta x_0 2\pi}
\, . 
\label{infinite_delta_n}
\end{equation}
This is useful as it can provide an estimate of how many eigenstates
must be included in a given expansion. In the limit that 
$n_0 >> \Delta n >> 1$, we can use this expression (relating
$b \propto \Delta x_0 \propto L/\Delta n$)
in Eqn.~(\ref{approximate_expansion}),  which then reduces to the standardly
used Gaussian expression coefficients in Eqn.~(\ref{gaussian_components}).

For most cases we consider in this section, we will also use
\begin{equation}
\alpha = \frac{1}{10\sqrt{2}}
\quad
\mbox{so that}
\quad
\Delta x_0 = 0.05 << L
\quad
,
\quad
\Delta p_0 = 10
\quad
\mbox{and}
\qquad
\Delta n = \frac{10}{\pi} \approx 3
\, .
\label{infinite_initial_choices}
\end{equation}
This choice of parameters also implies that the other relevant
time in the problem, the spreading time, $t_0$, defined by
\begin{equation}
t_0 
= m\hbar \alpha^2 
= \frac{m\hbar}{2 \Delta p_0^2}
= \frac{2m \Delta x_0^2}{\hbar}
\, , 
\qquad
\mbox{gives}
\qquad
\frac{T_{rev}}{t_0} = \frac{2}{\pi} \left(\frac{L}{\Delta x_0}\right)^2
\end{equation}
and $T_{rev}/t_0 >> 1$ for any wave packet sufficiently localized
to be contained in the well.
While we will typically focus on wave packets characterized by $n_0>>1$,
which is what is typically obtained in experimental realizations exhibiting
wave packet revivals, we start for simplicity with $p_0=0$ packets.

\subsubsection{Zero-momentum wave packets}
\label{subsec:zero_momentum}

We can make immediate use of Eq.~(\ref{approximate_expansion}) 
by considering zero
momentum ($p_0 = 0$) wave packets; this case corresponds to placing
a particle  `at rest' inside the infinite well potential. 
 For such cases, the only natural periodicity in the
problem is the revival time in Eq.~(\ref{infinite_well_revival}), 
because there is no corresponding classical periodic motion. 
In this special case, the expression for the $a_n$ in 
Eq.~(\ref{approximate_expansion}) simplifies even further to
\begin{equation}
a_n = \sqrt{\frac{4 b \pi}{L \sqrt{\pi}}} 
\,
e^{-b^2 n^2 \pi^2/L^2}
\sin\left(\frac{n \pi x_0}{L}\right)
\end{equation}
which shows that for several special values of 
$x_0$ in the well,
a number of the expansion coefficients will vanish for obvious
symmetry reasons. 

For example, for $x_0/L = 1/2$, all of the even ($n=2,4,6,\ldots$)
 coefficients are zero and the only non-vanishing terms in the expansion 
are the odd ones ($n = 2k+1$) which have energies of the form
\begin{equation}
E_{n} = \frac{\hbar^2 \pi^2}{2m L^2} (2k+1)^2
= E_{0} (4k^2 + 4k +1)
= E_0 + 8 E_0 \left[\frac{k(k+1)}{2} \right]
\,.
\label{zero_momentum_middle}
\end{equation}
The first term in Eq.~(\ref{zero_momentum_middle}) 
contributes only to the same overall phase 
of the time-dependence of each term. The second term is of the form
$8E_0$ times an integer and leads to revival times that are 8 times
{\it shorter} than the standard $T_{rev} = 2\pi \hbar/E_0$ in
Eq.~(\ref{infinite_well_revival}). We illustrate this behavior in 
Fig.~\ref{fig:zero_case}(a) where we also note that moving 
slightly away from $x_0/L = 1/2$ (as in Fig.~\ref{fig:zero_case}(b))
removes these exact sub-revivals,
while the required full revival is still present. (This was observed
in Ref.~\cite{1d_fractional}
for any even parity eigenstate in the infinite well.)

Similarly, for the cases of $x_0/L = 1/3$, 2/3, the $a_n$ with $n=3k$ 
vanish, leading to special exact revivals at multiples of $T_{rev}/3$
for these two initial locations and this behavior is also
shown in Fig.~\ref{fig:zero_case}(c) where we find exact revivals at these
shorter time intervals.
For $x_0/L = 0.8$ (still far enough away from the walls to be considered
reliably in this approximation), we also notice large partial (but
not exact) revivals at $0.4T_{\rm rev}$ and $0.6T_{\rm rev}$ for
similar reasons (because $\sin(4n\pi/5)$ vanishes for $n=5k$.)
(We note that it is also possible to construct
odd-parity wave packets that have different patterns of special
revivals at other locations in the well.) In the same plots, we also
indicate the values of the $a_n$ expansion coefficients and note the
patterns of vanishing $a_n$ values for the cases of special symmetry.
For these $p_0=0$ states, the distribution of $n$-values is far from
Gaussian.

We are more interested in semi-classical wave packets with $p_0 \neq 0$
and we next show, in Fig.~\ref{fig:add_momentum}, the effect of 
`turning on' momentum values for an $x_0 = L/2$ wave packet. 
For $p_0 = 0$ (top line, (a)), 
we have the special pattern of exact revivals at 
multiples of $T_{rev}/8$ noted above, due to the vanishing of
the even expansion coefficients (shown in the corresponding 
$|a_n|^2$ versus $n$ plot in the right column). For a small, 
non-zero value ($p_0 = 3\pi \approx \Delta p_0$, 
Fig.~\ref{fig:add_momentum}(b)), 
only the exact revival at $T_{\rm rev}$ remains,
because the even expansion coefficients are no longer forced to
vanish. The autocorrelation function decreases somewhat more
rapidly from its initial value than in the $p_0=0$ case,
because the particle is now slowly moving away from its initial
position, in addition to spreading out.

For still larger values of momentum, such as $p_{0} = 40\pi$ 
(Fig.~\ref{fig:add_momentum}(c)), 
we see obvious evidence for the classical periodicity and
the first appearance of fractional revivals
\cite{1d_fractional} at rational fraction multiples of
$T_{\rm rev}$, as in Sec.~\ref{subsec:revivals_and_fractional}. 
The corresponding $a_n$
now exhibit a more obvious Gaussian shape, with a spread, $\Delta
n$, which closely approximates the result in 
Eqn.~(\ref{infinite_delta_n}).  
For even larger momentum values (see the $p_{0} = 400\pi$ case
in Fig.~\ref{fig:add_momentum}(d), for example), 
the classical period becomes much shorter
than any obvious fractional revival time scale, and the shape of
the expansion coefficient distribution is unchanged (same $\Delta
n$), but simply shifted to higher values of $n$,  as expected.

In what follows, we will typically use this last case, $n_0 = 400$,
in our visualizations. This implies that the appropriate relative
time scales are
\begin{equation}
\frac{T_{rev}}{T_{cl}} = 2n_0 = 800
\qquad
\quad
\mbox{and}
\qquad
\quad
\frac{t_0}{T_{cl}} = n_0 \pi \left(\frac{\Delta x_0}{L}\right)^2
= \pi
\end{equation}
so that much of the fractional revival structure is obvious, while a
number of classical periods are present before significant spreading
occurs.

\subsubsection{Short-term, quasi-classical propagation}
\label{subsec:infinite_classical}

We expect the short-term propagation of wave packets in the semi-classical 
limit we study to share many properties with both the classical system, 
as well as with the quantum-mechanical free-particle in 
Sec.~\ref{subsec:free_particle}. For
example, well before the initial wave packet nears one of the infinite
wall barriers, we expect the wave packet to propagate with time-dependence
as in Eqn.~(\ref{gaussian_free_particle_position}),
 and with a decreasing auto-correlation function
(Eqn.~(\ref{free_particle_autocorrelation})) described by 
\begin{equation}
|A_{free}(t)|^2 
=
\frac{1}{\sqrt{1+(t/2t_0)^2}}
\exp\left[-2\alpha^2 p_0^2 \frac{(t/2t_0)^2}{(1+(t/2t_0)^2)}\right]
\, .
\label{infinite_free_auto}
\end{equation}
The quasi-periodic wave packet in this confining potential must return
to something like its original state at $t = T_{cl}$, so we expect also
suspect that the behavior near that point will  be approximately described 
by $A_{free}(T_{cl}-t)$. 
For later convenience, we will also find it useful to single out
the dispersive pre-factor above as the `envelope' and define
\begin{equation}
|A_{env}(t)|^2
= 
\frac{1}{\sqrt{1+(t/2t_0)^2}}
\,.
\label{infinite_free_dispersion}
\end{equation}
To see to what extent the free-particle autocorrelation function is
relevant, we plot in Fig.~\ref{fig:new_first} $|A(t)|^2$ over the first
classical period, using Eqn.~(\ref{gaussian_free_particle_position}) 
for the Gaussian wave packet, and increasing numbers ($N$) of eigenstates 
used in the expansion (solid curves.) 
For comparison, we also plot $|A_{free}(t)|^2$ (dashed) and 
$|A_{free}(T_{cl}-t)|^2$
(dot-dashed) and find that they are indeed good representations of the 
numerically evaluated result for the Gaussian packet over much of the
early and late parts of the classical period. The agreement improves
as increasing numbers of states used in the expansion can more 
approximately reproduce the exponential suppression predicted by
Eqn.~(\ref{infinite_free_auto}). We also plot (as diamonds) 
values of $|A(t)|^2$ from the generic expression given by Nauenberg
\cite{nauenberg} for the short-term quasi-classical behavior,
discussed in Appendix~\ref{appendix:time_scales},  and note that especially
for this case where there are no higher time scales, it is an excellent
approximation. 

Over longer  times, we know that the wave packet will spread significantly
and that $A(t),\overline{A}(t)$ will not be exactly periodic, 
but will decrease due to the dispersion, presumably described by 
Eqn~(\ref{infinite_free_dispersion}). To confirm this, we plot both
$|A(t)|^2$ and $|\overline{A}(t)|^2$ over the first $15$ classical periods
(corresponding to almost $15T_{CL}/t_0 = 15/\pi \approx 5$ spreading times)
in Fig.~\ref{fig:second}, along with the `envelope' function in 
Eqn.~(\ref{infinite_free_dispersion}) and note that the initial decrease
in the maximum values of the autocorrelation and anti-correlation functions
is indeed well described by this prescription.

Continuing in this vein, we next examine, in Fig.~\ref{fig:approach}, 
both $|A(t)|^2$ and $|\overline{A}(t)|^2$,
now evaluated at integral multiples of $T_{cl}$ over the first $150$ classical
periods. In this case we note that the dispersive prediction of
Eqn.~(\ref{infinite_free_dispersion}) is a good approximation for over
$40$ periods, while the anti-correlation function (evaluated at $t=nT_{cl}$) 
slowly grows from its exponentially suppressed value at $t=0$ and eventually
crosses the $|A(t)|$ curve. For comparison, we have plotted (horizontal
dotted line) the value of $|A(t)|^2$ which might be expected if the
wave packet were an incoherent sum of eigenstates, with no correlations,
given by
\begin{equation}
|A(t)|^2 \Longrightarrow
|A_{inc}|^2 
\equiv 
\sum_{n} |a_n|^4 
= \frac{1}{\Delta n 2\sqrt{\pi}}
\approx 0.089
\qquad (\mbox{for $\Delta n = 10/\pi$)}
\, .
\label{infinite_incoherent}
\end{equation}
The relationship between $A(t)$ and $\overline{A}(t)$ at some of the first
few fractional revival times is also obvious

\subsubsection{Revivals, fractional revivals, and mirror revivals}
\label{subsec:infinite_revivals}

The longer term structure of both $A(t)$ and $\overline{A}(t)$,
showing evidence for 
revivals, fractional revivals, and mirror revivals,  is illustrated
in Fig.~\ref{fig:all_autos}; only the interval $(0,T_{rev}/2)$ is shown
as the plot is symmetric about $T_{rev}/2$ for this system.
Many of the locations of
possible low-lying fractional revivals are shown as vertical dashed
lines, as well as the incoherent value, $|A_{inc}|$, from
Eqn.~(\ref{infinite_incoherent}). We can then examine the detailed
behavior of $A(t),\overline{A}(t)$ near various specific fractional
revival times. 

We begin, in Fig.~\ref{fig:frac_half}, 
with $t=T_{rev}/2$, the $p/q = 1/2$ (or mirror or anti-revival) 
time,  where we know 
that the wave packet will reform (Eqn.~(\ref{infinite_mirror_position})) 
on the opposite side of the well, with momentum in the opposite direction  
(Eqn.~(\ref{infinite_mirror_momentum}).) The upper plots 
in Fig.~\ref{fig:frac_half} show the
behavior of $|A(nT_{cl})|^2$ over the entire revival time (upper left)
evaluated at integral values of $T_{cl}$,
with the mirror revival indicated by the arrow, 
as well as the typical behavior of both $|A(t)|^2$ (solid)
and $|\overline{A}(t)|^2$ (dashed) near $T_{rev}$ (or $t=0$ as well) 
in the upper right corner.
The lower right shows the autocorrelation and anti-correlation functions
for one classical period on either side of $T_{rev}/2$, showing how the
anti-correlation function is unity at $T_{rev}/2$, with $A(t)$ being
$T_{cl}/2$ out of phase with it, as described very generally in 
Eqn.~(\ref{one_half_revival_function}).
Finally, the position-space probability density
is plotted (lower left) where the initial wave packet was located
at $x_0/L = 2/3$ (solid), and was given positive momentum (illustrated by the
arrow), while the value at $t=T_{rev}/2$ (dashed) is
indeed centered at the mirror location ($L-x_0$) and a short time later
($\Delta t = T_{cl}/20$, dotted curve) the packet is clearly moving to the
left, consistent with Eqn.~(\ref{infinite_mirror_momentum}).

We next turn our attention, in Fig.~\ref{fig:frac_third},
 to an example of a fractional revival at $T_{rev}/3$. In this case, the
plot of $|\psi(x,t)|^2$ in the lower left shows the pattern of three
`mini' versions (dashed) of the original wave packet (solid), moving in
a highly correlated manner (note the time-development at a time $T_{cl}/20$ 
later, dotted curve, indicated by the arrows.) This is also reflected in the 
plots of
$A(t)$ in the region around $t=T_{rev}/3$ where there is a periodicity of
$T_{cl}/3$ due to the `mini' packets (labeled $a,b,c$) being correlated
with the initial wave form, and the same shorter periodicity is also apparent
in the behavior of $\overline{A}(t)$, as predicted in 
Eqn.~(\ref{one_third_revival_function}).

The case of the $T_{rev}/4$ revival, shown in Fig.~\ref{fig:frac_quarter},
is interesting since in this case the maximal values of 
$|A(t)|$, $|\overline{A}(t)|$ near this fractional revival are identical 
($1/2$) and they are both in phase (see the lower right of
Fig.~\ref{fig:frac_quarter}.) Once again, the two 'mini' packets
are highly correlated in position-space (one packet reforming at $x_0$,
the other at $L-x_0$ as befits a mirror revival) and in momentum-space
(with their momenta in opposite directions) just as in 
Eqn.~(\ref{one_quarter_revival_function}).

Finally, if we look at many other times at random which are not near
a fractional revival, as in Fig.~\ref{fig:frac_collapse}, we see much
smaller values of $|A(t)|,|\overline{A}(t)|$, typically consistent with the
incoherent limit of Eqn.~(\ref{infinite_incoherent}), and position-space
probability densities (as in the lower left of Fig.~\ref{fig:frac_collapse})
which exhibit far less obvious structure, being much more consistent with
rapid oscillations about a uniform or `flat' classical probability 
distribution, namely $P(x) = 1/L$ (denoted here by
the two horizontal arrows.) This is typical of a collapsed phase where the
wave packet is described more by an incoherent sum of energy eigenstates,
with the probability density averaging about a semi-classical value,
as in Eqn.~(\ref{collapsed_wave_function}), which in this case is the 
trivial constant value.

Since the phenomena of wave packet revivals (and fractional revivals) 
is first and foremost about
the important phase relationships between the energy eigenstates, as well
as between the individual 'mini'-packets, it is instructive to examine both 
the real and imaginary parts of $A(t)$, both over the long term revival
time scales, as well as near fractional revivals or during the collapsed
phase.

 We begin by examining in Fig.~\ref{fig:newflower} both the
real and imaginary parts of $A(t)$ as well as a parametric plot 
(in an Argand diagram) over half a revival time. The plots show
$A(nT_{cl})$ sampled at multiples  of the classical period (just as in 
Figs.~\ref{fig:frac_half} - \ref{fig:frac_collapse}), illustrating the
intricate phase relationship present. The dashed circle corresponds to
the value of $|A|^2 = 1/2$ and we indicate the location of a $1/4$
fractional revival showing how the autocorrelation function `lingers'
tangentially  during the fractional revival period. Close-up versions
of the fractional revivals at $T_{rev}/4$ and $T_{rev}/3$ are shown
(in the same format) in Figs.~\ref{fig:quarterflower} and 
\ref{fig:thirdflower} respectively, where the obvious phase correlations
between the 'mini'-packets are clearly present. For the case of the
$T_{rev}/4$ fractional revival, the approximate wavefunction in 
Eqn.~(\ref{one_quarter_revival_function})
implies that the autocorrelation function near that time can be 
generally written in the form
\begin{equation}
A(t\approx T_{rev}/4) =
\frac{1}{\sqrt{2}}
\left[e^{i\pi/4} A_{cl}(t) + e^{-i\pi/4} A_{cl}(t+T_{cl}/2)
\right]
\label{quarter_argand_correlation}
\end{equation}
where we define
\begin{equation}
A_{cl}(t) \equiv \int_{-\infty}^{+\infty} \psi^*_{cl}(x,t)\,\psi_{cl}(x,0)
\,dx
\end{equation}
and we see that this form is 
consistent with the visualization in Fig.~\ref{fig:quarterflower}.

For the $T_{rev}/3$ case in Fig.~\ref{fig:thirdflower}, the 
wavefunction in Eqn.~(\ref{one_third_revival_function}) 
can be used to write
\begin{equation}
A(t\approx T_{rev}/3)
= \left(iA_{cl}(t) + e^{-2\pi i/3}
\left[
A_{cl}(t+T_{cl}/3) 
+
A_{cl}(t+2T_{cl}/3) 
\right]
\right)/\sqrt{3}
\label{third_argand_correlation}
\end{equation}
which describes the magnitude/phase relationships seen in 
Fig.~\ref{fig:thirdflower}. 

Finally, the behavior of these quantities during a typical time,
$T^{*}$, in the collapsed phase is shown in Fig.~\ref{fig:randomflower}, 
where in this case  the dotted circle corresponds to the incoherent value 
$|A_{inc}|^2 = \sum |a_n|^4$.

\subsubsection{Expectation value analysis}
\label{subsec:infinite_expectation}

While an analysis of the short-term quasi-classical and long-term revival
structure of wave packets using autocorrelation function methods
is often the most directly comparable to important experimental observables, 
the visualization of the time-dependence of quantum wave packets through 
their expectation values, both in position- and momentum-space, can 
also be a valuable tool \cite{observables}
for understanding many of the effects arising in the quantum
mechanical time evolution of wave functions. For example, the time-evolution
of the uncertainty principle product, $\Delta r \cdot \Delta p_r$,
has been used \cite{yeazell_2}, \cite{yeazell_fractional} to distinguish 
between regimes of classical versus non-classical behavior in Rydberg atoms.

In this section, using the
same parameter values as in earlier analyses, we examine the expectation
values, $\langle x \rangle_t$ and $\langle p \rangle_t$, 
and uncertainties, $\Delta x_t$ and $\Delta p_t$, for Gaussian wave
packets in the infinite well \cite{robinett_infinite_well}, 
both for short times (several classical
periods) and over the entire revival time. 

Once again, comparisons to both the
classical limits and the results of the free-particle wave packet in 
Sec.~\ref{subsec:free_particle} 
can be useful. For example, the time-dependence of a classical
particle in the infinite well potential, starting with velocity $+v_0$,
at the center of the well, is shown in Fig.~\ref{fig:x_p_classical}, with 
cusps in $x(t)$ at the `bounces' at the infinite walls, and a discontinuous
change in momentum (from $\pm mv_0$ to $\mp mv_0$) at the same points.
We might expect somewhat similar behavior, softened by quantum effects, at
least at early times for the quantum problem. For comparison, we recall
that the free-particle
Gaussian wave packet in Eqns.~(\ref{gaussian_free_particle_momentum}) 
and (\ref{gaussian_free_particle_position}) is characterized by
\begin{subequations}
\begin{eqnarray}
\langle x \rangle _t = p_0t/m + x_0
\qquad
& 
,
& 
\qquad
\Delta x_t = \Delta x_0 \sqrt{1+(t/t_0)^2}
\label{infinite_x_quantities}
\\
\langle p \rangle_t = p_0
\qquad
&
,
&
\qquad
\Delta p_t = \Delta p_0
\, .
\label{infinite_p_quantities}
\end{eqnarray}
\end{subequations}

The short-term, quasi-classical time-development can be seen 
in Fig.~\ref{fig:x_short} for the position-space variables and
in Fig.~\ref{fig:p_short} for the momentum-space quantities,
over the first 10 classical periods (approximately 3 spreading times.)
The classical looking behavior of $\langle x \rangle_t$ is apparent,
but with the maximal values near odd integral multiples of
$T_{cl}/2$ (corresponding to the `hits' on the walls) becoming smaller
and smaller, as the wave packet spreads, becomes wider and finds it
increasingly harder to get near the wall. The
results for $\Delta x_t$ are also consistent with expectations, namely
there are small `dips' in $\Delta x_t$ at $t = (2n+1)T_{cl}/2$ arising
from the `compression' of the packet 
\cite{bounce_anatomy}
as it strikes the wall, superimposed
on a uniform increase (the dotted line) consistent with the free-particle
spreading in Eqn.~(\ref{infinite_x_quantities}).

In our exemplary cases, the initial wave packet is characterized by positive 
momentum $p_0 = 400\pi \approx 1260$ and we see distinct evidence of 
`flips' between $\pm p_0$, at times consistent with the classical
`bounces' in the $\langle p \rangle_t$ plot at the 
top of Fig.~\ref{fig:p_short}. The initial momentum spread
from Eqn.~(\ref{infinite_initial_choices}) is $\Delta p_0 = 10$
and we expect this to stay roughly constant between successive collisions
with the wall. During the collision times, however, the spread arising from
the momentum-space probability density is dominated not by the intrinsic
spread of a single peak, but rather by the `distance' between the two
peaks at $p = \pm p_0$. In the limit where $P(p)$ can be approximated by
two highly peaked features at $\pm p_0$, namely
\begin{equation}
P(p) = \frac{1}{2}\left[\delta(p-p_0) + \delta(p+p_0)\right]
\end{equation} 
the corresponding spread in momentum is actually closer to 
$\Delta p_{cl} = p_0$,  and this type of behavior is also evident from 
Fig.~\ref{fig:p_short}.

For the autocorrelation function analysis, we found that the approach
to the collapsed state could be discussed in terms of the decrease of
$|A(t)|$ from its highly correlated value of unity near classical periods
to something closer to the incoherent value of $|A_{inc}|$ given by
Eqn.~(\ref{infinite_incoherent}). In more physical terms, we expect 
the increasingly wide quantum wave packet to approach a collapsed
state which is uniformly spread over the entire well, 
consistent with the classical probability density $P_{cl}(x) = 1/L$. 
For that distribution, we have
position spread is given by 
\begin{equation}
\langle x \rangle_{cl} = \frac{L}{2}
\qquad
,
\qquad
\langle x^2 \rangle_{cl} = \frac{L}{3}
\qquad
\mbox{so that}
\qquad
\Delta x_{cl} = \frac{L}{\sqrt{12}} \approx 0.288L
\, . 
\label{infinite_well_flat_values}
\end{equation}

To visualize this approach to the collapsed state, we plot in 
Fig.~\ref{fig:delta_x_approach}, the time-dependent position spread,
$\Delta x_t$ corresponding to three different values of $\Delta x_0$
and note that they all `saturate' at $\Delta x_{flat} = L/\sqrt{12}$,
at different times, due to the difference in spreading times 
($t_0 = 2m(\Delta x_0)^2/\hbar$). In each case, the bound state
values initially follow the free-particle prediction of 
Eqn.~(\ref{infinite_x_quantities}) for short times. The time scale
for this collapse can be estimated by equating
\begin{equation}
\frac{L}{\sqrt{12}} \equiv  \Delta x_{cl} = \Delta x_{(t=T_{coll})}
= \Delta x_0 \sqrt{1 + (T_{coll}/t_0)^2}
\end{equation}
giving
\begin{equation}
T_{coll} = t_0 \left(\frac{L}{\sqrt{12} \Delta x_0}\right)
= \frac{1}{\sqrt{3}} \left(\frac{mL \Delta x_0}{\hbar}\right)
= \sqrt{\frac{\pi T_{rev}t_0}{24}}
\label{original_infinite_well_collapse_time}
\end{equation}
which defines another time scale, intermediate between $t_0$ and
$T_{rev}$.

The same type
of approach to the values appropriate for a collapsed state can also
be seen (in Fig.~\ref{fig:all_approach}) 
for $\langle x \rangle_t$ (approaching $\langle x\rangle_{cl} = L/2$),
for $\langle p \rangle_{cl} = 0$ (equal admixtures of left and right 
moving states), 
and for $\Delta p_{cl} = + p_0$ (same reason) over the
same time scales, and with the same dependence on the spreading time
(via its dependence on the initial width.)

Finally, the behavior of selected average values over the entire
revival time, illustrating both the revival and mirror revival times,
as well as 'close up' views at intermediate times, are shown
in Figs.~\ref{fig:x_long}, \ref{fig:delta_x_long}, and
\ref{fig:p_long}. The plots are shown for times given by
$t = (n+1/8)T_{cl}$ (solid) and
$(n+5/8)T_{cl}$ (dashed) to help distinguish standard versus mirror
revivals. The magnitude of excursions from the classical value
of $\langle x \rangle_{cl} = L/2$ for the collapsed state
are shown on an expanded scale in Fig.~\ref{fig:x_long}, which includes
a fractional revival at $t = T_{rev}/6 = 133T_{CL}$ and 
$t = T_{rev}/5 =160T_{cl}$. 

For the time-dependent spread in position,
$\Delta x_t$, in Fig.~\ref{fig:delta_x_long}, 
we see that $\Delta x_t$ returns to the
initial value, $\Delta x_0$, at $T_{rev}$ (full revival) and
$T_{rev}/2$ ($p/q = 1/2$ or mirror revival) as expected, but also 
at $T_{rev}/4$
and $3T_{rev}/4$. These last two instances are special cases and are 
due to the choice of $x_0 = L/2$ for the initial wave packet 
(compared to $x_0 = 2L/3$ as shown
in Fig.~\ref{fig:frac_quarter}) where the two 'mini' packets are superimposed
at the same location (the midpoint of the well), but moving with
opposite momenta. This is consistent with the $\langle p \rangle_t$
and $\Delta p_t$ values at the same times in Fig.~\ref{fig:p_long}.

We note that the observed agreement of a number of expectation values and
spreads with results obtained from the classical probability distributions
(as in Eqn.~(\ref{infinite_well_flat_values})) during times much longer 
than the spreading time, while consistent with the approach to a 
purely `flat' distribution, can also be thought of, during periods of 
fractional revivals,  as being due to the correlated behavior of a 
number of `mini-packets' or 'clone' wave packets, which can give 
similar results \cite{force_paper}.

\subsubsection{Phase-space picture of fractional revivals using
the Wigner function}
\label{subsubsec:phase_space_wigner}

Another useful way to visualize the correlated position- and 
momentum-space structure of wave packets in a phase-space type picture
is by using the Wigner quasi- or pseudo probability density \cite{wigner}
defined by 
\begin{eqnarray}
W(x,p;t)
& \equiv &
\frac{1}{\pi \hbar}
\int_{-\infty}^{+\infty}
\psi^{*}(x+y,t)\,\psi(x-y,t)\,e^{2ipy/\hbar}\,dy \\
& = &
\frac{1}{\pi \hbar}
\int_{-\infty}^{+\infty}
\phi^*(p+q,t)\, \phi(p-q,t)\, e^{-2iqx/\hbar}\,dq
\, .
\end{eqnarray}
(The Wigner distribution for Gaussian wave packet solutions in the
infinite square well has been discussed in 
Ref.~\cite{belloni_doncheski_robinett} to which we
refer the reader for many calculational and visualization details.)
The general properties of the Wigner distribution have been discussed 
in a number of accessible reviews \cite{tatarskii} - \cite{weyl_review},
where, among other features, it is noted that integration of $W(x,p;t)$ 
over one variable or the other is seen to give the correct marginal 
quantum mechanical probability distributions for $x$ and $p$ separately, 
since
\begin{eqnarray}
\int_{-\infty}^{+\infty}
W(x,p;t)\, dp & = & |\psi(x,t)|^2 = P_{QM}(x,t) \\
\int_{-\infty}^{+\infty}
W(x,p;t)\, dx & = & |\phi(p,t)|^2 = P_{QM}(p,t) 
\,.
\end{eqnarray}
The Wigner function is also easily shown to be real, but need not, 
however, be positive definite, hence the name quasi- or 
pseudo-probability density.

The Wigner distribution for a standard time-dependent Gaussian free-particle
solution of the form in Eqns.~(\ref{gaussian_free_particle_momentum}) 
or (\ref{gaussian_free_particle_position}) is easily obtained, 
using standard Gaussian integrals, and is given by
\begin{equation}
W^{(G)}(x,p;t) = \frac{1}{\pi \hbar}
\, 
e^{-\alpha^2 (p-p_0)^2}
\, 
e^{-(x-x_0-pt/m)^2/\beta^2}
\label{free_particle_gaussian_wigner}
\end{equation}
where $\beta \equiv \hbar \alpha$; in this case, the ultra-smooth Gaussian 
form does give a positive $W(x,p;t)$, as has been discussed in the
literature \cite{hudson}.

Another result which will prove useful 
in what follows is the expression for the Wigner function for the case of a 
linear combination of two 1D Gaussians, characterized by different values 
of $x_0$ and $p_0$. For example, if we assume that at some instant of time 
we have
\begin{eqnarray}
\psi^{(A,B)}(x) & = & \gamma \psi_{(G)}(x;x_A,p_A) + 
\delta  \psi_{(G)}(x;x_B,p_B)
 \\
&  = & 
\gamma
\left[
\frac{1}{\sqrt{\beta\sqrt{\pi}}}
e^{-(x-x_A)^2/2\beta^2}e^{ip_A(x-x_A)/\hbar}
\right]
+ 
\delta
\left[
\frac{1}{\sqrt{\beta\sqrt{\pi}}}
e^{-(x-x_B)^2/2\beta^2}e^{ip_B(x-x_B)/\hbar}
\right] \nonumber 
\, ,
\end{eqnarray}
the corresponding Wigner function is given by
\begin{eqnarray}
P_{W}^{(A,B)}
& = &  \frac{1}{\pi \hbar}
\left[ 
|\gamma|^2
e^{-\alpha^2(p-p_A)^2}e^{-(x-x_A)^2/\beta^2}
+ 
|\delta|^2
e^{-\alpha^2(p-p_B)^2}e^{-(x-x_B)^2/\beta^2} \right. 
\label{two_gaussian_wigner}\\
& &
\qquad 
\left.
+ 
2e^{-\alpha^2(p-\overline{p})^2}e^{-(x-\overline{x})^2/\beta^2}
Re\left\{\gamma \delta^* e^{i(x_A p_B - x_B p_A)/\hbar}
\,
e^{-i(x_A-x_B)(p-\overline{p})/\hbar}
\,
e^{i(p_A-p_B)x/\hbar}
\right\}
\right]
\nonumber
\end{eqnarray}
where
\begin{equation}
\overline{x} \equiv 
\frac{x_A+x_B}{2}
\qquad
\mbox{and}
\qquad
\overline{p}
\equiv 
\frac{p_A+p_B}{2}
\, .
\label{average_values}
\end{equation}
In this case, the Wigner function is characterized by two smooth `lumps' 
in phase space, corresponding to the values of $(x_A,p_A)$ and $(x_B,p_B)$ 
of the individual Gaussians, but also by an oscillatory term, centered at a 
point in phase space defined by the average of these values; the
oscillations are either in $x$ (if $p_A \neq p_B$), 
$p$ (if $x_A \neq x_B$) or both. (See Ref.~\cite{ballentine} for a similar
result for the case when $p_A = p_B = 0$; see also Ref.~\cite{schleich_book} 
for a related expression  involving coherent states.)

For a general, time-dependent wave packet constructed from energy 
eigenstates of the form in Eqn.~(\ref{general_expansion_in_eigenstates}), 
the Wigner distribution is given by 
\begin{eqnarray}
W^{(\psi)}(x,p;t)
& \equiv & 
\frac{1}{\pi \hbar}
\int_{-\infty}^{+\infty}
\psi^{*}(x+y,t)\,\psi(x-y,t)\,e^{2ipy/\hbar}\,dy 
\nonumber \\
& = &
\sum_{m=1}^{\infty}
\sum_{n=1}^{\infty}
[a_m]^*a_n \, e^{i(E_m-E_n)t/\hbar}\,
\left[
\frac{1}{\pi \hbar}
\int_{-\infty}^{+\infty}
u_m(x+y)\,u_n(x-y)\,e^{2ipy/\hbar}\,dy \right] 
\nonumber \\
& \equiv & 
\sum_{m=1}^{\infty}
\sum_{n=1}^{\infty}
[a_m]^*a_n \, e^{i(E_m-E_n)t/\hbar}\,
P_{W}^{(m,n)}(x,p) 
\label{time_dependent_wigner}
\end{eqnarray}
where, in general, we must calculate both diagonal ($m=n$) and 
off-diagonal ($m\neq n$) terms of the form 
\begin{equation}
W^{(m,n)}(x,p) 
= \frac{1}{\pi \hbar}
\int_{-\infty}^{+\infty}
u_m(x+y)\,u_n(x-y)\,e^{2ipy/\hbar}\,dy
\label{off_diagonal}
\end{equation}
where we have assumed that the individual position-space bound state
eigenfunctions, $u_n(x)$, can be made purely real. The off-diagonal 
Wigner terms are, however, not real, but do satisfy 
\begin{equation}
\left[W^{(m,n)}(x,p)\right]^* 
=
W^{(n,m)}(x,p)
\, . 
\end{equation}

For the infinite square well, the diagonal ($m=n$) terms
are given by 
\begin{equation}
W^{(n,n)}(x,p) = \frac{1}{\pi \hbar}
\int u_{n}(x+y)\,u_{n}(x-y)\,e^{2ipy/\hbar}\,dy
\label{wigner_eigenstate}
\end{equation}
and the limits of integration are determined by the restriction that the 
$u_{n}(x \pm y)$ in Eqn.~(\ref{1d_eigenfunctions}) are non-vanishing 
only in the range $(0,L)$ and they must therefore simultaneously satisfy 
the requirements
\begin{equation}
0 \leq x+y \leq L
\qquad
\mbox{and}
\qquad
0 \leq x-y \leq L
\, .
\end{equation}
This leads to upper and lower bounds for the integral over $y$ in
Eqn.~(\ref{wigner_eigenstate}) which depend on $x$ via
\begin{eqnarray}
-x \leq y \leq +x
& \qquad & 
\mbox{for $0\leq x \leq L/2$} \\
-(L-x) \leq y \leq +(L-x)
& \qquad & 
\mbox{for $L/2\leq x \leq L$}
\, . 
\label{half_intervals}
\end{eqnarray}
Thus, over the left-half of the allowed $x$ interval, $(0,L/2)$, 
we have
\begin{eqnarray}
W^{(n)}(x,p) 
& = & 
\frac{1}{\pi \hbar} \int_{-x}^{+x} 
\left[\sqrt{\frac{2}{L}}\sin\left(\frac{n\pi(x+y)}{L}\right)\right]
\, 
\left[\sqrt{\frac{2}{L}}\sin\left(\frac{n\pi(x-y)}{L}\right)\right]
\, e^{2ipy/\hbar}\, dy  \nonumber \\
& = & 
\left(\frac{2}{\pi \hbar L}\right)
\left\{
\frac{\sin[2(p/\hbar - n\pi/L)x]}{4(p/\hbar - n\pi/L)}
+
\frac{\sin[2(p/\hbar + n\pi/L)x]}{4(p/\hbar + n\pi/L)}
\right. 
\label{isw_wigner_distribution} \\
& &
\qquad \qquad \qquad \qquad \qquad
\left. 
-
\cos\left(\frac{2n\pi x}{L}\right)
\,
\frac{\sin(2px/\hbar)}{(2p/\hbar)}
\right\}
\nonumber
\, ,
\end{eqnarray}
while over the right-half of the interval, $(L/2,L)$, 
one simply makes the replacement $x \rightarrow L-x$. 
This form has been derived before \cite{casas} - \cite{weyl_review}, 
although in at least one reference it is written
in terms of Bessel functions ($j_0(z)$) which somewhat obscures its
simple derivation.

For a general wave packet solution for the infinite square well, we also
require the off-diagonal terms in Eqn.~(\ref{off_diagonal}). Using the
position-space eigenstates in Eqn.~(\ref{1d_eigenfunctions}), over the
interval $(0,L/2)$ we find that
\begin{eqnarray}
W^{(m,n)}(x,p) & = & 
\frac{1}{\pi \hbar} \int_{-x}^{+x} 
\left[\sqrt{\frac{2}{L}}\sin\left(\frac{m\pi(x+y)}{L}\right)\right]
\, 
\left[\sqrt{\frac{2}{L}}\sin\left(\frac{n\pi(x-y)}{L}\right)\right]
\, e^{2ipy/\hbar}\, dy  \nonumber \\
& = &
\frac{1}{\pi \hbar}
\left[
e^{+i(m-n)\pi x/L}\,\frac{\sin[(2p/\hbar + (m+n)\pi/L)x]}{2pL/\hbar + (m+n)\pi}
\right. \nonumber \\ 
& & 
\qquad \qquad
\left. 
+
e^{-i(m-n)\pi x/L}\,\frac{\sin[(2p/\hbar - (m+n)\pi/L)x]}{2pL/\hbar - (m+n)\pi}
\right. \label{off_diagonal_wigner} \\
& & 
\qquad \qquad
\left.
-
e^{+i(m+n)\pi x/L}\,\frac{\sin[(2p/\hbar + (m-n)\pi/L)x]}{2pL/\hbar + (m-n)\pi}
\right. \nonumber \\
& &
\qquad \qquad
\left.
-
e^{-i(m+n)\pi x/L}\,\frac{\sin[(2p/\hbar - (m-n)\pi/L)x]}{2pL/\hbar - (m-n)\pi}
\right] \nonumber 
\end{eqnarray}
and it is easy to check that this result reduces to the expression
in Eqn.~(\ref{isw_wigner_distribution}) when $m=n$. In order to extend this
to the interval $(L/2,L)$,  it is important to note that the substitution
$x \rightarrow L-x$ should be made {\it only} in those terms arising from 
the integration over $dy$, namely, the $\sin[(2p/\hbar \pm (m\pm n)/L)x]$
terms.

Using these expressions, the expansion in Eqn.~(\ref{time_dependent_wigner}),
and the Gaussian expansion coefficients in Eqn.~(\ref{approximate_expansion}) 
for the $a_n$, 
we can evaluate the time-dependent Wigner function  for the infinite well.
As examples, we show in Figs.~\ref{fig:wigner_third} and 
\ref{fig:wigner_quarter},
$W(x,p;t)$ versus $(x,p)$ at two fractional revival times, namely 
$T_{rev}/3$ and $T_{rev}/4$, for direct 
comparison to the position-space probability densities in 
Figs.~\ref{fig:frac_third} and \ref{fig:frac_quarter} for the same values. 
(The same initial values of $x_0$ are used in 
each case, but we use $n_0 = 40$ here for ease of visualization.) 
The results are consistent with the expression in 
Eqn.~(\ref{two_gaussian_wigner}) with smooth isolated Gaussian `lumps' and 
oscillatory cross-terms; for example, in Fig.~\ref{fig:wigner_third},
the cross-term between the two Gaussians which have the same momentum
($-40\pi\hbar/L \approx -126$ in this case, so $p_A = p_B$), but different 
central locations ($x_A \neq x_b$) is seen to be oscillatory only in the 
$p$ variable. These views make clearer the Schr\"{o}dinger-cat like behavior
of the split wave packet at fractional revival times.

\subsubsection{Quantum carpets}
\label{subsec:infinite_quantum_carpets}

A semi-classical visualization of the initial spreading of bound
state wave packets in the infinite well was provided by Born
\cite{born}, \cite{born_2} who examined the diverging classical 
trajectories of particles with differing initial velocities 
(momenta) in $(x,t)$ plots where such classical paths can be identified
with individual `world-lines'. An example of such a plot is shown
in Fig.~\ref{fig:new_kinzel},  which illustrates how the spreading can lead
to an  almost uniform probability density, giving rise to 
calculations such as in Eqn.~(\ref{infinite_well_flat_values}). 

The corresponding quantum mechanical picture of such behavior can be 
obtained by
examination of the quantum mechanical position-space probability density
as a function of time, given by plots of $P(x,t) = |\psi(x,t)|^2$ versus
$(x,t)$.  As examples, we show in Fig.~\ref{fig:first_cycle}, the
probability density over the first classical period, comparing it to
the more interesting structures present over the same time interval
starting at the $T_{rev}/3$ fractional revival, as seen 
in Fig.~\ref{fig:third_cycle}.

To visualize these  over even longer time scales, we note that
because of the periodicity of the system given by the 
revival time, it suffices to plot $P(x,t) = |\psi(x,t)|^2$ over the 
two-dimensional
$(x,t)$ space given by $(0,L)$ and $(0,T_{rev}/2)$, such as 
in Fig.~\ref{fig:quantum_carpets},  which were first produced by
Kinzel \cite{quantum_kinzel}; such plots have come to be known 
as {\it quantum carpets}. 
The rich structure of ridges and canals apparent in such plots,
which are clearly correlated with the spatio-temporal structure of
revivals and fractional revivals, have been discussed in a number of
approaches \cite{lamb}, \cite{quantum_kinzel} - \cite{carpets_wigner},
including making use of a traveling wave decomposition of the 
wavefunction or using Wigner quasi-probability densities.

One useful approach \cite{simple_carpets}, \cite{even_more_carpets}
to the visualization of this pattern formation,
based on an expansion of the wavepacket in traveling waves,
begins by writing the stationary state solutions in the form
\begin{eqnarray}
\psi_{n}(x,t) & = &
\left[\sqrt{\frac{2}{L}} \sin\left(\frac{n\pi x}{L}\right)\right]
e^{-iE_nt/\hbar} \nonumber \\
& = &
\left[\frac{1}{i\sqrt{2L}}\left( e^{in\pi x/L} - e^{-in\pi x/L}
\right)\right]
e^{-2\pi i n^2 t/T_{rev}}
\end{eqnarray}
so that
\begin{equation}
\psi(x,t) = \sum_{n=1}^{\infty} a_n \psi_{n}(x,t)
\end{equation}
for any more general time-dependence state, including localized wave
packets. The probability density can then be written as 
\begin{equation}
P(x,t) 
= \psi^{*}(x,t) \psi(x,t)
= P_{cl}(x,t) + P_{qc}(x,t)
\end{equation}
where one defines 
{\it classical} ($P_{cl}$)
and 
{\it quantum carpet} ($P_{qc}$)
contributions
by 
\begin{eqnarray}
P_{cl}(x,t) = \left( \frac{1}{2L}\right)
\sum_{n,m=1}^{\infty}
a_m^* a_n
\left\{
e^{i(n-m)\pi[x/L + 2(n+m)t/T_{rev}]}
+
e^{-i(n-m)\pi[x/L - 2(n+m)t/T_{rev}]}
\right\} 
\label{classical_contribution}
\\
P_{qc}(x,t) = \left( \frac{-1}{2L}\right)
\sum_{n,m=1}^{\infty}
a_m^* a_n
\left\{
e^{i(n+m)\pi[x/L + 2(n-m)t/T_{rev}]}
+
e^{-i(n+m)\pi[x/L - 2(n-m)t/T_{rev}]}
\right\} 
\, . 
\label{quantum_contribution}
\end{eqnarray}

The contributions to the {\it classical} component 
($P_{cl}$) are dominated by $(x,t)$
locations at integral multiples of $[x/L \pm 2(n+m)t/T_{rev}]$. For
localized wave packets, the $n,m$ values will be peaked near $n_0$
and using the fact that $T_{rev} = 2n_0 T_{cl}$, 
the factors in the exponents can be approximated $[x/L \pm 2t/T_{cl}]$ 
which are therefore similar to those seen in Fig.~\ref{fig:new_kinzel} 
for classical trajectories.
Since $T_{cl} << T_{rev}$, or equivalently, since $n,m>>1$, 
these world-lines, when plotted over the ranges $(0,L)$ and $(0,T_{rev})$,
have very `flat' slopes and, also taking into account their diverging paths, 
soon cover the $(x,t)$ plane almost uniformly, as illustrated  in 
Fig.~\ref{fig:worldline}, providing a `background'.

The {\it quantum carpet} ($P_{qc}$) terms, on the other hand, have dominant 
contributions from terms in the exponentials of the 
form $[x/L \pm 2(n-m)t/T_{rev}]$. The $(n-m)$ terms will only contribute
significantly for $n,m$ values for which 
$|n-m| \leq {\cal O} (\Delta n)$.
This implies that the world-line slopes for
these factors will cover a significant fraction of the entire $(x,t)$
'plane', in a few `bounces', with two typical cases shown in 
Figs.~\ref{fig:worldline} (b) and (c); the interference of such terms with 
the {\it classical} background can then give rise to the observed patterns of 
ridges and canals.

For initial wave packets which are more highly localized (smaller 
$\Delta x_0$, larger $\Delta n$), more obvious features can be
resolved \cite{harter}, similarly to the dependence seen in 
Fig.~\ref{fig:new_three}. The relative `sharpness' of the features
can also be seen to arise from the differing pre-factors in the 
exponentials.
For example, the $(n-m)$ pre-factors in the $P_{cl}$ terms 
will be dominated by small values of $|n-m|$, so that relatively large
changes in the corresponding $[x\pm 2(n+m)t/T_{rev}]$ terms will still
contribute significantly, enhancing the uniformity of the classical
contributions; on the other hand, for the $P_{qc}$ terms, the $(n+m)$ 
pre-factors are large (${\cal O}(2n_0)$), so that small changes in the 
accompanying $[x\pm 2 |n-m|t/T_{rev}]$ factors will rapidly lead to 
cancellations, helping to define the 'contrast' seen in the quantum
carpet images.

It has also been observed that the 2D (in $(x,t)$)
probability  density patterns observed here are similar to what have
been described as `scars' \cite{heller_1} in
spatially two-dimensional ($(x,y)$) quantum mechanical systems. We also 
note that similar techniques have been used to
examine other systems which also exhibit exact revivals \cite{carpets}.

\subsection{Variations on the infinite well}
\label{subsec:infinite_variations}

Given the pedagogical familiarity and simplicity of the infinite
well potential, a number of variations on this system have been studied
and we briefly mention  three of them.

We have noted that the important time scales for wave packet propagation
are determined by the energy eigenvalues through 
Eqns.~(\ref{classical_period}), 
(\ref{quantum_revival}),
and (\ref{quantum_superrevival}), 
so it is clear that systems with identical (isospectral) 
or closely related energy spectra will have similar patterns of
classical periodicity, wave packet collapse, and revival.
For example, pairs of systems defined by superpartner potentials 
which are related by one-dimensional supersymmetry 
\cite{supersymmetry} - \cite{cooper_susy} have energy spectra which are
identical except for the ground state energy and would be expected to
exhibit almost the same pattern of revival behavior, at least 
for $n_0 >> 1$. The construction of superpartner potentials \cite{witten}
starts with the assumption of a 1D Hamiltonian  with a zero-energy ground 
state, namely one such that
\begin{equation}
\hat{H}_{(-)} \psi_{0}(x) = 
\left( 
-\frac{\hbar^2}{2m} \frac{d^2}{dx^2} + V_{(-)}(x) 
\right) \psi_{0}(x) = 0
\, . 
\end{equation}
Since $\psi_{0}(x)$ is assumed known, one can define a superpotential
for the problem by the identification 
\begin{equation}
W(x) \equiv  - \frac{\hbar}{\sqrt{2m}} \left(\frac{\psi_{0}'(x)}{\psi_{0}(x)}
\right)
\, . 
\end{equation}
The pair of systems  defined by the Hamiltonians
\begin{equation}
\hat{H}_{(\pm)}
\equiv 
-\frac{d^2}{dx^2} + V_{(\pm)}(x) 
\equiv
-\frac{d^2}{dx^2} + 
\left\{\left[W(x)\right]^2 \pm \frac{\hbar}{\sqrt{2m}}W'(x)
\right\}
\end{equation}
can then be shown to have the same energy level spectrum, 
with $E_{n}^{(+)} = E_{n+1}^{(-)}$, except that the zero-energy 
ground state of $\hat{H}_{(-)}$ ($E_{0}^{(-)}=0$) has no counterpart 
in $\hat{H}_{(+)}$. The wavefunctions in the two systems are related
by generalized raising and lowering operators.

The standard infinite well potential of Eqn.~(\ref{infinite_well_potential}) 
can be put into the form of $\hat{H}_{(-)}$ by subtracting the appropriate 
zero-point energy to write
\begin{equation}
V_{(-)}(x) = V(x) - \frac{\hbar^2 \pi^2}{2mL^2}
\qquad
\mbox{with}
\qquad
\psi_{0}^{(-)}(x) = \sqrt{\frac{2}{L}}\sin\left(\frac{\pi x}{L}\right)
\end{equation}
with quantized energies given by
\begin{equation}
E_{n}^{(-)} = \frac{\hbar^2 \pi^2}{2mL^2} n(n+2)
\, .
\end{equation}
The superpotential is then given by
\begin{equation}
W(x) = - \frac{\hbar}{\sqrt{2m}} \frac{\pi}{L} 
\cos\left(\frac{\pi x}{L}\right)
\end{equation}
and the superpartner potential of the infinite well potential is given
by
\begin{equation}
V_{(+)}(x) = \frac{\hbar^2 \pi^2}{2mL^2}
\left[2\csc^2\left(\frac{\pi x}{L}\right) -1 \right]
\end{equation}
so that large $n$ wave packets in this potential will have the same
pattern of semi-classical periodicity, and exact revivals, as for the
infinite well. We note that this potential is a special case of
a larger class of P\"{o}schl-Teller potentials  and the authors of 
Ref.~\cite{klauder} have constructed coherent states 
(see also Refs.~\cite{nieto_1} - \cite{nieto_3}) for such
systems, making using of their $SU(1,1)$ symmetries, while the revival 
structure in P\"{o}schl-Teller and Rosen-Morse \cite{carpets} and other
related potentials \cite{torre} have also been considered. Revivals and
fractional revivals in other $SU(1,1)$ symmetric systems have also 
been discussed \cite{banerji}.
 
We note in passing that the same approach can be used with the harmonic
oscillator potential, defining
\begin{equation}
V_{(-)}(x) = \frac{m \omega^2}{2} x^2 - \frac{\hbar \omega}{2}
\end{equation}
but trivially returns the supersymmetric partner potential
\begin{equation}
V_{(+)}(x) = \frac{m \omega^2}{2} x^2 + \frac{\hbar \omega}{2}
\end{equation}
and the raising/lowering operator formalism is the standard one seen in
textbooks. The use of a phenomenological  supersymmetry involving atomic
energy levels,  has been discussed \cite{bluhm_supersymmetry} 
as it relates to the revivals and fractional revivals of Rydberg atoms and 
radial squeezed states.

The finite well has been studied by several groups 
\cite{other_finite_well}, \cite{finite_well}
with an eye towards
providing a more realistic model system, focusing on the appearance of
superrevival time scales since the energy eigenvalue spectrum is no
longer exactly quadratic. Aronstein and Stroud \cite{finite_well_formula}
have provided a useful
description of the bound state energy eigenvalue spectrum which is helpful
for the determination of both the global energy spectrum as well as
local approximations, useful for wave packet construction.
Finally, the addition of a $\delta$-function to the infinite well problem
is a staple of the pedagogical literature \cite{delta_background}, 
and its effects on the revival structure of wave packet development 
has been discussed in Ref.~\cite{dirac_delta}.

\subsection{The quantum bouncer}
\label{subsec:quantum_bouncer}

\subsubsection{Energy eigenfunctions and eigenvalues}
\label{subsubsec:energy_eigenfuctions_and_eigenvalues}

One of the more  familiar systems of classical mechanics, the `bouncer',
has been considered in the context of quantum mechanical wave packet
propagation and revivals. This system, defined by the potential
\begin{equation}
      V(z)  = \left\{ \begin{array}{ll}
               +\infty & \mbox{for $z\leq 0$} \\
               Fz & \mbox{for $z>0$}
                                \end{array}
\right.
\, ,
\end{equation}
corresponds to a particle under the influence of a constant force $F$
(such as gravity), but with an infinite wall at $z=0$ from which the
particle bounces. While this system has long been a popular non-trivial 
example in the pedagogical literature and in collections of problems
in quantum mechanics, advances in experimental techniques
\cite{advances_bouncing}, using reflection from laser-induced evanescent 
waves, have allowed the observation of such `bouncing' atoms 
\cite{bouncing_1}, \cite{bouncing_2} and even the quantum mechanical
interference between different atomic trajectories \cite{bouncing_3}.

The classical periodicity for this system is simply
twice the time a particle takes to fall from its maximum height, $z_0$,
and is given by
\begin{equation}
T_{cl} = 2 \sqrt{\frac{2mz_0}{F}}
\end{equation}
and several groups \cite{milburn_and_chen} - \cite{robinett_bouncer}
have considered different aspects of the time-development of (Gaussian)
quantum mechanical wave packets, initially centered at $z_0$, including 
their subsequent revival behavior. 
(The effect on the bouncer of the application of a periodic external 
driving force, especially on the structure of revivals, has also been 
discussed \cite{driven_gravity}, \cite{saif_driven}.)

The Schr\"{o}dinger equation for this system can be written as
\begin{equation}
- \frac{\hbar^2}{2m} \frac{d^2 \psi(z)}{d z^2} + Fz \psi(z)
= E\psi(z)
\end{equation}
and the change of variables
\begin{equation}
z = \rho y +\sigma
\, , 
\qquad
\mbox{where}
\qquad
\rho = \left(\frac{\hbar^2}{2mF}\right)^{1/3}
\qquad
\mbox{and}
\qquad
\sigma = \frac{E}{F}
\, , 
\end{equation}
reduces it to the familiar Airy equation
\begin{equation}
\frac{d^2 \psi(y)}{d y^2} = y \psi(y)
\, .
\end{equation}
The condition that the wavefunction be well-behaved as $z/y \rightarrow
+\infty$ excludes the divergent $Bi(y)$ solution, while the boundary
condition at the infinite wall, $\psi(z=0) = 0$, implies that
\begin{equation}
Ai\left(-\frac{\sigma}{\rho}\right) = 0 =
Ai(-y_n)
\end{equation}
and the quantized energy eigenvalues are determined by the zeros ($y_n$) 
of the well-behaved $Ai(y)$ solution. (This problem is discussed in a
variety of textbooks, but the first analysis using the Schr\"{o}dinger 
equation seems to have been done by Breit \cite{breit_bouncer}.)

The energy spectrum then has the form
\begin{equation}
E_n = y_n \left(\frac{\hbar^2 F^2}{2m}\right)^{1/3}
\, . 
\end{equation}
The Airy function zeros ($y_n$) 
can be derived from standard handbook results, 
but the energies in the large $n$ limit relevant for wave
packet construction can also be easily obtained from a WKB approximation.
The WKB quantization condition is
\begin{equation}
\int_{0}^{z_{0}} \sqrt{2m(E_n -Fz)}\,dz 
= \left( n + \frac{1}{2} + \frac{1}{4}\right) \hbar \pi
\end{equation}
where we use the appropriate matching coefficients, 
$C_L = 1/2$ for the infinite wall, and
$C_R = 1/4$ for the linear barrier. The resulting energies are given by
\begin{equation}
E_n = \left(\frac{3\pi}{2}(n+3/4)\right)^{2/3}
\left(\frac{\hbar^2 F^2}{2m}\right)^{1/3}
\qquad
\quad
\mbox{where $n=0,1,2,...$,}
\end{equation}
which agrees with the Airy function analysis \cite{quantum_bouncer} to 
this order in $n$. 
(We note that experimental evidence for the quantized nature of the 
bound state spectrum of neutrons in the Earth's gravitational field
\cite{neutron_bound_states} has recently appeared, motivating additional 
studies of wave packet dynamics \cite{neutron_oscillations} of such systems.)
This expression can also be used to associate the central
value of $n_0$ in the eigenstate expansion with the initial position, $z_0$,
via
\begin{equation}
Fz_0 = E_{initial} = \left(\frac{3\pi}{2}(n_0+3/4)\right)^{2/3}
\left(\frac{\hbar^2 F^2}{2m}\right)^{1/3}
\label{identification}
\end{equation}
which will prove useful.

\subsubsection{Classical period and revival time}
\label{subsubsec:classical_period_and_revival_time}

The classical periodicity of the system, as encoded in  the energy
eigenvalue spectrum, is given by
\begin{equation}
T_{cl} = \frac{2\pi \hbar}{|E'(n_0)|}
=
\frac{3\pi \hbar (n_0+3/4)}{E(n_0)}
= 
2 \sqrt{\frac{2mz_0}{F}}
\end{equation}
(where we have used the $z_0/n_0$ identification in 
Eqn.~(\ref{identification})) and is consistent with the classical result.
As an example of the semi-classical, short-term 
time-development of a Gaussian
wave packet in this potential, we show in Fig.~\ref{fig:bouncer_classical} 
the position- and momentum-space probability densities for such a packet with
\begin{equation} 
z_0 = 25
\qquad
\mbox{and}
\qquad
\Delta z_0 = 1
\label{other_bouncer_parameters}
\end{equation}
using the parameters
\begin{equation}
\hbar = 1
\quad
,
\quad
2m = 1
\quad
, 
\quad
F = 1
\label{bouncer_parameters}
\end{equation}
so that $T_{cl} = 10$. 
On the left of Fig.~\ref{fig:bouncer_classical}, the position-space
probability is initially seen to spread in a manner which is numerically 
consistent with Eqn.~(\ref{accelerating_values}), 
while the calculated position
value $\langle z \rangle_t$ (solid curve) agrees well with the classical
expectation for the trajectory (dashed) except, of course, for the 
cusp at the 'bounce'.  The packet exhibits the standard `interference'
pattern during the collision with the wall \cite{bounce_anatomy},
\cite{andrews}, \cite{bouncing_bob}
at the `bounce', and then reforms into something like the
initial packet (compare to the dotted initial packet superimposed
on the $t=10$ case),  only wider.

For the momentum-space distributions (shown on the right of 
Fig.~(\ref{fig:bouncer_classical})), we
also see features of both the classical motion and the uniformly 
accelerated wave packet, as in Eqn.~(\ref{accelerating_momentum_values}). 
The expectation value of momentum $\langle p \rangle_t$,
calculated from $|\phi(p,t)|^2$ and plotted as the solid curve, is 
once again consistent with the classical trajectory (dashed curve), 
except near the
discontinuous, impulsive change in momentum values at the `bounce'. The
shape of the momentum-distribution follows the form expected from
Eqn.~(\ref{accelerating_momentum_shape}), namely uniform translation with no
change in shape, from $t=1 \rightarrow  t=3$ and then again from
$t=8 \rightarrow  t = 10$, that is, during the time when it is not in 
collision with the wall, but with a definite final change in shape, 
compared to the initial $|\phi_0(p)|^2$ superimposed on the $t=10$ result, 
resulting from the collision with the wall. 
The dotted vertical lines indicate the values of $p=0$, and  also the 
classically expected minimum and maximum values of momentum given by
$\pm p_M = \pm \sqrt{2mE} = \pm \sqrt{2mFz_0}$. The change in shape
can be understood, in great part,  from purely classical arguments 
\cite{quantum_bouncer}, \cite{robinett_bouncer}.

Using Eqn.~(\ref{quantum_revival}), the revival time is then given by
\begin{equation}
T_{rev} = \frac{4\pi \hbar}{|E''(n_0)|}
=
\frac{16 m z_0^2}{\pi \hbar}
\label{bouncer_revival_time}
\end{equation}
which can be compared to the similar result for the infinite well
in Eqn.~(\ref{infinite_well_revival}) 
which obviously has the same dimensions, but a differing
dependence on the initial energy. Gea-Banacloche \cite{quantum_bouncer}
examined the explicit
time-dependence of Gaussian wave packets for the `quantum bouncer' and noted
that packets returned to their classical periodicity, but half a period
out of phase, at a time half that of Eqn.~(\ref{bouncer_revival_time}),
which he described as the revival time and this would correspond to a
$p/q = 1/2$ revival in the language of Averbukh and Perelman 
\cite{perelman_1}; discussions of the structure of fractional revivals 
\cite{milburn_and_chen} in this system  have also appeared.  
The probability density at $t=0$ and near the first two
(complete) revivals are shown in Fig.~\ref{fig:bouncer_collapse},
showing the revival structure becoming increasingly approximate as the
effect of higher order ($(n-n_0)^3/T_{super}$) terms becomes important.

As noted in Sec.~\ref{subsec:period_and_collapse}, 
during the collapsed phase, the probability
density will be consistent with an incoherent sum of the individual
probability densities, as in Eqn.~(\ref{collapsed_wave_function}). 
We observe this behavior
at the bottom of Fig.~\ref{fig:bouncer_collapse} where we plot 
$|\psi(x,T^{*})|^2$ at a time
$T^{*}$ which is not close to any obvious fractional revival time. 
The anti-correlation
between the `wiggliness' of the wavefunction and its magnitude,
familiar from semi-classical discussions of stationary states or from 
the WKB approximation, is apparent. For additional comparison, we plot 
a purely classical probability density, given by
\begin{equation}
P_{cl}(z) = \frac{2}{\tau}\frac{1}{\sqrt{E-V(z)}}
\, ,
\end{equation}
based on simple ``{\it How much time, $dt$, 
does the particle spend in a given position bin, $dx$?}'' arguments or,  
equivalently, given by the pre-factor of a WKB wavefunction. 
In this case, the classical distribution is given by
\begin{equation}
P_{cl}(z;z_0) = \frac{1}{2\sqrt{z_0(z_0-z)}}
\label{bouncer_classical}
\end{equation}
and is shown in Fig.~\ref{fig:bouncer_collapse}(b) as the dotted curve.
The visualization of the time-dependent expectation values in both the
short-term semiclassical  and the long-term revival phases have been
discussed in Ref.~\cite{robinett_bouncer}, 
where, for example. the classical position-space 
average values, using Eqn.~(\ref{bouncer_classical}),  are given by
\begin{equation}
\langle z \rangle = \frac{2z_0}{3}
\qquad
,
\qquad
\langle z^2 \rangle = \frac{8z_0^2}{15}
\qquad
\mbox{and}
\qquad
\Delta z = \frac{2z_0}{\sqrt{45}} \approx 0.3 z_0
\end{equation}
and quantum wave packets collapse to near these values during much of
the time between revivals.

\subsection{2D rotor and related systems}
\label{subsec:2d_rotor}

\subsubsection{Two dimensional free quantum rotor}
\label{subsubsec:2d_free_quantum_rotor}

The problem of a particle confined to a circle of radius $L$, but 
otherwise free to rotate, defines the free quantum rotor, and is
described by the Schr\"{o}dinger equation
\begin{equation}
- \frac{\hbar^2}{2I} \frac{d^2 \psi_{m}(\theta)}{d \theta^2}
= E_m \psi_{m}(\theta)
\end{equation}
where $I\equiv \mu L^2$ is the moment of inertia. (We will occasionally
denote the particle mass by $\mu$ to avoid confusion with standard notation
for angular quantum numbers when appropriate.) 
The quantized energies and normalized wavefunctions are then given by
\begin{equation}
E_m = \frac{\hbar^2 m^2}{2I}
\qquad
\mbox{and}
\qquad
\psi_{m}(\theta) = \frac{1}{\sqrt{2\pi}} e^{im\theta}
\end{equation}
for $m=0, \pm 1, \pm 2, ...$. The solutions with $+m,-m$ for
$|m|\neq 0$ are doubly degenerate, corresponding to the equivalence
of clockwise and counter clockwise motion.  The angular wavefunctions
can also be written in the form
\begin{equation}
      \psi_{m}(\theta) = \left\{ \begin{array}{ll}
               1/\sqrt{2\pi} & \mbox{for $m=0$} \\
               \cos(m\theta)/\sqrt{\pi}& \mbox{for $m>0$} \\
               \sin(m\theta)/\sqrt{\pi} & \mbox{for $m>0$} \\
                                \end{array}
\right.
\label{alternate_angular_wavefunctions}
\end{equation}
which displays the pattern of degeneracies in a different way.

The classical system would
have an energy, classical frequency, and period associated by
\begin{equation}
E = \frac{1}{2} I \omega^2
\qquad
\mbox{and}
\qquad
T_{cl} = \frac{2\pi}{\omega} =
2\pi \sqrt{\frac{I}{2E}}
\, .
\end{equation}
The corresponding quantum mechanical period requires
\begin{equation}
|E'_{m}| = \frac{\hbar^2m}{I} = \hbar \sqrt{\frac{2E_m}{I}}
\end{equation}
so that
\begin{equation}
T_{cl} = \frac{2\pi \hbar}{|E'_m|} = 2\pi /\sqrt{2E_m/I}
\end{equation}
as expected. 

For this system, with purely quadratic dependence on the single quantum
number, the revival time is given by
\begin{equation}
T_{rev} = \frac{4\pi I}{\hbar}
= \frac{4\pi \mu L^2}{\hbar}
\label{rotor_revival_time}
\,. 
\end{equation}
This system can, of course, be thought of as a free particle with
periodic boundary conditions, to be compared to 1D infinite well 
(with reflecting boundary conditions) which has a very similar energy
spectrum. Localized (angular) wave packets can then be constructed with
many of the same properties as those in the 1D infinite well, including
zero (angular) momentum states with shorter revival times, and fractional
revivals at $p/q$ multiples of $T_{rev}$. The related problem of quantum
diffusion \cite{diffusion}
on a circular one-dimensional lattice has been found to have
similar revival-like behavior, while a propagator approach 
\cite{propagator_1}, \cite{propagator_2} for the 1D problem 
has also shown explicit evidence for fractional revivals. The time
evolution of the quantum rotor has also been examined in an interesting
way in Ref.~\cite{schleich_factorize_numbers}.

\subsubsection{Quantum pendulum}
\label{subsubsec:quantum_pendulum}

A related problem, with a far richer structure of classical time-dependence
and quantum energy eigenvalue spectra, is the quantum pendulum, defined
by the potential
\begin{equation}
V(\theta) = -V_0 \cos(\theta)
\end{equation}
where $V_0 = mgl$ for a pendulum under the influence of gravity, or
$V_0 = qEL$ for a point charge $q$ acted on by a constant electric
field, $E$. This problem was first studied by Condon \cite{condon} 
in the early days of quantum theory, and is a staple of the pedagogical 
literature, appearing as an example of perturbation theory in many textbooks
and collections of problems.
The corresponding Schr\"{o}dinger equation can be written in
the form
\begin{equation}
- \frac{\hbar^2}{2I} \frac{d^2 \psi(\theta)}{d \theta^2}
-V_0 \cos(\theta) \psi(\theta)
= E_m \psi(\theta)
\, .
\end{equation}
This problem is of interest since it reduces to the free-rotor case in the
high energy limit ($E_n >> +V_0$), while it has the harmonic oscillator
case 
in the limit of small oscillations, with predictable anharmonic corrections
derivable from the $\cos(\theta)$ term. The classical problem is also
more interesting as the periodicity can be evaluated in both limits,
as well as in general case (using elliptic integrals), with the limiting 
case of $E \approx +V_0$ defining the separatrix where the classical period
diverges, namely the `stuck on top' point.

The quantum mechanical problem can be written as one of the familiar 
equations of mathematical
physics, namely Mathieu's equation. Recently, this problem has been
studied both in the context of understanding the connections between
the classical periodicity and the quantum energy eigenstates, but also
in terms of the revival behavior. The authors of 
Ref.~\cite{robinett_pendulum} use fourth-order
 perturbation theory and numerical evaluation of the energy eigenvalues
to discuss the classical periodicity in the low-energy, high-energy,
and separatrix  limits, where the classical divergence in $T_{cl}$ is
`softened' by quantum effects, while also discussing the energy 
dependence of the quantum revival and super-revival times. For example,
the lowest-order anharmonic correction to the energies in the low-energy
limit is given by
\begin{equation}
E_n^{(1)} = \frac{\hbar^2}{32I}(2n^2 + 2n+1)
\end{equation}
which gives a (non-infinite) revival time
\begin{equation}
T_{rev} = \frac{2\pi \hbar}{|E_n''|/2}
= \frac{32\pi I}{\hbar}
\end{equation}
which is independent of $V_0$ and $8$ times larger than the revival
time for the high-energy, pure-rotor limit in 
Eqn.~(\ref{rotor_revival_time}).

Revivals in 3D rotational systems (with energy eigenvalues proportional
to $l(l+1)$) have also been studied \cite{bluhm_ajp}, \cite{clones},
\cite{auzinsh}.

\section{Two- and three-dimensional quantum systems}
\label{sec:2d_and_3d_systems}

A number of integrable two-dimensional infinite well or quantum billiard
geometries lend themselves to the study of quantum revival behavior 
in systems with several quantum numbers, and we focus here on three 
polygonal  billiard footprints, namely the square ($N=4$), equilateral
triangle ($N=3$), and circular ($N\rightarrow \infty$) infinite wells.
Discussions of time-dependent wave packet solutions of the first and
third cases go back to at least de Broglie \cite{debroglie} and also
provide useful examples of the connections of the short-term 
time-development of wavepackets to classical periodic orbits.

We note that evidence of wave packet revivals has  been presented from 
numerical simulations of quantum systems with classically chaotic 
behavior \cite{chaotic_revivals}, with specific examples being 
illustrated for the stadium billiard.

\subsection{Two dimensional infinite well and variations}
\label{subsec:2d_infinite_well}

For the two-dimensional infinite square well 
(with dimensions $L_x \times L_y = L \times L$),
the problem simplifies to two copies of a single 1-D infinite well because
of the separability of the potential. For example, 
the energy eigenvalues, $E(n_x,n_y)$,  
and position-space eigenstates, $w_{(n_x,n_y)}(x,y)$,  
are given by
\begin{equation}
E(n_x,n_y) = \frac{\hbar ^2 \pi^2 (n_x^2+n_y^2)}{2m L^2}
\qquad
\quad
\mbox{and}
\qquad
\quad
w_{(n_x,n_y)}(x,y)  = u_{(n_x)}(x) u_{(n_y)}(y)
\label{2d_energies}
\end{equation}
where $n_x,n_y = 1,2,3,...$ are the appropriate quantum numbers and
the $u_{n}(x)$ are given by Eqn.~(\ref{1d_eigenfunctions}).
The two non-vanishing revival times are given by 
Eqn.~(\ref{two_revival_times})
and are simply related to each other via
\begin{equation}
T_{rev}^{(n_x)} = \frac{4 mL^2}{\hbar \pi} = T_{rev}^{(n_y)}
\label{2d_revival_time}
\end{equation}
with no cross-term ($T_{rev}^{(n_x,n_y)}$) present. 
Therefore, the quantum revival structure is 
very simply related to that of the 1-D infinite well, including the 
possibilities of special `symmetric' revivals for zero-momentum 
(${\bf p}_0 = (0,0)$) wave packets at particular locations, such as 
for initial values of 
$(x_0,y_0) = (L/2,L/2)$ or $(L/3,2L/3)$, as well as the same rich
structure of fractional revivals. 
For rectangular infinite wells with commensurate 
($L_x \times L_y$, $L_x/L_y = p/q$) or incommensurate ($L_x/L_y \neq p/q$) 
sides,  the structure of the revival times may be more complex and 
interesting examples have been given in Refs.~\cite{bluhm_2d}
and \cite{agarwal_2d}.

The pattern of closed or periodic orbits is also interesting as the
path lengths for closed orbits in the 2-D square billiard
can be readily deduced from simple geometric arguments 
and are given by
\begin{equation}
L(p,q) = 2L\sqrt{p^2 + q^2}
\label{2d_billiard_orbits}
\end{equation}
where $2p,2q$ count the number of `hits' on the horizontal and vertical
walls respectively,  before returning to the starting point in phase space.
The corresponding classical periods for such closed trajectories are 
given by
\begin{equation}
T_{cl}^{(po)} = \frac{L(p,q)}{v_0}
\label{2d_closed_orbit_periods}
\end{equation}
where $v_0$ is the classical speed. Such orbits can be produced by 
point particles in the 2-D billiard, starting from any initial location,
$(x_0,y_0)$,  inside the box, provided they are `pointed' appropriately, 
namely in the
$\tan(\theta) =q/p$ direction. The values of 
\begin{equation}
\theta = \tan^{-1}(q/p)
\qquad
\mbox{and}
\qquad
T_{cl}^{(po)}/\tau = \sqrt{p^2+q^2}
\label{special_conditions}
\end{equation} (where $\tau \equiv 2L/v_0$ is the
period for the simplest, `back-and-forth' closed trajectory) for many of 
the low-lying cases are tabulated in Ref.~\cite{blueprint}.

The condition for periodic orbits in Eqn.~(\ref{two_classical_periods})
can  be implemented very easily in this case, and we will examine this
example in detail.  For such closed orbits to occur we require that
\begin{equation}
p\left(\frac{2m L^2}{\hbar \pi n_x}\right) 
= p T_{cl}^{(n_x)}
= q T_{cl}^{(n_y)} 
= q\left(\frac{2mL^2}{\hbar \pi n_y}\right) 
\end{equation}
or $n_y = n_x (q/p)$. Substituting this back into 
Eqn.~(\ref{2d_energies}), as well as equating the quantized total energy, $E(n_x,n_y)$, 
with the classical kinetic energy, gives
\begin{eqnarray}
\frac{1}{2} m  v_0^2  
\longleftrightarrow
 E(n_x,n_y) 
& = & \frac{\hbar^2 \pi^2}{2m L^2} \left( n_x^2 + n_y^2\right) \nonumber \\
& = & \frac{\hbar^2 \pi^2}{2m L^2} \left( n_x^2 + n_x^2
\left( 
\frac{q}{p}\right)^2 
\right) \\
& = &
\frac{\hbar^2 \pi^2}{2m L^2} \left[\frac{n_x^2 (p^2+q^2)}{p^2}\right]
\nonumber 
\end{eqnarray}
or
\begin{equation}
n_x = \left(\frac{m L v_0}{\hbar \pi} \right)\frac{p}{\sqrt{p^2+q^2}}
\qquad
\mbox{and}
\qquad
n_y = \left(\frac{m L v_0}{\hbar \pi} \right) \frac{q}{\sqrt{p^2+q^2}}
\end{equation}
so that
\begin{equation}
T_{cl}^{(po)} = pT_{cl}^{(n_x)} = p \left(\frac{2m L^2}{\hbar \pi n_x}\right)
= \frac{2L \sqrt{p^2+q^2}}{v_0}
\label{square_box_relation}
\end{equation}
which is consistent with the purely classical, and geometrical,  result from
Eqn.~(\ref{2d_closed_orbit_periods}).

We illustrate the short-term time-dependence of such wave packets, 
through plots of $|A(t)|^2 = |A_x(t)A_y(t)|^2$ versus $t$,  
in Fig.~\ref{fig:square_well_short}. 
The results shown there are for wave packets characterized
by initial positions $(x_0,y_0) = (L/2,L/2)$ 
and initial momenta given by $(p_{0x}, p_{0y}) =
(p_0\cos(\theta), p_{0}\sin(\theta))$ where we use $p_{0}
= 400\pi$ and vary $\theta = \tan^{-1}(p_{0y}/p_{0x})$; 
we also use $\Delta x_0 = \Delta y_0 = 0.05$ 
and the same physical parameters as in Eqn.~(\ref{physical_parameters}). 
With these values, the classical period (for the simplest back-and-forth 
motion),
$\tau = 2L/v_0$, the spreading time, 
$t_0$ (from Eqn.~(\ref{spreading_time})), 
and the revival time, $T_{rev}$ (from Eqn.~(\ref{2d_revival_time})), 
are given numerically by
\begin{eqnarray}
\tau  & = & 
 \frac{(2\mu) L}{p_0} = \frac{1}{400\pi} \approx 0.8 \times 10^{-3}, \\
t_0 & = &
 \left(\frac{2\mu}{\hbar}\right) \Delta x_0^2 = (0.05)^2 = 2.5 \times 10^{-3}
\\
T_{rev} & =&  \frac{4\mu L^2}{\hbar \pi} = \frac{2}{\pi} \approx 0.64
\, . 
\end{eqnarray}
The wave packet is seen to exhibit a reasonable number of classical periods
before significant spreading occurs, with the revival time scale being
much larger than both; the locations of classical closed or
periodic orbits are denoted by stars, with the closed orbit patterns
indicated for some of the simpler 2D trajectories.

\subsection{Isosceles ($45^{\circ}-45^{\circ}-90^{\circ}$) triangle billiard}
\label{subsec:isoceles_billiard}

The energy eigenvalues and wavefunctions for a special 2-D 
triangular billiard `footprint' can be easily derived 
\cite{other_berry} - \cite{robinett_jmp} from those
of the two-dimensional square infinite well solutions in
Eqn.~(\ref{2d_energies}).  The standard results for the 2-D square well
\begin{equation}
E(n_x,n_y) = \frac{\hbar^2 \pi^2}{2m L^2} \left( n_x^2 + n_y^2\right)
\qquad
\mbox{and}
\qquad
w_{(n_x,n_y)}(x,y) = u_{(n_x)}(x) u_{(n_y)}(y)
\end{equation}
hold for any integral $n_x,n_y\geq 1$ and
for $n_x=n_y$ there is a single state, while for $n_x \neq n_y$,
there is a two-fold degeneracy. Linear combinations of these 
solutions can be written in form 
\begin{eqnarray}
w_{(n,m)}^{(-)}(x,y) & = & \frac{1}{\sqrt{2}} \left[u_{(n)}(x)u_{(m)}(y) - u_{(m)}(x)u_{(n)}(y)\right] 
\qquad
(m \neq n) 
\label{triangle_1}
\\
w_{(n,m)}^{(+)}(x,y) & = & \frac{1}{\sqrt{2}} \left[u_{(n)}(x)u_{(m)}(y) + u_{(m)}(x)u_{(n)}(y)\right] 
\qquad
(m \neq n) 
\label{triangle_2}
\\
w_{(n,n)}^{(o)}(x,y) & = & u_{(n)}(x)u_{(n)}(y) 
\label{triangle_3}
\end{eqnarray}
which have the same energy degeneracy, but exhibit different patterns
of nodal lines.
These alternative forms  are  useful since they  allow one to discuss 
the energy eigenvalues and 
eigenfunctions of the $45^{\circ}-45^{\circ}-90^{\circ}$ triangle 
billiard formed by `folding' the square along a diagonal, 
since the $w_{(n,m)}^{(-)}(x,y)$ now satisfy the 
appropriate boundary condition along the new hypotenuse 
as one can easily see from Eqn.~(\ref{triangle_1}) that 
$w_{(n,m)}^{(-)}(x,y=x) = 0$, by construction.
The allowed eigenvalues for this case are still given by 
\begin{equation}
E(n_x,n_y) = \frac{\hbar^2 \pi^2}{2m L^2} (n_x^2 + n_y^2)
\quad
\longrightarrow
\quad
E(n,\tilde{n}) = \frac{\hbar^2 \pi^2}{2m L^2} (n^2 + \tilde{n}^2)
\end{equation}
but now with only a single $(n_x,n_y) \rightarrow (n,\tilde{n})$ 
state allowed, with corresponding wavefunctions given by
Eqn.~(\ref{triangle_1}), but multiplied by $\sqrt{2}$ to account for the
different normalization needed in the smaller area billiard.

The single common revival time in the $45^{\circ}-45^{\circ}-90^{\circ}$ 
triangle billiard is then still given by Eqn.~(\ref{2d_revival_time}) 
and localized wave packets can also  be constructed, now using the 
appropriately normalized analogs of the wavefunctions in 
Eqn.~(\ref{triangle_1}), once again, provided they are kept away from any 
of the infinite wall boundaries. 

The structure of the classical closed or periodic orbits in this case
is the same as for the square billiard, since all of the standard $(p,q)$
orbits in the 2-D square well are simply reflected off the new diagonal
wall (along the hypotenuse),  giving rise to the same allowed orbits
as in Eqn.~(\ref{special_conditions}). 
The only new feature in the semi-classical propagation
of such wave packet solutions in this `folded' geometry
is the existence  of a special isolated closed orbit 
\cite{blueprint}, \cite{robinett_jmp} at 
$135^{\circ}$ (one which bisects the $90^{\circ}$
right angle, bouncing normally off the hypotenuse, and 
returning to the corner) 
which has path lengths which are multiples of 
$(\sqrt{1^2 + 1^2})L/2 = \sqrt{2}L/2$, namely half that of the standard
$(p,q) = (1,1)$ features. When we construct wave packets using
parameters appropriate to this geometry, we see twice as many features
in the $A(t)$ plot for the $(1,1)$ case because of this special classical
closed orbit, but otherwise reproduce the results shown in 
Fig.~\ref{fig:square_well_short}.
Extensions to 3D cubic or rectangular parallelepiped billiard systems
are also possible.

\subsection{Equilateral triangle billiard}
\label{subsec:equilateral_triangle}

\subsubsection{Energy eigenvalues and eigenfunctions}
\label{subsubsec:equi_energies_and_eigenfunctions}

It is perhaps an under-appreciated fact that 
the energy eigenvalues and position-space wave functions for a particle 
in an equilateral ($60^{\circ}-60^{\circ}-60^{\circ}$) 
triangular infinite well (or billiard) of side 
$L$ are available in closed form. They have been discussed by
many authors, in a variety of different contexts, 
using complementary methods of derivation
\cite{math_methods} - \cite{blinder_version}
and more recently in the context of wave packet revivals
\cite{robinett_annals}. For definiteness in what follows, 
we will assume such a triangular billiard with vertices 
located at 
$(0,0)$, $(L/2,\sqrt{3}L/2)$, and $(-L/2,\sqrt{3}L/2)$ and we will
denote the particle mass by $\mu$ to avoid confusion with standardly used
notation for various quantum numbers. With this notation, 
the resulting energy spectrum is given by
\begin{equation}
E(m,n) = \frac{\hbar^2}{2\mu L^2} \left(\frac{4 \pi}{3}\right)^2
\left( m^2 + n^2 - mn\right)
\label{equilateral_energy_eigenvalues}
\end{equation}
for integral values of $m,n$,  with the restriction that $m \geq 2n$.  
(In what follows, we will use the notation of 
Ref.~\cite{richens_and_berry} except for a minor relabeling, 
for the energies and 
wavefunctions.) For the case of $m > 2n$, there are 
two degenerate states with different symmetry properties 
which can be written in the forms
\begin{eqnarray}
w_{(m,n)}^{(-)}(x,y) & = &
\sqrt{\frac{16}{L^2 3\sqrt{3}}}
\left[
\sin\left(\frac{2\pi (2m-n)x}{3L}\right) 
\sin\left(\frac{2\pi ny}{\sqrt{3}L}\right) 
\right. \nonumber  \\ 
& & 
\qquad \qquad
- 
\sin\left(\frac{2\pi (2n-m)x}{3L}\right) 
\sin\left(\frac{2\pi my}{\sqrt{3}L}\right) 
\label{odd_wavefunctions}
\\
& &
\qquad \qquad 
\left.
- 
\sin\left(\frac{2\pi (m+n) x}{3L}\right)
\sin\left(\frac{2\pi (m-n) y}{\sqrt{3}L}\right)
\right] \nonumber 
\end{eqnarray}
and
\begin{eqnarray}
w_{(m,n)}^{(+)}(x,y) & = &
\sqrt{\frac{16}{L^2 3\sqrt{3}}}
\left[
\cos\left(\frac{2\pi (2m-n)x}{3L}\right) 
\sin\left(\frac{2\pi ny}{\sqrt{3}L}\right) 
\right. \nonumber \\ 
& & 
\qquad \qquad 
- 
\cos\left(\frac{2\pi (2n-m)x}{3L}\right) 
\sin\left(\frac{2\pi my}{\sqrt{3}L}\right) 
\label{even_wavefunctions}  \\
& &
\qquad \qquad 
\left.
+
\cos\left(\frac{2\pi (m+n) x}{3L}\right)
\sin\left(\frac{2\pi (m-n) y}{\sqrt{3}L}\right)
\right] \, .
\nonumber 
\end{eqnarray}
One can confirm by direct differentiation that they satisfy the
2-D Schr\"odinger equation, with the energy eigenvalues in 
Eqn.~(\ref{equilateral_energy_eigenvalues}), 
as well as the appropriate boundary 
conditions. 
Extending earlier results, we have here also included the correct 
normalizations,  since we are, of course, interesting in expanding
Gaussian wave packets in such eigenstates.

For the special case of $m=2n$,  there is a single non-degenerate state 
for each value of $n$, given by 
\begin{equation}
w_{(2n,n)}^{(o)}(x,y) = 
\sqrt{\frac{8}{L^2 3\sqrt{3}}}
\left[
2\cos\left(\frac{2\pi nx}{L}\right) 
\sin\left(\frac{2\pi n y}{\sqrt{3}L}\right)
- \sin\left(\frac{4 \pi ny}{\sqrt{3}L}\right)
\right] \, .
\label{special_wavefunctions}
\end{equation}
Clearly these states satisfy
\begin{equation}
w_{(m,n)}^{(\pm)}(-x,y) = 
\pm w_{(m,n)}^{(\pm)}(x,y)
,
\qquad
\quad
w_{(m,n)}^{(o)}(-x,y) = 
+ w_{(m,n)}^{(o)}(x,y)
\end{equation}
and the $w_{(m=2n,n)}^{(\pm)}(x,y)$ states also satisfy 
\begin{eqnarray}
w_{(m=2n,n)}^{(+)}(x,y) & =& \sqrt{2} w_{(2n,n)}^{(o)}(x,y) \\
w_{(m=2n,n)}^{(-)}(x,y) & =& 0 \, . \nonumber
\end{eqnarray} 
The pattern of energy level degeneracies, and wavefunction symmetries
is thus very similar to that for the 2-D square
billiard, especially when the solutions  for that problem are written
in the form in Eqns.~(\ref{triangle_1}), (\ref{triangle_2}), and
(\ref{triangle_3}). (See Ref.~\cite{liboff_polygon}
for a discussion of such pairwise degeneracies in 2D polygonal billiards.)

\subsubsection{Classical periodicity and revival times}
\label{subsubsec:equi_period_and_revival}

Turning now to the time-dependence of wave packets in this geometry,
the long-term, revival time scales in this two quantum number system
are given by Eqns.~(\ref{two_revival_times}) and
(\ref{third_revival_time}) as 
\begin{equation}
T_{rev}^{(m)} = \frac{2\pi \hbar}{|\partial^2 E/\partial m^2|/2}
,
\qquad
T_{rev}^{(n)} = \frac{2\pi \hbar}{|\partial^2 E/\partial n^2|/2}
,
\qquad
\mbox{and}
\qquad
T_{rev}^{(m,n)} = \frac{2\pi \hbar}{|\partial^2 E/\partial m \partial n|}
\end{equation}
which all give the single common revival time 
\begin{equation}
T_{rev}^{(m)} 
= 
T_{rev}^{(n)} 
=
T_{rev}^{(m,n)} 
\equiv
T_{rev}
= \frac{9 \mu L^2}{4\hbar \pi}
\label{equilateral_triangle_revival_time}
\end{equation}
and exact revivals (just as for the 2-D square well) and fractional revivals
are present in this system, with a single revival time guaranteed for any 
and all possible initial 
wave packets. Thus, both the $N=3$  and $N=4$ polygonal billiards (the
equilateral triangle and square) exhibit similar and simple energy 
eigenvalues and exact quantum revivals. Just as with the
1D and 2D infinite wells, we note that 
special, shorter-time scale revivals are also possible in the equilateral
triangle case for zero-momentum ($(p_{0x},p_{0y}) = (0,0)$) 
wave packets initially centered at 
`symmetric' locations within the triangular billiard, such as at the 
geometric center, and half-way down a bisector \cite{robinett_annals}.

The short-term, classical periodicity of quantum wave packets in
this geometry can also be determined from calculations of
\begin{eqnarray}
T_{cl}^{(m)}
& = & 
\frac{2\pi \hbar}{|\partial E/\partial m|}
 = 
\left[\frac{9\mu L^2}{4\hbar \pi} \right]\frac{1}{(2m-n)}
\\
T_{cl}^{(n)} 
& = & 
\frac{2\pi \hbar}{|\partial E/\partial n|}
=
\left[\frac{9\mu L^2}{4\hbar \pi} \right] \frac{1}{(2n-m)}
\, . 
\label{new_shorty}
\end{eqnarray}
The condition leading to closed orbits can then be written as
\begin{equation}
\frac{(2m-n)}{(2n-m)} = \frac{T_{cl}^{(n)}}{T_{cl}^{(m)}}
= \frac{p}{q}
\qquad
\quad
\mbox{or}
\quad
\qquad
n = m \left(\frac{2p+q}{2q+p} \right)
\, .
\end{equation}
If we substitute this condition into the energy spectrum in 
Eqn.~(\ref{equilateral_energy_eigenvalues}), as well as equating 
the quantum energies with 
the classical kinetic energy, $\mu v_0^2/2$ (where $v_0$ is 
the classical speed),  we are led to the association
\begin{equation}
\frac{1}{2} \mu v_0^2 
\longleftrightarrow
E(m,n) = \left(\frac{16 \pi^2}{9}\right) \left(\frac{\hbar^2}{2\mu L^2}\right)
\left[ m^2 + m^2\left(\frac{2p+q}{2q+p}\right)^2 
- m^2 \left(\frac{2p+q}{2q+p}\right) \right]
\end{equation}
or
\begin{equation}
\left(\frac{2\mu v_0L}{4\pi \hbar}\right)^2
= m^2 \left[\frac{3(p^2+pq + q^2)}{(2q+p)^2} \right]
\, . 
\end{equation}
This implies that 
\begin{equation}
m = \left(\frac{2\mu v_0L}{4\pi \hbar} \right) \frac{(2q+p)}{\sqrt{3}\sqrt{p^2 + pq + q^2}}
\qquad
\mbox{and}
\qquad
n = \left(\frac{2\mu v_0L}{4\pi \hbar} \right) \frac{(2p+q)}{\sqrt{3}\sqrt{p^2 + pq + q^2}}
\, .
\end{equation}
The period for classical, closed/periodic orbits is then given by 
\begin{equation}
T_{cl}^{(po)} = pT_{cl}^{(m)}
= \frac{L \sqrt{3}\sqrt{p^2 + pq + q^2}}{v_0}
=
\frac{L(p,q)}{v_0}
\label{closed_orbit_period}
\end{equation}
where 
\begin{equation}
L(p,q) = d(p,q) = L\sqrt{3}\sqrt{p^2 + pq + q^2}
\end{equation}
are the corresponding path lengths for periodic orbits. 
The possible classical closed or periodic orbits can be derived
from a geometric construction (involving tiling of the 2-D plane, 
just as for the square case)  giving this
same result. (See Refs.~\cite{liboff_polygon} 
and \cite{other_polygon}
for discussions of the derivation of
energy eigenvalues and eigenfunctions for convex plane polygonal
billiards using tiling methods.)

We note that the special cases of $2m=n$ and $2n=m$
correspond to  $q=0$ and $p=0$ respectively, both of which give
$L(p,q) = \sqrt{3}L$. In these cases, one of the classical periods 
from Eqn.~({\ref{new_shorty}) formally goes to infinity and this can
be understood classically from the corresponding path length, which
corresponds to periodic, `back and forth' motion from one corner, 
along a bisector, to the opposite side, but with no repetition in
the complementary direction. (See also Ref.~\cite{richens_and_berry} for 
a derivation of the quantized energies from which this effect can 
be also inferred.)

Because of the (relatively) simple form of the allowed wavefunctions
in Eqns.~(\ref{odd_wavefunctions}) and (\ref{even_wavefunctions}), 
involving trigonometric functions, just as for the 1-D infinite well,
one can evaluate the expansion coefficients for any 2-D Gaussian
wave packet  by extending
the region of integration from the (finite) triangular region to the
entire 2-D space, so long as the initial wave packet is well away from
any of the walls. The required Gaussian-type
integrals are of the forms
\begin{eqnarray}
\int_{-\infty}^{+\infty}\, e^{ip_0(x-x_0)/\hbar}\, 
e^{-(x-x_0)^2/2b^2}\,\cos\left(\frac{Cx}{a}\right)\,dx
& = &
\frac{b\sqrt{2\pi}}{2}
\left[
e^{iCx_0/a} e^{-b^2(C/a + p_0/\hbar)^2/2} \right.
 \\
& & 
\qquad
\quad
\left. 
+
e^{-iCx_0/a} e^{-b^2(-C/a + p_0/\hbar)^2/2}
\right]
\nonumber
 \end{eqnarray}
and
\begin{eqnarray}
\int_{-\infty}^{+\infty}\, e^{ip_0(x-x_0)/\hbar}\, 
e^{-(x-x_0)^2/2b^2}\,\sin\left(\frac{Cx}{a}\right)\,dx
& = &
\frac{b\sqrt{2\pi}}{2i}
\left[
e^{iCx_0/a} e^{-b^2(C/a + p_0/\hbar)^2/2} \right.
 \\
& & 
\qquad
\quad
\left. 
-
e^{-iCx_0/a} e^{-b^2(-C/a + p_0/\hbar)^2/2}
\right] 
\nonumber
\end{eqnarray}
with similar expressions for $y$-integrations. The resulting closed form
expressions for the expansion coefficients can be used in calculations
of the auto-correlation function to illustrate
the long-term revival structure of wave packets, as well as the
short-term, semi-classical propagation giving rise to closed orbits
of the form in in Eqn.~(\ref{closed_orbit_period}); 
for example, the analog of Fig.~\ref{fig:square_well_short}
for the equilateral triangle billiard have been presented in 
Ref.~\cite{robinett_annals}.

A `folding' of the equilateral ($60^{\circ}-60^{\circ}-60^{\circ}$) 
triangle along a bisector yields another special triangular geometry,
namely a $30^{\circ}-60^{\circ}-90^{\circ}$ right triangle. The energy
eigenvalues and eigenfunctions for this case can also be trivially obtained 
from those of Eqn.~(\ref{odd_wavefunctions}) as they satisfy the new boundary
condition along the `fold'. Such a system will have the same energy
eigenvalues as in Eqn.~(\ref{equilateral_energy_eigenvalues}) 
(but with only one $(m,n)$ combination, with $m > 2n$,  possible) and the 
same common revival time, $T_{rev}$, as in 
Eqn.~(\ref{equilateral_triangle_revival_time}). Wave packet solutions
can also be constructed (remembering to include an additional factor of
$\sqrt{2}$ to account for the normalization difference) from the
$w_{(m,n)}^{(-)}(x,y)$ in Eqn.~(\ref{odd_wavefunctions}). 

Finally, we note
that Liboff \cite{liboff_hexagon} has displayed a subset of the wavefunctions 
and energy eigenstates for the hexagonal ($N=6$ regular polygon) which 
can be seen to also give exact revival behavior.

\subsection{Circular billiard and variations}
\label{subsec:circular_billiard}

\subsubsection{Energy eigenvalues and eigenfunctions}
\label{subsubsec:circular_eigenvalues_and_eigenfunctions}

We finally turn our attention to the $N \rightarrow \infty$ limit of 
the $N$-sided regular polygonal billiard, namely the circular infinite well,
which has been explored \cite{robinett_circular} in terms of quantum revivals.
The potential for this problem can be defined by 
\begin{equation}
      V_C(r) = \left\{ \begin{array}{ll}
               0 & \mbox{for $r<R$} \\
               \infty  & \mbox{for $r\geq R$}
                                \end{array}
\right.
\end{equation}
and we once again use the notation $\mu$ for the particle mass.
The (unnormalized) solutions of the corresponding 2-D Schr\"odinger equation 
are given by
\begin{equation}
 u_{(m)}(r,\theta) = J_{|m|}(kr) e^{im\theta}
\end{equation}
where the quantized angular momentum values are given by $L_z = m \hbar$
for $m=0, \pm 1, \pm 2,...$ and  the $J_{|m|}(kr)$ are the (regular) 
Bessel functions of order $|m|$.  The angular eigenstates can also
be written in the form in Eqn.~(\ref{alternate_angular_wavefunctions}).

The wavenumber, $k$,  is related to the  energy via 
$k = \sqrt{2\mu E/\hbar^2}$ and the energy eigenvalues are
quantized by application of the boundary conditions at the infinite wall
at $r=R$, namely $J_{|m|}(z=kR) =0$. The quantized energies are then given by
\begin{equation}
E_{(m,n_r)} = \frac{\hbar^2 [z_{(m,n_r)}]^2}{2\mu R^2}
\end{equation}
where $z_{(m,n_r)}$ denotes the zeros of the Bessel function of order
$|m|$,  and $n_r$ counts the number of radial nodes. 
A general time-dependent state in this system can be written in the
form
\begin{equation}
\psi(r,\theta,t) = \sum_{n_r=0}^{\infty}
\sum_{m=-\infty}^{+\infty} a_{(m,n_r)} 
\left[N_{(m,n_r)} J_{|m|}(k_{(m,n_r)}r) \right] \, 
\left(\frac{e^{im\theta}}{\sqrt{2\pi}}\right)           \, 
e^{-iE_{(m,n_r)}t/\hbar}
\label{circular_expansion_in_eigenstates}
\end{equation}
where the radial normalization factors are determined by insisting that
\begin{equation}
\int_{0}^{R} \left[ N_{(m,n_r)} J_{|m|}(k_{(m,n_r)}r)\right]^2\,r\,dr
= 1
\, . 
\end{equation}

The energy spectrum
is doubly degenerate for $|m| \neq 0$ corresponding to the equivalence
of clockwise and counter-clockwise ($m>0$ and $m<0$) motion. 
We therefore see a pattern of degeneracies very similar to that of both the
square and equilateral triangle wells, with two  equal energy states
for each $(|m|>0,n_r)$ value, and a single one for each $(m=0,n_r)$.
Because the quantum number dependence  of the energy eigenvalues is the
determining factor in the structure of wave packet revivals, we need to
examine  the $m,n_r$ dependence of the $E_{(m,n_r)} \propto [z_{(m,n_r)}]^2$.

As a first approximation, one can look at the large $z$ behavior of the 
Bessel function solutions for fixed values of $|m|$, 
namely
\begin{equation}
J_{|m|}(z) 
\longrightarrow 
\sqrt{\frac{2}{\pi z}}
\cos\left(z - \frac{|m|}{2}- \frac{\pi}{4}\right) + \cdots
\,\,\, .
\end{equation}
With this approximation, the zeros are  given by
\begin{equation}
z_{(m,n_r)} -\frac{|m|}{2} - \frac{\pi}{4} \approx  (2n_r+1)\frac{\pi}{2}
\qquad
\mbox{or}
\qquad
z_{(m,n_r)} \approx  \left(n_r+\frac{|m|}{2}+\frac{3}{4}\right)\pi 
\, . 
\label{first_approximation}
\end{equation}
If this result were exact, the quantized energies would depend on two quantum
numbers, in at most a quadratic manner, and there would be exact wave packet
revivals, just as for the 2-D square or equilateral triangle billiards.
However, there are important corrections to this first-order formula
which means that the Bessel function zeros are not given by exact integral
values: however, a useful approximation in the large quantum number limit
relevant for wave packet expansions can be obtained in a straightforward and 
accessible way  by use of the WKB approximation. 

The two-dimensional Schr\"{o}dinger equation for the radial wavefunction
can be written in the form
\begin{equation}
- \frac{\hbar^2}{2\mu}\left( \frac{d^2R(r)}{dr^2}
+ \frac{1}{r} \frac{dR(r)}{dr}\right) + \frac{\hbar^2 m^2}{2\mu r^2}
R(r) = ER(r)
\end{equation}
and to recast this into a one-dimensional equation suitable for the WKB
approximation we write $u(r) \equiv R(r)/\sqrt{r}$ to obtain
\begin{equation}
-\frac{\hbar^2}{2\mu} \frac{d^2u(r)}{dr^2} + \frac{\hbar^2(m^2-1/4)}{2\mu r^2}
u(r) = Eu(r)
\, .
\end{equation}
In order to obtain WKB wavefunctions with the correct behavior (same
phase for large $r$ as the exact solutions), one makes
use of the Langer transformation \cite{wkb_approximation},
\cite{wkb_approximation}, \cite{langer} - \cite{froman}
which effectively replaces $m^2 -1/4$ with $m^2$. This substitution
is valid for all but $s$-states,  which must be treated differently
\cite{berry_two_dimensions}, \cite{trost}.

In this approach, we note that in the radial direction the particle moves 
freely up to the infinite wall at $r=R$, but is subject 
to an effective centrifugal
potential given by $V_{eff}(r) = L_z^2/2\mu r^2 = (m\hbar)^2/2\mu r^2$. 
The classical particle cannot penetrate this centrifugal barrier and has an 
associated inner radius or distance of closest approach, $R_{min}$,
given by
\begin{equation}
V_{eff}(R_{min}) = \frac{m^2\hbar^2}{2\mu R_{min}^2} = E
\qquad
\mbox{or}
\qquad
R_{min} = \frac{|m|\hbar}{\sqrt{2\mu E}}
\, . 
\end{equation}
We can also write this in the useful form
\begin{equation}
R_{min} = \frac{|m|R}{z}
\qquad
\mbox{where}
\qquad
E \equiv \frac{\hbar^2 z^2}{2\mu R^2}
\label{useful_form}
\end{equation}
more directly in terms of the desired dimensionless variable $z$,
which is equivalent to the energy eigenvalue.

The WKB quantization condition on the radial variable, $r$,
is then given by
\begin{equation}
\int_{R_{min}}^{R} k_r(r)\,dr = (n_r + C_L + C_R) \pi
\qquad
\mbox{where}
\qquad
k_r(r) \equiv  \sqrt{\frac{2\mu E}{\hbar^2}} \sqrt{1 - \frac{R_{min}^2}{r^2}}
\, . 
\end{equation}
The matching coefficients  are given by 
$C_L = 1/4$ and $C_R = 1/2$ which are appropriate for 'linear' boundaries 
(at the inner centrifugal barrier) and 'hard' or 'infinite wall' boundaries 
(such as at $r=R$), respectively. 
The WKB energy quantization condition for the quantized energies, 
in terms of $n_r$ explicitly and $|m|$ implicitly, via both the $E$ and 
$R_{min}$ terms, can then be written in the form
\begin{equation}
\sqrt{\frac{2\mu E}{\hbar^2}} 
\int_{R_{min}}^{R} \sqrt{1 - \frac{R_{min}^2}{r^2}} \,dr
= (n_r + 3/4)\pi
\, . 
\label{wkb_condition}
\end{equation}
(This result was obtained earlier in Ref.~\cite{robnik}. We also note
that an improved approximation for $m\neq 0$ eigenvalues has also been
obtained through the use of periodic orbit theory \cite{wunner}.) 
The integral 
on the left can be evaluated  in the form
\begin{eqnarray}
\int_{R_{min}}^{R} \frac{\sqrt{r^2 -R_{min}^2}}{r} \,dr
& = &
\sqrt{R^2 - R_{min}^2} - R_{min}\sec^{-1} \left(\frac{R}{R_{min}}\right) \\
& = & R \left[\sqrt{1-x^2} - x \sec^{-1}(1/x) \right] \nonumber
\end{eqnarray}
where we define $x \equiv R_{min}/R = |m|/z$. 
This result can be expanded for small values of $x$ (i.e., $R_{min}/R << 1$
or $|m|/z << 1$) to obtain
\begin{equation}
\sqrt{1-x^2} - x\sec^{-1}(1/x) = 1 - \frac{\pi x}{2} + \frac{x^2}{2}
+ \frac{x^4}{24} + \frac{x^6}{80} + \frac{5x^8}{896} + \cdots
\,\,\, . 
\end{equation}
The WKB quantization condition in Eqn.~(\ref{wkb_condition})
can then be written, in terms of $z$,  in the form
\begin{equation}
z\left(1 - \frac{\pi}{2}\frac{|m|}{z} + \frac{m^2}{2z^2} + 
\frac{m^4}{24z^4} + \cdots \right)
= (n_r+3/4)\pi
\end{equation}
If we keep only the first two terms on the left-hand side, we find that
\begin{equation}
z \approx  (n_r +|m|/2+3/4)\pi \equiv  z_{0}(m,n_r)
\label{lowest_order}
\end{equation}
which is the lowest-order result obtained directly from the limiting form 
of the wavefunction. 

To improve on this result, we simply keep successively higher order terms, 
solving iteratively at each level of approximation using a lower-order result 
for $z$, and we find the much improved approximation
\begin{eqnarray}
z_{(m,n_r)} & = &  z_0(m,n_r) - \frac{m^2}{2z_0(m,n_r)}
- \frac{7}{24} \frac{m^4}{[z_0(m,n_r)]^3} 
- \frac{83}{240} \frac{m^6}{[z_0(m,n_r)]^5} 
\nonumber  \\
& &
\qquad
\qquad
\qquad
- \frac{6949}{13440}\frac{m^8}{[z_0(m,n_r)]^7} + \cdots
\label{zero_expansion}
\end{eqnarray}
which we have confirmed numerically is an increasingly good approximation,
especially for $n_r >> 1$. For the study of wave packet revivals, we only
require the energy eigenvalue dependence on $m,n_r$ to second order, but
higher order terms such as those above might  be useful for super-revivals
and even longer-term time-dependence studies.

For the special case of $m=0$, this WKB technique only returns the
zero-th order result, $z = z(0,n_r)$, and the authors of 
Ref.~\cite{robinett_circular}, motivated  by the form of the expansion in 
Eqn.~(\ref{zero_expansion}), numerically fit the first $50$ lowest-lying 
$m=0$ zeros to a similar form and find the result
\begin{equation}
z_{(0,n_r)} = z_0(0,n_r) + \frac{1}{8z_0(0,n_r)} + \cdots
\, 
\label{zero_zero_approximation}
\, .
\end{equation}
which gives a much improved  approximation. 
Friedrich and Trost \cite{trost} have obtained an improved WKB
approximation for the 2D circular well for the $m=0$ case which,
when expanded in terms of $z_{(0,n_r)}$, gives the first two terms
in Eqn.~(\ref{zero_zero_approximation}), so we will use that next-to-leading
result in what follows.

Using Eqns.~(\ref{zero_expansion}) and (\ref{zero_zero_approximation}),
we can evaluate the energy eigenvalues to quadratic order in $n_r,m$ in
order to probe the revival structure of wave packets. For the special 
case of $m=0$, we find that
\begin{equation}
E_{(0,n_r)} = \frac{\hbar^2 [z_{(0,n_r)}]^2}{2\mu R^2}
= \frac{\hbar^2 \pi^2}{2\mu R^2}\left[\left(n_r+\frac{3}{4}\right)^2 + 
\frac{1}{4\pi^2}\right]
\label{zero_energies}
\end{equation}
while for the more general case with $m\neq 0$, we find
\begin{equation}
E_{(m,n_r)} = \frac{\hbar^2 [z_{(m,n_r)}]^2}{2\mu R^2}
= \frac{\hbar^2 \pi^2}{2\mu R^2}
\left[\left(n_r + \frac{|m|}{2} + \frac{3}{4}\right)^2 - \frac{m^2}{\pi^2}
\right]
\, . 
\label{other_energies}
\end{equation}

The fact that these energies depend on non-integral values of the
effective quantum numbers is reminiscent of the case of Rydberg
wave packets in alkali-metal atoms due to quantum defects 
\cite{bluhm_defects},  \cite{other_defects}
and methods similar to those used there might
prove useful. In what follows, however, we simply examine the time-dependence 
of typical $m=0$ and $m \neq 0$ wave packets directly.

\subsubsection{Wave packets and time scales}
\label{subsubsec:wave_packets_and_time_scales}

We begin by focusing on the special case of zero-momentum wave packets,
centered at the origin, namely with vanishing values of
$(p_{0x},p_{0y})$ and $(x_0,y_0)$. 
For such states, where only the $m=0$ eigenstates contribute, we can write
the energy eigenvalues from Eqn.~(\ref{zero_energies}) in the form
\begin{eqnarray}
E(n_r) & = & \frac{\hbar^2\pi^2}{2\mu R^2}
\left[ \left(n_r+\frac{3}{4}\right)^2 + \frac{1}{4\pi^2}
+ {\cal O}\left(\frac{1}{(n_r+3/4)^2}\right) \right]
\nonumber  \\
& \approx  & 
\frac{\hbar^2\pi^2}{32\mu R^2}\left[ 8n_r(2n_r+3) + 
\left(9 + \frac{4}{\pi^2}\right)\right] 
\label{constant_piece}\\
& = & \frac{\hbar^2 \pi^2}{4\mu R^2} 
\left[ l(n_r) + \left(\frac{9}{8} + \frac{1}{2\pi^2}\right)\right] \nonumber
\end{eqnarray}
where $l(n_r) \equiv n_r(2n_r+3)$ is an integer 
(neither even nor odd in general).
The last term in the square brackets is independent of $n_r$ and will make
the same, constant, overall phase contribution to the autocorrelation
function,  so we focus on the $l(n_r)$ term. Since this integer has no
special evenness/oddness properties, its contribution to the phase of
each $|a_{(n,n_r)}|^2$ term in Eqn.~(\ref{circular_expansion_in_eigenstates})
will be identically unity at a revival time given by
\begin{equation}
\left(\frac{\hbar^2 \pi^2}{4\mu R^2}\right) \frac{T_{rev}^{(m=0)}}{\hbar}
=  2\pi
\qquad
\mbox{or}
\qquad
T_{rev}^{(m=0)} = 4\left[\frac{2\mu R^2}{\hbar \pi}\right] \equiv
4T_0
\end{equation}
Thus, at integral multiples of $4T_0$, we expect nearly perfect revivals
because of the almost regularly spaced structure of the $m=0$ Bessel function
zeros. At these recurrences, we can also predict  the overall phase
corresponding to the last term in Eqn.~(\ref{constant_piece}), namely
\begin{equation}
e^{-i\hbar^2\pi^2/4\mu R^2(4T_0) (9/8+1/2\pi^2)}
= e^{-2\pi i (9/8 +1/2\pi^2)}
=
e^{-2\pi i} e^{-2\pi i(1/8+1/2\pi^2)}
\equiv e^{-i\pi F}
\label{extra_phase_1}
\end{equation}
where $F = 1/4 + 1/\pi^2 \approx 0.351$.

To investigate these predictions numerically, we have used an initial
Gaussian wave packet consisting of a product of two forms as 
in Eqn.~(\ref{just_initial_gaussian}) with the specific  values
\begin{equation}
2m\, , \, \hbar\, , \,  R = 1
\qquad
\mbox{and}
\qquad
b = \frac{1}{10\sqrt{2}}
\qquad
\mbox{so that}
\qquad
\Delta x_0 = \Delta y_0 = 0.05
\label{numerical_values}
\, . 
\end{equation}
The expansion coefficients $a_{(m,n_r)}$ in 
Eqn.~(\ref{circular_expansion_in_eigenstates}) are calculated by numerical
evaluation of the required `overlap' integrals of the initial state
with the eigenstates.

Using the expansion coefficients for this state, we plot the modulus squared
of the autocorrelation function, $|A(t)|^2$, in the bottom plots of both
Figs.~\ref{fig:circular_revivals_1} and \ref{fig:circular_revivals_2}, 
with time
`measured' in units of $T_0$. The almost exact
revival structure at integral multiples of $4T_0$ is evident. As a further
check, we can evaluate the phase of $A(t)$ at each revival and find that
to an excellent approximation it is given by $-nF\pi$, as in 
Eqn.~(\ref{extra_phase_1}). 

We next move away from the special case of the zero-momentum, central
wave packet by considering individually the case of $x_0 \neq 0$ and
 $p_{0y} \neq 0$ (but not both). In each case,  the average angular momentum
of the state is still vanishing, 
but $m\neq 0$ values of the expansion coefficients are now required.
We must now use the more general case for the energies, which to 
second order in $m\neq 0, n_r$, are given by Eqn.~({\ref{other_energies})
as
\begin{eqnarray}
E_{(m,n_r)} & = & \frac{\hbar^2 \pi^2}{2\mu R^2} 
\left[ \left(n_r + \frac{|m|}{2} + \frac{3}{4}\right)^2 
- \frac{m^2}{\pi^2} \right] \nonumber \\
& = & \frac{\hbar^2 \pi^2}{32 \mu R^2}
\left[ (16n_r^2 + 24n_r + 16|m|n_r) + 4|m|(|m|+3) - \frac{16 m^2}{\pi^2}
\right] \\
\label{approximate_energies}
& = & \frac{\hbar^2 \pi^2}{32 \mu R^2}
\left[ 8\tilde{l}(n_r) + 8\overline{l}(|m|) - \frac{16 m^2}{\pi^2} +9 \right]
\nonumber
\end{eqnarray}
where 
\begin{equation}
\tilde{l}(n_r) \equiv n_r(2n_r +3 +2|m|)
\qquad
\mbox{and}
\qquad
\overline{l}(|m|) \equiv |m|(|m|+3)/2
\end{equation}
are both integers, again, with no special even or oddness properties.
We can then write these energies in the form
\begin{equation}
E_{(m,n_r)} = \frac{2\pi \hbar}{4T_0}\left[\tilde{l}(n_r) + \overline{l}(|m|)
- \frac{2m^2}{\pi^2} + \frac{9}{8} \right]
\, . 
\end{equation}
At integral multiples of the $m=0$ revival time, $t_N = N(4T_0)$, 
the first two terms give $e^{-N(2\pi i)} = 1$ phases to each $(m,n_r)$ 
term in the autocorrelation function, while the last term gives an overall, 
$(m_,n_r)$-independent  phase,  just as in the $m=0$ case.  
The other term, however,  gives a contribution
\begin{equation}
e^{-(2\pi i) (m^2 N) (2/\pi^2)}
\end{equation}
which depends on $m$ explicitly and which therefore eliminates the
revivals, increasingly so, as the wave packet is dominated by $m \neq 0$
terms. However, because of a seeming numerical accident, at integral
multiples of $5T^{(m=0)}_{rev} = 20 T_0$, we recover approximate revivals
due to the fact that $5 \times (2/\pi^2) = 1.013$. We thus find approximate
revivals for the more general $m \neq 0$ case given by
$T_{rev}^{(m \neq 0)} = (\pi^2/2) T_{rev}^{(m=0)} \approx 5T_{rev}^{(m=(0))}$.

This effect is illustrated in more detail in 
Figs.~\ref{fig:circular_revivals_1}
and 
\ref{fig:circular_revivals_2}  where we
plot $|A(t)|^2$ versus $t$ as we move from the central, zero-momentum
wave packet by first moving away from the origin ($x_0 \neq 0$, 
in Fig.~\ref{fig:circular_revivals_1})
or having non-zero momentum values ($p_{0y} \neq 0$, 
in Fig.~\ref{fig:circular_revivals_2}.)
In each case, as we increase the parameter ($x_0$ or $p_{0y}$),
 we necessarily
include more and more $|m| \neq 0$ eigenstates. For even a small mix of
such states, the $T_{rev}^{(m=0)}$ revival periods at most integral
multiples of $4T_0$ disappear, while evidence for the more general
$T_{rev}^{(m \neq 0)} = 20 T_{0}$ revivals remains evident.

We note that this 'lifting' of a seemingly 'accidental' degeneracy in the 
pattern of revival times is somewhat similar to the special case of a 
zero-momentum Gaussian wave packet in a 2D square or equilateral triangular 
billiard, initially placed at the center.

This pattern of revival times depending on two distinct quantum numbers
is also somewhat reminiscent of that encountered in a rectangular billiard
with differing sides of length $L_x,L_y$ where if the sides are
incommensurate one would expect a less elaborate revival structure. Since the
revival times typically scale as $T_{rev} \propto L^2$, the appearance of a
$\pi^2$ scale factor which can  give rise to very close to  an integer ratio 
$10/\pi^2 \approx 1$ (to within $1.3\%$) is appropriate; in this case, 
the  relevant length scales for the radial quantum number and azimuthal
quantum numbers are most likely multiples of $R$ and  $2\pi R$ respectively,
so that relative factors of $\pi^2$ in the revival times can appear naturally. 

The presence of the $\Delta m \neq 0$ revivals becomes increasingly
less obvious as the average angular momentum is increased away from zero
(with both $x_0$ and $p_{0y}$ now non-vanishing), 
since the  required energy eigenvalues are in a region of large
$|m|/z$ where the lowest-order approximation (from Eqn.~(\ref{lowest_order}))
of evenly spaced $z$ values becomes worse. We also note that even with
$\langle \hat{L} \rangle = 0 $, as we increase  $p_{0y}$, the spread
in $m$ values required also increases, 
so that the overall number of states required to reproduce the initial
Gaussian, and which have to `beat' against each other appropriately, 
increases as well,
making revivals more difficult to produce.

\subsubsection{Variations on the circular billiard}
\label{subsubsec:circular_variations}

The discussion of the circular billiard can be easily extended to
the case of the half-circular footprint, with the addition of an
infinite wall along any diameter. The resulting energy eigenvalue
spectrum is obtained in a similar way as in the cases of the 
square and equilateral triangle billiards when `cut' along a diagonal.
The angular wavefunctions for the half-circle problem can be obtained
from the form in Eqn.~(\ref{alternate_angular_wavefunctions}) by 
choosing only the $\sin(m\theta)$ ($m>1$) forms 
which vanish on the new boundary.
Thus, the energy eigenvalue spectrum consists of one copy of the
$m\neq 0$ energies of the full well, very similar to the cases
encountered in Sec.~\ref{subsec:isoceles_billiard} and 
at the end of Sec.~\ref{subsec:equilateral_triangle}. 
However, because purely $m=0$ wave packets are  no longer possible, 
the existence of recognizable quantum revivals is less obvious 

While we have focused on the long-term revival structure of
quantum wave packets in the circular well, the short-term,
semi-classical propagation leading to closed orbits in this geometry
can also be studied using the same methods as in 
Secs.~\ref{subsec:2d_infinite_well}
 and
\ref{subsec:equilateral_triangle} 
 and we present such an analysis in Appendix~\ref{appendix:wkb_circular}.
The 2D annular billiard, where an additional
infinite wall at $r=R_{in} < R$ is added,  can be studied with the
same WKB methods used here, and we discuss that case in 
Appendix~\ref{appendix:wkb_circular} as well.

The problem of the spherical billiard, with an otherwise free
particle confined to a circular region of radius $R$, is an obvious
extension, with the (unnormalized) solutions in spherical coordinates 
given by
\begin{equation}
\psi(r,\theta,\phi) = j_{l}(k_{(l,n_r)} r)\, Y_{(l,m)}(\theta,\phi)
\end{equation}
and the quantized energies determined by the zeros of the regular 
{\it spherical} Bessel functions via
\begin{equation}
E_{(l,n_r)} = \frac{\hbar^2}{2\mu R^2} 
\left[z_{(l,n_r)})\right]^2
\qquad
\mbox{where}
\qquad
j_{l}(z_{(l,n_r)}) 
= 0
\, . 
\end{equation}
For such a central potential, the energy eigenvalues do not depend on
the $m$ quantum number, so the corresponding classical period
($T_{cl}^{(m)}$) and revival times ($T_{rev}^{(m)},T_{rev}^{(n_r,m)},
T_{rev}^{(l,m)})$ decouple from the problem. For the special case
of spherically symmetric ($l=0$) solutions, one has
\begin{equation}
j_{0}(z) \propto \frac{\sin(z)}{z}
\end{equation}
and the $l=0$ eigenvalues are exactly quadratic in the $n_r$ quantum
number; this implies that initially central (${\bf r}_0 = (0,0,0)$)
Gaussian wave packets with vanishing initial momentum
(${\bf p}_0 = (0,0,0)$) will exhibit exact revival behavior, with no
higher order time scales present, while all other packets, which
necessarily include  $l\neq 0$ components, will have less obvious 
revival behavior.

Finally, the addition of a single infinite wall along any radius
(not diameter) of the 2D circular billiard (adding a `baffle') can be 
analyzed in detail \cite{robinett_baffle}, with the result that 
half-integral values of the 2D angular momentum ($m = 1/2, 3/2, ...$) 
are allowed. Then, noting that the 2D ($J_{m}(z)$) and 3D ($j_{m}(z)$) 
Bessel functions are related by
\begin{equation}
j_{m}(z) = \sqrt{\frac{\pi}{2z}} \, J_{m+1/2}(z)
\, , 
\end{equation}
we see that wave packets constructed from only $m=1/2$ eigenstates in this
geometry will also exhibit purely quadratic dependence on the radial quantum 
number and also have exact revival behavior.

\section{Experimental realizations of wavepacket revivals}
\label{sec:experimental_results}

The existence of revival and fractional revival behavior in
quantum bound states, first found numerically in simulations of
Rydberg atoms \cite{parker}, has led to a number of experimental tests
in atomic, molecular, and other systems. We briefly review some of the
experimental evidence for quantum wave packet revivals, while noting that
excellent reviews of wave packet physics \cite{rydberg_review} 
- \cite{new_fielding}, \cite{averbukh_review} 
have appeared elsewhere. In addition, we discuss other
experimental realizations of quantum revivals found in the occupation
probability of a two-state atom system in a quantized electromagnetic
field, described by the Jaynes-Cummings model, and in the behavior of
the macroscopic wave function of Bose-Einstein condensates. We also discuss
revival-like behaviors which arise in various optical phenomena.

\subsection{Atomic systems}

As mentioned above, early formal studies \cite{lowell_brown} - \cite{ghosh}
of the construction of Coulomb wave packets were expanded upon by a number of 
authors \cite{yeazell_angular} - \cite{circular_1}, 
including suggestions for experimental production of localized
wave packets. Since excellent reviews \cite{rydberg_review} -
\cite{new_fielding}, \cite{averbukh_review} exist on this subject, 
we limit ourselves to a few comments.

The production of localized Rydberg wave packets makes use of the fact that
the large $n$ energy levels in such atoms are very closely spaced; for
example, the energy difference between successive Rydberg levels for
$n_0 = 50$ ($\Delta n = \pm 1$) is approximately $\Delta E_n = 
2 \times 10^{-3}\,eV$. Short-duration ($\Delta t \sim \tau_{pump}$) laser 
pulses necessarily have a range of energies ($\Delta E = \hbar/\Delta t$) 
and can therefore simultaneously excite a number of states (experiments have 
been done where wavepackets containing of order $3-10$ states have 
been produced, 
corresponding to roughly $\Delta n = 1.5-5$ in our notation.) 
One-electron-like atoms such as potassium and rubidium
have ionization potentials of order $4\,eV$, requiring laser wavelengths
of roughly this order to excite states to $E_n = -E_0/n_0^2 \leqnew 0$ 
for $n_0 >> 1$.
As examples, several classic experiments \cite{yeazell_2} (\cite{meacher}, 
\cite{wals}) used potassium (rubidium) atoms which have ionization potentials
given by $4.341\,eV$ ($4.177\,eV$) and produced excited Rydberg states with 
$n_0 = 65$ ($n_0 = 62$, $n_0 = 46.5$) using single photon excitation 
corresponding to 
laser wavelengths of $2858$ \AA \,($297\, nm$); an earlier experiment 
\cite{wolde} also used rubidium, but utilized two-photon excitation with 
$\lambda = 594\,nm$, and contained only $2-3$ component states.

The time-development of the wave packet can be subsequently probed with
a second laser pulse, as a function of the time delay from the first pulse.
A classical description \cite{yeazell_2} of the process involves noting 
that the rate of energy absorption from the laser pulse by the wave packet 
is given by 
$R(t) = {\bf J}(t) \cdot {\bf E}(t)$ where ${\bf J}(t) = e{\bf v}(t)$ 
is the classical electron current and ${\bf E}(t)$ is the time-dependent
electric field. This rate is large (small) near the inner (outer) Keplerian 
turning point, where the electron speed is big (small), so that when 
the electron is near the nucleus, energy from the `probe' laser can be 
efficiently absorbed, resulting in an increased probability of ionization
(which is the observable signal), a method first proposed by Alber, 
Ritsch, and Zoller \cite{generation}. Other experiments \cite{wals} have 
used different techniques \cite{ramsey_fringes} to monitor the dynamic 
behavior of the packets. 

For systems described by the Coulomb-like spectrum in 
Eqn.~(\ref{coulomb_energies}), such as large-$n$ Rydberg states where
$Z$ is effectively unity, the classical period will be given by
\begin{equation}
T_{cl}^{(Coul)}(n_0) = \frac{2\pi \hbar}{|E'(n_0)|}
=
(1.52\times 10^{-16} \,s) (n_0)^3
\approx 20\,ps \, (n_0/50)^3
\, . 
\end{equation}
and the corresponding revival time is then given by
\begin{equation}
T_{rev}^{(Coul)} = (2n_0/3) T_{cl}^{(Coul)}
\approx 
670 \,ps \, (n_0/50)^4
\, . 
\label{coulomb_revival_time}
\end{equation}
For example, the original numerical predictions of Parker and Stroud
\cite{parker} used an effective $n_0 = 85$ which gives
\begin{equation}
T_{cl} = 93.5\,ps
\qquad
\mbox{and}
\qquad
T_{rev} = 5.3\,ns
\end{equation}
so that the $p/q = 1/2$ revival shown in Fig.~\ref{fig:perelman}
is indeed observed to occur at roughly $T_{rev}/2 = 2.7\, ns$ as
seen in their simulations (where Averbukh and Perelman \cite{perelman_1}
added the appropriate labeling to the original data,
as in Fig.~\ref{fig:perelman}.)

In one of the early  experimental observations of fractional revivals
by Yeazell  and Stroud \cite{yeazell_fractional}, 
the initial wave packet was excited with $n_0 = 72$ (and including
roughly $5$ states, or $\Delta n \sim 2.5$) which gives
\begin{equation}
T_{cl} = 57\,ps
\qquad
\mbox{and}
\qquad
T_{rev} = 2.7 \,ns
\, .
\end{equation}
In the data from Ref.~\cite{yeazell_fractional}, reproduced here in
Fig.~\ref{fig:yeazell_stroud}, that  initial periodicity is clearly
visible, and a fractional revival, with local periodicity roughly
half that of $T_{cl}$, corresponding to a $p/q=1/4$ fractional revival 
at $T_{rev}/4 \approx 680\,ps$ is also apparent (note the two closely 
spaced vertical dashed lines); similar results were 
obtained by Meacher {\it et al.} \cite{meacher}.

Experimental data on rubidium from Ref.~\cite{wals}, shown in 
Fig.~\ref{fig:wals}, corresponding to $n_0 = 46.5$ and 
$\Delta n \approx 3$ ($6.5$ states), gives $T_{cl} = 15\,ps$, 
and a $p/q = 1/2$ revival is evident, with the initial local classical 
periodicity, 
at $T_{rev}/2 = 237\,ps$ (which is labeled there as $T_{rev}$.)  
The inset shows similar data from the same experiment, 
but obtained with $n_0 = 53.3$ and almost $10$ states excited in the 
wavepacket expansion, where fractional revivals up to order $T_{rev}/7$ 
are now visible, due in part to the larger value of $\Delta n$ used, 
and connections such as those in Eqn.~(\ref{resolve_features}). 
Observations of fractional revivals in two-electron atoms 
such as calcium \cite{calcium} have also been reported.

In contrast to most studies where localized electron wave packets
are excited by short optical pulses, it has also been shown possible to
generate Rydberg wave packets using THz frequency half cycle pulses
\cite{THz_wavepackets}. While the short-term Kepler orbit motion and 
longer-term revival behavior of packets produced in this way are comparable,
the THz wavepacket is initially delocalized and only becomes localized after
half a revival time.

States where the effects of the inner atomic core are important are often
described by  quantum defects, where the effective principal 
quantum number is given by $n^{*} = n - \delta(n,l)$,
 with an angular-momentum quantum number dependent correction 
characterizing the effect of the inner core electrons. The effects of
quantum defects on the structure of wave packet revivals has been examined in
Refs.~\cite{bluhm_defects} and \cite{other_defects}, while its effect
on the detailed recurrence spectrum has been documented \cite{wals}
experimentally.

The time-evolution  of localized  electron packets in an external electric 
field (Stark wave packets) \cite{alber_stark} has been studied experimentally
\cite{noordam_1} - \cite{bucksbaum}. The effect of the field on the
energy spectrum is most easily determined by solving the Schr\"{o}dinger
equation using parabolic
coordinates \cite{bethe_landau} where the principal quantum number, 
$n$, 
can be written in terms of two parabolic quantum numbers, $n_1,n_2$, 
and the azimuthal (magnetic) quantum number, $m$ in the form 
\begin{equation}
n = n_1+n_2+|m|+1
\, .
\label{parabolic_quanta}
\end{equation}
The linear (in applied field) correction to the energy spectrum due 
to an external
electric field of the form $V(z) = -eFz$ can be written 
\cite{bethe_landau} as 
\begin{equation}
E^{(1)}(k,n) = \frac{3}{2}nk[eFa_0]
\end{equation}
where $k\equiv n_1-n_2$ and $a_0 = \hbar^2/me^2$ is the Bohr radius.
For fixed $n$, the spacing between adjacent energy levels is given by
$\Delta E^{(1)}  = 3n[eFa_0]$, due to the form in 
Eqn.~(\ref{parabolic_quanta}) which implies $\Delta k = 2$ jumps,
which is therefore similar to the even spaced levels of an oscillator
spectrum. 
The classical periodicity associated with
the $k$ quantum number will then be given by Eqn.~(\ref{two_periods}), 
suitably modified to read,
\begin{equation}
T_{cl}^{(k)} = \frac{2\pi \hbar}{2|\partial E/\partial k|}
= \frac{2\pi \hbar}{3n[eFa_0]}
= 
\frac{2.6\,ps}{n[F/(100\,V/cm)]}
\label{stark_periods}
\end{equation}
and experimental observations of up to $10$ periods have been
reported \cite{noordam_1} - \cite{bucksbaum}. Since one is here concerned
with only the classical periodicity, the application of variations of
periodic orbit theory have also proved useful \cite{other_fielding} - 
\cite{wals_periodic_orbit}.
The dynamics of Stark wave packets above the field-ionization threshold
\cite{above_threshold} have also been measured experimentally, observing
that part of the electron wave function returns to the core,
before escaping over the saddle point, with classical periodicities
consistent with Eqn.~(\ref{stark_periods}). 

Novel Stark wave packets consisting of an $H^{+}-H^{-}$ ion pair
have been produced \cite{heavy_rydberg} in which the energy states
are scaled from the simple  hydrogen results by
\begin{equation}
\mu \sim m_e
\quad
\longrightarrow
\quad
\mu \sim \frac{M_p}{2}
\end{equation}
since it is now the nuclear motion which is relevant; the classical 
periodicity in Eqn.~(\ref{stark_periods}) is increased by roughly $3$ orders 
of magnitude and data for this system \cite{heavy_rydberg} show good,  
but not perfect, agreement with that prediction.

The longer-term,  revival structure of Stark wave packets, in the case 
where the time-dependent states contain components of both varying
$n$ and $k$ have been analyzed in Ref.~\cite{bluhm_stark}.  
The purely Coulomb revival time in
Eqn.~(\ref{coulomb_revival_time}) is unchanged, but there is a non-vanishing 
cross-revival time given by
\begin{equation}
T_{rev}^{(n,k)} = \frac{2\pi \hbar}{3[eFa_0]}
= nT_{cl}^{(k)}
\end{equation}
and the revival (and fractional revival) structure of Stark wave packets
differs in interesting ways from the free Rydberg case.

\subsection{Molecular systems}

The vibrational states of diatomic systems, described by anharmonic
potentials with unequally spaced energy levels, constitute another
physical system in which wave-packet excitations have been prepared.
The subject has been reviewed in Ref.~\cite{averbukh_review}, 
where it is rightly pointed out that the excitation of localized 
packets in such molecular systems can be easier because of the 
one-dimensional nature of the vibrational degree of freedom, 
versus the 3D nature of hydrogen-like atoms. Vibrational
wave packet motion has been seen in a number of systems,
including 
$I_2$ \cite{molecule_1}, \cite{molecule_1p},
$NaI$ \cite{molecule_2},
and 
$ICl$ \cite{molecule_3}, often with a large number of classical
periods apparent. 

For example, the recurrence of the semi-classical periodicity of a 
vibrational wave packet at a longer revival time in the sodium dimer 
$Na_2$ was exhibited in Ref.~\cite{baumert}, with a hierarchy of classical 
periodicity versus revival times given by $T_{cl} \sim 300\,fs$ and 
$T_{rev} \sim 47\,ps$. More recent experiments on $Br_2$ \cite{vrakking} 
have presented evidence for fractional revivals (of order $p/q=1/2$ and 
$1/4$) in such vibrational wave packets, as have experiments on $I_2$ 
\cite{stolow}. 

An especially illustrative method of visualizing the appearance of revivals 
(and fractional revivals) in such systems involves the calculation of a 
time-windowed Fourier transform power spectrum, or spectrogram, 
$S(\omega, \tau)$; this is  basically a two-dimensional map of frequency 
content of the packet versus time delay. The experimentally obtained time 
delay scan signal, $s(t)$, is convoluted with a window function, $g(t)$ to 
obtain
\begin{equation}
S(\omega,\tau) = \int_{0}^{\infty}\, s(t)\, g(t-\tau)\, e^{i\omega t}
\,dt
\label{spectrogram_definition}
\end{equation}
where the window function is often chosen to be an Gaussian of the form
$g(t) = \exp(-t^2/t_0^2)$; it  is typical to plot $\ln(|S(\omega,\tau)|^2)$
versus $(\omega,\tau)$. An example of such a plot (taken from 
Ref.~\cite{vrakking}) is shown in Fig.~\ref{fig:stolow_1}. The dark
regions near $\tau = 0$ and $\tau \approx 88\,ps$ and $f = 95\,cm^{-1}$
correspond to the initial wave packet and the $p/q=1/2$ revival, while
the features along $2f= 190\,cm^{-1}$ (i.e, at half the classical periodicity)
and structures at $3f$ show evidence of fractional revivals.

Revivals and fractional revivals in $NaK$ systems for
different isotope-selected samples \cite{heufelder} have shown evidence 
for fractional
revivals of up to order $p/q=1/6$, as well as being able to distinguish 
differing classical periods and revival times for different isotopes; 
isotope-selective studies of $K_2$ \cite{rutz_isotope} have also appeared.
A novel application of such vibrational wave packet revival behavior 
has been demonstrated in the laser separation of isotopes. Standard
methods of isotope separation \cite{standard_methods} involve 
gaseous diffusion and centrifugation which, in turn, rely on the
differences in isotopic masses. More modern laser separation
methods \cite{modern_methods} make use of the isotopic shifts in
various atomic or molecular lines. The excitation of vibrational
wave packets in diatomic molecules with differing isotopes can yield
quantum revival behavior which depends on the detailed structure of
the vibrational eigenfrequencies (and anharmonicities) which then 
determine the long-term free evolution and revival times. The difference
in revival time can yield spatial separation of the two species and this
effect has been demonstrated experimentally \cite{isotope_1}, \cite{isotope_2}
and patented \cite{isotope_3}.

Observations of revival behavior (up to order $p/q=1/8$) in triatomic
molecules have  been reported \cite{triatomic}, while other proposals 
\cite{meyer} to use differences in revival times 
in more complex molecules, ones with several vibrational degrees of freedom,
 have also appeared.

Evidence for coherence in the time-development of rotational wave packets
goes back to at least 1975 \cite{heritage} with observations of 
short-duration birefringence in $CS_2$ vapor with periodicity of order
$40\,ps$. The theoretical background and many experimental realizations of 
such molecular structures arising from coherent production of rotational
wave packets has been nicely reviewed by Felker and Zewail \cite{felker}.
Evidence for revival behavior in the rotational behavior of molecular 
wave packets has also been presented \cite{molecule_1p}, \cite{seideman} - 
\cite{other_vrakking} and reviewed \cite{other_pruna}, while
more recent experiments \cite{bartels}, \cite{comstock} have made use 
of the revivals in such molecular rotational wavepackets to manipulate 
the form (phase and spectral content) of ultrashort
laser pulses. The revival time is determined by the difference in energy
eigenvalues of the rotational states making up the coherent packet. Using
a standard notation for the rotational eigenvalues, $E_J = BJ(J+1)$, 
the relevant differences are given by
\begin{equation}
\Delta E_J = E_{J+N} - E_{J}
=
B[2JN +N(N+1)]
= 2B[JN +N(N+1)/2]
\label{rotational_eigenvalues}
\end{equation}
where two states differ by $N$ rotational quanta. For molecules with no
special symmetries, the $JN$ and $N(N+1)/2$ values are integers (neither
even nor odd) and the time to return to original state will be determined
by
\begin{equation}
\frac{\Delta E_{J}  T_{rev}}{\hbar} = 2\pi
\qquad
\mbox{or}
\qquad
T_{rev} = \frac{\pi \hbar}{B}
\end{equation}
with shorter revival times possible for molecules of specific symmetries
where only certain rotational states are allowed \cite{heritage}.
Predictions of cross-revivals \cite{cross_revivals} due to 
vibration-rotation coupling and discussions of the wavepacket dynamics of 
rotational quantum states of $C_{60}$ \cite{c_60} have also appeared.

\subsection{Jaynes-Cummings model}

One of the most frequently discussed fully quantum mechanical models 
of the interaction of a two-level atom with a single mode of the quantized 
electromagnetic field was proposed by Jaynes and Cummings 
\cite{jaynes_and_cummings}. The Hamiltonian for this system is given by
\begin{equation}
\hat{H} = \frac{\hbar \omega_0}{2} \hat{\sigma}_3
+ \hbar \omega \hat{a} \hat{a}^{\dagger}
+ \hbar \lambda \left( \hat{\sigma}_{+} \hat{a} + \hat{a}^{\dagger}
\hat{\sigma}_{-}\right)
\end{equation}
where $\hat{a}^{\dagger},\hat{a}$ are the raising and lowering operators 
for the boson field mode of frequency $\omega$, 
and the $\hat{\sigma}_{+},
\hat{\sigma}_{-},\hat{\sigma}_{3}$ are the Pauli matrices representing
the two-state system, with $\hbar \omega_0$ being the energy difference
between the two levels. Discussions of this system have routinely
appeared in reviews of the subject \cite{stenholm_review}
- \cite{barnett_review} and the model is known to have analogs 
in many other areas of physics.

Applied to a two-level atom in a resonant cavity with $n$ photons, 
the system will undergo Rabi oscillations with frequency given by
\begin{equation}
\Omega_n = (n\lambda^2 + \Delta^2/4)^{1/2}
\qquad
\mbox{with}
\qquad
\Delta = \omega_0 - \omega
\end{equation} 
which simplifies in the de-tuning limit when $\omega \longrightarrow 0$.
For that case, the probability that the system is in the excited state 
is given by
\begin{equation}
P_{e,n}(t) = \frac{1}{2} \left( 1 + \cos(2\sqrt{n}\lambda t) \right)
\end{equation}
while for a system with a distribution of photons, the solution is 
averaged over the initial probability distribution, $p_n$,  to give
\begin{equation}
P_{e}(t)  = \sum_{n=0}^{\infty} p_n\, P_{e,n}(t)
= 
\frac{1}{2} + \frac{1}{2}
\sum_{m=0}^{\infty} p_n\, \cos(\sqrt{n} \lambda t)
\, . 
\end{equation}
For an initial coherent state distribution, one has $p_n$ given
by the Poisson distribution, yielding 
\begin{equation}
P_{e}(t) = \frac{1}{2} + \frac{e^{-|\alpha|^2}}{2} \sum_{n=0}^{\infty}
\frac{|\alpha|^{2n}}{n!}\, \cos(2 \sqrt{n} \lambda  t)
\label{summation}
\end{equation}
and the average value of $n$ is given by $\overline{n} = |\alpha|^2$,
while the spread is $\Delta n = |\alpha|$.
This expression for $P_{e}(t)$ 
has many obvious similarities to the autocorrelation
function, $A(t)$,  for wave packets, namely highly localized expansion
coefficients (when $\overline{n} = |\alpha|^2 >> |\alpha| = \Delta n >> 1$) 
and oscillatory terms which are not purely harmonic,
 so it is not surprising that some  aspects of the short- and long-term 
behavior of Eqn.~(\ref{summation}) have features in common with $A(t)$.

The dynamics of this system exhibit Rabi oscillations with a frequency
centered at $\Omega_{\overline{n}}$, but with a dephasing given by a
Gaussian envelope, $\exp(-(\lambda t)^2/2)$, first derived 
in Refs.~\cite{cummings} and \cite{meystre},  and later improved  
upon \cite{eberly_1} for $\Delta \neq 0$.  For longer times, 
using familiar physical reasoning (or more formal expansions about
$n-\overline{n}$), one can see that at the time that neighboring terms
in Eqn.~(\ref{summation}) acquire a common $2\pi$ phase difference, 
one expects the Jaynes-Cummings 
summation to return to close to its $t=0$ behavior, exhibiting revivals,
this time due to the quantized nature of the electromagnetic field.
The condition for this to occur is 
\begin{equation}
2\Omega_{\overline{n}+1} T_{rev} - 2 \Omega_{\overline{n}}T_{rev}
\equiv \phi_{\overline{n}+1} - \phi_{\overline{n}} = 2\pi
\end{equation}
yielding a  revival time given by
\begin{equation}
T_{rev} = \frac{2\pi \sqrt{\lambda^2 \overline{n} + \Delta^2/4}}{\lambda^2}
\qquad
\longrightarrow
\qquad
\frac{2\pi \sqrt{\overline{n}}}{\lambda}
\qquad
\mbox{as $\Delta=0$} 
\, . 
\label{jaynes_revival_time}
\end{equation}
This time scale, given by how long it takes the phases from  consecutive 
frequencies to catch up to each other, is more akin to the classical
periodicity in Eqn.~(\ref{classical_period}) in the formalism we have 
used so far, but is routinely referred to as the revival time.

The result in Eqn.~(\ref{jaynes_revival_time}) was derived in 
numerical simulations by Eberly, Narozhny, and Sanchez-Mondragon 
\cite{eberly_1}, \cite{eberly_2}, but one should also note the much 
earlier work by J. Frahm \cite{frahm} who also demonstrated very similar
results in this system.
Eberly {\it et al.} also found evidence of subsequent revivals at integral 
multiples of $T_{rev}$, as well as accurate representations of the decreasing
magnitudes at $t = kT_{rev}$, and discussed as well the even longer-term 
approach to more irregular behavior, due to overlapping revivals 
\cite{eberly_3}. For example, they were able to demonstrate that the 
peak heights were bounded by the long-term limits (when $\Delta =0$) 
\begin{equation}
B(t) = \frac{1}{2} \pm 
\frac{1}{2} \frac{1}{(1 +\lambda^2 t^2/4\overline{n})^{1/4}}
\, . 
\label{suppression_factor}
\end{equation}

As an example of their solutions, we show in Fig.~\ref{fig:jaynes}
plots of $P_{e}(t)$ for $\overline{n} = 36$ and $\lambda = 0.01$.
We note the short-term Rabi oscillations, with Gaussian dephasing factor 
in Fig.~\ref{fig:jaynes}(a), as well as the presence of increasingly 
width revivals at integral multiples of $T_{rev}$ in 
Fig.~\ref{fig:jaynes}(b). (Plotted as a function of 
$\tau = \lambda t/\pi$, the revivals predicted by 
Eqn.~(\ref{jaynes_revival_time}) 
are expected at $\tau = k (2\sqrt{\overline{n}}) 
= 12k$ for this example.) The reappearance of the `classical' periodicity 
near the first two revivals ((c) and (d)) is also apparent. 
We note that there are  similarities in the spreading time formalism 
appearing in the envelope functions found by them with the formulae 
by Nauenberg \cite{nauenberg}: Fleischhauer and Schleich \cite{poisson}
have, in fact,  applied a Poisson summation approach to the evaluation
of Eqn.~(\ref{summation}) which makes this connection more apparent. 
The time-development of this system in phase space, in terms of 
the quasiprobability distribution of Cahill and Glauber 
\cite{cahill_and_glauber} (or Q-function), has been presented
by Eiselt and Risken \cite{eiselt_and_risken}.

The direct observation of these effects (with at least two obvious
revivals) was  first demonstrated  by
Rempe, Walther, and Klein \cite{rempe} using Rydberg atom states 
interacting with single field modes in a superconducting cavity 
constituting a one-atom maser \cite{one_atom_maser}. 
More detailed observations 
\cite{brune} have been able to extract discrete Fourier components 
of the Rabi oscillation time-dependence 
proportional to the square root of integers, as in 
Eqn.~(\ref{summation}),  providing very direct evidence of the
quantization of the electromagnetic field. (It should be noted that
Wright and Meystre \cite{wright_and_meystre} have examined in detail
the collapses and revivals in micromaser systems and found subtle
differences with those in the Jaynes-Cummings model.)

Theoretical work extending these results further to fractional 
revivals \cite{averbukh_jaynes}, \cite{gora_jaynes} and
to sub-Poisson photon distributions \cite{gora_jaynes} has also appeared,
as well as suggestions for using the revivals in the population inversion  
to measure novel phenomena \cite{cirac}; variations on these results
in an optical (as opposed to a microwave) cavity have also been discussed
\cite{quang}.
Studies have shown \cite{gea_jaynes} - \cite{moya}
that at half the revival time it is possible to obtain a 
entangled, Schr\"odinger-cat-like, atom-field state, similar in some ways
to the structure in Eqn.~(\ref{one_quarter_revival_function}) and
Fig.~\ref{fig:wigner_quarter}, but with no short-term periodicity 
(no Rabi oscillations), while extensions
to $N$-level atoms \cite{other_knight} have been observed to
give different behaviors.

Other similar models which are soluble in closed form have been
presented \cite{closed_form_1}, \cite{closed_form_2}, while realizations
of similar phenomena in the context of laser-driven excitation of
electronic transitions in diatomic molecules 
\cite{diatomic_revivals} have also been discussed.

Using harmonically bound ions, it has proved possible to create 
non-classical motional states of trapped atoms, including thermal, 
Fock, coherent and squeezed states of motion \cite{meekhof}, including
observations of the classical periodicity. The coupling between the
internal and motional states is in a regime which can be described by the
Jaynes-Cummings model, so that the evolution of the atomic state can
provide information on the number distribution of the motional state.

\subsection{Revivals in other systems}
\label{sec:other_systems}

\subsubsection{Atoms in optical lattices and Bose-Einstein condensates}
\label{subsubsec:bose_einstein_condensates}

Optical lattices, formed by the interference of multiple laser beams,
can act as local periodic potentials for atoms. Anharmonicities present
in the potential well can cause dispersive spreading of wave packets
formed in such systems, while dissipation can also arise from 
spontaneous emission, leading to loss of coherence. Localized 
coherent-state-like wave packets can be formed by suddenly shifting the
optical lattice and two groups
\cite{optical_1}, \cite{optical_2} have examined the interplay between
wave packet spreading, dissipation, and tunneling between adjacent wells
and their impact on wave packet revival times.

Motivated by the production of Bose-Einstein condensates, several
groups analyzed the time-dependence of the macroscopic wave function
\cite{bec_1} - \cite{bec_4}  and found periodic collapses and revival 
behavior. The energy of the external potential experienced by the atoms
could be modeled as being due to the interaction energy between the atoms,
giving an effective Hamiltonian of the form
\begin{equation}
\hat{H} = \frac{U_0\hat{n}(\hat{n}-1)}{2}
\label{bec_energies}
\end{equation}
where $\hat{n}$ is the number operator for atoms in the confining potential,
and $U_0$ is determined by the inter-particle interactions through the 
$s$-wave scattering length, $a$, and the ground state wavefunction, $w(x)$,
to be $U_0 = (4\pi \hbar^2 a/m) \int|w({\bf x})|^4 \, d{\bf x}$. 

In this case, the initial coherent state excitations can be of  the form
\begin{equation}
|\alpha \rangle_{t} = \left\{
e^{-|\alpha|^2/2}
\sum_{n=0}^{\infty} \frac{\alpha^n}{\sqrt{n!}}\, e^{-iU_0n(n-1)t/2\hbar}
\right\}
|n\rangle
\label{bec_coherent_state}
\end{equation}
where $|\alpha|^2 = N$ and $|\alpha| = \Delta N$; the time-dependence
has at most a quadratic dependence on $n$. Using by now
familiar techniques, we see that since $n(n-1)/2$ is an integer (neither
even nor odd), there is a common revival time given by 
$T_{rev} = 2\pi \hbar/U_0$ at which $|\alpha \rangle_{t+T_{rev}}
= |\alpha\rangle_{t}$. This effect was confirmed experimentally for a 
Bose-Einstein condensate \cite{bec_revivals} confined to a three-dimensional 
optical lattice, where the collapse and (approximate) revivals of the number
of atoms in the coherent state was monitored.

We note that the macroscopic
matter field, $\psi = \langle \alpha_{t}|\hat{a} | \alpha \rangle_{t}$, can
be written in the form \cite{bec_first} 
\begin{equation}
\psi(t) = \alpha 
\exp
\left\{ -|\alpha|^2
\left[
\left(1-\cos\left(\frac{2\pi t}{T_{rev}}\right)
 +
i\sin\left(\frac{2\pi t}{T_{rev}}\right)\right)
\right]
\right\}
\label{bec_dynamic}
\end{equation}
which is very similar in form to the autocorrelation function,
$A(t)$, in Eqn.~(\ref{sho_autocorrelation}) 
for the minimum-uncertainty, coherent-state like solution of the 
harmonic oscillator.
The collapse time for this system can also be derived 
\cite{bec_first}, \cite{bec_1} - \cite{bec_4} from the short-term
time-dependence of Eqn.~(\ref{bec_dynamic}) giving $T_{coll} \propto
T_{rev}/|\alpha| = T_{rev}/\Delta N$. 

The dynamical behavior of $|\alpha\rangle_{t}$ can also be examined
in the context of fractional revival 
behavior by evaluating the time-dependence
(especially the interplay between the $n$ and $n^2$ terms in the 
exponentials) of each term  at $p/q$ multiples of $T_{rev}$. For example, 
at $T_{rev}/2$, each term in the coherent state expansion contains a term 
of the form
\begin{eqnarray}
e^{i\pi n/2}\, e^{-i\pi n^2/2}
& = & 
e^{i\pi n/2}
\left[\frac{1}{\sqrt{2}}
\left(e^{-i\pi/4} + e^{+i\pi/4}
e^{-in\pi}\right) \right] \nonumber \\
& = & 
\frac{1}{\sqrt{2}} \left( e^{-i\pi/4} e^{i\pi n/2} + 
e^{+i\pi/4} e^{-i\pi n/2} \right)
\, , 
\end{eqnarray}
using the expansion in Eqn.~(\ref{second_trick}). When used in the
evaluation of $|\alpha\rangle_{t}$ one finds
\begin{equation}
|\alpha \rangle_{t=T_{rev}/2} = \frac{1}{\sqrt{2}}
\left( e^{-i\pi/4} |e^{+i\pi/2}\alpha\rangle
    + e^{+i\pi/4} |e^{-i\pi/2}\alpha\rangle
\right)
\label{bec_cat_state}
\end{equation}
which is another Schr\"{o}dinger-cat like superposition, with two 
distinct states rotated by $90^{\circ}$ from the initial $|\alpha\rangle$
state.  The structure of these coherent states are visualized in a standard 
way by defining the overlap of $|\alpha\rangle_{t}$ with an arbitrary 
coherent state, $|\beta\rangle$,  via
\begin{equation}
\langle \beta | \alpha\rangle_{t}
= e^{-|\alpha|^2/2}\, e^{-|\beta|^2/2}
\sum_{n=0}^{\infty} \frac{\alpha^n (\beta^{*})^n}{n!} 
e^{-i\pi n(n-1)t/T_{rev}}
\end{equation}
and then plotting
\begin{equation}
P[Re(\beta),Im(\beta);t]= 
P(\beta;t) \equiv |\langle \beta |\alpha\rangle_{t}|^2
\label{beta_plot}
\end{equation}
and several examples are shown in Fig.~\ref{fig:coherent_states}; for
example, the case of Fig.~\ref{fig:coherent_states}(c) corresponds to
the state in Eqn.~(\ref{bec_cat_state}). Related ways of visualizing 
such coherence phenomena include the use of Wigner quasi-probability 
distributions (as in Ref.~\cite{perelman_1} for fractional wave
packet revivals or Figs~\ref{fig:wigner_third} and 
\ref{fig:wigner_quarter})  
and the so-called $Q$ function \cite{schleich_book}, 
\cite{scully_quantum_optics} in quantum optics.

\subsubsection{Revivals and fractional revivals in optical systems}
\label{subsubsec:optical_analogs}

The quantized structure of important physical properties (energy eigenvalues 
for quantum wave packets, or the quantized EM field for Jaynes-Cummings 
systems) are not limited to purely quantum phenomena.
The existence of full revivals, mirror revivals (with a reformation of
the original coherence, but out of phase), and fractional revivals,
due to the discrete nature of classical wave systems, has been observed 
experimentally in several optical phenomena. In most cases, the
observation of revivals in the spatial distributions of light is referred
to as {\it self-imaging} (for an excellent review of this general topic, 
see Ref.~\cite{patorski}) and we will focus on two examples, the 
Talbot effect and self-imaging in waveguides, as they are the most
analogous to wave packet revivals in their mathematical structure and 
analyses.

In 1836 H. F. Talbot \cite{talbot} (a co-inventor of photography, with
Daguerre) illuminated a diffraction grating
with a small (coherent) white light source and examined the resulting 
transmitted light with a magnifying glass. He noted recurring patterns of 
colored bands, repeating themselves as the lens was moved further away from 
the grating. Rayleigh \cite{rayleigh} correctly interpreted these as 
resulting from the interference of the diffracted beams, in what would now 
be called the near-field regime. Subsequent work showed that the 
self-images of the original periodic structure, illuminated in plane wave
approximation, with monochromatic light, would appear at multiples of a 
distance $L_{T} = 2a^2/\lambda$ where $a$ is the grating distance and 
$\lambda$ is the wavelength. The detailed mathematical analysis showed
that the same pattern would also recur at $z = L_{T}/2$, but shifted in
space by half the spatial periodicity (the analog of a mirror revival.)
Higher order fractional revivals, at rational multiples of $L_{T}$
($z = pL_{T}/q$) consist of  $q$ copies of the original grating, separated
by $a/q$. In the context of self-imaging, the revivals (and mirror revivals)
are often called Fourier images, while the fractional versions are referred
to as Fresnel images \cite{fresnel}. The clear physical and mathematical 
analogies to the structure of wave packet revivals, and their intricate 
dependence on number theoretic identities, has been extensively discussed 
\cite{berry_fractional} - \cite{carpets_of_light} and the Talbot effect 
\cite{factorize_1}, along with corresponding effects in wave packet revivals 
\cite{schleich_factorize_numbers} \cite{harter}, have been put forward 
as a novel way to factorize numbers.

The Talbot effect is not limited to purely electromagnetic waves, but has
also been discussed for matter waves in atom optics \cite{rohwedder}
and observed experimentally with relatively light atoms 
\cite{atomic_talbot_1} - \cite{atomic_talbot_3} as well as with 
large molecules such as $C_{60}$ \cite{atomic_talbot_4}.
Fractional revivals, i.e., higher order Talbot fringes, of up to
$7$-th order have been observed \cite{atomic_talbot_3}, which is 
coincidently quite similar to the fractional revival resolution obtained 
with the best wave packet studies \cite{wals}.

The formation of images by {\it phase coincidences} in optical
waveguides is another example of such self-imaging processes. The
prediction that real optical images could form in planar optical
waveguides was made by Bryngdahl \cite{first_waveguide}, was then
demonstrated in a series of elegant experiments by Ulrich and 
collaborators \cite{ulrich_waveguide} and resulted in several
patents \cite{talbot_patents}.
For this system, the recurrence length is given by $L = 4n_f W_a^2/\lambda$ 
where $W_a$ is the effective aperture size in this slab geometry (including 
penetration depth effects) and $n_F$ is the index of refraction of the 
dielectric slab. Multiple/fractional image formation \cite{ulrich_waveguide},
\cite{metal_waveguide} has also been observed, and the phenomenon 
has also been demonstrated  with $x$-ray waveguides \cite{xray}. 
The planar waveguide geometry corresponds to the one-dimensional infinite 
well, with $W$ corresponding to the well size, and the length along the
waveguide corresponding to time. 
The ray tracing visualization of the intensity patterns in this geometry 
leading to fractional revivals (see especially Ref.~\cite{heaton} for a 
nice example) are therefore quite similar to the semi-classical pictures 
discussed by Born \cite{born}, \cite{born_2} for trajectories in the 
infinite square well, leading to quantum carpets. 

The  application of this effect for use
in novel optical beam splitters has been advocated \cite{heaton}.
The generalization of this self-imaging
method to two-dimensional waveguides of rectangular cross-sections
\cite{ulrich_other} has also been demonstrated and fiber optic geometries
of equilateral triangle and hexagonal cross-sections 
(as in Sec.~{\ref{subsec:equilateral_triangle}), have also been discussed.

Finally, we note that 
while observations of wavepacket revival behavior has been mainly 
limited to atomic and molecular systems, the possible use of localized 
low-energy 
wave packet excitations \cite{edge} localized at the edge of a 2D electron 
system (edge magnetoplasmons) to probe the dynamical properties of the 
integer and quantum Hall effects has also been proposed.

The study of classical and quantum mechanical systems subject to
external periodic forces, focusing on issues related to chaotic behavior,
has a rich literature. A number of such studies have focused on the
appearance of revival-type behaviors \cite{driven_gravity}, 
\cite{saif_driven}, \cite{forced_1} - \cite{forced_4} in such systems.

\section{Discussion and conclusions}
\label{sec:discussion}

The connections between the energy eigenvalue spectrum of a quantum
bound state system and the classical periodicities of the system have 
been a standard subject in quantum theory since the first
discussions of the correspondence principle by Bohr. 

Some semiclassical techniques, such as periodic orbit theory, can 
connect the quantum energy spectrum with classical closed or periodic orbits,
but often do so in a way  which does not exhibit the time-dependence
of quantum wave packets. Truly dynamic  observations of localized
quantum wave packets, exhibiting the classical periodicities expected
in such semi-classical limits,  have become possible with observations
in atomic, molecular, and other systems, with analog behaviors seen in
atom-field, BEC, optical, and other systems as well. 
The medium-term collapse of localized quantum states also present in such
systems is familiar from simple examples of spreading wave packets which 
were constructed during the early days of the development of quantum
theory. It can be analyzed quite generally for such
bound state systems and has been observed  in a number of systems as well.

The truly novel observation that such wave packets can relocalize  and
once again exhibit the classical periodicity, first observed in numerical
simulations,  has now been widely confirmed by experiments in a large number 
of different physical systems including atomic (Coulomb and Stark effect) 
systems using both electronic and nuclear states, and vibrational and 
rotational states in molecules. Besides being a fundamental realization of 
the discrete nature of quantum bound states, the simple time-dependence
of such eigenstates, and interference effects, the 
phenomenon of quantum packet revivals has been increasingly used in
the development of modern quantum control experiments,  to assist in 
the shaping of specific quantum states, and is likely to remain an important
aspect of the production of specific target states in the future. 
Given the relative simplicity of the quantum mechanical background which is 
needed to understand many aspects of these effects, and the connections to
a wide variety of other revival phenomena (in optics and elsewhere), it is 
likely that this important manifestation of the time-development of 
quantum mechanical bound states could easily find a place in the 
undergraduate and graduate curricula, and this review can be seen as one 
step towards that end.

\vskip 2cm

\noindent
{\bf Acknowledgments}

We thank M. Doncheski for fruitful and enjoyable collaborations on
many projects. We are very grateful to I. Averbukh, M. Belloni,
R. Bluhm, H. Fielding,
A. Kosteleck\'{y}, I. Marzoli, W. Schleich, C. Stroud, 
W. van der Zande, and D. Villeneuve for helpful comments and communications.
Some of the original work of 
the author cited here was supported, in part, by the National Science 
Foundation under Grant DUE-9950702.

\appendix
\section{Energy eigenvalues in power law potentials}
\label{appendix:power_law_wkb}

The energy eigenvalue spectrum in the family of power-law potentials
\cite{sukhatme} -- \cite{robinett_power_law} given by
\begin{equation}
V_{(k)}(x) \equiv V_0 \left|\frac{x}{L}\right|^k
\label{appendix_power_law_potential}
\end{equation}
can be obtained for large $n$ (which is the situation encountered in
wave packet revivals) by the use of the WKB approximation. In this
case one writes
\begin{equation}
\int_{-x_0}^{+x_0} \sqrt{2m(E_n^{(k)} - V_{(k)}(x)}) \, dx
= \left(n + C_L + C_R\right) \hbar \pi
\end{equation}
where
\begin{equation}
\pm x_0 = \pm \left(\frac{E_n}{V_0}\right)^{1/k}L
\end{equation}
are the classical turning points.
The matching coefficients have been discussed in detail in 
Ref.~\cite{wkb_approximation} and are given by $C_L, C_R = 1/4$ for smooth 
potentials, but with $C_L, C_R \rightarrow 1/2$ for the case of
infinite walls (i.e., in the limit when $k \rightarrow \infty$).
The integrals can be done using standard handbook results to give
\begin{eqnarray}
E_n^{(k)} & = &
\left[(n+C_L+C_R) \frac{\hbar \pi}{2L\sqrt{2m}}
V_0^{1/k}
\frac{\Gamma(1/k+3/2)}{\Gamma(1/k+1) \Gamma(3/2)}
\right]^{2k/(k+2)} \nonumber \\
& = & 
{\cal E}^{(k)} (n+C_L+C_R)^{2k/(k+2)} 
\end{eqnarray}
which does reproduce both the oscillator ($k=2$) and infinite well
$k \rightarrow \infty$) limits, for large $n$, 
and even gives the appropriate scaling 
for the Coulomb ($k=-1$) case. The same approach can also be used for
`half' well potentials, where an infinite barrier at $x=0$ is introduced,
as for the case of the `quantum bouncer' 
in Sec.~\ref{subsec:quantum_bouncer}, for appropriate values of $C_L,C_R$.

\section{General time scales}
\label{appendix:time_scales}

The general expansion in Eqn.~(\ref{expansion_in_time}) defines three time 
scales, namely the classical periodicity, the revival, and superrevival 
times, all of which we have discussed in detail.
In this Appendix, we describe two other relevant time scales, as well
as another quite general approach to visualizing the short-term classical
periodicity and subsequent spreading. 

Nauenberg \cite{nauenberg} noted that,  for the Gaussian coefficients in 
Eqn.~(\ref{gaussian_components}),
the autocorrelation function could be written (in the notation used here)
as 
\begin{equation}
A(t) = \langle \psi_t | \psi_0 \rangle
= \sum_{n=0}^{^\infty}
\left[\frac{e^{-(n-n_0)^2/2\Delta n^2}}{\Delta n \sqrt{2\pi}}
\right]
e^{i[E(n_0)t + E'(n_0)(n-n_0)t + E''(n_0)(n-n_0)^2/2]/\hbar}
\end{equation}
to second order in $(n-n_0)$. Using the Poisson formula,
\begin{equation}
\sum_{q=0}^{\infty}
\, f(q) 
= \sum_{m=-\infty}^{+\infty}\, 
\left[ \int_{0}^{+\infty}
\, f(q)\, e^{2\pi i qm}\, dq \right]
+ \frac{f_0}{2}
\, , 
\end{equation}
he found he could write $A(t)$ in the suggestive form
\begin{equation}
A(t)
= \frac{e^{iE(n_0)t/\hbar}}{2\pi \Delta n
\sqrt{\alpha(t)}} \sum_{m=-\infty}^{+\infty}
e^{-(m-t/T_{cl})^2/2 \alpha(t)}
\label{nauenberg_formula}
\end{equation}
where
\begin{equation}
\alpha(t) \equiv \frac{1}{4\pi^2}\left(\frac{1}{(\Delta n)^2}
+ \frac{4\pi i t}{T_{rev}}\right)
\end{equation}
and $T_{cl}$ and $T_{rev}$ are defined, in the notation used here,
 by Eqns.~(\ref{classical_period}) and (\ref{quantum_revival}).
This form exhibits, in a quite general way, the periodicity of the
autocorrelation function, with the summation of exponentials centered
at integral multiples of $T_{cl}$. It also exhibits the spreading (due
to dispersion) apparent in the time-dependent width given by
\begin{equation}
\Delta (t) = 
\frac{1}{2\pi \Delta n}
\left( 1 + \left(\frac{4\pi \Delta n^2 t}{T_{rev}}\right)^2\right)^{1/2}
= 
\frac{1}{2\pi \Delta n}
\left(1+\left(\frac{t}{T_s}\right)^2\right)^{1/2}
\end{equation}
where
\begin{equation}
T_s \equiv  \frac{T_{rev}}{4\pi (\Delta n)^2}
\, .
\end{equation}

This type of spreading is similar to that due to the pre-factor terms in
Eqns.~(\ref{free_particle_autocorrelation}) and 
(\ref{accelerating_autocorrelation}),
which were written in terms of the familiar
spreading times, $t_0$, in the free-particle (infinite well) or
uniform acceleration (quantum bouncer) cases. In order to exhibit this
connection more generally, we use the power-law potential in 
Eqn.~(\ref{appendix_power_law_potential})
and the results in Appendix~\ref{appendix:power_law_wkb}. 
For example, we can write
the quantized energies (in the large $n$ limit) for the general 
$V_{(k)}(x)$ potential in the form
\begin{equation}
\frac{p_n^2}{2m} = E_{n}^{(k)} = {\cal E}^{(k)} \, n^{2k/(k+2)}
\label{energy_momentum_identification}
\end{equation}
which we also equate to the maximum value of momentum, $p_n$.
In these systems, we have 
\begin{equation}
T_{rev}^{(k)} = \frac{4\pi \hbar}{{\cal E}^{(k)}} n^{4/(k+2)} 
\left|\frac{(k+2)^2}{2k(k-2)} \right|
\, . 
\end{equation}
For an initial Gaussian wave packet of width $\Delta p_n$, 
the spreading time can be written in the form
\begin{equation}
t_0^{(k)} = \frac{m\hbar}{2\Delta p_n^2}
\end{equation}
and we can use the identification  
in Eqn.~(\ref{energy_momentum_identification})
to equate
\begin{equation}
\Delta p_n = \sqrt{2m{\cal E}^{(k)}} 
\, n^{-2/(k+2)} \left(\frac{k}{k+2}\right)
\Delta n
\end{equation}
which gives
\begin{equation}
t_0^{(k)} = \left(\frac{\hbar}{4 {\cal E}^{(k)}}\right)
 \frac{n^{4/(k+2)}}{(\Delta n)^2}
\left(\frac{k+2}{k}\right)^2
\, .
\end{equation}
Thus, for this family of potentials, we have the relation
\begin{equation}
\frac{T_{rev}^{(k)}}{t_0^{(k)}} = 8\pi \left|\frac{k}{k-2}\right| (\Delta n)^2
\end{equation}
and this ratio is equal to $8\pi (\Delta n)^2$ for both $k=1$
(quantum bouncer, uniform acceleration) and $k \rightarrow \infty$
(infinite well, free particle). We then note that the spreading time,
$T_{s}$, given by the Nauenberg formula for these cases, 
is calculated to be
\begin{equation}
T_{s}^{(k)} = \frac{T_{rev}^{(k)}}{4\pi (\Delta n)^2}
= 2t_0
\end{equation}
which agrees with the results in Eqn.~(\ref{free_particle_autocorrelation}) 
and (\ref{accelerating_autocorrelation}).

Nauenberg also pointed out that the wavepacket spreads sufficiently that
{\it quantum self-interference} occurs on a time scale of (in our 
notation) $T_{rev}/\Delta n$, marking the approach the collapsed state.
The notation of a {\it collapse time} has also been studied in the
context of the infinite \cite{robinett_infinite_well}
well and the quantum bouncer \cite{quantum_bouncer}, with quite 
different approaches. For the infinite well, the time required for 
the various expectation values to approach their semi-classical limits
was given approximately by Eqn.~(\ref{original_infinite_well_collapse_time})
which can be written in the form
\begin{equation}
T_{coll} = \frac{T_{rev}}{4\sqrt{12} \Delta n}
\label{infinite_well_collapse_time}
\end{equation}
while for the quantum bouncer, a direct examination of the phase differences
between various terms in the eigenstate expansion led to the result
\begin{equation}
T_{coll} \sim \frac{T_{rev}}{(8/\pi) \Delta n}
\label{quantum_bouncer_collapse_time}
\end{equation}
(where we use our notation for $T_{rev}$,  since the one described in
Eqn.~(42) of Ref.~\cite{quantum_bouncer} is more rightly associated with the
$p/q = 1/2$ revival in the language of Averbukh and Perelman.)
A third, quite general measure of the time taken to collapse to the
incoherent state would be to use the spreading ({\it dispersive}) pre-factors
in Eqn.~(\ref{nauenberg_formula}),  or the similar ones in 
Eqns.~(\ref{free_particle_autocorrelation})
and (\ref{accelerating_autocorrelation}),  and define the collapse time
as how long it takes for the envelope of $|A(t)|^2$ to decrease 
to $|A_{inc}|^2$, namely when 
\begin{equation}
\frac{2t_0}{T_{coll}}
\approx 
\frac{1}{\sqrt{1+(T_{coll}/2t_0)^2}}
= |A(T_{coll})|^2 \equiv |A_{inc}|^2 = \frac{1}{\Delta n 2\sqrt{\pi}}
\end{equation}
which gives
\begin{equation}
T_{coll} = \left(4\sqrt{\pi} \Delta n\right) t_0
= \frac{T_{rev}}{(2\sqrt{\pi}) \Delta n}\left(\frac{k-2}{k}\right)
\end{equation}
which gives  the same general form as the explicit analyses resulting in
Eqns.~(\ref{infinite_well_collapse_time}) and
(\ref{quantum_bouncer_collapse_time}).
 We note that the earliest time at which a 
possible  fractional revival which might be resolvable (over the 
incoherent background) is given by the condition
\begin{equation}
\left|A\left(\frac{T_{rev}}{q}\right)\right|^2 = \frac{1}{q} 
= \frac{1}{\Delta n 2\sqrt{\pi}}
\qquad
\mbox{at the time}
\qquad
T_{early} = \frac{T_{rev}}{q} = \frac{T_{rev}}{\Delta n 2\sqrt{\pi}}
\end{equation}
so that this time scale is also singled out as one `comes down' in time
from $T_{rev}$, as well as from 'going up' in time from $t=0$.
We note that Fleischhauer and Schleich \cite{poisson} made similar use of
the Poisson summation formula to obtain improved approximate expressions
for the Jaynes-Cummings sum in Eqn.~(\ref{summation}).

Finally, the presence of the $T_{rev}$ term provides the dispersion
necessary for the decay to the collapsed state. For the harmonic oscillator
system, when $T_{rev} \rightarrow \infty$, we would expect to reproduce
something like the result of Eqn.~(\ref{sho_autocorrelation}). Using
Eqn.~(\ref{nauenberg_formula}), with $T_{rev} \rightarrow \infty$, we
find that
\begin{equation}
A(t) = e^{iE(n_0)t/\hbar}
\sum_{m=-\infty}^{+\infty}
e^{-(m-t/T_{cl})^2 (2\pi \Delta n)^2/2}
\, .
\end{equation}
Near $t\approx 0$, this reduces to
\begin{equation}
A(t) = e^{iE(n_0)t/\hbar} e^{- (\Delta n)^2 \omega^2 t^2/2}
\label{compare_autos}
\end{equation}
since $2\pi/T_{cl} = \omega$. The exact result in 
Eqn.~(\ref{sho_autocorrelation}) was derived for $\beta = \beta_0$,
but arbitrary $x_0,p_0$, and using Eqns.~(\ref{sho_total_energy})
(to evaluate $\langle E \rangle$) and (\ref{sho_deltan})
(to evaluate $\Delta n$), we find that the oscillator result
for $\omega t << 1$ can be written in the form 
\begin{equation}
A(t) = e^{+i\langle E \rangle t/\hbar}
e^{-(\Delta n)^2 (\omega t)^2/2}
\end{equation}
which agrees with Eqn.~(\ref{compare_autos}) in that limit. More
generally,  the two formulae agree very well for $|t-kT_{cl}|<<T_{cl}$,
i.e., near any multiple of the classical period.

\section{Fractional ($p/q$) revivals for even $q$}
\label{appendix:fractional_revivals}

The derivations of Averbukh and Perelman \cite{perelman_1}
for the temporal structure of fractional wave packet
revivals at $t=pT_{rev}/q$ for odd values of $q$ was discussed in 
Sec.~\ref{subsec:revivals_and_fractional} and we extend the analysis
here. 
For case of even $q$ values (implying $p$ is odd, since $p,q$ are assumed to
be relatively prime), the periodicity in $l$ required to satisfy
Eqn.~(\ref{periodic_condition}) is given by
\begin{eqnarray}
l = q   & \qquad & \mbox{for $q=2 \,\,\,(mod \,4)$} \\
l = q/2 & \qquad & \mbox{for $q=4 \,\,\,(mod \,4)$} 
\end{eqnarray}
and we can treat each case separately. 

For the first case, we note that the recursion relation in 
Eqn.~(\ref{recursion_relation}) 
will connect $b_r$ values with even and odd $r$ separately. Since
$q=2$ ($mod \, 4$), $q/2$ will be an odd integer and we can rewrite
the expression for $b_0$,  using the relabeling $\overline{n} = n + q/2$,
to write
\begin{eqnarray}
b_0 & = & \frac{1}{l} \sum_{n} e^{-2\pi i p n^2/q} \nonumber \\ 
& =  & 
\frac{1}{l} \sum_{\overline{n}} e^{-2\pi i p (\overline{n}+q/2)^2/q} 
\nonumber \\
& = &
\frac{1}{l} \sum_{\overline{n}} 
e^{-2\pi i p \overline{n}^2/q}
\, e^{-2\pi i p \overline{n}}
\, e^{-\pi i p q/2}  \\  
& = & 
- \frac{1}{l} \sum_{\overline{n}} e^{-2\pi i p \overline{n}^2/q} 
\nonumber \\
& = & -b_0 \nonumber
\end{eqnarray}
since
\begin{equation}
e^{-2\pi i p \overline{n}} = 1
\qquad
\mbox{and}
\qquad
e^{-\pi i p q/2} 
= \left[ e^{-\pi i q/2}\right]^p
= (-1)^{p}
\end{equation}
because $q/2$ is an odd integer and $p$ is necessarily odd if $q$ is even.
Since $b_0$ vanishes, all of the even $b_r$ do as well, and only the
$q/2$ $b_r$ values with odd $r$ are non-vanishing, leading to $q/2$
`clones' or 'mini-packets' near the fractional revival time in these cases.

An explicit example is for the half-revival time, $T_{rev}/2$ where $p=1$, 
$q=2$, and $l=q/2=1$ where one can explicitly find that
\begin{eqnarray}
b_0 & = & \frac{1}{2}\sum_{k=0}^{1} e^{-2\pi ik^2/2} = \frac{1}{2}(1-1) = 0 \\
& \mbox{and} & \nonumber \\
b_1 & =  &\frac{1}{2}\sum_{k=0}^{1} e^{2\pi i (k/2-k^2/2)} = 
\frac{1}{2}(1+1) = 1 
\end{eqnarray}
and we obtain the result of Eqn.~(\ref{one_half_revival_function}). 
The next case in this series is for multiples of $T_{rev}/6$,  which 
turn out to be similar to the $p/q = 1/3,2/3$ cases.

Finally, for the case of $q=4\, (mod\, 4)$, the periodicity in 
Eqn.~(\ref{periodic_condition})
is given by $l=q/2$ and the analysis proceeds in a similar fashion. 
For example, for the case of $T_{rev}/4$, we have $p=1$, $q=4$, $l=q/2=2$, 
and the various expansion coefficients in Eqn.~(\ref{fourier_expansion}) 
are
\begin{eqnarray}
b_0 & = &\frac{1}{2} \sum_{0}^{1} e^{-2\pi i  k^2/4}
= \frac{1}{2}(1-i) = \frac{1}{\sqrt{2}} e^{-i\pi /4} \nonumber \\
b_1 & = &\frac{1}{2} \sum_{0}^{1} e^{2\pi i (k/4- k^2/4)}
= \frac{1}{2}(1-1) = 0 \\ 
b_2 & = & \frac{1}{2} \sum_{0}^{1} e^{2\pi i (2k/4- k^2/4)}
= \frac{1}{2}(1+i) = \frac{1}{\sqrt{2}} e^{+i\pi /4} \nonumber 
\end{eqnarray}
which explains the result of Eqn.~(\ref{one_quarter_revival_function}) 
in a more `turnkey' fashion.

We note that Bluhm and Kosteleck\~{y} \cite{bluhm_long_term} have
obtained similar results for
the algebra of the complex phases arising from the {\it superrevival}
terms of the form $\exp(-2\pi i p k^3/q)$. They have also extended
this  formalism of fractional revival analysis to the case of systems 
with two or more quantum numbers \cite{bluhm_2d}.

\section{The `inverted' oscillator}
\label{appendix:inverted_oscillator}

For the case of the so-called `inverted' oscillator, 
the general wave packet solution in Eqn.~(\ref{sho_general_case}), for
example, can be directly carried over using the identifications in
Eqn.~(\ref{runaway_identifications}) to obtain the `runaway' wavepacket,
with probability density given by
\begin{equation}
|\psi(x,t)|^2
= \frac{1}{\sqrt{\pi} |B(t)|}
\exp\left[
-\frac{(x-x_0\cosh(\tilde{\omega} t) 
- p_0 \sinh(\tilde{\omega} t)/m\tilde{\omega})^2}{|B(t)|^2}
\right]
\end{equation}
with
\begin{equation}
\langle x \rangle_t = x_0 \cosh(\tilde{\omega} t) + 
\frac{p_0 \sinh(\tilde{\omega} t)}{m\tilde{\omega}}
\qquad
\quad
\mbox{and}
\qquad
\quad
\Delta x_t = \frac{|B(t)|}{\sqrt{2}}
\end{equation}
where
\begin{equation}
|B(t)| = \sqrt{\beta^2 \cosh^2(\tilde{\omega} t) 
+ (\hbar/m\tilde{\omega} \beta)^2
\sinh^2(\tilde{\omega} t)}
\,.
\end{equation}
Just as for the harmonic oscillator, 
the expression for $A(t)$ for the general case is cumbersome,
so we only examine it for one specific case as an example, namely the
case where $\beta = \beta_0 = \sqrt{h/m\tilde{\omega}}$. 
This situation no longer corresponds to a constant width wave packet, since
\begin{equation}
\Delta x_t \longrightarrow \frac{\beta_0}{\sqrt{2}}
\sqrt{\cosh^2(\tilde{\omega} t) + \sinh^2(\tilde{\omega} t)}
\end{equation}
increases exponentially, as the individual momentum 
components comprising the wave packet quickly diverge in $p$-space. 
For the case of $x_0=0$, we have the general expression
\begin{equation}
A(t) = \frac{1}{\sqrt{\cosh(\tilde{\omega} t)}}
\exp\left[
\left(
\frac{p_0^2}{2m\tilde{\omega} h}
\right)
\left\{
\frac{\cosh(\tilde{\omega} t)-1 + i\sinh(\tilde{\omega}t)(2\cosh(\tilde{\omega} t) -1)}
{\cosh(\tilde{\omega} t)(\cosh(\tilde{\omega} t) - i\sinh(\tilde{\omega} t))}
\right\}
\right]
\end{equation}
In the limit when $t >> 1/\tilde{\omega}$, the hyperbolic functions
both approach $\exp(\tilde{\omega}t)/2$ and we have the limiting case
\begin{equation}
A(t) \longrightarrow
\frac{1}{\sqrt{\exp(\tilde{\omega}t)/2}}
\exp\left[- \frac{p_0^2}{2m\tilde{\omega} \hbar}(1-i)\right]
\, . 
\end{equation}
The exponential (`dynamical') suppression once again is seen to 
`saturate', as in the free-particle case, and for the same reason, namely 
that both $x(t) - x_0$ and $\Delta x_t$ have the same large $t$ behavior. 
The resulting modulus is given by
\begin{equation}
|A(t)|^2
\longrightarrow
2e^{-\tilde{\omega} t}
\exp\left[
- \frac{p_0^2}{m\tilde{\omega} \hbar}
\right]
\end{equation}
which still becomes exponentially small, but now 
due to the (`dispersive') prefactor.
If one also has $x_0 \neq 0$, the expression above includes an
additional factor of $\exp(-x_0^2/\beta_0^2)$ (similar to that
in Eqn.~(\ref{free_particle_A}),  with no cross-term involving 
$x_0$ times  $p_0$.

\section{The full and annular circular wells: WKB energy eigenvalues, 
classical periods, and closed orbits}
\label{appendix:wkb_circular}

While we focus on the information about wave packet revival times
encoded in the energy eigenvalue spectrum of quantum systems, it is
also interesting to see how the pattern of closed (or periodic) orbits
supported in a number of simple 2D quantum billiards  arise
from the connections between the classical periods in systems with
more than two quantum numbers. This is especially true since most of
the experimentally observed data from 2D circular billiard systems 
\cite{microstructures}, \cite{atom_optics} have involved measurements
which are  relevant for short-term, quasi-classical ballistic propagation. 
Such closed orbits are also the ones of relevance to periodic orbit
theory \cite{microwave_1} - \cite{microwave_3} analyses  
of such billiard systems.

We have illustrated these connections using the explicit expressions
for $E(n,m)$ for both the square and equilateral triangle billiard
and we now turn to the cases of the circular and annular infinite wells.
(The arguments presented here are adapted and extended from 
Ref.~\cite{robinett_circular}, but see also Ref.~\cite{balian_and_bloch}
for discussions of the circular billiard in the context of what has
come to be known as periodic orbit theory.)
For the circular well, we first note that the allowed closed orbits
are characterized by two integers $(p,q)$, where $q$ counts the number
of `net revolutions' the trajectories make before closing on themselves,
while $p$ counts the number of `hits' on the circular walls. For consistency,
one must have $p \geq 2q$ and when $p,q$ have common factors, say
$\overline{p} = np$ and $\overline{q} = nq$, one simply has an $n$-fold
recurrence of a basic (or primitive) closed path.  A number of low-lying
trajectories of this type are shown in Ref.~\cite{balian_and_bloch}. 
The total path length for
one classical period in any of these closed orbits is given by
\begin{equation}
L(p,q) = 2pR \sin\left(\frac{\pi q}{p}\right)
\label{circular_path_lengths}
\end{equation}
so that the classical period is simply
\begin{equation}
T(p,q) = \frac{L(p,q)}{v_0}
\end{equation}
where $v_0$ is again the classical speed of the point particle inside
the billiard. The minimum distance to the origin for any of these
trajectories (distance of closest approach) is given by
\begin{equation}
R_{min} = R \cos\left(\frac{\pi q}{p}\right)
\, . 
\label{circular_closest_approach}
\end{equation}

For the quantum case, the two appropriate quantum numbers give rise
to classical periods given by
\begin{equation}
T_{cl}^{(n_r)} \equiv \frac{2\pi \hbar}{|\partial E/\partial n_r|}
\qquad
\mbox{and}
\qquad
T_{cl}^{(m)} \equiv \frac{2\pi \hbar}{|\partial E/\partial m|}
\end{equation}
and the two periods can beat against each other to produce the
classical periodicity ($T_{cl}^{(po)}$) for closed or periodic orbits 
if they satisfy 
\begin{equation}
pT_{cl}^{(n_r)} = T_{cl}^{(po)} = q T_{cl}^{(m)}
\end{equation}
again, with $p\geq2q$ for this geometry. 

We can then use this formalism to understand how these conditions can
give rise to the classical expressions for the path lengths and 
minimum radii in 
Eqns.~(\ref{circular_path_lengths})
and (\ref{circular_closest_approach}). 
Instead of using the approximate
expression in Eqn.~(\ref{other_energies}) for the $(m,n_r)$-dependent
energies, we make use of the WKB condition in 
Eqn.~(\ref{wkb_condition}), namely
\begin{equation}
\sqrt{\frac{2\mu E}{\hbar^2}} 
\int_{R_{min}}^{R} \sqrt{1 - \frac{R_{min}^2}{r^2}} \,dr
= (n_r + 3/4)\pi
\end{equation}
where the appropriate $C_L+C_R = 1/2 + 1/4 = 3/4$ factor corresponds to
infinite wall boundary conditions at $r=R$ and `linear' boundary
conditions at the inner `turning point', 
\begin{equation}
R_{min} = \frac{|m|\hbar}{\sqrt{2\mu E}}
\qquad
\mbox{or}
\qquad
R_{min} = \frac{|m|R}{z}
\end{equation}
which defines the useful parameter $z$. 

We then simply take partial derivatives of both sides with respect to
$n_r$ and $m$ respectively. We thus obtain the conditions
\begin{eqnarray}
\sqrt{\frac{\mu}{2\hbar^2}}
\left[ \int_{R_{min}}^{R} \frac{dr}{\sqrt{E - m^2\hbar^2/2\mu r^2}} \right]
\left|\frac{\partial E}{\partial n_r} \right| & = & \pi \\
\sqrt{\frac{\mu}{2\hbar^2}}
\left[ \int_{R_{min}}^{R} \frac{dr}{\sqrt{E - m^2\hbar^2/2\mu r^2}} 
\left(
\left|\frac{\partial E}{\partial n_r}\right|
- \frac{|m|\hbar^2}{\mu r^2}\right) \right]
& = & 0
\, . 
\end{eqnarray}
The condition to be satisfied for periodic orbits can then be written
as
\begin{equation}
\frac{q}{p} = \frac{T_{cl}^{(n_r)}}{T_{cl}^{(m)}}
= \frac{|\partial E/ \partial m|}{|\partial E/\partial n_r|}
= \left(\frac{|m|\hbar}{\pi \sqrt{2\mu E}} \right)\left[\int_{R_{min}}^{R}
\frac{dr}{r\sqrt{r^2 - R_{min}^2}}\right]
\end{equation}
Evaluating the integral and using $R_{min} \equiv |m|\hbar/\sqrt{2\mu E}$,
we find that
\begin{equation}
\frac{q}{p} =\frac{1}{\pi} \sec^{-1}\left(\frac{R}{R_{min}}\right)
\qquad
\mbox{or}
\qquad
R_{min}(p,q) \equiv R_{min} = R \cos\left(\frac{\pi q}{p}\right)
\end{equation}
as the condition on periodic orbits, as expected. 
 To find the classical period for such closed orbits, we note that
\begin{eqnarray}
T_{cl}^{(po)} = p T_{cl}^{(n_r)} = 
p\left( \frac{2\pi \hbar }{|\partial E/\partial n_r|}\right)
& =&  \left(2p \sqrt{R^2 - R_{min}^2} \right) \sqrt{\frac{\mu}{2E}} 
\nonumber \\
& = &  \frac{[2pR\sin(\pi q/p)]}{v_0} = \frac{L(p,q)}{v_0}
\end{eqnarray}
where we identify $v_0 = \sqrt{2E/\mu}$ with the classical speed
which reproduces the purely geometric result of 
Eqn.~(\ref{circular_path_lengths}).

This type of analysis can be easily generalized to the case of an 
annular billiard, a circular system with infinite walls at both
the outer radius $R$ as well as at an inner radius $R_{0}
\equiv fR$ (where $0<f<1$), defined by the potential
\begin{equation}
      V_A(r) = \left\{ \begin{array}{ll}
               0 & \mbox{for $R_{0}< r< R$} \\
               \infty  & \mbox{for $r\leq R_{0}$ and $r\geq R$}
                                \end{array}
\right.
\, . 
\end{equation}
This system has also been studied in the context of periodic orbit theory
\cite{annular_1}. The appropriate quantum wave functions are given by
\begin{equation}
\psi(r,\theta) = (\alpha J_{|m|}(kr) + \beta Y_{|m|}(kr))
\left(e^{im\theta}/\sqrt{2\pi}\right)
\end{equation}
where one now includes the `irregular' (divergent at the origin)
$Y_{|m|}(kr)$ since the inner wall now guarantees that $r>0$. 
Application of the boundary conditions at $r=R$ and $r=R_0=fR$
yields the corresponding energy eigenvalue condition 
\begin{equation}
J_{|m|}(kR)
Y_{|m|}(kR_{inner})
-
J_{|m|}(kR_{inner})
Y_{|m|}(kR)
= 0
\end{equation}
which can be solved numerically just as easily as in the purely circular 
case.

In terms of the connection between the classical closed orbits and the
quantum periods, we note that the same periodic orbits
described by Eqns.~(\ref{circular_path_lengths}) and 
(\ref{circular_closest_approach}) are still allowed, so long as
\begin{equation}
\frac{R_{min}}{R} = \cos\left(\frac{\pi q}{p}\right) \geq f
\equiv \frac{R_{0}}{R}
\end{equation}
as shown in Fig.~\ref{fig:both_circle_and_annular}. 
Another set of orbits is allowed, also characterized by
the same set of integers $(p,q)$, which bounce  off the inner wall, an
example of which is also shown in Fig.~\ref{fig:both_circle_and_annular}.
For this geometry, the corresponding path lengths are always at least
as large as as those in Eqn.~(\ref{circular_path_lengths}), and are given by
\begin{equation}
\tilde{L}(p,q) = 2pR\sqrt{1 + f^2 - 2f\cos(\pi q/p)}
\, . 
\label{annular_path_lengths}
\end{equation}
The two classes of periodic orbits and path lengths coalesce in the
limit that $f \rightarrow f_{max} = \cos(\pi q/p)$ and then both
disappear from the allowed set of paths. We know that the effective
distance of closest approach is an important parameter in this geometry,
and for the case of the annular ring, we can define $R_{min} = zR$ as
before, but we now have
\begin{equation}
z = f \frac{\sin(\pi q/p)}{\sqrt{1 + f^2 -2f\cos(\pi q/p)}}
\label{annular_inner}
\end{equation}
as illustrated in Fig.~\ref{fig:annular_closest}. 
We note that $z \leq f$ (as expected) for
these trajectories which bounce off the inner wall, with the limiting
case (`just touching') arising when $f=f_{max} = \cos(\pi q/p)$.

The WKB analysis for this geometry proceeds as in 
Sec.~\ref{subsubsec:circular_eigenvalues_and_eigenfunctions}, with the 
lower integration limit changed from $R_{min}$ to $R_0$, and the WKB 
`matching coefficient' changed to reflect the fact that the inner boundary 
is now also of the `infinite wall type'. These changes give
\begin{equation}
\sqrt{\frac{2\mu E}{\hbar^2}} 
\int_{R_{0}}^{R} \sqrt{1 - \frac{R_{min}^2}{r^2}} \,dr
= (n_r + 1)\pi
\end{equation}
with $R_{min} = zR$ with $f<z$ as noted above. 
The implicit differentiation proceeds as before and the condition
for closed orbits becomes
\begin{equation}
\frac{q}{p}
= \frac{|\partial E/\partial m|}{|\partial E/\partial n_r|}
= \sqrt{\frac{2\mu}{E}} \left(\frac{|m| \hbar}{2\mu \pi}\right)
\left[\int_{R_0}^{R} \, \frac{dr}{r \sqrt{r^2-R_{min}^2}} \right]
\, . 
\end{equation}
The integral can be done exactly and using $R_{min} = zR$ and
$R_{0} = fR$, we obtain
\begin{equation}
\frac{\pi q}{p} = \sec^{-1}\left(\frac{1}{z}\right)
- \sec^{-1}\left(\frac{f}{z}\right)
\end{equation}
as the condition for closed orbits, or
\begin{equation}
\sec^{-1}\left(\frac{1}{z}\right)
= \frac{\pi q}{p}  +
\sec^{-1}\left(\frac{f}{z}\right)
\end{equation}
This can be inverted to give
\begin{equation}
z = \cos\left(\frac{\pi q}{p}\right)\left(\frac{z}{f}\right)
 - 
\sin\left(\frac{\pi q}{p}\right) \sqrt{1- \frac{z^2}{f^2}}
\end{equation}
which can be solved for $z$ to yield
\begin{equation}
z = f \frac{\sin(\pi q/p)}{\sqrt{1 + f^2 -2f\cos(\pi q/p)}}
\label{almost_last_equation}
\end{equation}
which is the appropriate condition for these `inner touching' closed
orbits. The corresponding classical periods, $T_{cl}^{(po)}$, 
which then give the corresponding path lengths,  are given by
\begin{eqnarray}
T_{cl}^{(po)} & = &  p T_{cl}^{(p)} 
= p \left(\frac{2\pi \hbar}{|\partial E/\partial n_r|} \right)
\nonumber \\
& = & p \left[\sqrt{\frac{2\mu}{E}}\right]
\left[\int_{R_0}^{R} \frac{r\,dr}{\sqrt{r^2 - R_{min}^2}}\right] 
\\
& = &
p\left[\frac{2R}{v_0}\right]\left(\sqrt{1-z^2} -\sqrt{f^2 - x^2}\right)
\nonumber \\
& = & 
\frac{2pR\sqrt{1 + f^2 - 2f\cos(\pi q/p)}}{v_0}
 \nonumber
\end{eqnarray}
where we have used the relation in Eqn.~(\ref{almost_last_equation}) 
and the classical
connection $E = \mu v_0^2/2$ and recover the expected result. 

The spherical billiard can also be discussed in the same context, using
these methods. The centrifugal barrier term, in a WKB expansion,
is obtained from Eqn.~(\ref{wkb_condition}) by the substitution
$l(l+1) \rightarrow l(l+1)+1/4 = (l+1/2)^2$, the so-called Langer 
modification \cite{langer} - \cite{froman}.

\clearpage

\begin{figure}
\epsfig{file=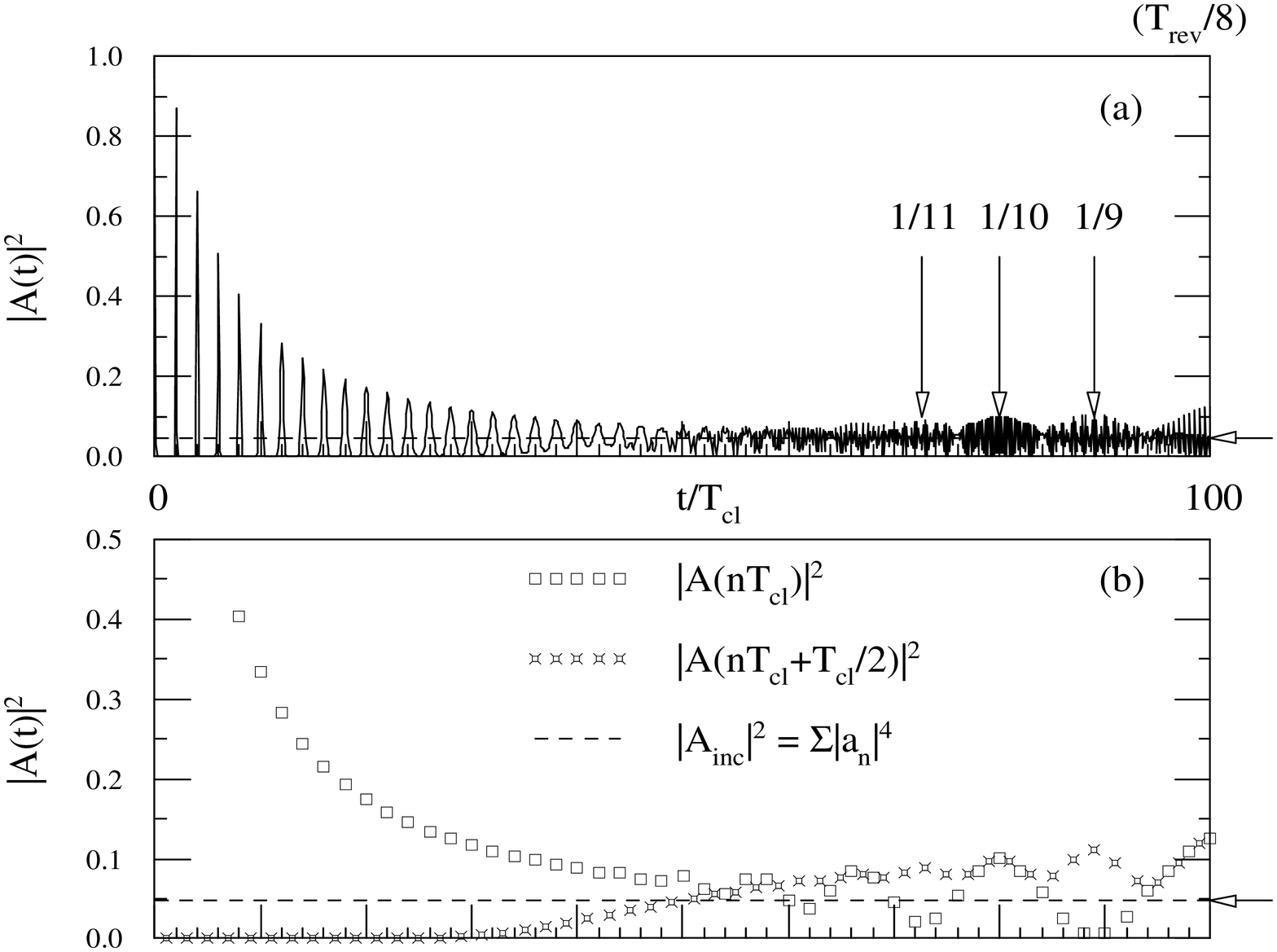,width=12cm,angle=270}
\caption{Plot of the autocorrelation function, $|A(t)|^2$, over the
first 100 classical periods, for the model system described by
Eqns.~(\ref{gaussian_components}) and (\ref{anharmonic}) and Case A in 
Table I. 
The value of $|A(t)|^2$ at multiples of the classical period (squares) and
at half a period out of phase (stars) is shown in the bottom (b), 
compared to the `incoherent' value of $|A_{inc}|^2 = \sum_{n} |a_n|^4
= 1/(\Delta n 2\sqrt{\pi}) \approx 0.047$ in Eqn.~(\ref{incoherent_value}),
shown as the dotted horizontal lines (and indicated by arrows.) 
Locations of fractional revivals are indicated by vertical arrows.
\label{fig:collapse}}
\end{figure}

\begin{figure}
\epsfig{file=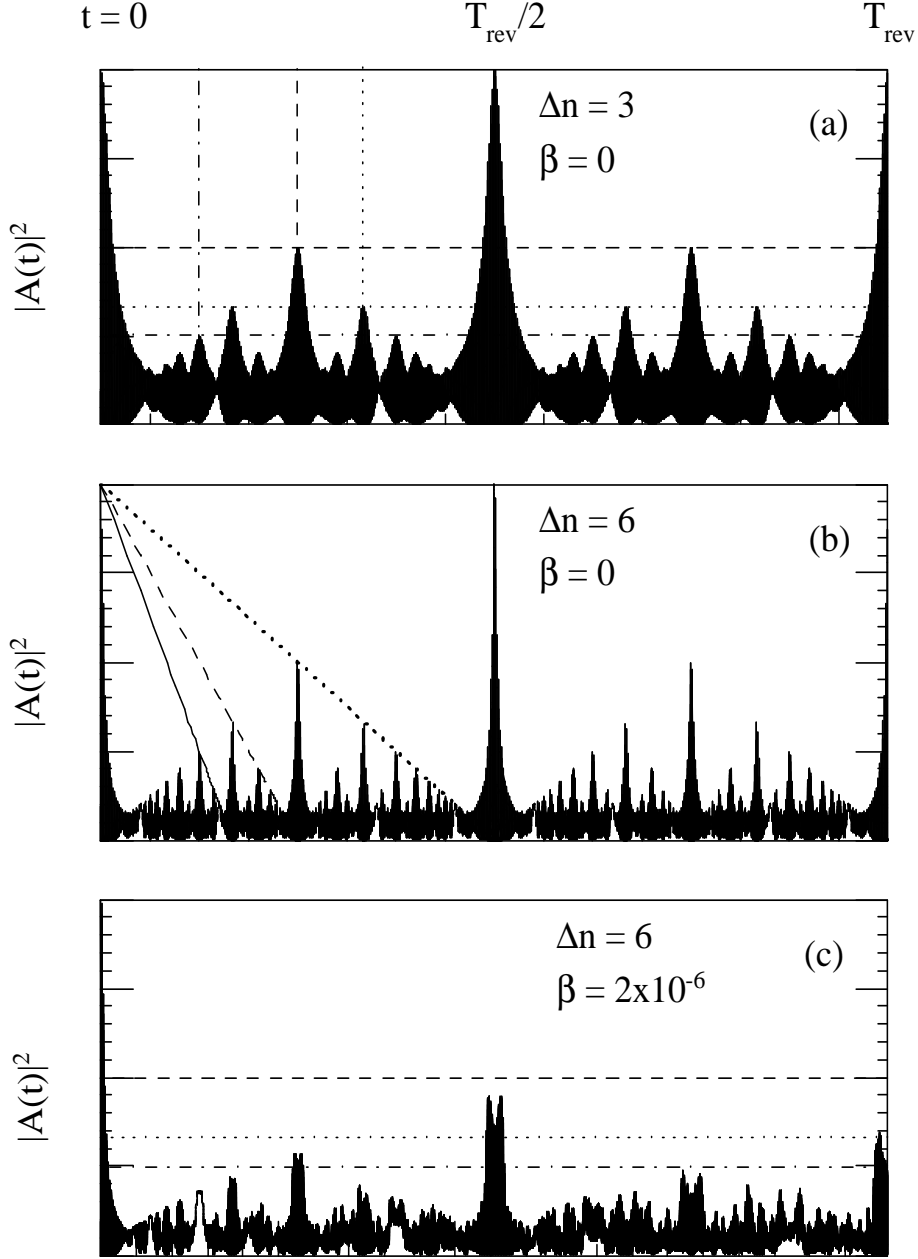,width=12cm,angle=0}
\caption{Plot of the autocorrelation function, $|A(t)|^2$, over the interval
$(0,T_{rev})$ for the energy spectrum in Eqn.~(\ref{anharmonic}) 
and the Gaussian distribution of energy eigenstates in 
Eqn.~(\ref{gaussian_components}) with $n_0 = 400$ and $\alpha = 1/800$ 
and three different sets of parameters: 
(a) $\Delta n = 3$, and $\beta = 0$,
(b) $\Delta n = 6$, and $\beta = 0$,
and
(c) $\Delta n = 6$, and $\beta = 2 \times 10^{-6}$. 
The horizontal lines indicate several fractional values of $|A|^2=1$,
namely $1/2$ (dashed), $1/3$ (dotted), and $1/4$ (dot-dash), at rational
fractions of $t/T_{rev}$ given by $1/4$, $1/3$, and $1/8$ respectively.
\label{fig:new_three}}
\end{figure}

\clearpage

\begin{figure}
\epsfig{file=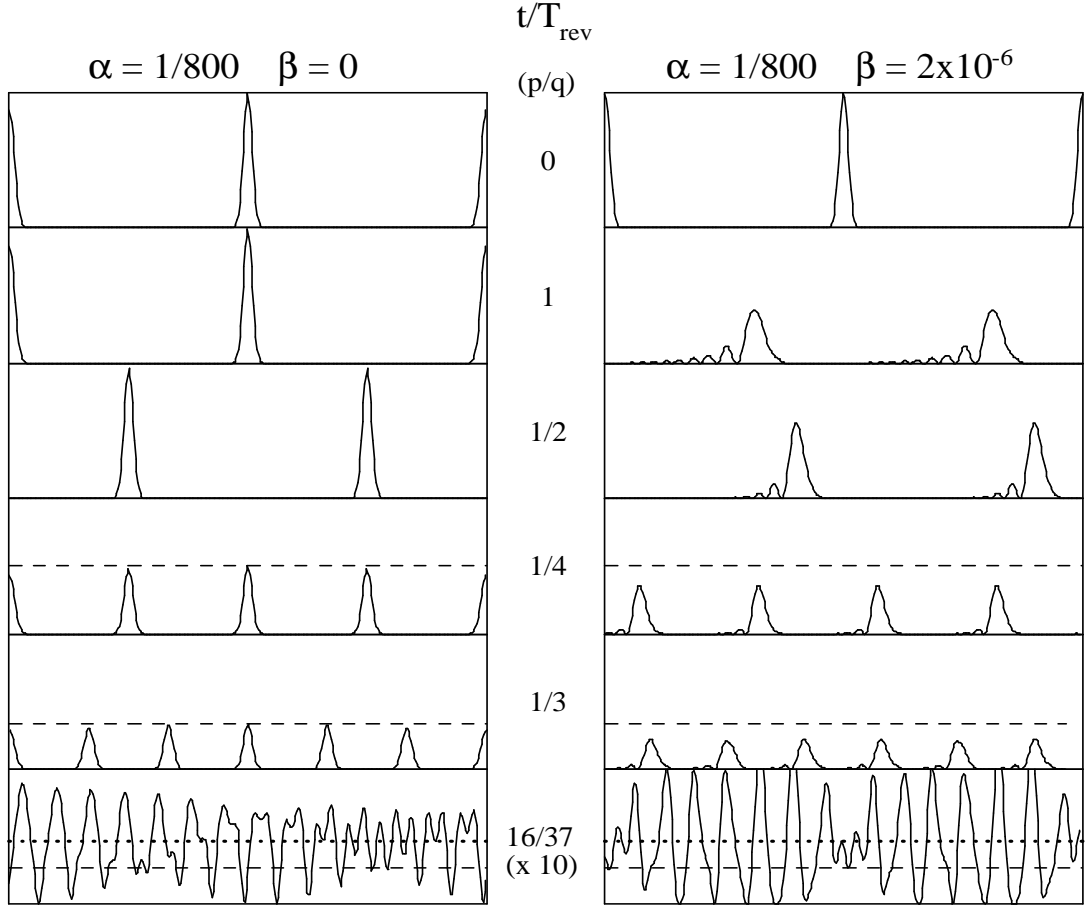,width=12cm,angle=270}
\caption{Plot of the autocorrelation function, $|A(t)|^2$,
near various full or fractional revivals, i.e. over the intervals 
$(pT_{rev}/q-T_{cl},pT_{rev}/q + T_{cl})$ for various values of
$p/q$ and for the parameters
in Eqns.~(\ref{gaussian_components}) and (\ref{anharmonic}) 
and Table~I. The values at the bottom
are multiplied by $10$ to illustrate the oscillations about the
constant incoherent value $|A_{inc}|^2$ (shown as the dotted horizontal
line) for times not near resolvable fractional revivals. For comparison,
the constant value of $|A|^2 = p/q = 1/37$ is indicated by the horizontal
dashed line to compare a large $q$ fractional revival versus a collapsed
or incoherent state.
\label{fig:compare}}
\end{figure}

\clearpage

\begin{figure}
\epsfig{file=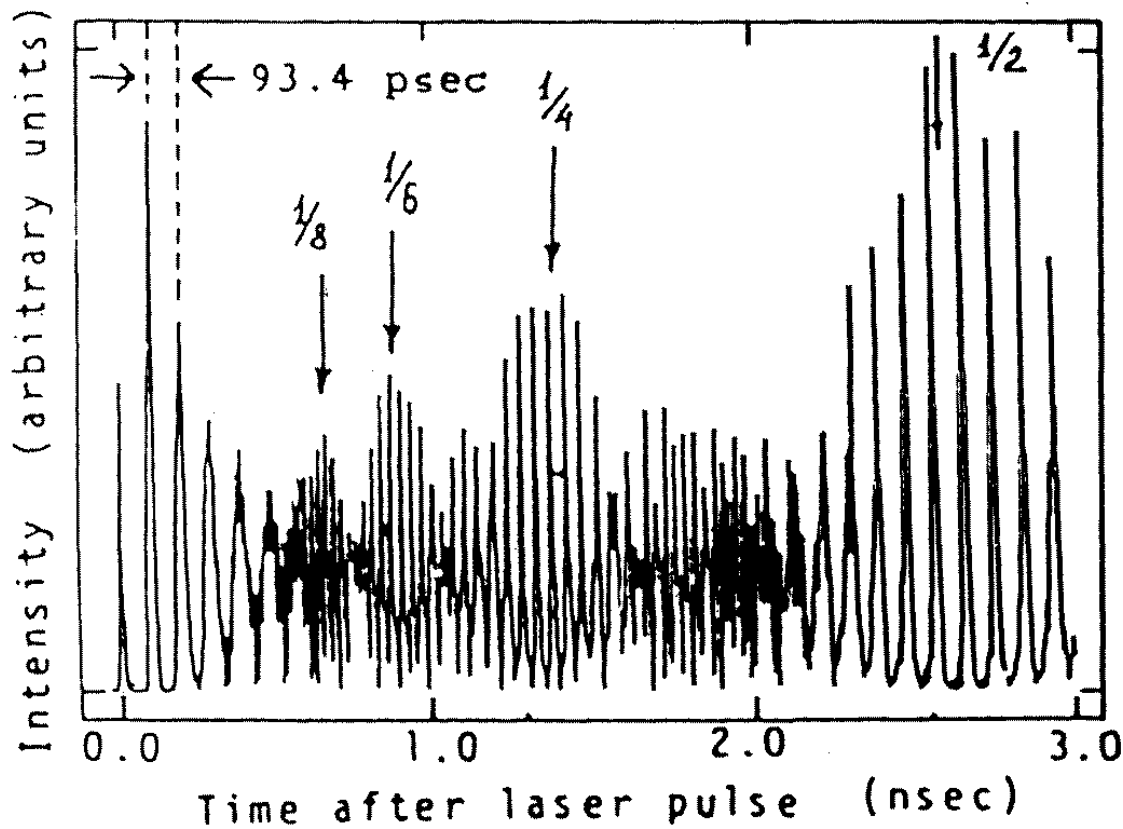,width=12cm,angle=0}
\caption{Numerical calculations by Parker and Stroud (originally from 
Ref.~\cite{parker}) for the intensity of the ionization probability for 
atoms excited by short laser pulses,  showing numerical evidence for revival
behavior, as explained in Ref.~\cite{perelman_1}. The additional notations, 
identifying the locations of the fractional revivals,
 were added by Averbukh and Perelman. (Reprinted from
Ref.~\cite{perelman_1}.)
\label{fig:perelman}}
\end{figure}

\begin{figure}
\epsfig{file=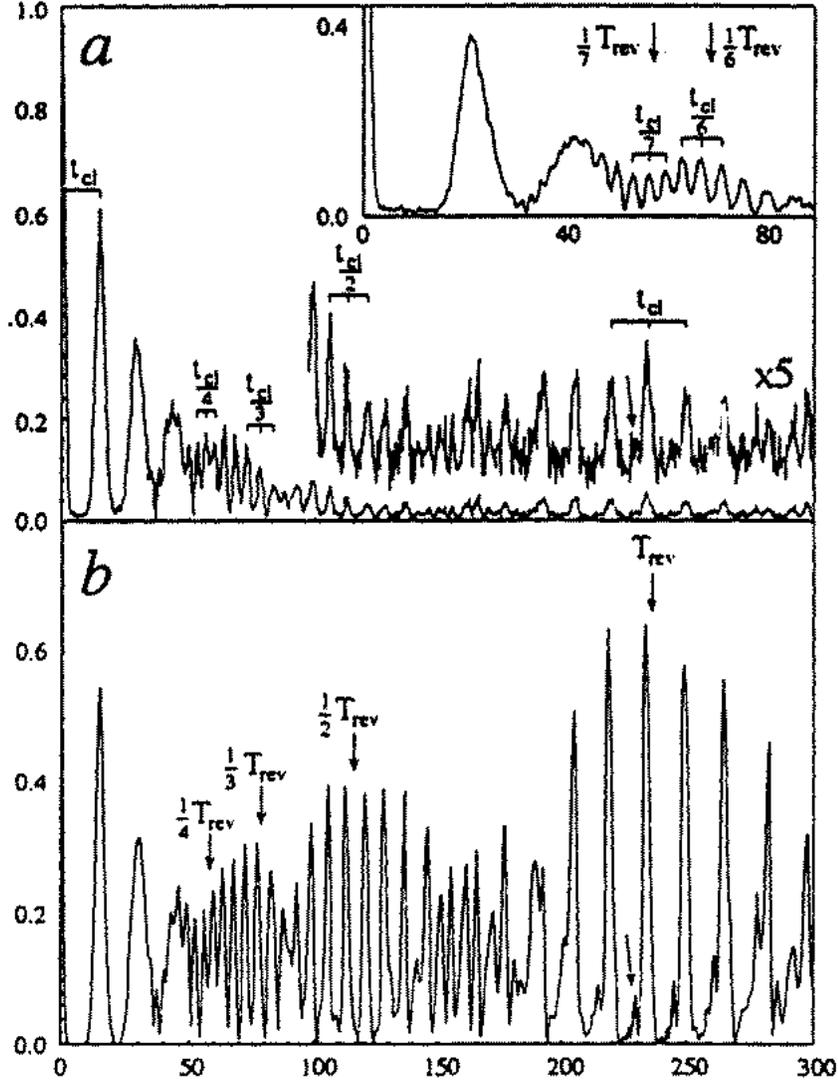,width=12cm,angle=0}
\caption{Recurrence spectra for the excitation of a Rydberg
electron wave packet from Ref.~\cite{wals} showing fractional revivals 
of up to order $T_{rev}/7$. The top plot shows measurements taken
using rubidium atoms, with $n_0 = 46.5$ and roughly $6.5$ states excited, 
while the inset shows data for $n_0 = 53.5$ and $9.9$ states, showing the
improved resolution as one increases $\Delta n$. The bottom plot
shows the calculated time-dependent transition probabilities
for the same system. We note that some of the identifications with various 
fractional revivals differ from the notation used by Averbukh and Perelman 
\cite{perelman_1}. (Reprinted from Ref.~\cite{wals}.)
\label{fig:wals}}
\end{figure}

\clearpage

\begin{figure}
\epsfig{file=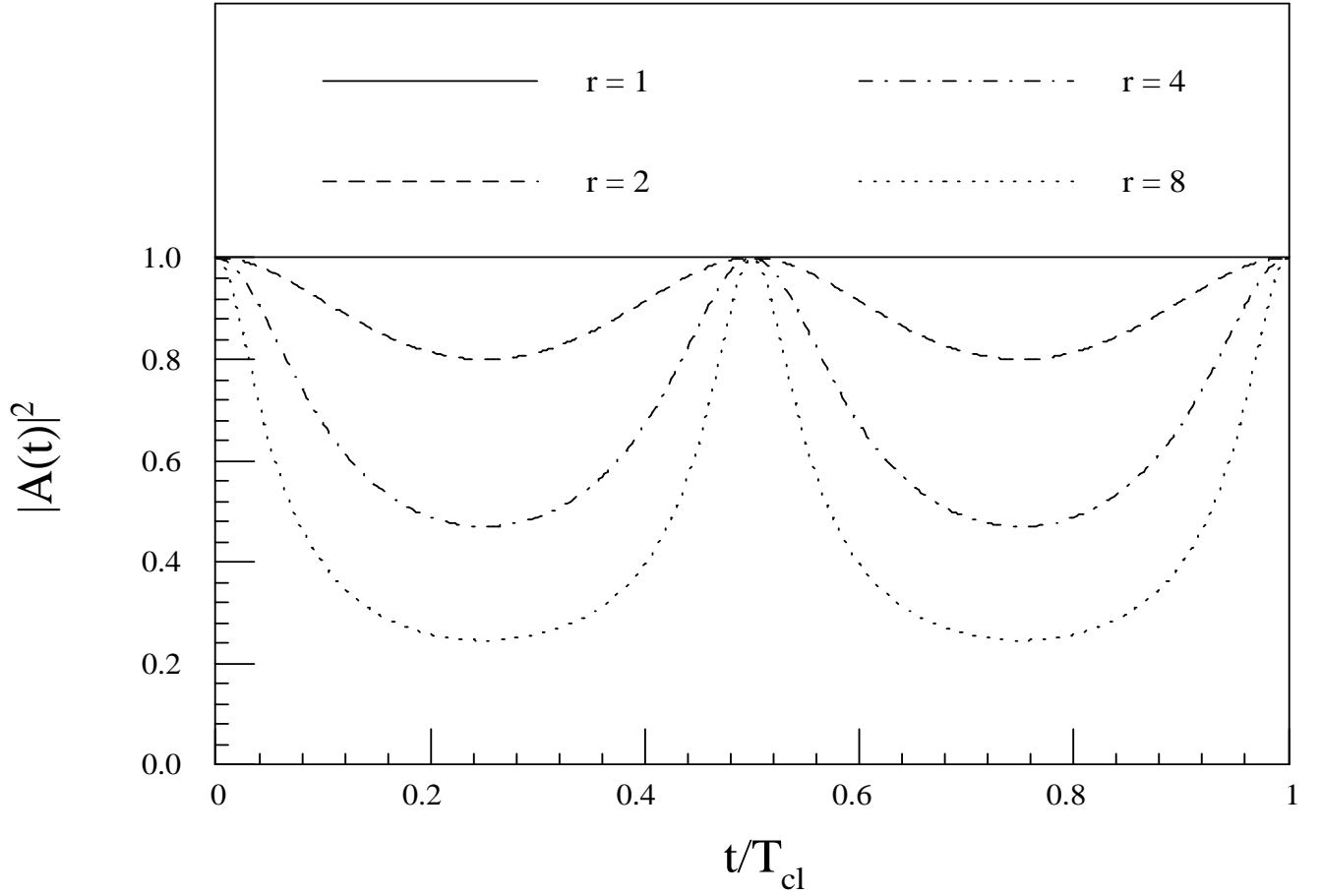,width=12cm,angle=270}
\caption{Plot of the autocorrelation function, $|A(t)|^2$, over one
classical period, for the `pulsating' Gaussian wave packets solutions
in the harmonic oscillator, with $x_0,p_0 = 0$,  
described by Eqn.~(\ref{special_case}), for various values of 
$r = \beta^2/\beta_0^2$. Note that the plots are
invariant under $r \rightarrow 1/r$. For $r=1$ the solution reduces
to the ground state energy eigenstate, with trivial time dependence,
and $|A(t)|=1$.
\label{fig:pulsate_auto}}
\end{figure}

\clearpage

\begin{figure}
\epsfig{file=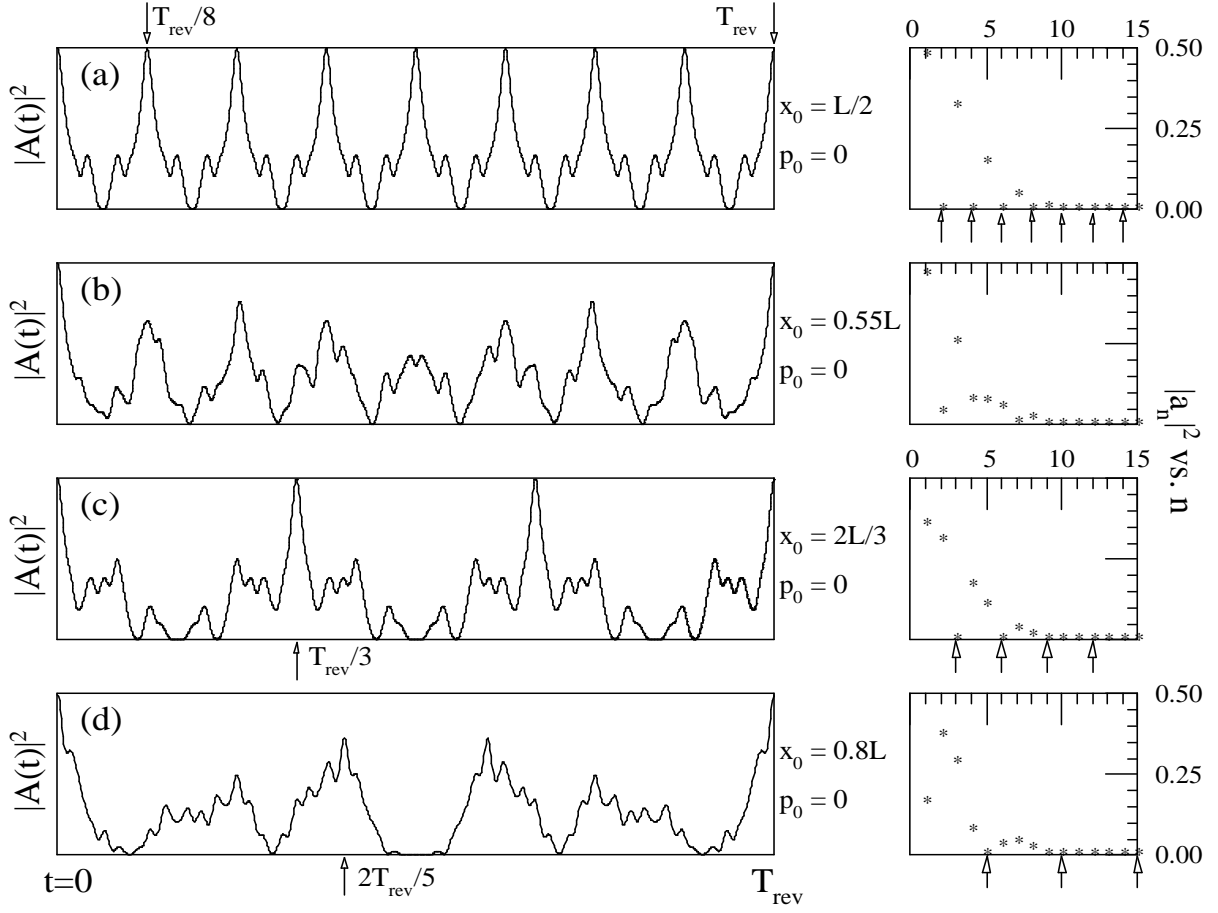,width=12cm,angle=270}
\caption{Plots of the autocorrelation function, $|A(t)|^2$ over one
revival time, for zero-momentum Gaussian wave packets in the infinite
well, for various values of the initial position, $x_0$. 
Note the special shorter time-scale
revivals for cases with additional symmetries, such as for 
(a) $x_0 = L/2$ ($T_{rev}/8$) and 
(b) $x_0 = 2L/3$ ($T_{rev}/3$). The corresponding
expansion coefficients from Eqn.~(\ref{approximate_expansion}) 
are shown alongside.
\label{fig:zero_case}}
\end{figure}

\begin{figure}
\epsfig{file=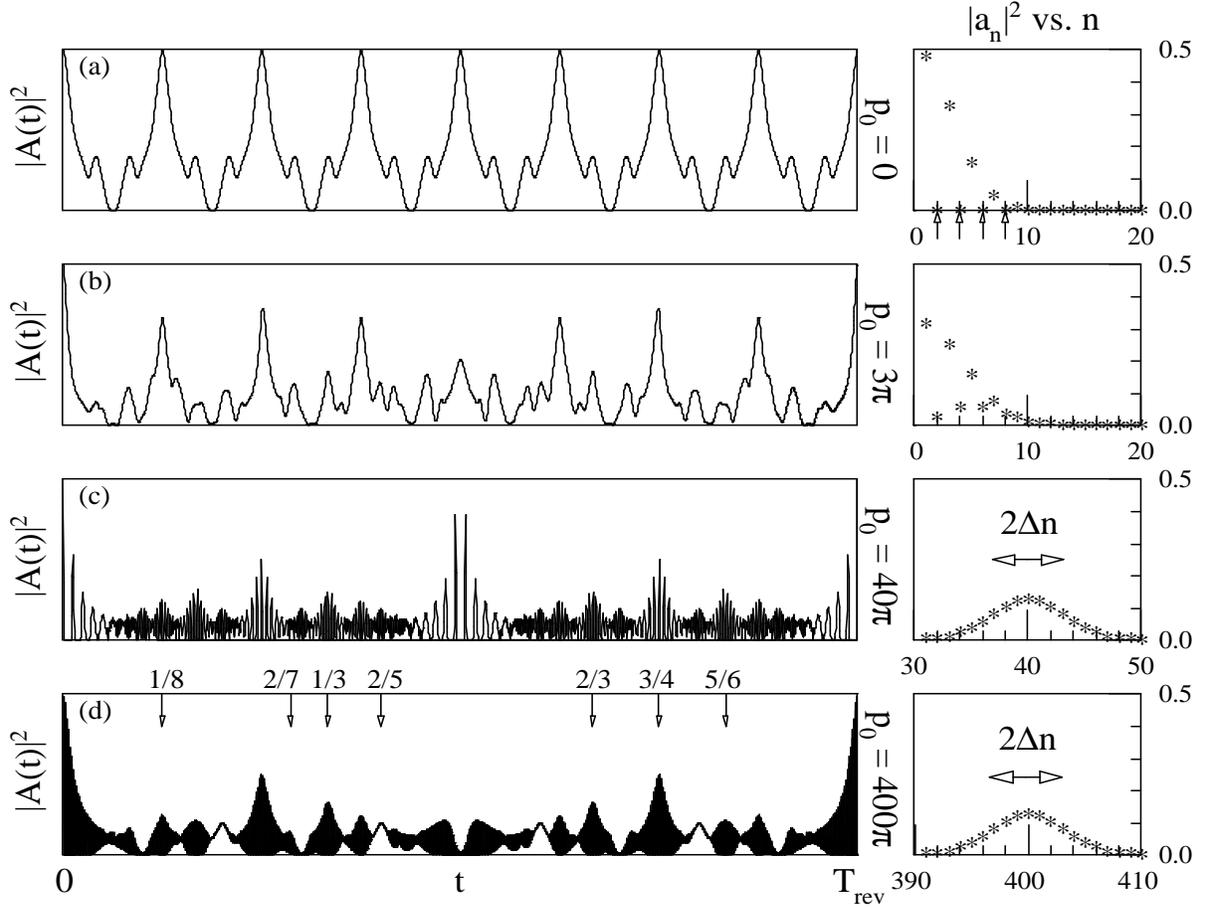,width=12cm,angle=270}
\caption{Same as Fig.~\ref{fig:zero_case}, but for $x_0 = L/2$
wave packets with increasing momentum values. Note the appearance of
the classical periodicity for $p_0 >> \Delta p_0$ for the bottom two
cases, where the $|a_n|$ versus $n$ distribution approaches the
expected Gaussian form in Eqn.~(\ref{gaussian_components}) for
$n_0 >> 1$. 
\label{fig:add_momentum}}
\end{figure}

\clearpage

\begin{figure}
\epsfig{file=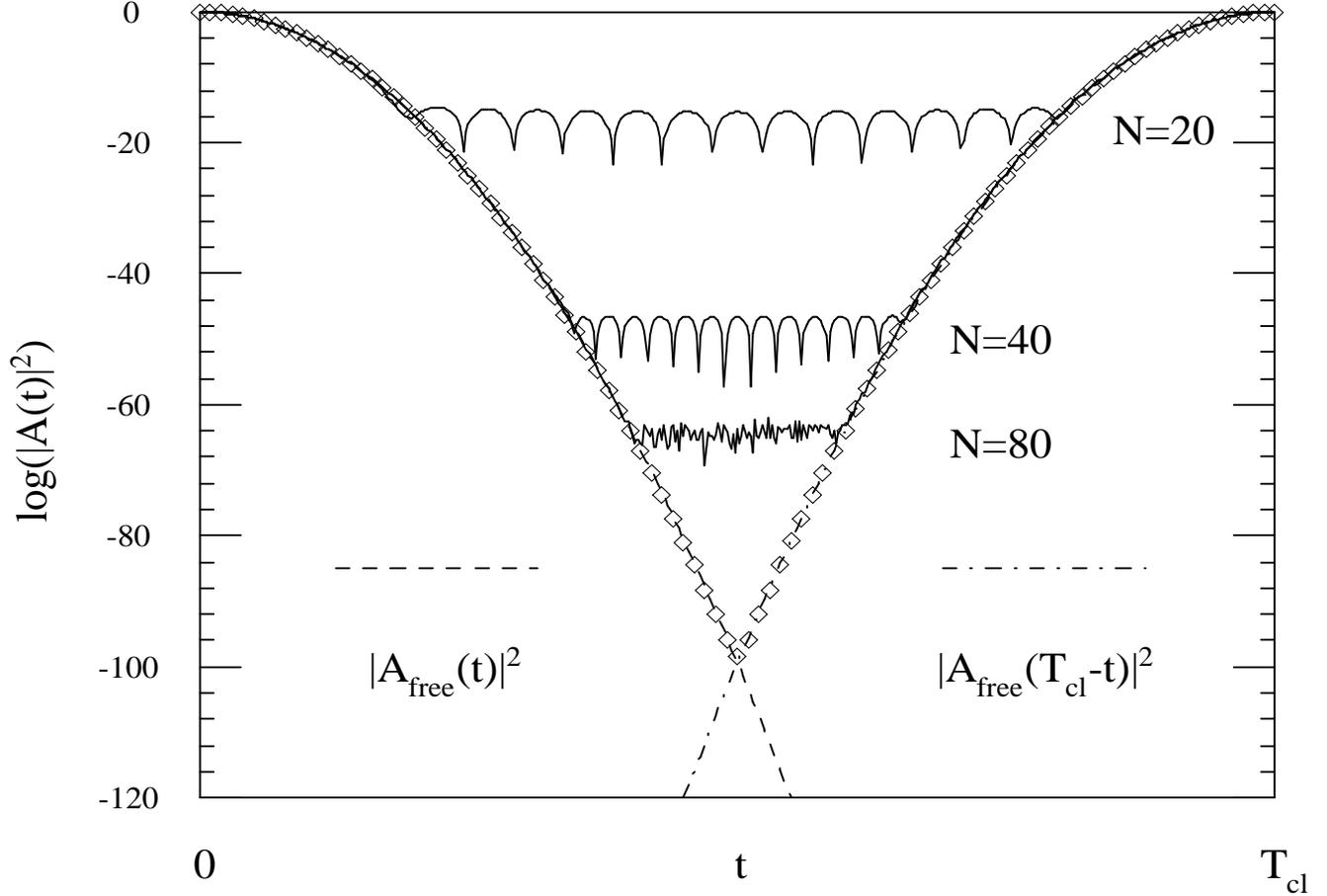,width=12cm,angle=270}
\caption{Plot of the auto-correlation function ($\log(|A(t)|^2)$
for the Gaussian wave packet in the 1D infinite well
for the first classical period, for
increasing values of the number of eigenstates ($N = 20,40,80$) 
(with $a_n$ given by Eqn.~(\ref{approximate_expansion}))
used in the expansion (solid curves). The free-particle autocorrelation 
function in 
Eqn.~(\ref{infinite_free_auto}) as a function of $t$ (dashed) and
$T_{cl}-t$ (dot-dash) are shown for comparison. The values plotted
as diamonds are  from the generic result given by Nauenberg in 
Eqn.~(\ref{nauenberg_formula}).
\label{fig:new_first}}
\end{figure}

\begin{figure}
\epsfig{file=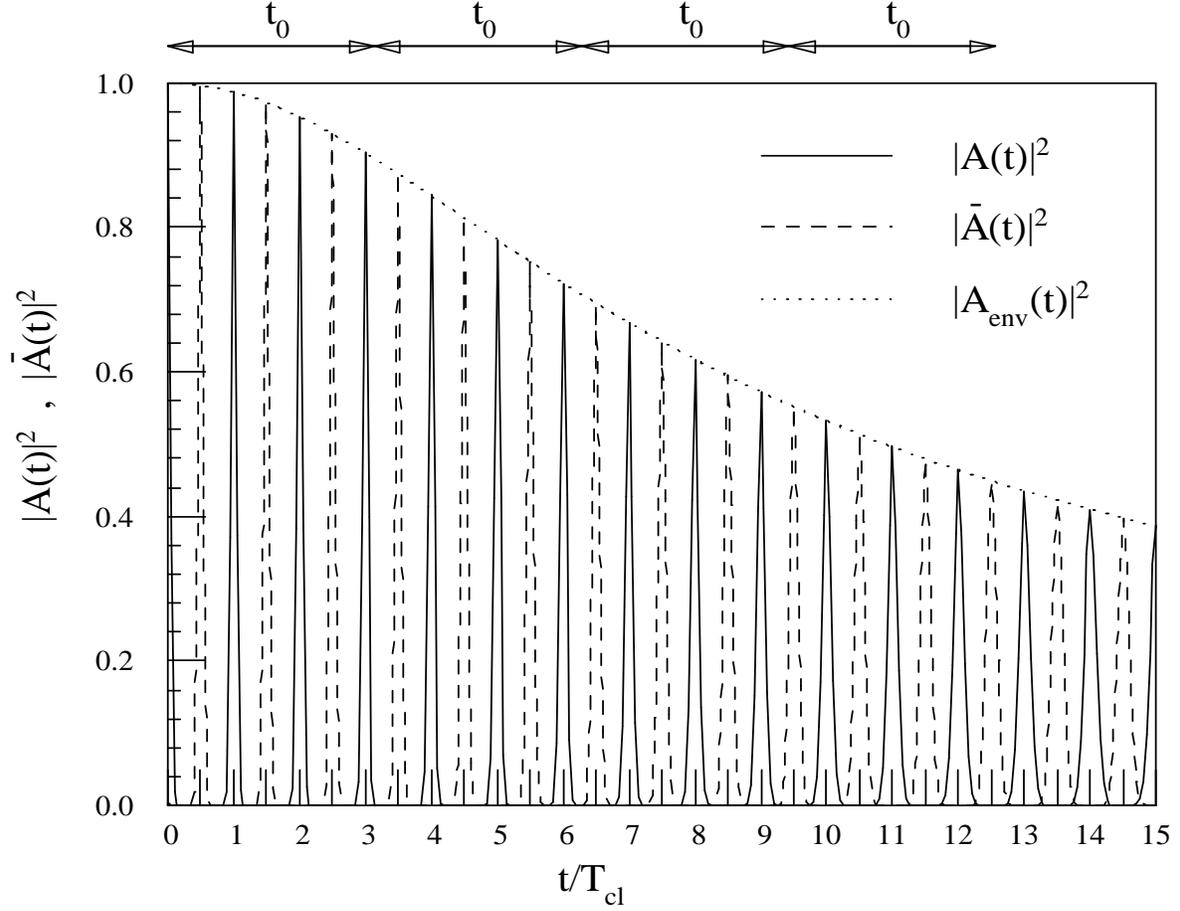,width=12cm,angle=270}
\caption{Plot of $|A(t)|^2$ (solid) and $|\overline{A}(t)|^2$ (dashed)
over the first $15$ classical periods, along with the dispersive part 
of the free-particle autocorrelation function, $|A_{env}(t)|^2$ (dotted),
in Eqn.~(\ref{infinite_free_dispersion}). The value of the spreading
times, $t_0$ are shown for comparison.
\label{fig:second}}
\end{figure}

\clearpage

\begin{figure}
\epsfig{file=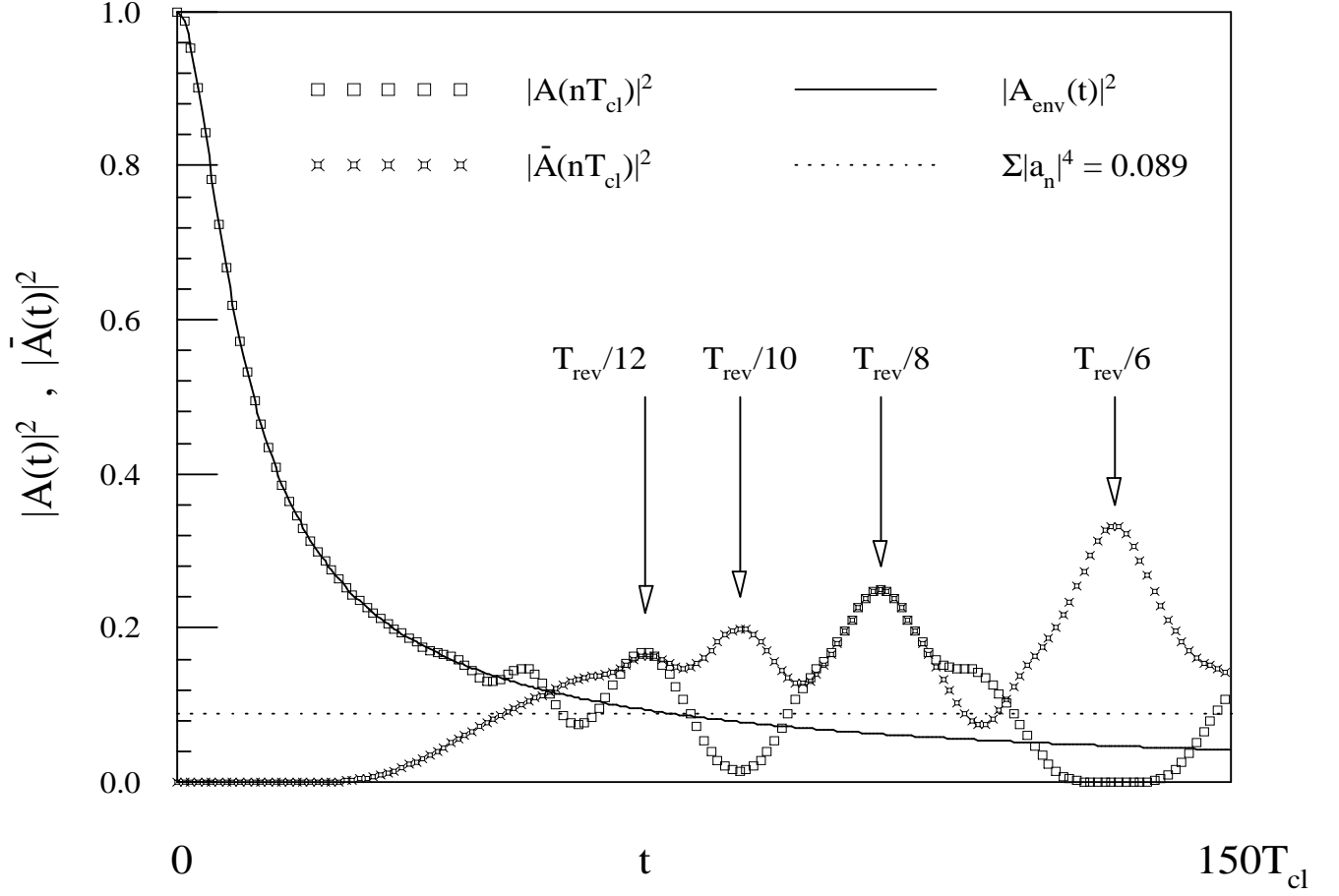,width=12cm,angle=270}
\caption{Plot of $|A(nT_{cl})|^2$ (square) and $|\overline{A}(nT_{cl})|^2$ 
(stars) over the first $150$ classical periods, along with the dispersive 
part  of the free-particle autocorrelation function, 
$|A_{env}(t)|^2$ (solid), in Eqn.~(\ref{infinite_free_dispersion}). 
The horizontal dotted line corresponds to the incoherent sum of uncorrelated
energy eigenstates, given by Eqn.~(\ref{infinite_incoherent}). 
\label{fig:approach}}
\end{figure}

\begin{figure}
\epsfig{file=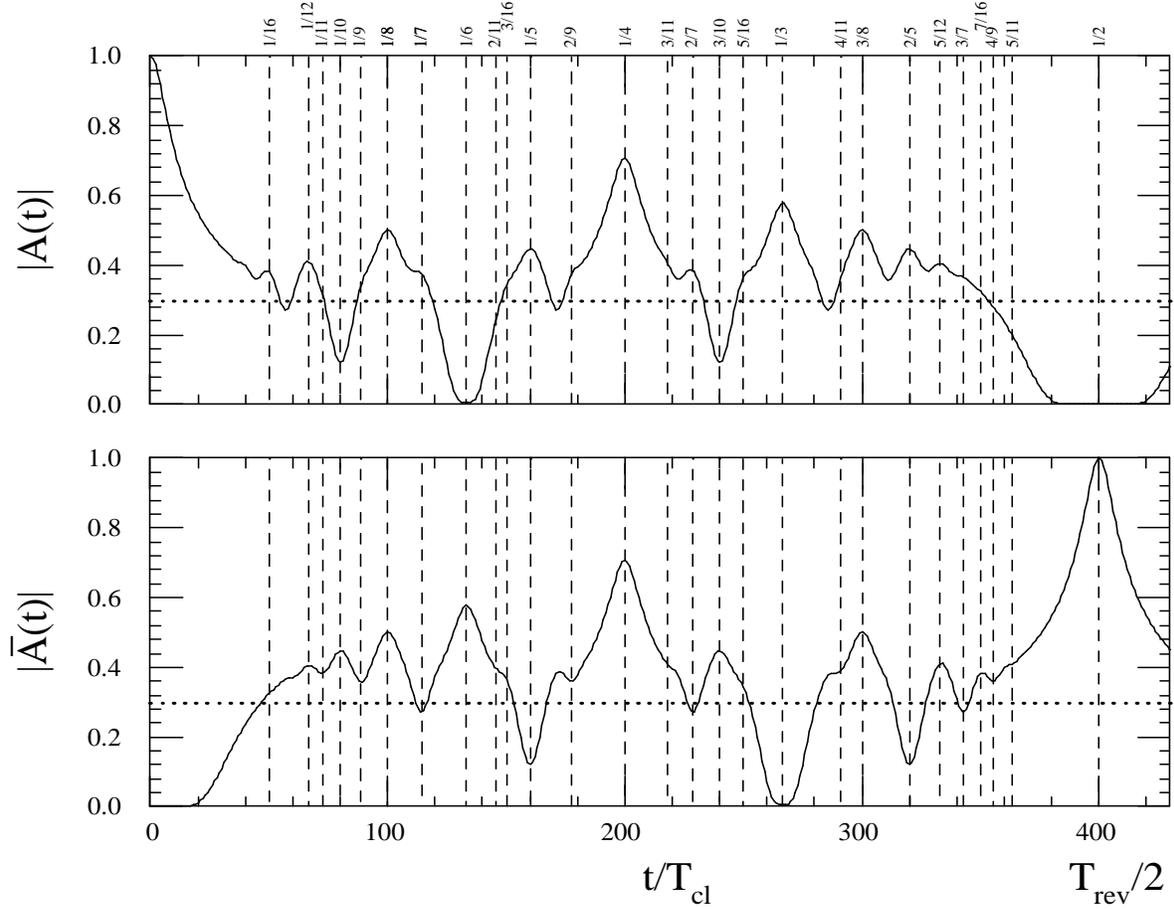,width=12cm,angle=270}
\caption{Plot of $|A(nT_{cl})|$ (top) and $|\overline{A}(nT_{cl})|$ 
(bottom) over half a revival time (just over $T_{rev}/2$.) A large number
of possible fractional revivals are denoted by vertical dashed lines,
while the value of $|A_{inc}|$ for an incoherent sum of eigenstates,
as in Eqn.~(\ref{infinite_incoherent}), is also shown for comparison.
A number
of the higher order fractional revivals are obvious. For example, the
revivals at $T_{rev}/2$ and $T_{rev}/6$ are obvious in the `out-of-phase' 
$\overline{A}(t)$ anti-correlation function, while that at
$T_{rev}/3$ is apparent in $A(t)$; the fractional revival at $T_{rev}/4$
is equally distinct in both plots, consistent with the form in 
Eqn.~(\ref{one_quarter_revival_function}).
\label{fig:all_autos}}
\end{figure}

\begin{figure}
\epsfig{file=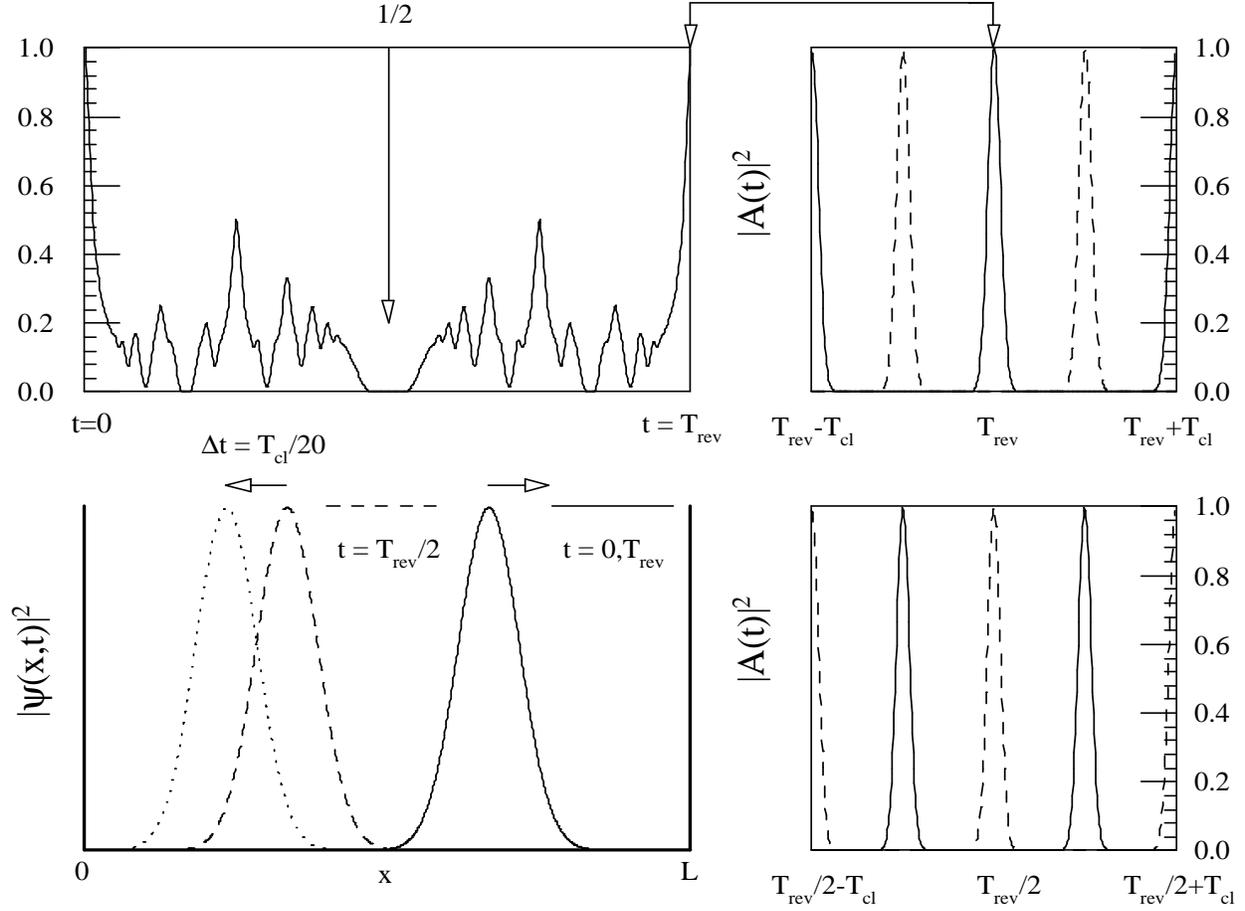,width=12cm,angle=270}
\caption{Plots of $|A(nT_{CL})|^2$ over one revival time (upper left)
and $|A(t)|^2,|\overline{A}(t)|^2$ over two classical periods, centered
about $T_{rev}$ (upper right). The autocorrelation and anti-correlation
functions near $t=T_{rev}/2$ (lower right) and the position-space
probabilities at $t=T_{rev}/2$ (dashed) and $T_{rev}/2+T_{cl}/20$
(dotted) are shown, illustrating the mirror revival, the wave packet
being reformed on the opposite side of the well from the $t=0$ packet
(solid) centered at $x_0/L = 2/3$.
\label{fig:frac_half}}
\end{figure}

\clearpage

\begin{figure}
\epsfig{file=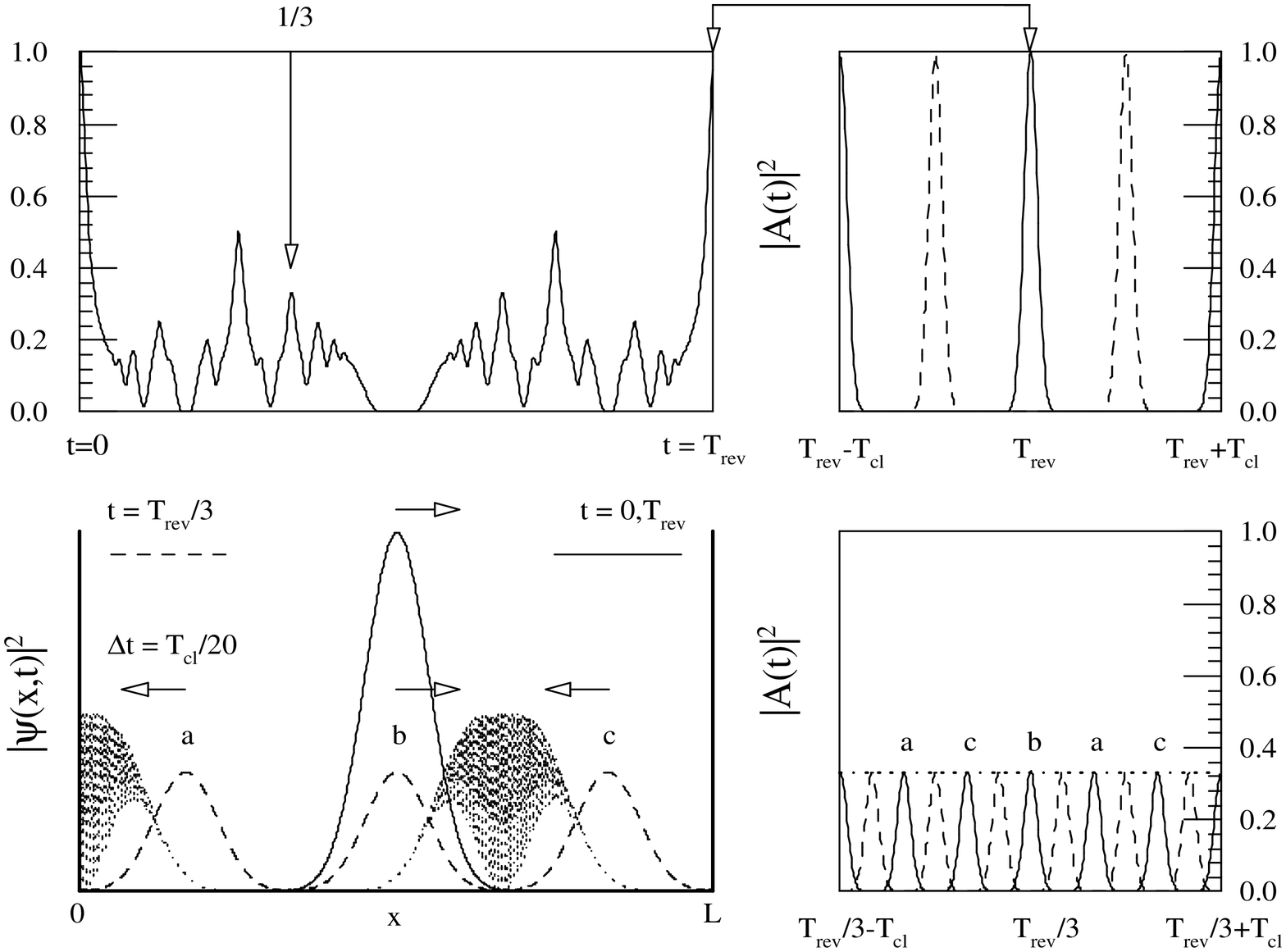,width=12cm,angle=270}
\caption{Same as Fig.~\ref{fig:frac_half}, except illustrating the
fractional revival at $T_{rev}/3$. The horizontal dotted line
corresponds to $|A(t)|^2 = 1/3$,  and $x_0/L = 1/2$ was used. The
same fractional revival is also visualized using the Wigner distribution
in Fig.~\ref{fig:wigner_third}.
\label{fig:frac_third}}
\end{figure}

\begin{figure}
\epsfig{file=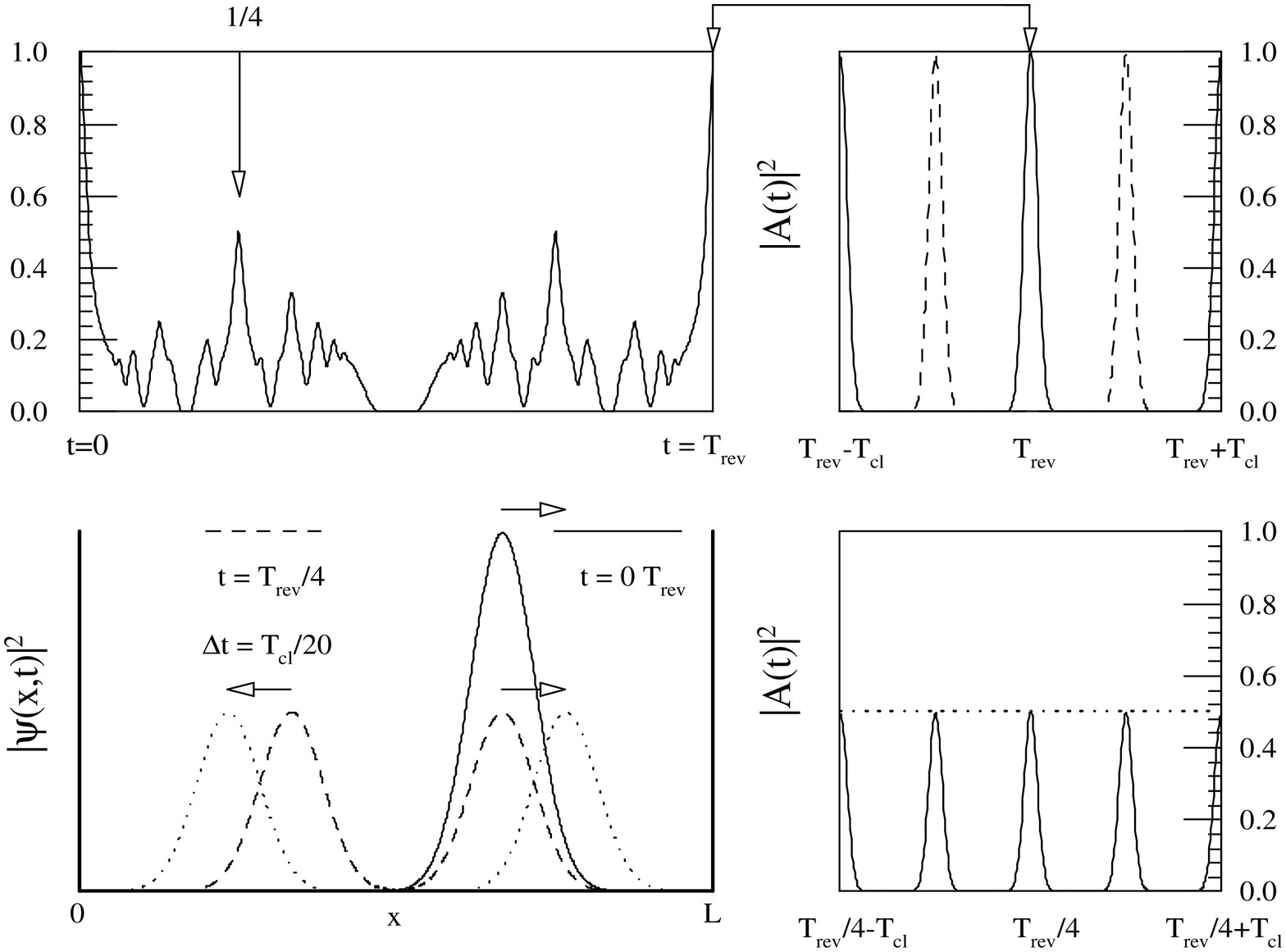,width=12cm,angle=270}
\caption{Same as Fig.~\ref{fig:frac_half}, except illustrating the
fractional revival at $T_{rev}/4$. The horizontal dotted line
corresponds to $|A(t)|^2 = 1/2$,  and $x_0/L = 2/3$ was used. The
same fractional revival is also visualized using the Wigner distribution
in Fig.~\ref{fig:wigner_quarter}.
\label{fig:frac_quarter}}
\end{figure}

\clearpage

\begin{figure}
\epsfig{file=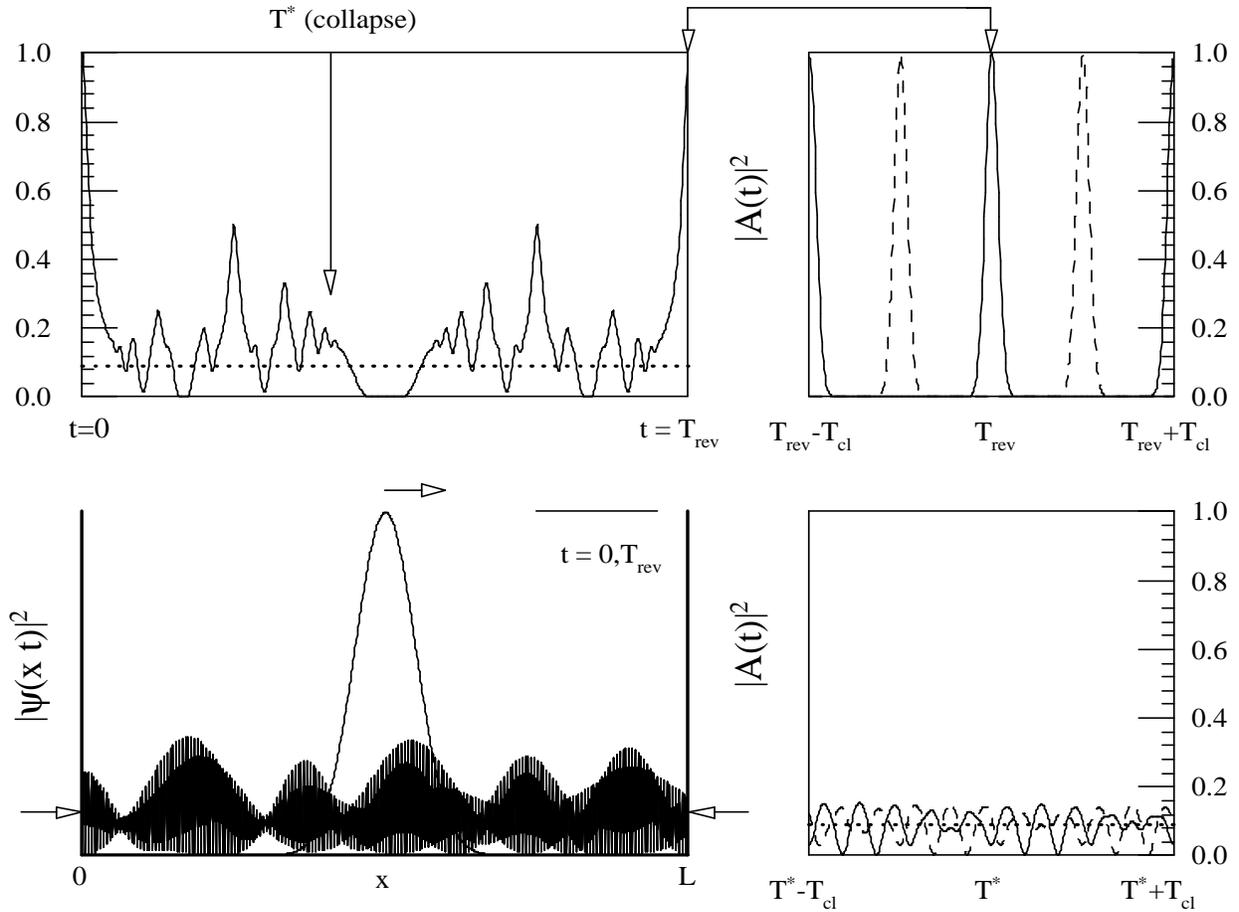,width=12cm,angle=270}
\caption{Same as Fig.~\ref{fig:frac_half}, except illustrating the
behavior of both the probability density (bottom left) and 
$A(t),\overline{A}(t)$ (bottom right) during the collapsed phase, where
$P(x,t) = |\psi(x,t)|^2$ is more consistent with the uniform or `flat' 
value of $P(x) = 1/L$ (which is indicated by the two horizontal arrows.) 
The values of $|A(t)|^2,|\overline{A}(t)|^2$ can be compared to the
incoherent value in Eqn.~(\ref{infinite_incoherent}) shown as the
horizontal dotted line.
\label{fig:frac_collapse}}
\end{figure}

\begin{figure}
\epsfig{file=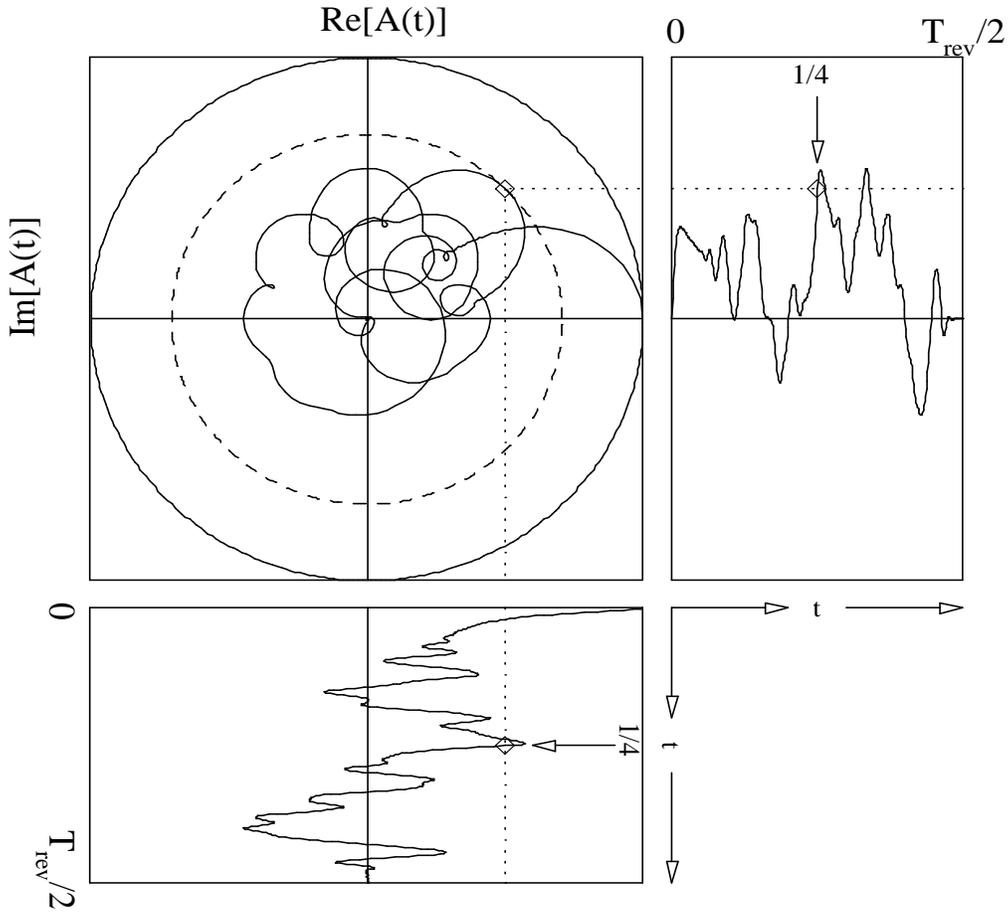,width=12cm,angle=270}
\caption{Plots of the real (bottom) and imaginary (right) parts of
$A(t)$ versus time (over half a revival time) as well as a parametric
plot (Argand diagram) of the same data. In this case, $A(t)$ is evaluated
at integral multiples of $T_{cl}$, as in Figs.~\ref{fig:frac_half} - 
\ref{fig:frac_collapse}. The solid circle corresponds to $|A(t)|^2=1$, 
while the dashed circle corresponds to $|A(t)|^2 = 1/2$; the location of 
a quarter-revival ($T_{rev}/4$) is denoted by the diamond, illustrating 
how at fractional  revivals the autocorrelation function 
(sampled at these intervals) is tangent to the circle.
\label{fig:newflower}}
\end{figure}

\begin{figure}
\epsfig{file=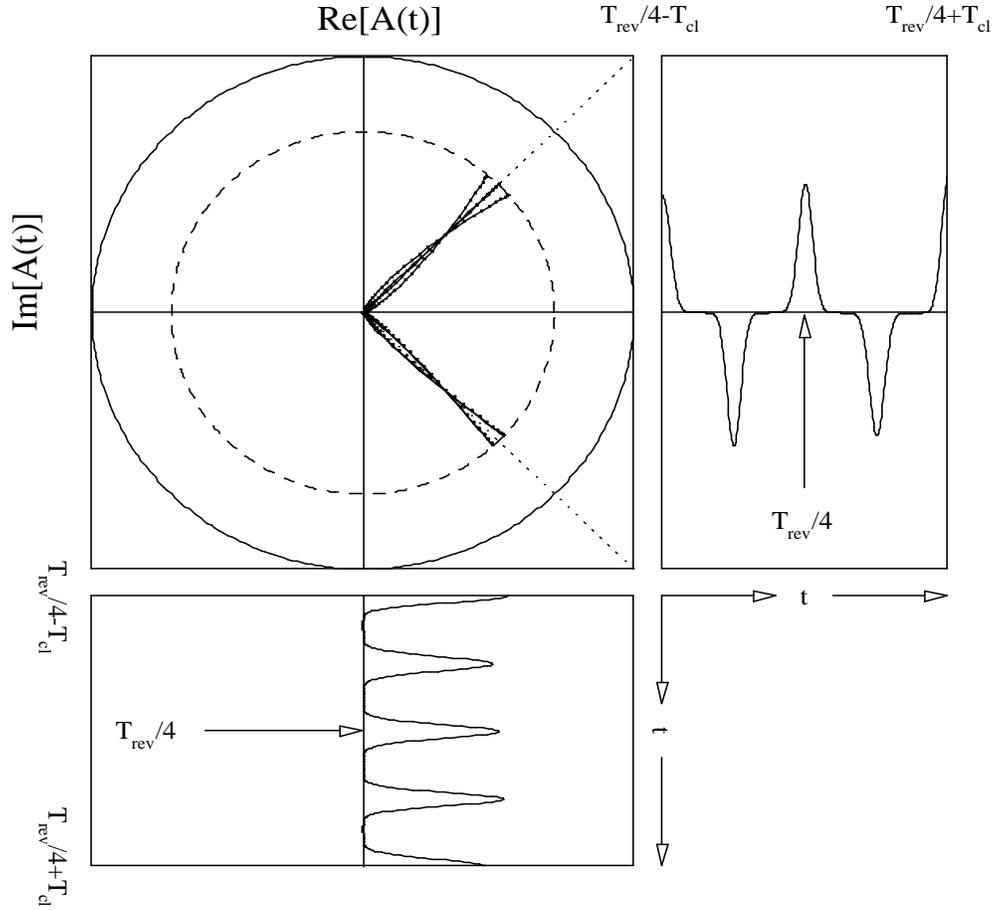,width=12cm,angle=270}
\caption{Same as Fig.~\ref{fig:newflower}, but for the 
time period $T_{cl}$ on either side of the $T_{rev}/4$
fractional revival to be compared to the general wavefunction
in Eqn.~(\ref{one_quarter_revival_function}) and the correlations predicted 
by Eqn.~(\ref{quarter_argand_correlation}). The dashed circle corresponds
to $|A(t)| = 1/\sqrt{2}$. The dotted lines correspond to the directions
$e^{+i\pi/4}$ and $e^{-i\pi/4}$.
\label{fig:quarterflower}}
\end{figure}

\clearpage

\begin{figure}
\epsfig{file=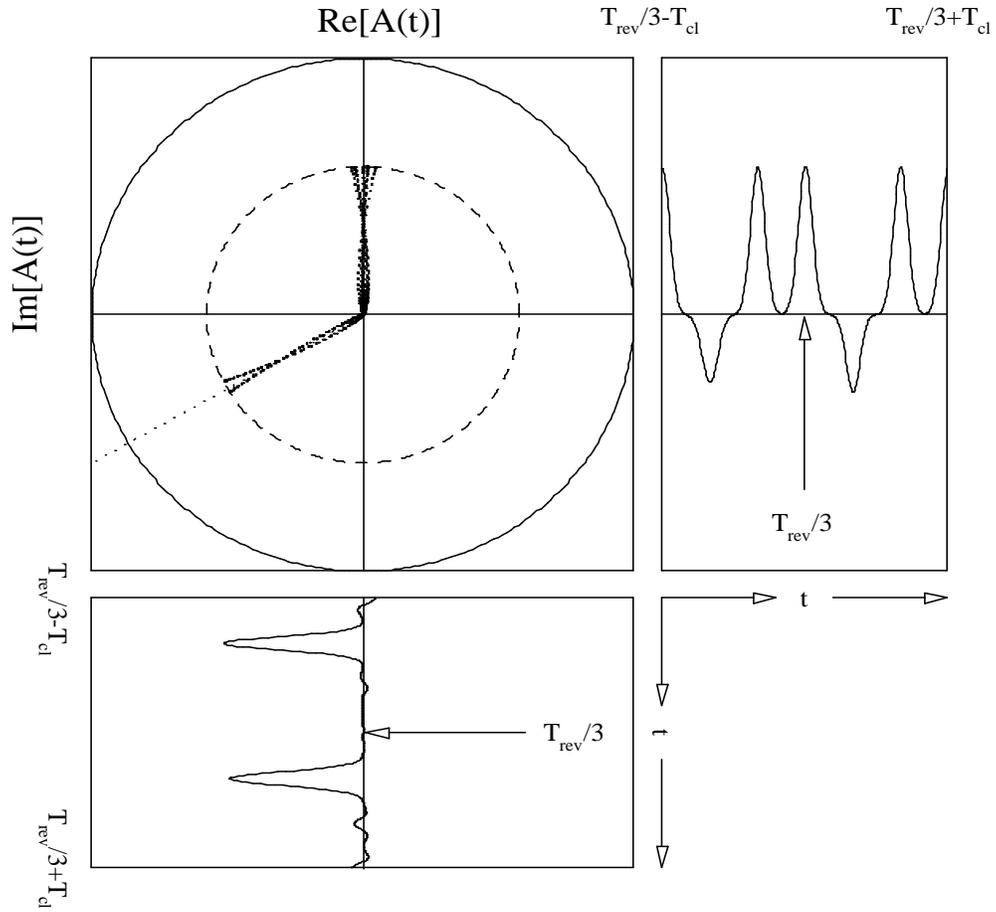,width=12cm,angle=270}
\caption{Same as Fig.~\ref{fig:newflower}, but for the 
time period $T_{cl}$ on either side of the $T_{rev}/3$
fractional revival and for all times, not just multiples of
$T_{cl}$. Note the highly correlated phase relationship 
consistent with the general results of 
Eqn.~(\ref{one_third_revival_function}) and in
Eqn.~(\ref{third_argand_correlation}). In this case, 
the dashed circle corresponds to $|A(t)| = 1/\sqrt{3}$. 
\label{fig:thirdflower}}
\end{figure}

\clearpage

\begin{figure}
\epsfig{file=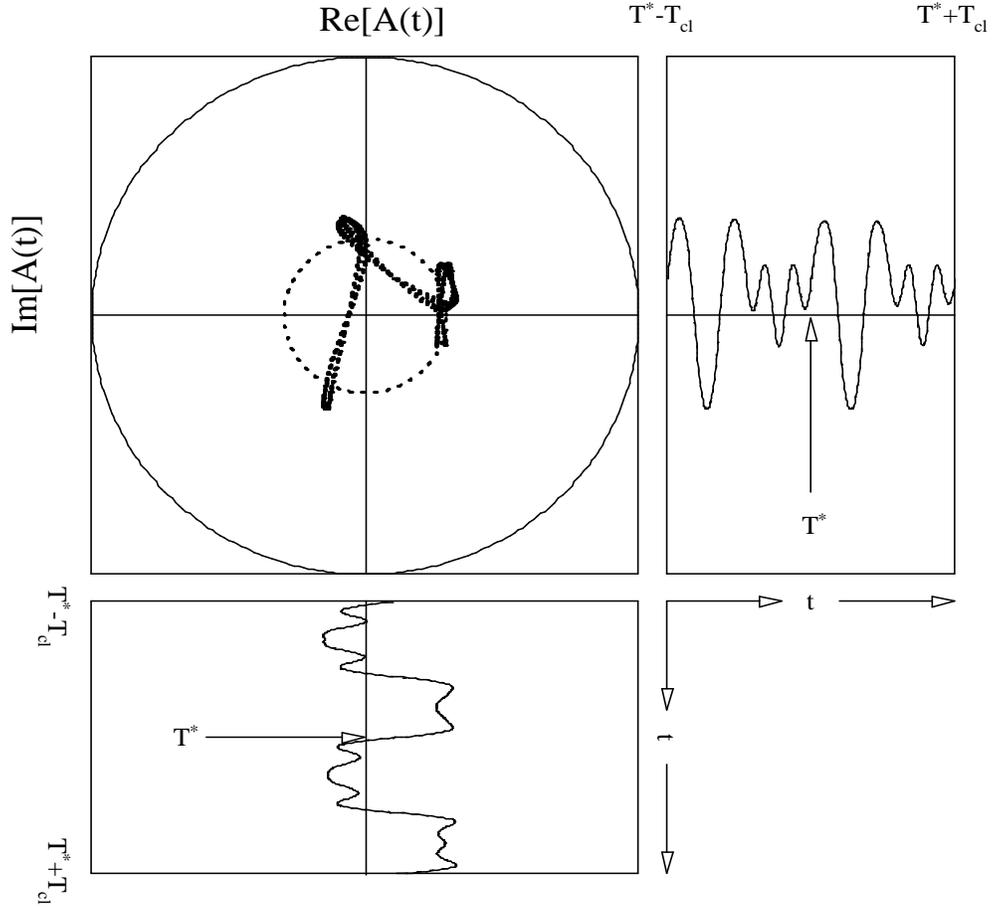,width=12cm,angle=270}
\caption{Same as Fig.~\ref{fig:newflower}, but for the 
time period $T_{cl}$ on either side of a typical time ($T^{*}$)
during the collapsed phase, not near any resolvable fractional revival.
 In this case, the dotted  circle corresponds
to $|A_{inc}|^2 = \sum |a_n|^4 = 1/\Delta n 2\sqrt{\pi}$,
typical of the incoherent
sum of many eigenstates, with little or no phase relationship obvious.
\label{fig:randomflower}}
\end{figure}

\begin{figure}
\epsfig{file=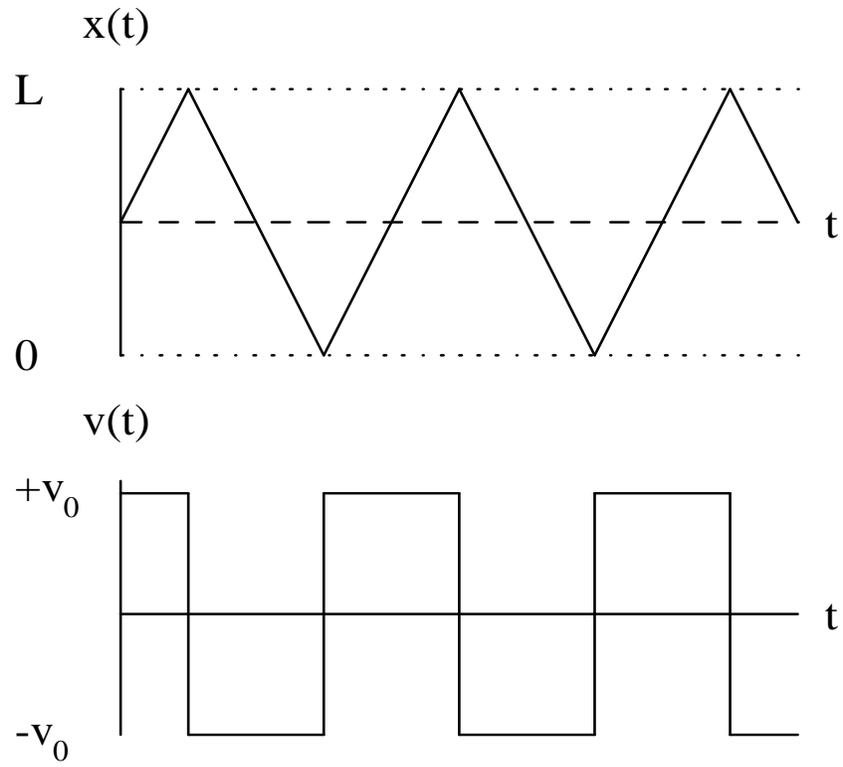,width=10cm,angle=270}
\caption{The classical motion of a particle in an infinite well potential.
\label{fig:x_p_classical}}
\end{figure}

\clearpage

\begin{figure}
\epsfig{file=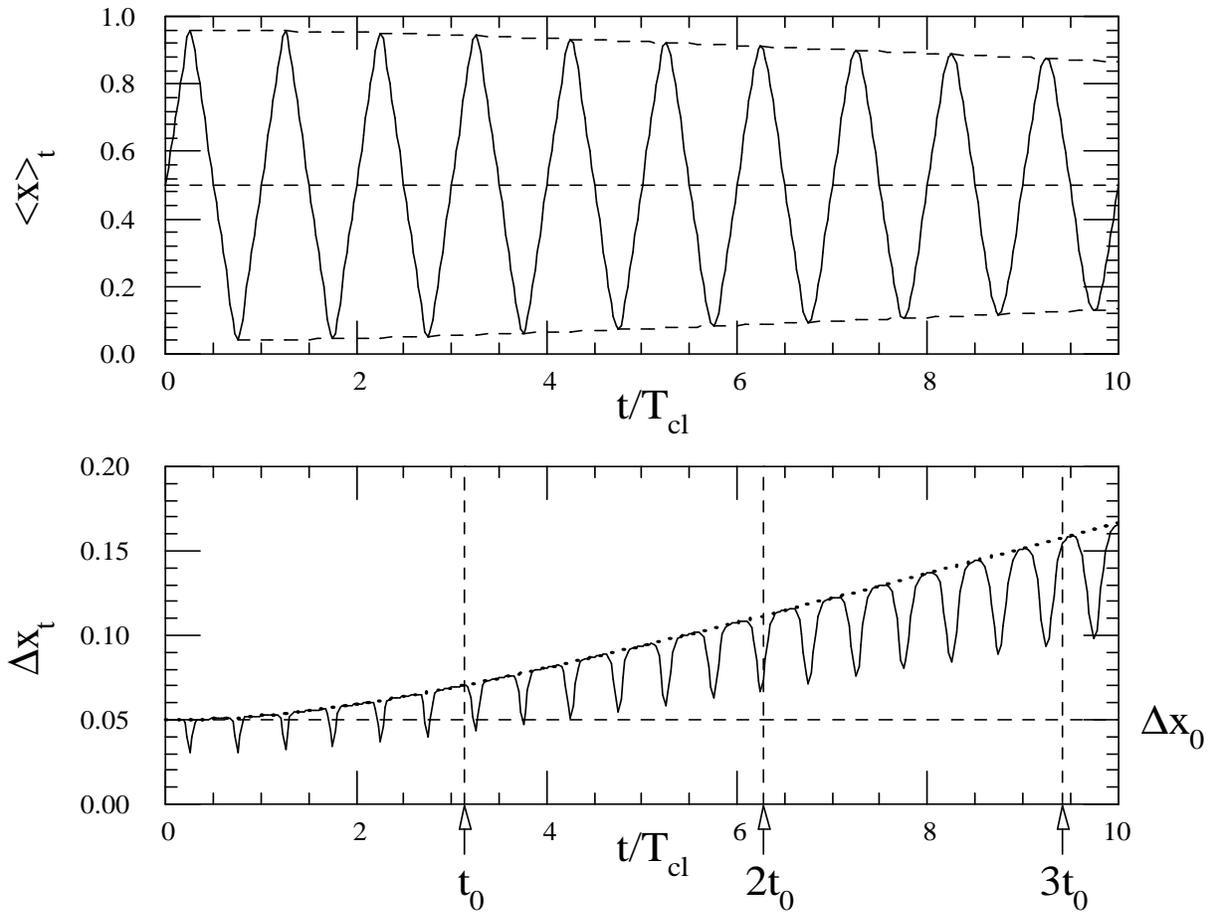,width=12cm,angle=270}
\caption{The expectation value, $\langle x \rangle_t$ (top), 
and uncertainty, $\Delta x_t$ (bottom), for the infinite Gaussian wave
packet over the first ten classical periods, measured in units of
$L$. The dotted line which forms the envelope for the
$\Delta x_t$ curve is the free-particle spread given by
Eqn.~(\ref{infinite_x_quantities}). (Reprinted from  
Ref.~\cite{robinett_infinite_well}.)
\label{fig:x_short}}
\end{figure}

\begin{figure}
\epsfig{file=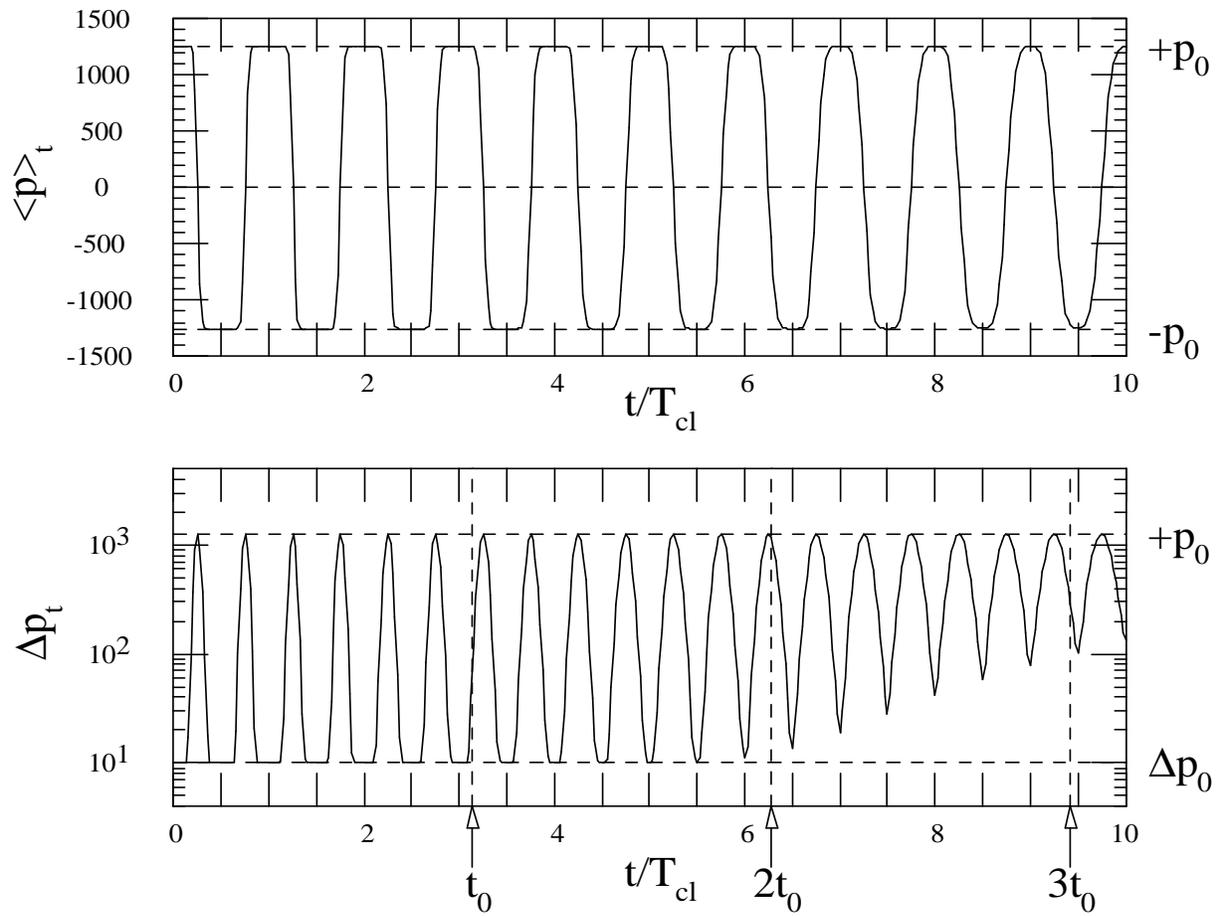,width=12cm,angle=270}
\caption{Same as for Fig.~\ref{fig:x_short}, but for 
$\langle p \rangle_t$ (top), and $\Delta p_t$ (bottom). 
(Reprinted from Ref.~\cite{robinett_infinite_well}.)
\label{fig:p_short}}
\end{figure}

\begin{figure}
\epsfig{file=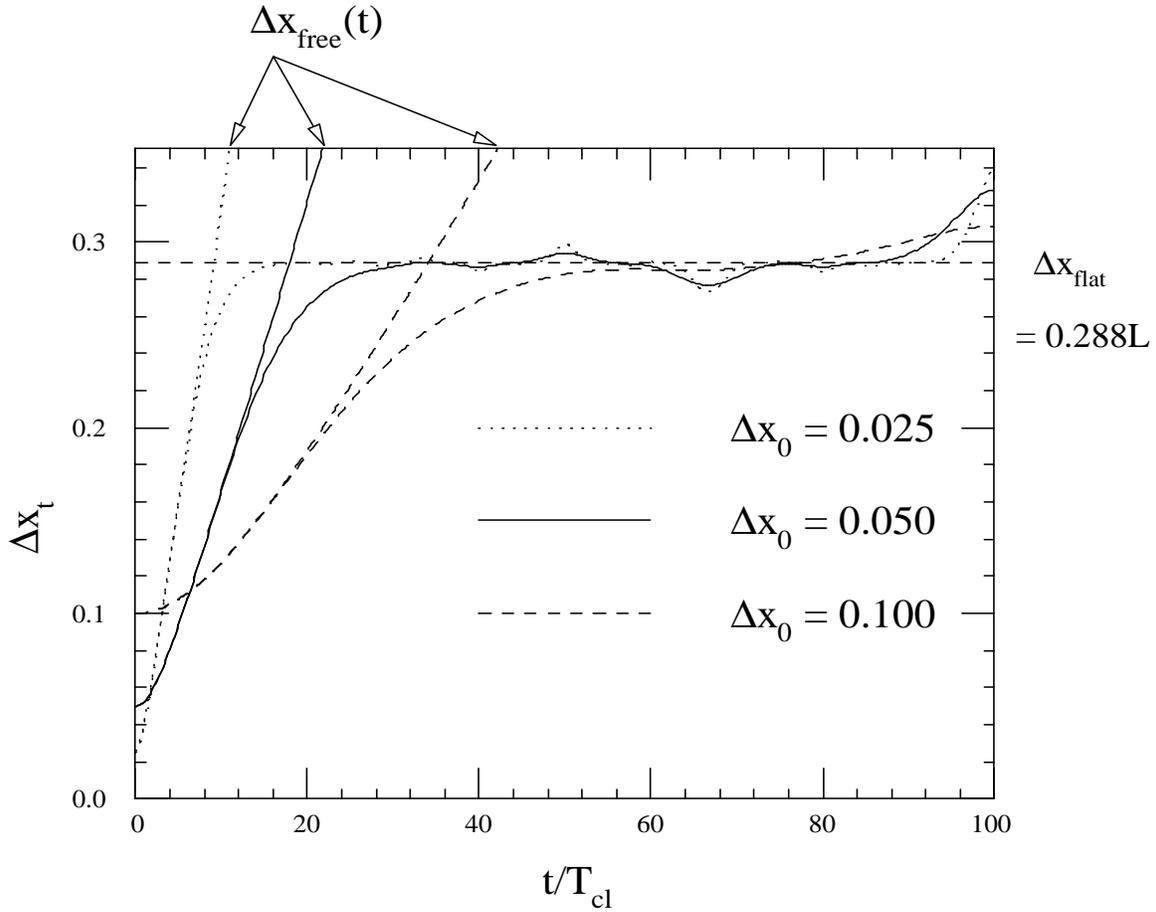,width=12cm,angle=270}
\caption{The position-spread, $\Delta x_t$ versus $t$, for
various initial width Gaussian states, and the approach to the
collapsed state, characterized by $\Delta x_{cl} = L/\sqrt{12}$,
evaluated at integral values of the classical period. 
(Reprinted from Ref.~\cite{robinett_infinite_well}.)
\label{fig:delta_x_approach}}
\end{figure}

\begin{figure}
\epsfig{file=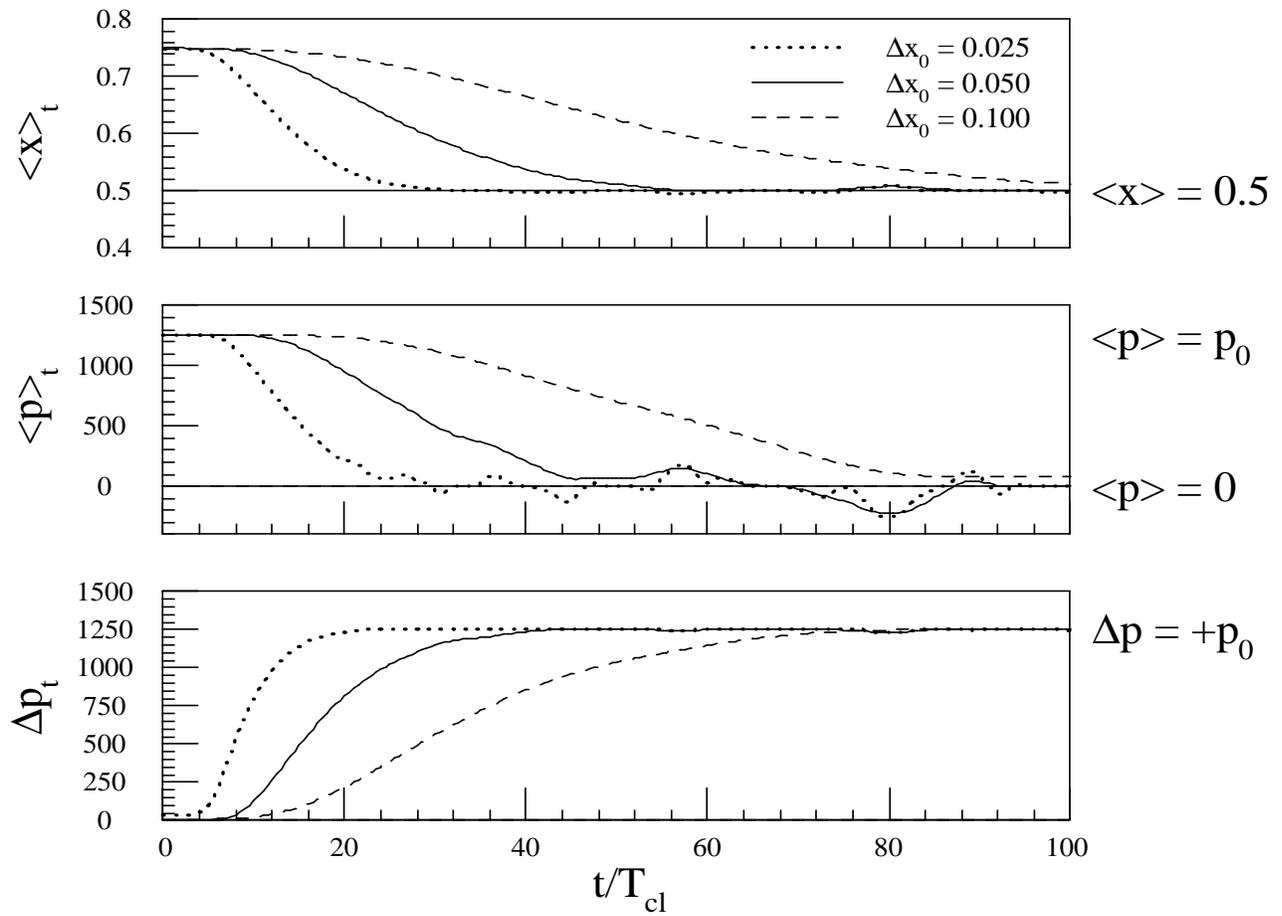,width=12cm,angle=270}
\caption{The same as Fig.~\ref{fig:delta_x_approach}, but for
$\langle x \rangle_t$ (top), $\langle p \rangle_t$ (middle), and
$\Delta p_t$ (bottom).
(Reprinted from Ref.~\cite{robinett_infinite_well}.)
\label{fig:all_approach}}
\end{figure}

\begin{figure}
\epsfig{file=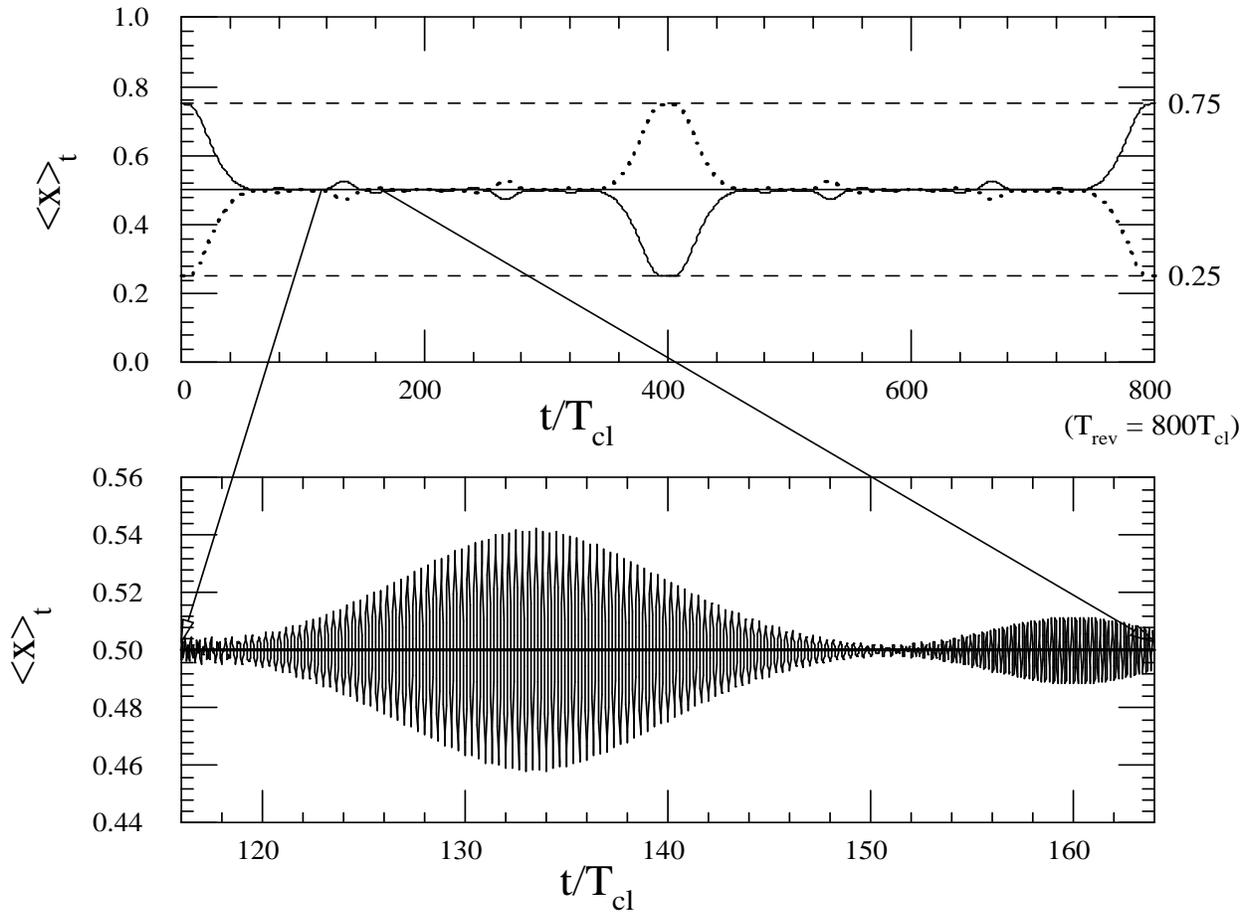,width=12cm,angle=270}
\caption{The expectation value of position $\langle x \rangle_t$
evaluated at $t = (n+1/8)T_{cl}$ (solid) and $t= (n+5/8)T_{cl}$
(dashed) over an entire revival time. The mirror revival, where the
wave packet is reversed, at $t = T_{rev}/2$ is apparent.
(Reprinted from Ref.~\cite{robinett_infinite_well}.)
\label{fig:x_long}}
\end{figure}

\begin{figure}
\epsfig{file=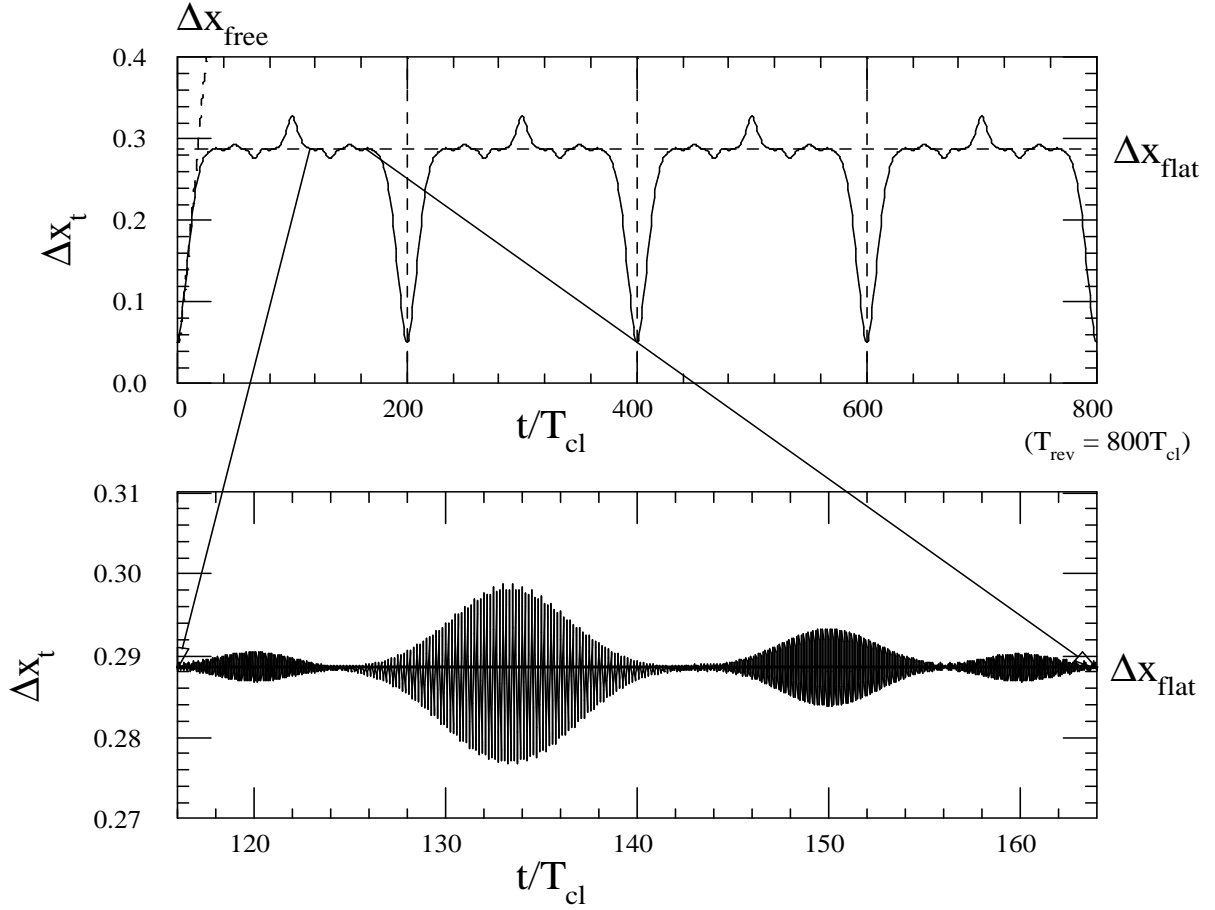,width=12cm,angle=270}
\caption{Same as for Fig.~\ref{fig:x_long}, but for
$\Delta x_t$ over one revival time. The return of $\Delta x_t$
to its original (small) value at $T_{rev}$ (revival time) and
$T_{rev}/2$ (mirror revival) are familiar, while the small values
at the $T_{rev}/4$ and $3T_{rev}/4$ fractional revivals 
are special features of an initial
state centered at $x_0 = L/2$. At those points, the wave packet consists
of two 'mini' packets, superimposed and hence of small width, but with
opposite momenta. (Reprinted from Ref.~\cite{robinett_infinite_well}.)
\label{fig:delta_x_long}}
\end{figure}

\begin{figure}
\epsfig{file=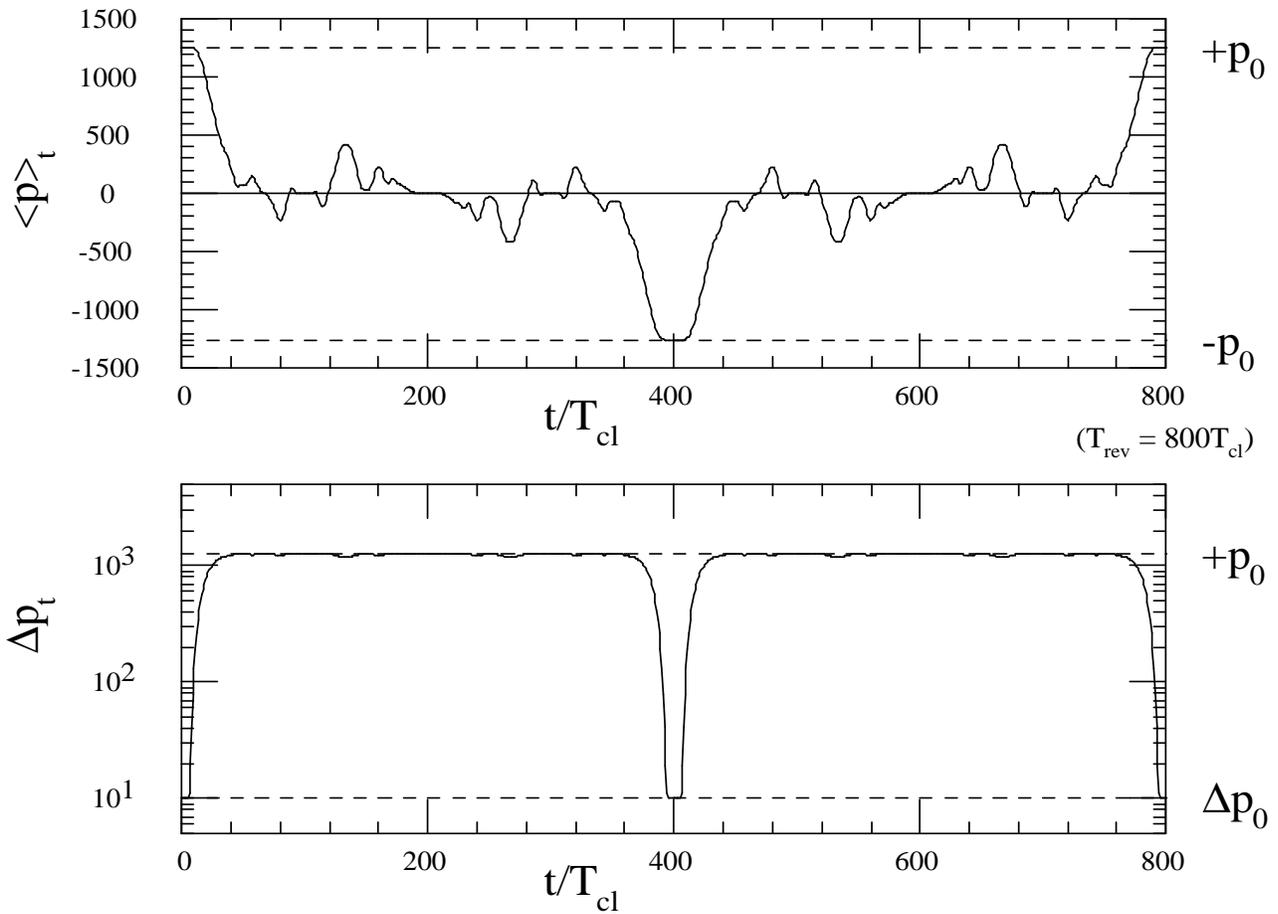,width=12cm,angle=270}
\caption{Same as for Fig.~\ref{fig:x_long}, but for
$\langle p \rangle_t$ and $\Delta p_t$. 
Note that there are no obvious deviations from
the collapsed value of $\Delta p_{cl} = +p_0$  at
$T_{rev}/4$ ($200T_{cl}$) or $3T_{rev}/4$ ($600T_{cl}$), 
even though $\Delta x_t$ returns to its
original (small) size. (Reprinted from Ref.~\cite{robinett_infinite_well}.)
\label{fig:p_long}}
\end{figure}

\clearpage

\clearpage
\begin{figure}
\epsfig{file=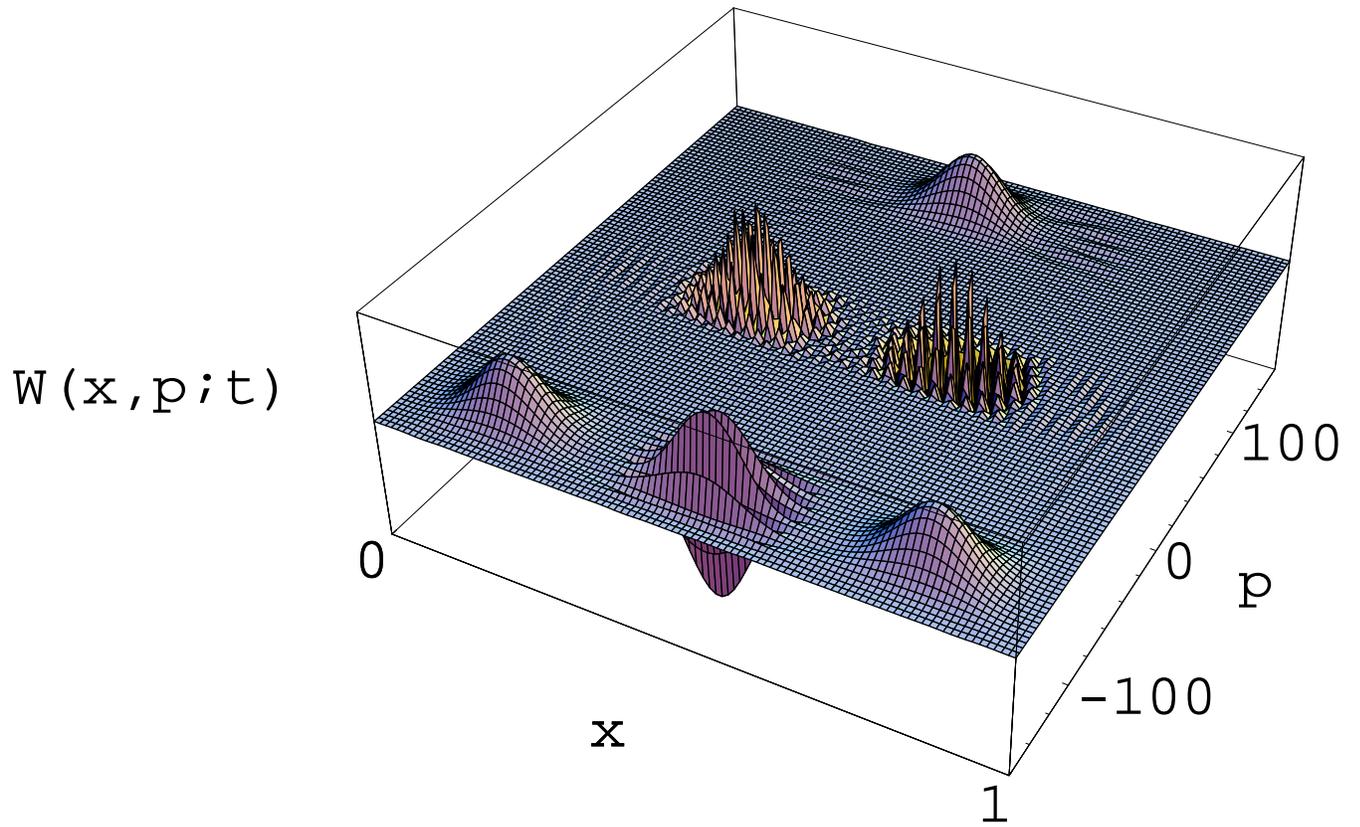,width=18cm,angle=0}
\caption{Phase-space structure of the $T_{rev}/3$ fractional
revival using the Wigner distribution. The parameters of 
Eqns.~(\ref{gaussian_components}) and (\ref{anharmonic}) are used, 
with $x_0/L = 1/2$, but with $n_0= 40$ used to make the oscillatory
`cross-terms' more obvious.
\label{fig:wigner_third}}
\end{figure}

\clearpage

\begin{figure}
\epsfig{file=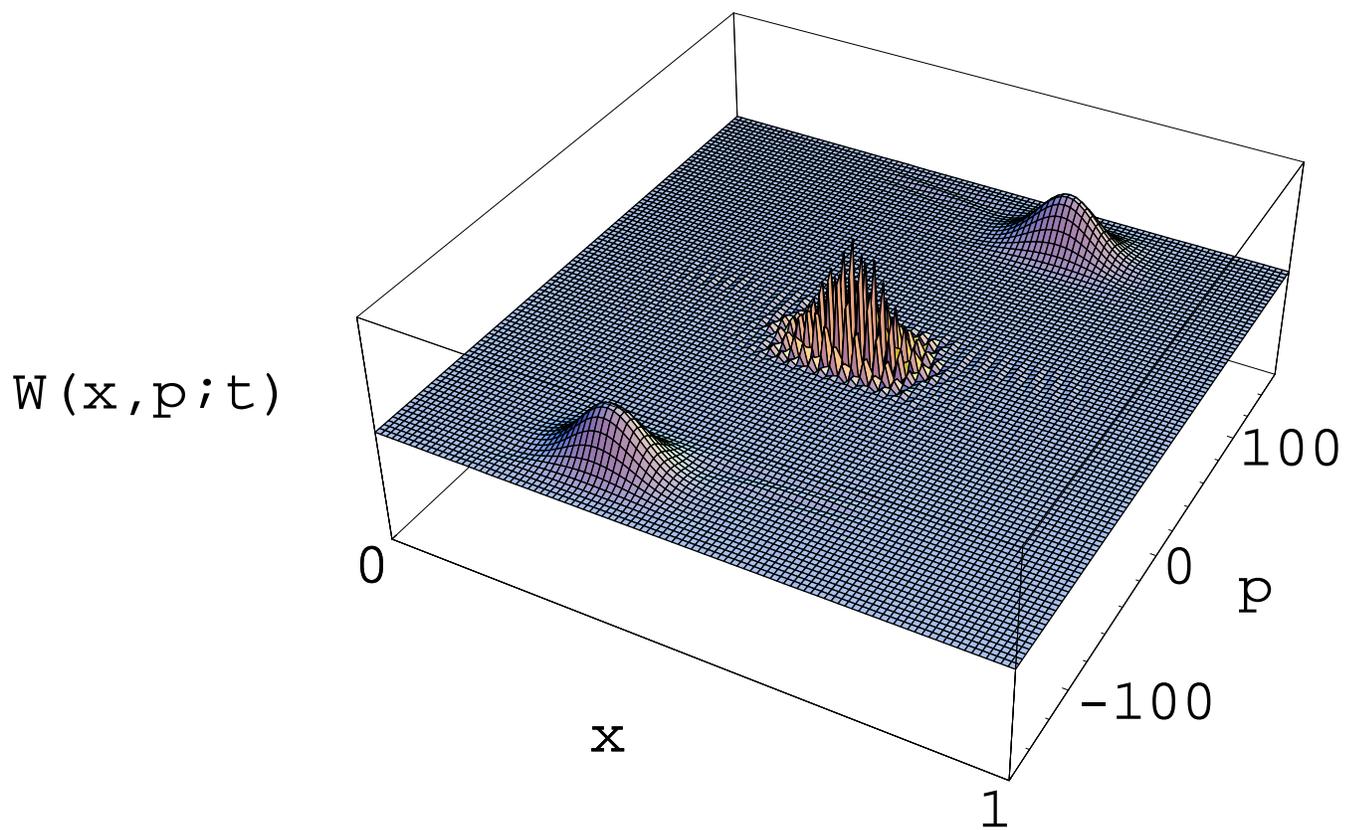,width=18cm,angle=0}
\caption{Same as Fig.~\ref{fig:wigner_third}, but for 
the fractional revival at $t=  T_{rev}/4$, where $x_0/L = 2/3$
is used.
\label{fig:wigner_quarter}}
\end{figure}

\clearpage

\begin{figure}
\epsfig{file=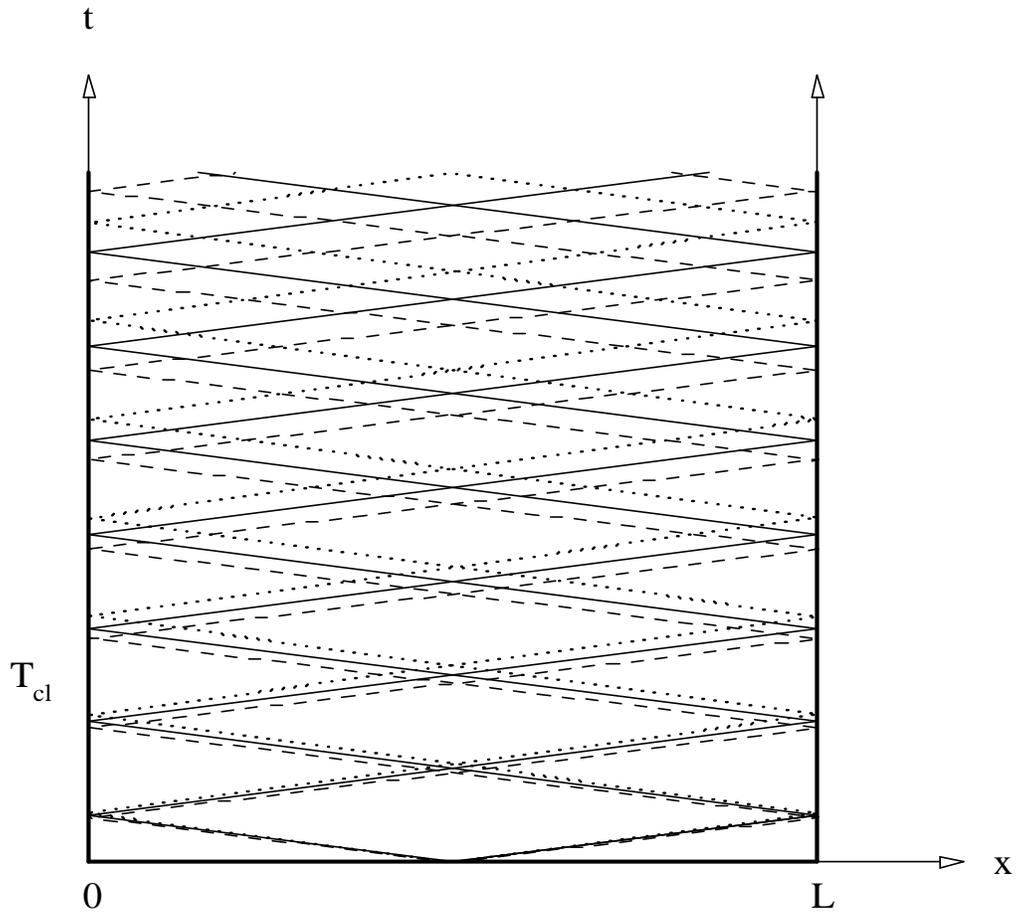,width=12cm,angle=270}
\caption{Plot of the classical trajectories (or world-lines), 
shown in the $(x,t)$ plane, of three point particles with slightly different 
speeds, illustrating wave  packet spreading in the infinite well with 
walls at $x=0,L$, as first discussed by Born \cite{born}, \cite{born_2}.
\label{fig:new_kinzel}}
\end{figure}

\clearpage

\begin{figure}
\epsfig{file=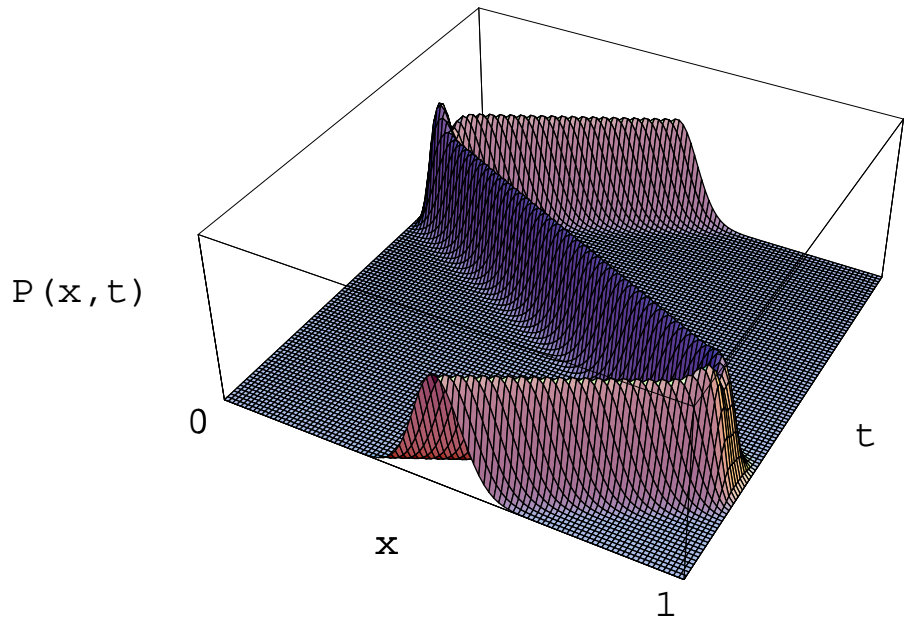,width=12cm,angle=0}
\caption{Plot of the position-space probability density, 
$P(x,t) = |\psi(x,t)|^2$, versus $x,t$ in the infinite well, for a
Gaussian wave packet, over the first classical period. The
parameters in Eqn.~(\ref{infinite_initial_choices}), with $n_0 = 400$,
 are used, so the infinite well is over the interval $x \in (0,1)$.
\label{fig:first_cycle}}
\end{figure}

\clearpage

\begin{figure}
\epsfig{file=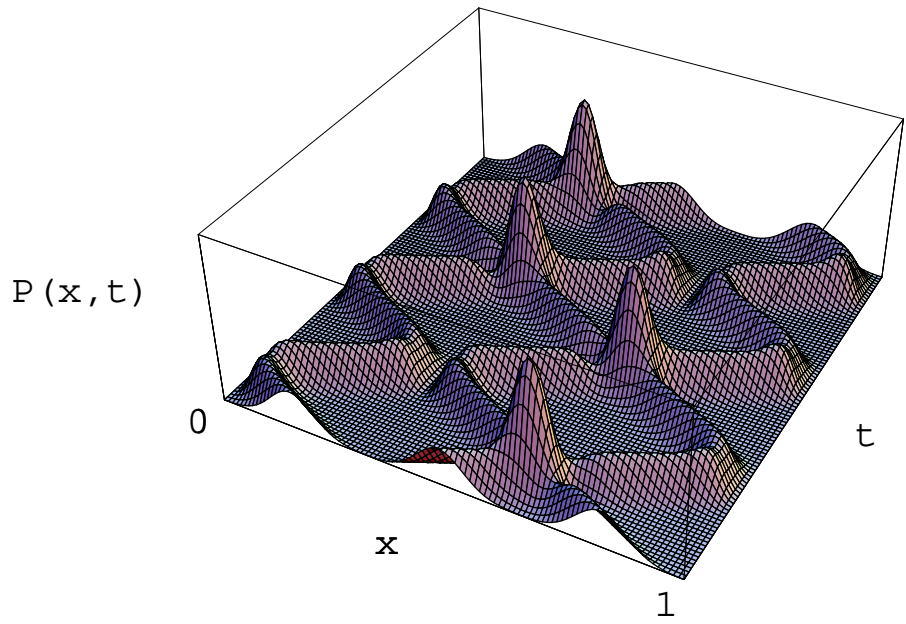,width=12cm,angle=0}
\caption{Same as Fig.~\ref{fig:first_cycle}, but over one
classical period, starting at $T_{rev}/3$. 
\label{fig:third_cycle}}
\end{figure}

\clearpage

\begin{figure}
\epsfig{file=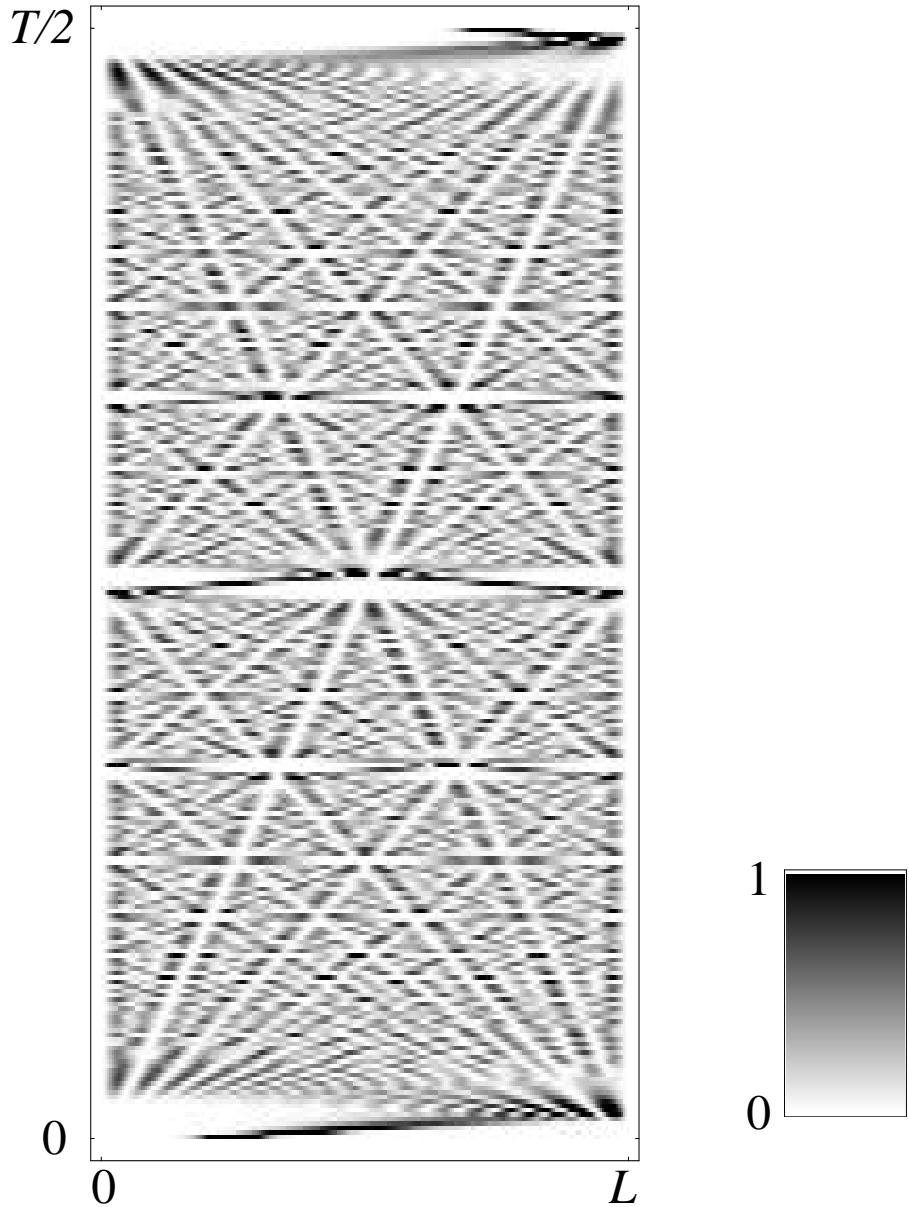,width=12cm,angle=0}
\caption{Contour plot of the probability density, $|\psi(x,t)|^2$ versus $x,t$
for the infinite well with $x \in (0,L)$ over the interval $(0,T_{rev}/2)$, 
with darker areas corresponding to higher probability, 
illustrating the pattern of ridges and canals described as a quantum carpet.
A Gaussian wavepacket with $n_0 = 15$, $x_0 = L/4$, and $\Delta x_0 = L/20$ 
is used. (Adapted from Fig.~1 of Ref.~\cite{carpets_wigner}, courtesy of 
I. Marzoli.)
\label{fig:quantum_carpets}}
\end{figure}

\clearpage

\begin{figure}
\epsfig{file=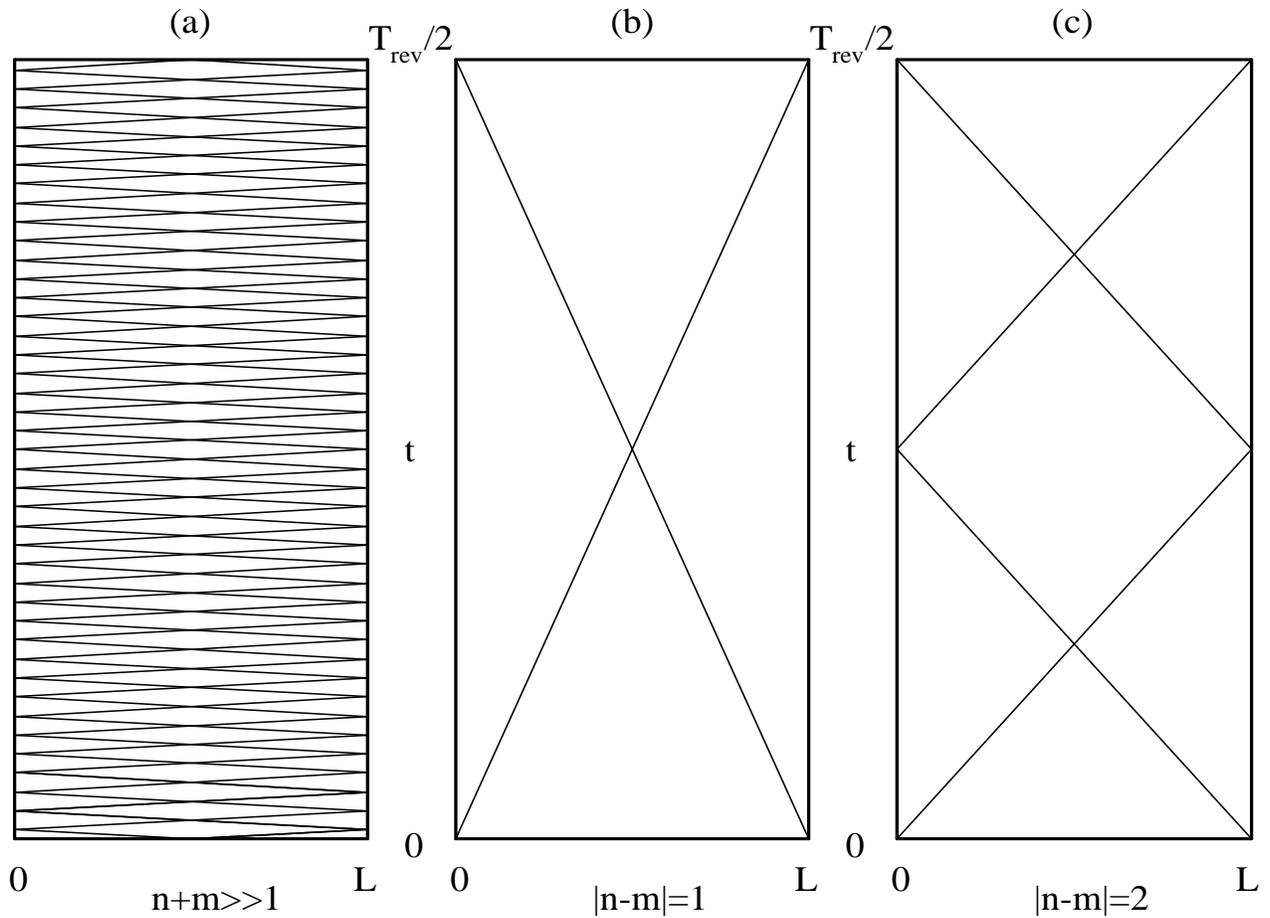,width=12cm,angle=270}
\caption{Plots of typical worldlines in the $(x,t)$ plane corresponding 
to the classical contribution to the probability density in 
Eqn.~(\ref{classical_contribution}) 
where $n+m>>1$ (a), and two cases from the {\it quantum carpet} terms 
in Eqn.~(\ref{quantum_contribution}) 
corresponding to $|n-m|=1$ (b) and $|n-m|=2$ (c).
\label{fig:worldline}}
\end{figure}

\clearpage

\begin{figure}
\epsfig{file=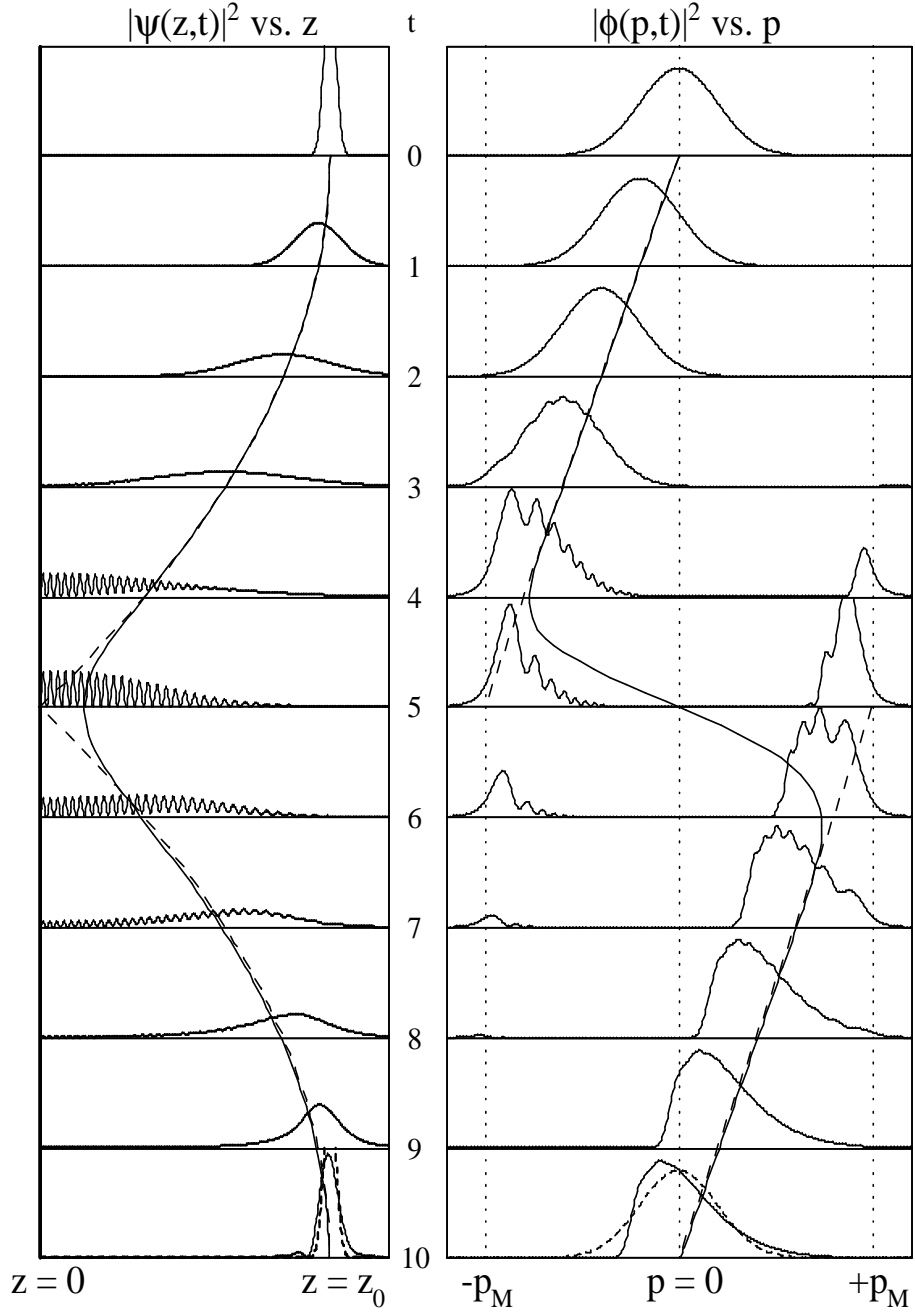,width=12cm,angle=0}
\caption{Gaussian wave packet solutions for the
quantum bouncer in position-space ($|\psi(z,t)|^2$ versus $z$, left)
and momentum-space ($|\phi(p,t)|^2$ versus $p$, right) for various times
during the first classical period. The solid curves represent the
time-dependent expectation values of position  ($\langle z \rangle_t$, left) 
and momentum ($\langle p \rangle_t$, right) for these solutions.
The similar dashed curves are the 
classical trajectories, $z(t)$ (left) and $p(t)$ (right),  superimposed.
The wave packet parameters in Eqns.~(\ref{other_bouncer_parameters}) and 
(\ref{bouncer_parameters}) are used.
For the momentum-space figure, the vertical dotted lines represent the values
$p=0$ and the classical extremal   values of momentum,  $\pm p_M = \pm 
\sqrt{2mFz_0}$. (Reprinted from Ref.~\cite{robinett_bouncer}.)
\label{fig:bouncer_classical}}
\end{figure}

\begin{figure}
\epsfig{file=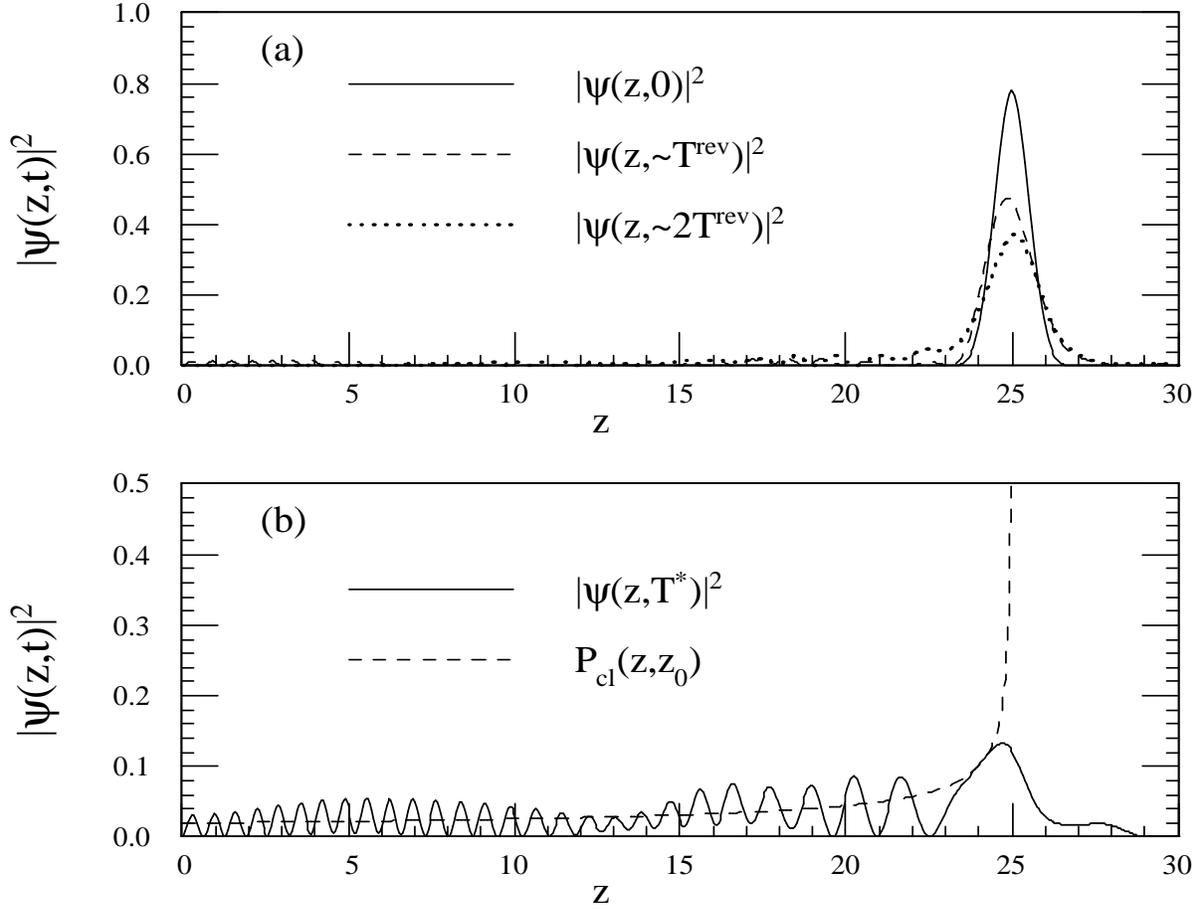,width=12cm,angle=270}
\caption{The top plot (a) shows the position-space probability density,
$|\psi(z,t)|^2$  for the initial Gaussian wave packet (solid), at
at times near the first (dashed) and second (dotted) revival times.
The bottom plot (b) shows $|\psi(z,T^{*})|^2$ at a typical time, $T^{*}$, 
not close to any
fractional revival, during the collapsed phase (solid), while the
dashed  curve is the classical probability density given
by Eqn.~(\ref{bouncer_classical}) (but note the change in scale.)
\label{fig:bouncer_collapse}}
\end{figure}

\begin{figure}
\epsfig{file=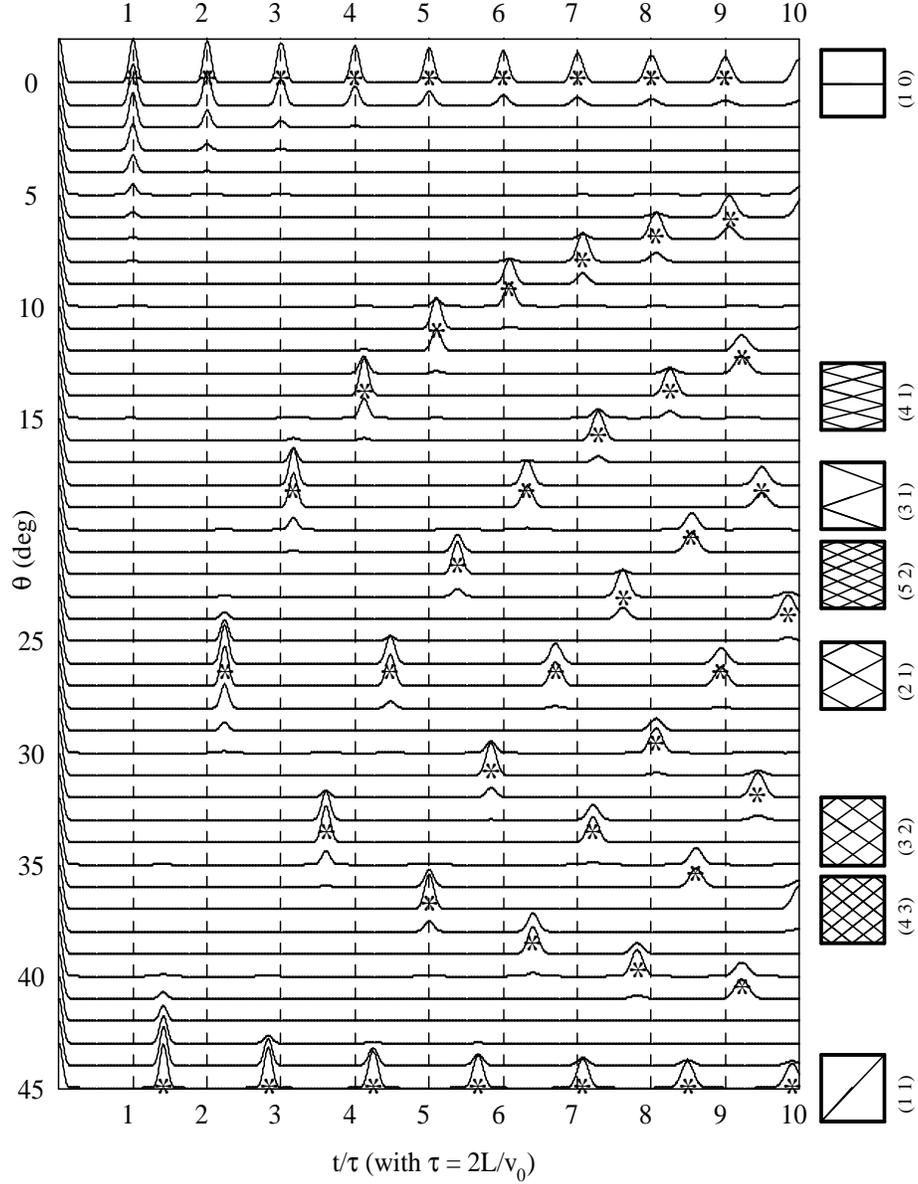,width=12cm,angle=0}
\caption{Plots of the auto-correlation function,
$|A(t)|^2$ versus $t$, for the 2-D square billiard. The plots are
over a time period equal to ten classical
`back-and-forth' periods, $10\tau$, where $\tau \equiv 2L/v_0$.
Plots for different values of the initial angle, $\tan(\theta)
= p_{0y}/p_{0x}$ are shown. The stars indicate the positions of
classical closed orbits (and recurrences) as given by 
Eqn.~(\ref{special_conditions}). (Reprinted from Ref.~\cite{blueprint}.)
\label{fig:square_well_short}}
\end{figure}

\begin{figure}
\epsfig{file=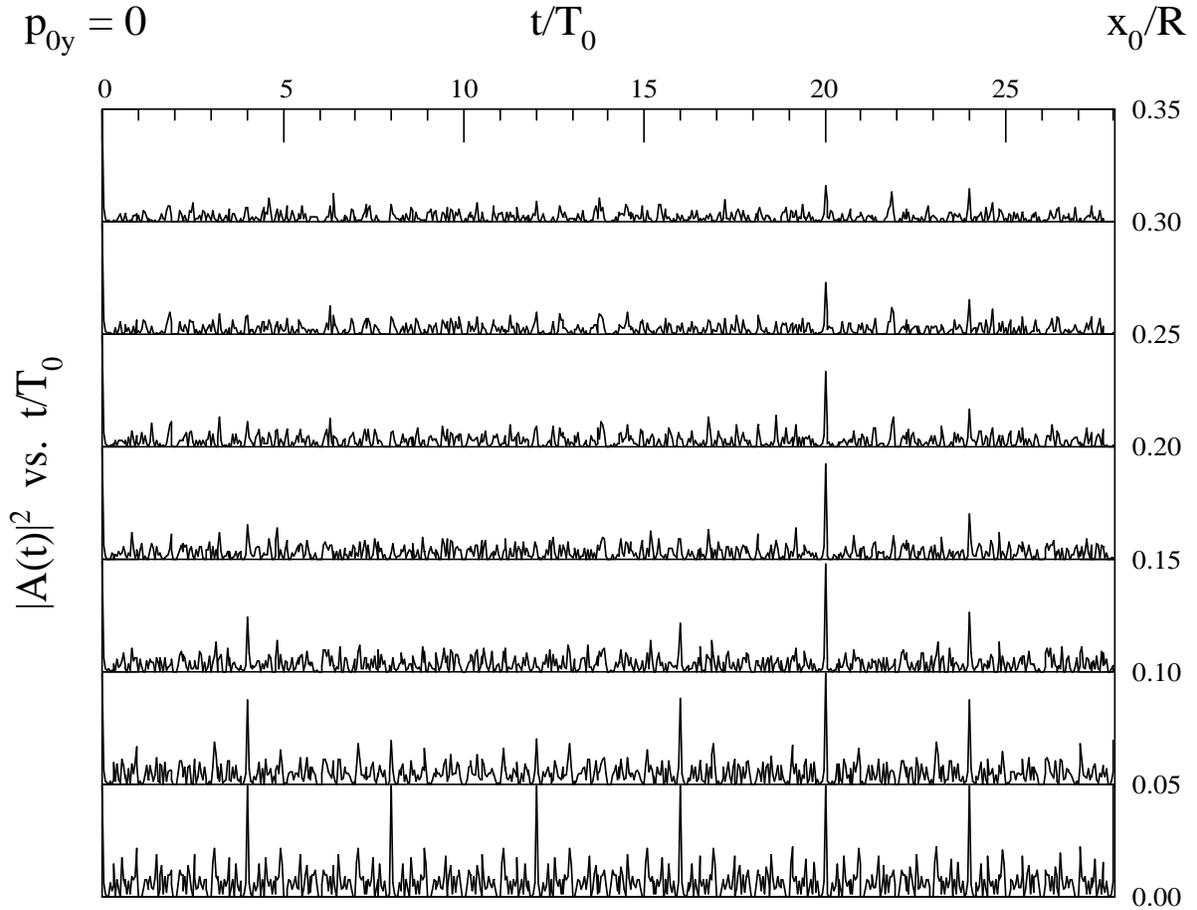,width=12cm,angle=270}
\caption{Plot of the autocorrelation function, $|A(t)|^2$
versus $t$, in units of $T_0 \equiv  2\mu R^2/\hbar \pi$, 
The numerical values
of Eqn.~(\ref{numerical_values}) are used, along with $y_0= 0$ 
and $p_{0x}=p_{0y} = 0$. The results for $|A(t)|^2$ versus $t$, 
 as one varies the $x_0$ of the  initial wave packet away from the
center of the circular billiard, are shown on horizontal axes.
(Reprinted from Ref.~\cite{robinett_circular}.)
\label{fig:circular_revivals_1}}
\end{figure}

\begin{figure}
\epsfig{file=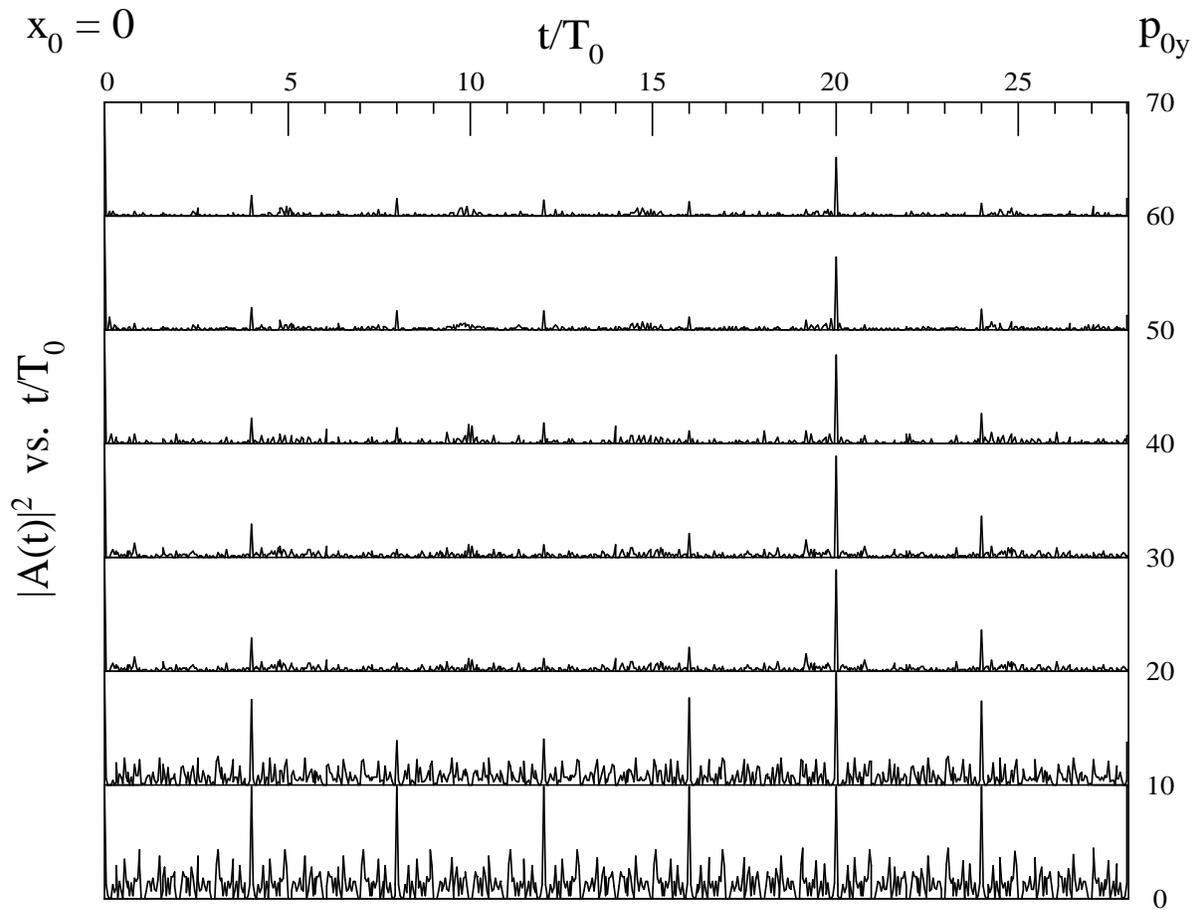,width=12cm,angle=270}
\caption{Same as Fig.~\ref{fig:circular_revivals_1}, 
but with $x_0=y_0=0$ and $p_{0x}=0$, as one increases $p_{0y}$.
(Reprinted from Ref.~\cite{robinett_circular}.)
\label{fig:circular_revivals_2}}
\end{figure}

\clearpage

\begin{figure}
\epsfig{file=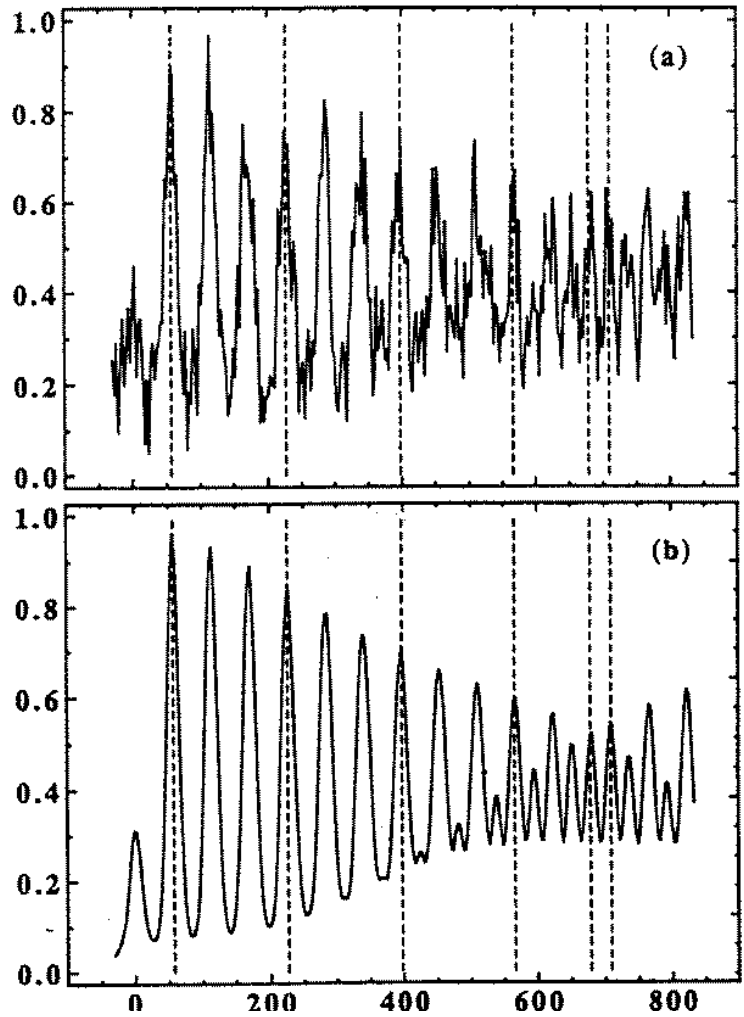,width=10cm,angle=0}
\vskip 1cm
\caption{Experimental (top) and theoretical (bottom)
photoionization signal from Rydberg atoms as a function of
the time delay after the initial probe pulse (a) compared with 
theoretical predictions (b). A fractional revival of order $p/q = 1/4$,
with half the classical periodicity, is apparent near $680\,ps$.
(Reprinted from Ref.~\cite{yeazell_fractional}.)
\label{fig:yeazell_stroud}}
\end{figure}

\clearpage

\begin{figure}
\epsfig{file=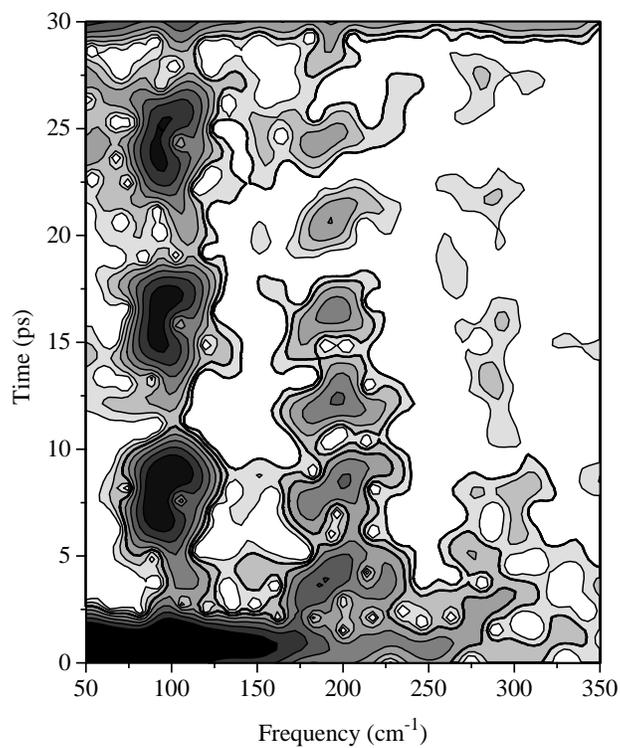,width=12cm,angle=0}
\caption{Log spectrogram contour plot of data from $Br_2$ showing
the spectral content of the observed signal as a  function of time
delay, defined by Eqn.~(\ref{spectrogram_definition}). Evidence for
full and fractional revivals at multiples of $f=95\,cm^{-1}$ are
clear. (Reprinted from Ref.~\cite{vrakking}, courtesy of D. Villeneuve.)
\label{fig:stolow_1}}
\end{figure}

\clearpage

\begin{figure}
\epsfig{file=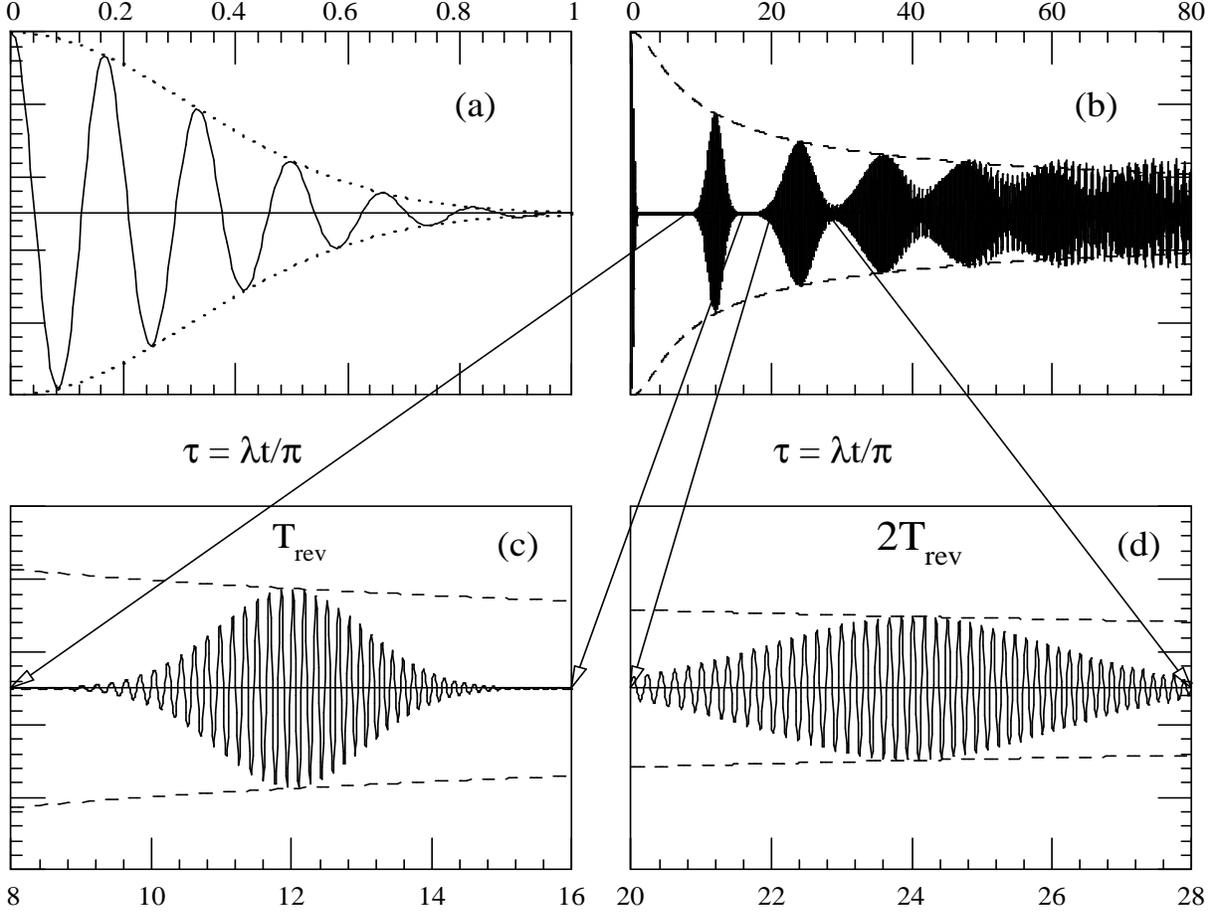,width=12cm,angle=270}
\caption{Plots of $P_{e}(t)$ versus scaled time ($\tau = \lambda t/\pi$)
from the solution in Eqn.~(\ref{summation})
over (a) the first few Rabi cycles and 
(b) over a number of revivals times, 
showing the first (c) and second (d) revivals in detail. Values of
$\overline{n} = 36$ and $\lambda = 0.01$ are used. The dotted
curve corresponds to the initial Gaussian de-phasing envelope
($\exp(-(\lambda t)^2/2$), while the dashed curves correspond to the
long-term suppression given by Eqn.~(\ref{suppression_factor}).
\label{fig:jaynes}}
\end{figure}

\clearpage

\begin{figure}
\epsfig{file=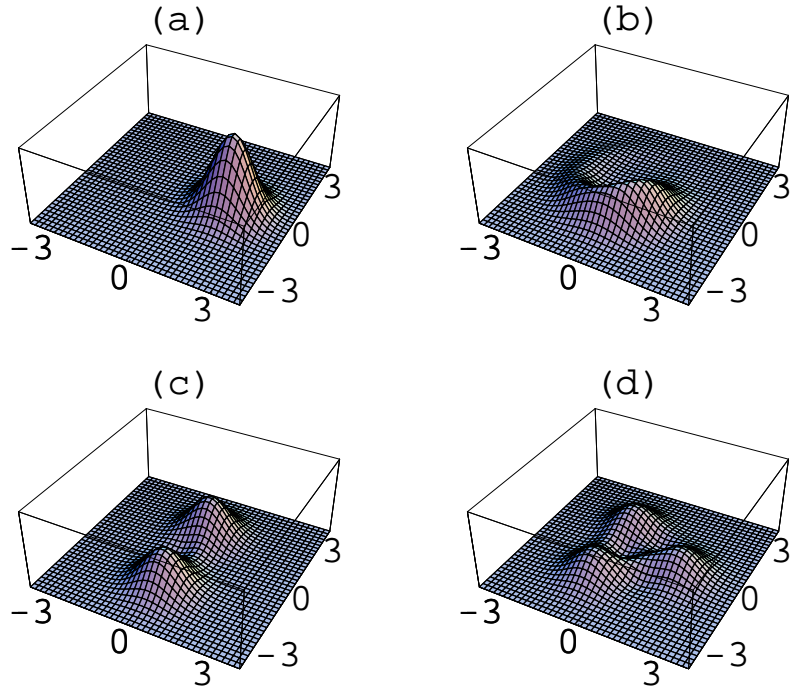,width=12cm,angle=0}
\caption{Plots of $P(\beta;t) = P(Re(\beta),Im(\beta);t)$ versus
$(Re(\beta),Im(\beta)$) from Eqn.~(\ref{beta_plot}). Values for
(a) $t=0$, (b) $0.1T_{rev}$, (c) $T_{rev}/2$, and (d) $T_{rev}/3$ are
shown. An initial coherent state with $\alpha = (3,0)$ is used for
illustration.
\label{fig:coherent_states}}
\end{figure}

\clearpage

\begin{figure}
\epsfig{file=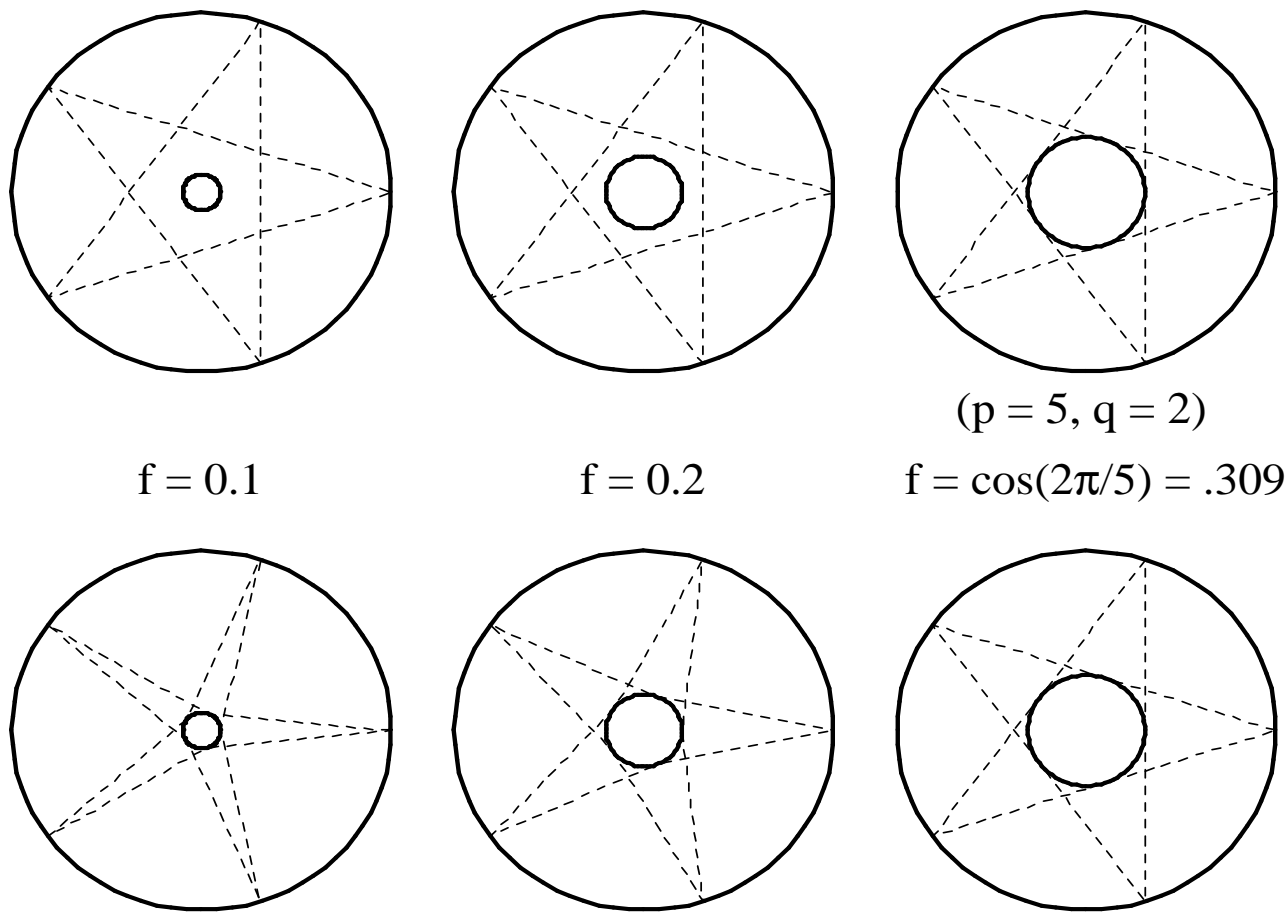,width=12cm,angle=270}
\caption{An example of two possible closed orbits in the annular
billiard, with $(p,q) = (5,2)$, corresponding to the
path lengths in Eqns.~(\ref{circular_path_lengths}) (top)
and Eqn.~(\ref{annular_path_lengths}) (bottom),
illustrating as well the critical value  of $f_{max} = \cos(\pi q/p)$,
beyond which neither orbit can be supported in this geometry.
\label{fig:both_circle_and_annular}}
\end{figure}

\clearpage

\begin{figure}
\epsfig{file=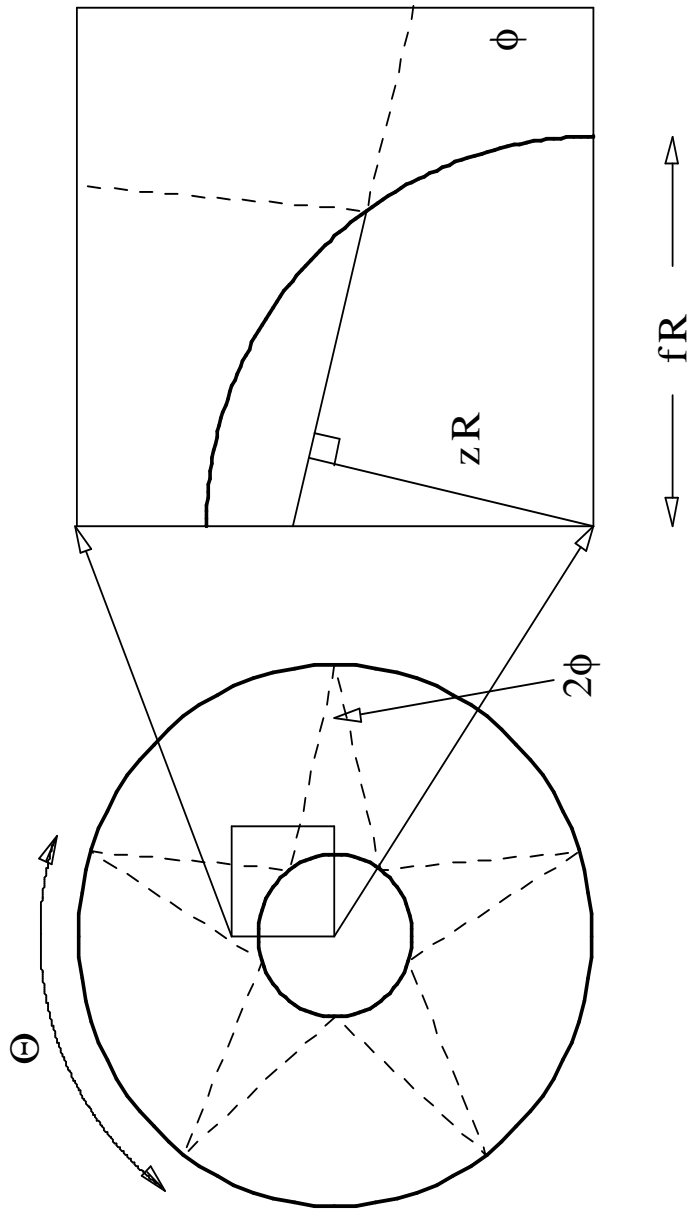,width=9cm,angle=0}
\caption{The geometry defining the effective distance of closest approach
for the `inner touching' closed orbits for the annular billiard, leading
to Eqn.~(\ref{annular_inner}).
\label{fig:annular_closest}}
\end{figure}

\end{document}